\documentclass[a4paper,11pt]{article}
\usepackage{jcappub}
\usepackage[utf8]{inputenc}

\usepackage{amsmath}
\usepackage{bm}
\usepackage{xcolor}
\usepackage{float}
\usepackage{multirow}
\usepackage{makecell}

\newcommand{\aap}{{Astron.~Astrophys.}}
\newcommand{\apjl}{{Astrophys.~J.~Lett.}}
\newcommand{\apjs}{{Astrophys.~J.~Supp.}}

\usepackage{graphicx}
\usepackage[large]{subfigure}
\usepackage{amssymb, amsmath}
\usepackage[amssymb]{SIunits}
\usepackage{epstopdf}
\usepackage{hyperref}
\usepackage{aas_macros}

\usepackage[mathscr]{eucal}	
\DeclareTextSymbol{\degre}{T1}{23}

\newcommand{\vl} { {\mbf{\ell}} }

\newcommand{\beq} {\begin{equation}}
\newcommand{\eeq} {\end{equation}}
\newcommand{\bal} {\begin{aligned}}
\newcommand{\eal} {\end{aligned}}

\newcommand{\mbf}[1]{\mbox{\boldmath$#1$}}
\newcommand{\fnl}{f_{\rm NL}}

\title{Photo-z outlier self-calibration in weak lensing surveys}

\author[a,b]{Emmanuel Schaan,}
\author[a,b]{Simone Ferraro}
\author[a,b,c]{and Uros Seljak}
\affiliation[a]{Lawrence Berkeley National Laboratory, One Cyclotron Road, Berkeley, CA 94720, USA}
\affiliation[b]{Berkeley Center for Cosmological Physics, UC Berkeley, CA 94720, USA}
\affiliation[c]{Department of Physics, University of California, Berkeley, CA 94720, USA}
\emailAdd{eschaan@lbl.gov}
\emailAdd{sferraro@lbl.gov}
\emailAdd{useljak@berkeley.edu}

\abstract{

Calibrating photometric redshift errors in weak lensing surveys with external data is extremely challenging.
We show that both Gaussian and outlier photo-z parameters can be self-calibrated from the data alone.
This comes at no cost for the neutrino masses, curvature and dark energy equation of state $w_0$, but with a $65\%$ degradation when both $w_0$ and $w_a$ are varied.

We perform a realistic forecast for the Vera Rubin Observatory (VRO) Legacy Survey of Space and Time (LSST) 3$\times$2 analysis, combining cosmic shear, projected galaxy clustering and galaxy - galaxy lensing.
We confirm the importance of marginalizing over photo-z outliers.
We examine a subset of internal cross-correlations, dubbed ``null correlations'', which are usually ignored in 3$\times$2 analyses. 
Despite contributing only $\sim 10\%$ of the total signal-to-noise, these null correlations improve the constraints on photo-z parameters by up to an order of magnitude.
Using the same galaxy sample as sources and lenses dramatically improves the photo-z uncertainties too.
Together, these methods add robustness to any claim of detected new Physics, and reduce the statistical errors on cosmology by $15\%$ and $10\%$ respectively.
Finally, including CMB lensing from an experiment like Simons Observatory or CMB-S4 improves the cosmological and photo-z posterior constraints by about $10\%$, and further improves the robustness to systematics.

To give intuition on the Fisher forecasts, we examine in detail several toy models that explain the origin of the photo-z self-calibration.
Our Fisher code LaSSI (Large-Scale Structure Information), which includes the effect of Gaussian and outlier photo-z, shear multiplicative bias, linear galaxy bias, and extensions to LCDM, is publicly available at
\url{https://github.com/EmmanuelSchaan/LaSSI}.
}

\begin{document}

\maketitle

\section{Introduction}

Modern weak gravitational lensing surveys aim at understanding cosmic acceleration, the properties of the dark matter and the masses of the neutrinos.
These experiments, including 
CFHTLenS\footnote{\url{https://www.cfhtlens.org/}}, KiDS\footnote{\url{http://www.astro-wise.org/projects/KIDS/}}, DES\footnote{https://www.darkenergysurvey.org/}, HSC\footnote{\url{https://www.naoj.org/Projects/HSC/}}, the Rubin Observatory Legacy Survey of Space and Time (LSST\footnote{\url{http://www.lsst.org}}), Euclid\footnote{\url{https://www.euclid-ec.org/}} and the Nancy Grace Roman Space Telescope (formerly WFIRST)\footnote{\url{https://roman.gsfc.nasa.gov/}}, 
probe the mass distribution in the Universe by detecting its gravitational lensing effect on the images of distant galaxies.
Disentangling this correlated percent shear ellipticity from the uncorrelated $20\%$ intrinsic galaxy ellipticity requires samples of one hundred million to one billion galaxies,
which can currently only be achieved by photometric surveys.
One major challenge is then to infer accurate photometric redshifts (``photo-z''), given only the galaxy brightnesses in a few photometric bands, their positions on the sky and any morphological information available.

Confidence in the photometric redshift estimation is crucial if the resulting cosmological parameters are to be trusted.
Demanding unbiased cosmological parameters, within the statistical error, places stringent requirements on the photo-z accuracy.
The DESC science requirements document \cite{Alonso2018} derives and lists these requirements for the 3$\times$2-point analysis, i.e. the joint analysis of cosmic shear (correlation of galaxy shear fields), galaxy - galaxy lensing (cross-correlation of galaxy number density and galaxy shear) and galaxy clustering (correlation of galaxy number density fields), for LSST after 10 years of data collection.
Generally, the galaxy catalog is split into a ``lens'' sample and a ``source'' sample.
The lenses are the galaxies whose number density is used as a tracer of the mass density, which causes the lensing. The lens sample is thus often also used as a ``tracer'' sample, used to measure the galaxy clustering.
The sources are the galaxies whose ellipticities are measured to estimate the shear.
In the analysis of \cite{Alonso2018}, the lens and tracer sample is different from the source sample, and the photometric redshift requirements on these two samples are distinct (requirements LSS1, LSS2, WL1 and WL2 in the DESC SRD) \cite{Alonso2018}. We list them in Tab.~\ref{tab:desc_srd_reqs}.
\begin{table}[h!!!]
\centering
\begin{tabular}{c|c|c|}
\cline{2-3}
&Tracer/lens sample & Source sample\\
\hline
\multicolumn{1}{|c|}{\makecell{Uncertainty on tomographic bin\\ mean redshift $\frac{\sigma\left( \langle z \rangle_\text{bin} \right)}{1 + \langle z \rangle_\text{bin}}$}} & $\leq 0.003$ & $\leq 0.001$ \\
\hline
\multicolumn{1}{|c|}{Individual photo-z scatter $\frac{\sigma_z}{1+z}$} & $0.03$ & $0.05$ \\
\hline
\multicolumn{1}{|c|}{Uncertainty on individual photo-z scatter $\frac{\sigma\left( \sigma_z \right)}{1 + z}$} & $\leq 0.03$ & $\leq 0.003$ \\
\hline
\multicolumn{1}{|c|}{Uncertainty on stellar fraction $f_\star$} & -- & $\leq 0.001$ \\
\hline
\end{tabular}
\caption{Photo-z requirements for the tracer/lens and source samples, for the 3 2-point analysis of LSST after 10 years of observation, as presented in the DESC SRD \cite{Alonso2018}.}
\label{tab:desc_srd_reqs}
\end{table}
Meeting such stringent requirements is needed in order to fully exploit the statistical power of LSST. 
However, achieving this level of systematic control poses a number of challenges.

The photo-z requirements above can be split into training and calibration requirements \cite{Newman2015}. 
Training consists in minimizing the scatter $\sigma_z$ and bias $\delta_z$ in individual galaxy photo-z, i.e. constructing the algorithm that returns a photo-z as close to the truth as possible, given only photometry in several bands. 
Once such an algorithm is chosen, calibration consists in characterizing as accurately as possible the performance of the algorithm, i.e. minimizing the uncertainties $\sigma\left( \sigma_z \right)$ and $\sigma\left( \delta_z \right)$ on the achieved scatter and bias.
Like in any physical experiment, obtaining a quality measurement requires the instrument to be precise (training of the photo-z algorithm), and this precision to be known (calibration of the photo-z algorithm).

Several physical effects limit photo-z training, and complicate the photo-z calibration.
Because galaxy spectra have a limited number of features, different galaxies at different redshifts may have the same exact fluxes in the observed photometric bands.
This ``type''-redshift degeneracy is a fundamental limit to photo-z training.
In principle, it should manifest itself by a multimodal posterior distribution for the galaxy redshift. 
This degeneracy can be alleviated by adding more bands to the photometric datasets \cite{Masters2017,Hildebrandt2018,Buchs2019}.
A second important physical effect is blending: if two superimposed galaxies are misidentified as one, the question of the redshift of the blended object becomes ill-defined.
The issue of blending can in principle be reduced by adding higher resolution imaging at the same depth, e.g. from space.
In practice, both effects typically lead to outliers or ``catastrophic failures'' in algorithms that output point estimates for the objects' redshifts.

A proper statistical description of these uncertainties requires an accurate prior on the galaxy types and true redshifts present in the photometric sample of interest.
Photo-z algorithms rely explicitly (template-fitting methods) or implicitly (machine learning methods) on such a prior, 
and inaccuracies in this prior result in biases in the inferred photo-z.
However, constructing such accurate priors for the galaxy types and redshifts present in the photometric sample is difficult.
This is typically achieved with spectroscopy, but spectroscopy at the same depth as the photometry is expensive,
and can therefore only be performed on a small subsample of the photometric catalog.
If a random subsample of the photometric catalog is followed up with spectroscopy, one needs to make sure that the subsample is representative of the overall mix of galaxy types and redshifts present in the photometric survey.
In particular, the spectroscopic follow up needs to cover a wide enough area, in order to limit cosmic variance. 
Indeed, a small patch of the sky may correspond to an overdense or underdense part of the Universe, in which the galaxy types would be systematically different from those in the overall photometric survey.
Critically, the spectroscopic success rate also needs to be extremely high, to guarantee that a certain galaxy population is not systematically missed by the spectroscopic follow up, leading to a biased measurement of the true redshift distribution.
Indeed, ref.~\cite{Newman2015} estimates that photo-z training for LSST requires 30k spectra over more than one square degree, split into 15 fields, with a redshift completeness of $\sim 90\%$.
The requirements for photo-z calibration are even more stringent, amounting to spectroscopy with $99.9\%$ completeness for 100,000 objects over $\sim 100$ deg$^2$ \cite{Newman2015} .
Achieving training alone was estimated to require 32 years of dedicated observation on a DESI-like instrument, implying that calibration would be even further out of reach.
Proposals have been made to alleviate these difficult requirements, by expanding the photometric coverage to reduce the type-redshift degeneracy \cite{Masters2017,Hildebrandt2018,Buchs2019}, by not following up a random subset of the photometric sample, but instead a subset that better samples the actual color space occupied by the photometric sample \cite{Buchs2019}, or by stacking spectra from shallower surveys with higher spectral resolution \cite{Padmanabhan2019}.

Assuming that photo-z training can be achieved, a promising approach for calibration is that of clustering redshifts.
This approach relies on an overlapping spectroscopic sample, for which true redshift distribution and galaxy bias as a function of redshift are known, and which is split into narrow spectroscopic redshift bins.
Each spectroscopic bin is then cross-correlated with the photometric sample.
This results in a cosmology-dependent determination of the product of the bias and the redshift distribution of the photometric sample $b_\text{phot}(z) \frac{dn_\text{phot}}{dz}$.
This method enjoys a high statistical power, and has been used in practice, e.g. \cite{2019arXiv190907412K} and references therein.
At the level of the LSST calibration requirements, the uncertainty in the redshift evolution of the photometric galaxy bias will limit the accuracy of the inferred redshift distribution \cite{2013MNRAS.433.2857M, Bernstein2010, 2015APh....63...81N, 2018MNRAS.481.2427C}.
Magnification bias will also have to be controlled.

In light of these challenges, it is natural to ask whether there exist additional methods to validate the photo-z calibration for LSST, or if the standard 3$\times$2 analysis can be slightly modified in order to alleviate the LSST requirements, for example by enabling self-calibration of the photo-z parameters.
If possible, such an approach would be very valuable to validate external photo-z calibrations and to give confidence in any potential discovery or tension in cosmological parameters from LSST.

In this paper, we perform various forecasts for the LSST 3$\times$2 point analysis, 
and ask to what extent the photo-z nuisance parameters can be self-calibrated internally.
We investigate the dependence of this self-calibration on the following assumptions:
\begin{itemize}
\item{\textbf{Catastrophic photo-z errors}}.
If the distributions of photo-z errors were Gaussian, photo-z calibration would be an easy task: for a photo-z scatter of $\sigma_z \sim 3-5\%$, the relative uncertainty on $\sigma_z$ would simply be $1/\sqrt{2 N_\text{spec}}$, where $N_\text{spec}$ is the number of spectroscopic redshifts available. Reaching a calibration of $\sigma\left( \sigma_z \right) \sim 3-0.3\%$ would require less than $200$ spectroscopic redshifts \cite{Newman2015}. This would be easily achievable.
Since the whole difficulty arises from the non-Gaussianity of the photo-z errors, we not only marginalize over the Gaussian core of the photo-z error distribution (shift $\delta_z$ and scatter $\sigma_z$), but we also marginalize over a highly flexible description of the photo-z outliers.
\item{\textbf{Including the ``null'' cross-correlations}}.
The DESC SRD \cite{Alonso2018} and the latest DES, HSC and KiDS 3$\times$2 analyses \cite{2017arXiv170609359K, 2019PhRvD..99b3508B, 2018PhRvD..98d3526A, 2019PhRvD.100b3541A, 2018MNRAS.476.4662V, 2018MNRAS.474.4894J} 
include the auto-correlations of the tracer galaxies but not their cross-correlations (although cross-correlations are included in photo-z outlier tests outside of the 3x2 analysis, e.g. \cite{Elvin-Poole2018}).
Indeed, these cross-correlations would be null in the absence of photo-z errors and magnification bias (and within the Limber approximation), and therefore only add a negligible contribution to the total signal-to-noise when photo-z errors are present. 
However, these low signal-to-noise cross-correlations actually contain precious information on the photo-z errors, which feed back to a better determination of the cosmological parameters.
This possibility is discussed in \cite{2013MNRAS.433.2857M,Elvin-Poole2018,Erben2009,Hildebrandt2009,Hildebrandt2009a,Schneider2006,Benjamin2010,Zhang2010, 2006JCAP...08..008Z}.
Because clustering cross-correlations have a lower signal than auto-correlations, and a similar level of systematics, one may worry about the feasibility of measuring them accurately. Recently, ref.~\cite{2020JCAP...03..044N} demonstrated that this can be done with HSC data.
Another set of ``null'' cross-correlations are provided by the galaxy-galaxy lensing measurements where the lens bin is at higher redshift than the source bin \cite{Prat2018, 2018PhRvD..98d3526A}. This also contains useful information about the photo-z outliers. 
We quantify the improvement in photo-z and cosmological parameters when these ``null'' cross-correlations are included.
\item{\textbf{Using the same galaxy samples for lenses and sources, or at least measuring the clustering of the source galaxies}}.
The DESC SRD \cite{Alonso2018} and the latest DES/HSC/KiDS 3$\times$2 analyses do not include the clustering of the source galaxies, only their shear. 
However, these galaxies also cluster, and in particular their clustering cross-correlations also contain valuable information about their photo-z errors, which we would like to include.
In the Limber approximation, for a given tomographic bin, clustering and lensing are two independent integrals of the redshift distribution: the former involves the galaxy bias, the second involves the lensing kernel.
We will show that measuring these two independent linear combinations of the same redshift distribution allows to solve for the redshift distribution, and therefore constrain the photo-z errors.
This can be done by using the same galaxy samples for lenses and sources, or by adding the clustering of the source sample to the data vector.
A number of systematic effects may hinder this approach; we discuss them in detail in the paper.
However, as mentioned in the DESC SRD, clustering redshifts are a promising way to achieve the LSST requirements for photo-z calibration.
If this is the case, then one will have to extract clustering information from the source galaxy sample, in cross-correlation with a spectroscopic survey. 
Replacing the spectroscopic reference sample by a photometric one was already considered in \cite{2018MNRAS.477.2196D}. 
We build upon this idea, and include the redshift information from the clustering of the source sample.

Furthermore, if this requires selecting a smaller, better controlled set of galaxies for the source sample than currently planned, this is likely acceptable:
indeed, Fig.~3 in \cite{2017MNRAS.470.2100K} shows that
the LSST constraints on dark energy are not shot noise or shape noise limited, and are thus not very sensitive to changes in the galaxy sample size. They are instead determined by the level of systematics control.
We thus investigate whether using the same samples for sources and lenses improves the photo-z and cosmological parameters enough to justify the potential challenges associated with this non-standard approach.
\item{\textbf{Cosmological parameter space}}.
Finally, we explore different cosmological parameter spaces, varying not only the dark energy equation of state, but also neutrino masses and curvature.
For instance, \cite{2018PhRvD..97l3544M} shows that perfect photo-z would improve the constraints on $w_0 $ and $w_a$, but would give a smaller improvement on neutrino masses and curvature. We investigate this question further, in the case where photo-z outliers are also modeled.
\item{\textbf{Inclusion of CMB lensing}}. 
We build on \cite{2017PhRvD..95l3512S} and \cite{Cawthon2018a} and ask whether CMB lensing helps with calibrating photo-z Gaussian uncertainties and outliers for LSST.
\item{\textbf{Magnification bias}}. 
We also include magnification bias in our analysis, as this is a concern for inferring redshift distributions from clustering.
However, we assume that the slope of the luminosity function is known, measured from the catalog. This is a reasonable assumption in practice, as one can quantify the magnification bias empirically, by artificially reducing the brightness of the galaxies in the catalog and asking which ones would still be detected.
\end{itemize}

We begin our study with a set of realistic Fisher forecasts for the LSST $3\times 2$ analysis, in order to quantitatively answer the questions above.
We then study simple toy models to gain intuition on how such a photo-z self-calibration works, and where the corresponding information comes from.
We discuss in detail the various additional systematics that may impact our results,
and finally summarize our conclusions.

\section{Realistic Fisher forecast for LSST and CMB S4 lensing}
\label{sec:full_fisher}

We perform a realistic Fisher forecast for the LSST $3\times 2$-point functions, i.e. clustering, galaxy-galaxy lensing and shear tomography.

\subsection{Forecast assumptions \& comparisons}

Following the DESC SRD \cite{Alonso2018}, we simulate an LSST-like survey over $14,300$ deg$^2$, i.e. a sky fraction of $f_\text{sky}=35\%$.

\subsubsection{Galaxy samples: shape noise, galaxy, shear and magnification biases}

We use the same galaxy sample as tracer, lens and source.
To be conservative, we follow the assumptions in the DESC SRD for their source sample, which is much smaller than their tracer sample, and with worse photo-z uncertainties.
There, the shear sample has an effective number density of 27 galaxies per squared arcminute, after applying the shear weights, as opposed to 48 for the gold sample used as tracer and lens sample in \cite{Alonso2018}.

Specifically, we assume a redshift distribution $dn/dz \propto z^2 e^{-\left( z / z_0 \right)^\alpha}$, with $z_0=0.11$ and $\alpha=0.68$, truncated at redshift 4 and normalized to $27$ galaxies per squared arcminute.
The mean redshift of the galaxy sample is 1.03.
We split the galaxy sample into 10 tomographic bins with equal number of galaxies per bin by applying sharp cuts in photo-z.
It may be interesting to further explore splitting the galaxy sample into more tomographic bins.
As suggested in \cite{2020MNRAS.491.3535T}, cosmological constraints keep improving until the tomographic bin size is reduced to the individual photo-z scatter, especially when cross-correlations are included.

We assume the shape noise per ellipticity component to be $\sigma_\epsilon = 0.26$,
and the fiducial shear multiplicative biases $m_i$ to be zero for all bins, although we marginalize over them in the forecast.

The galaxy bias is assumed to follow $b(z) = 0.95 / D(z)$, where $D(z)$ is the linear growth factor normalized to match $D(z) = 1/(1+z)$ in matter domination.
For each tomographic bin, we fit for one galaxy bias parameter $b_i$, such that $b(z) = b_i \times 0.95 / D(z)$.

We include the effect of magnification bias in the data and the model, but do not marginalize over it.
To do this, we assume that the slopes of the luminosity functions have been measured from the data, and follow the true values in Table~C.1 in \cite{Joachimi2010}.

\subsubsection{Highly flexible photo-z parameterization}

We parametrize the Gaussian photo-z uncertainties with a shift $\delta z_i$ and scatter $\sigma_{z_i} / (1+\langle z\rangle_i)$ for each tomographic bin, such that
\beq
\frac{dn_i^G}{dz} 
= 
\int_{z_\text{phot} \in \text{bin }i} dz_\text{ph}\;
\frac{dn_i}{dz_\text{ph}} \;
p(z | z_\text{ph}),
\text{ with }
p(z | z_\text{ph})
=
\frac{1}{\sqrt{2\pi \sigma_{z_i}^2}}
e^{
-\frac{\left( z - z_\text{ph} - \delta z_i \right)^2}{2 \sigma_{z_i}^2}}.
\eeq
We assume the fiducial Gaussian photo-z scatter to be $\sigma_{z_i} / \left(1+\langle z \rangle_i\right) = 0.05$, with no fiducial shifts $\delta z_i = 0$.
However, we marginalize over both shifts and scatters for the 10 tomographic bins (20 parameters).

\begin{figure}[h!!!]
\centering
\includegraphics[width=0.45\textwidth]{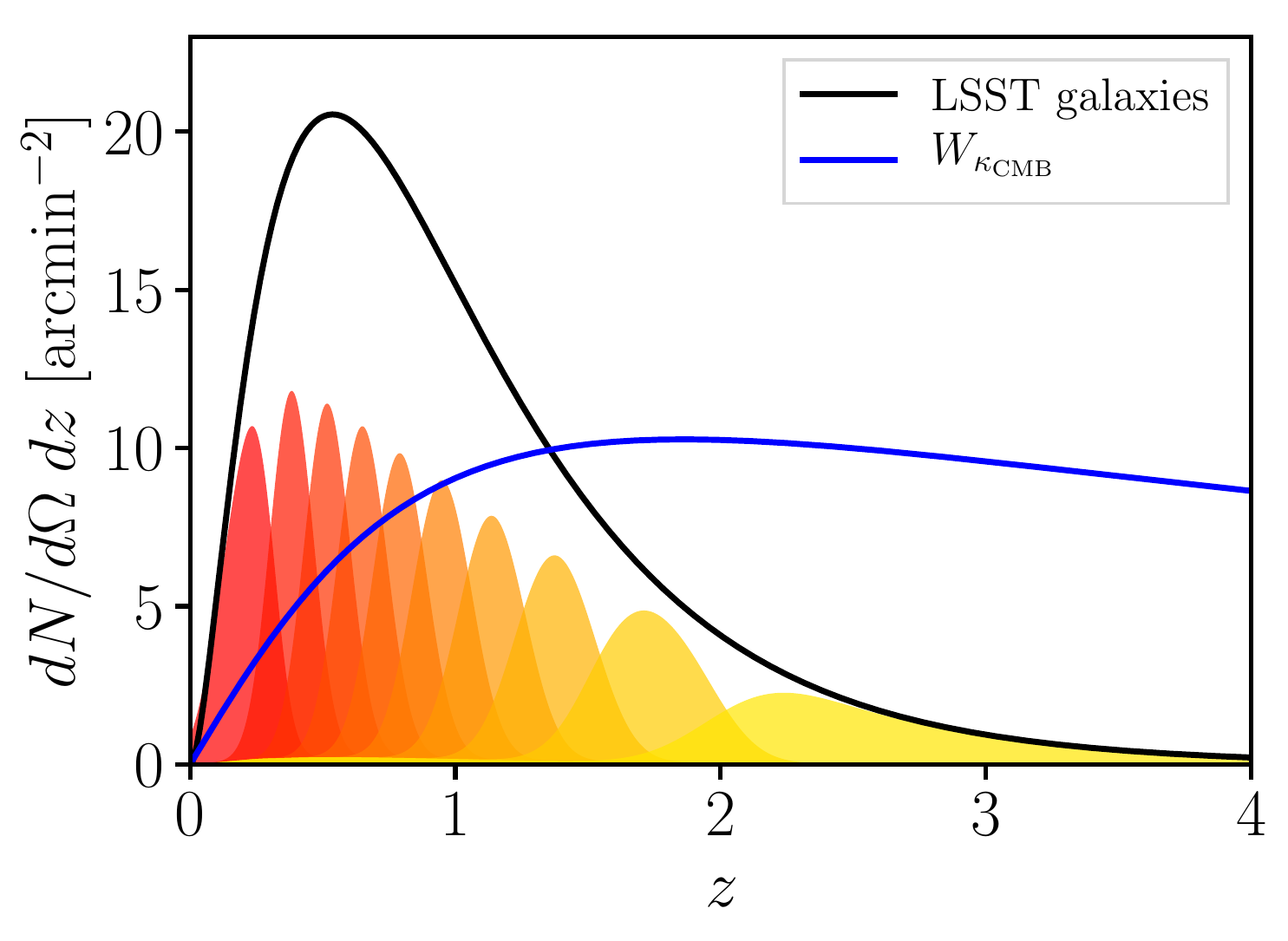}
\includegraphics[width=0.45\textwidth]{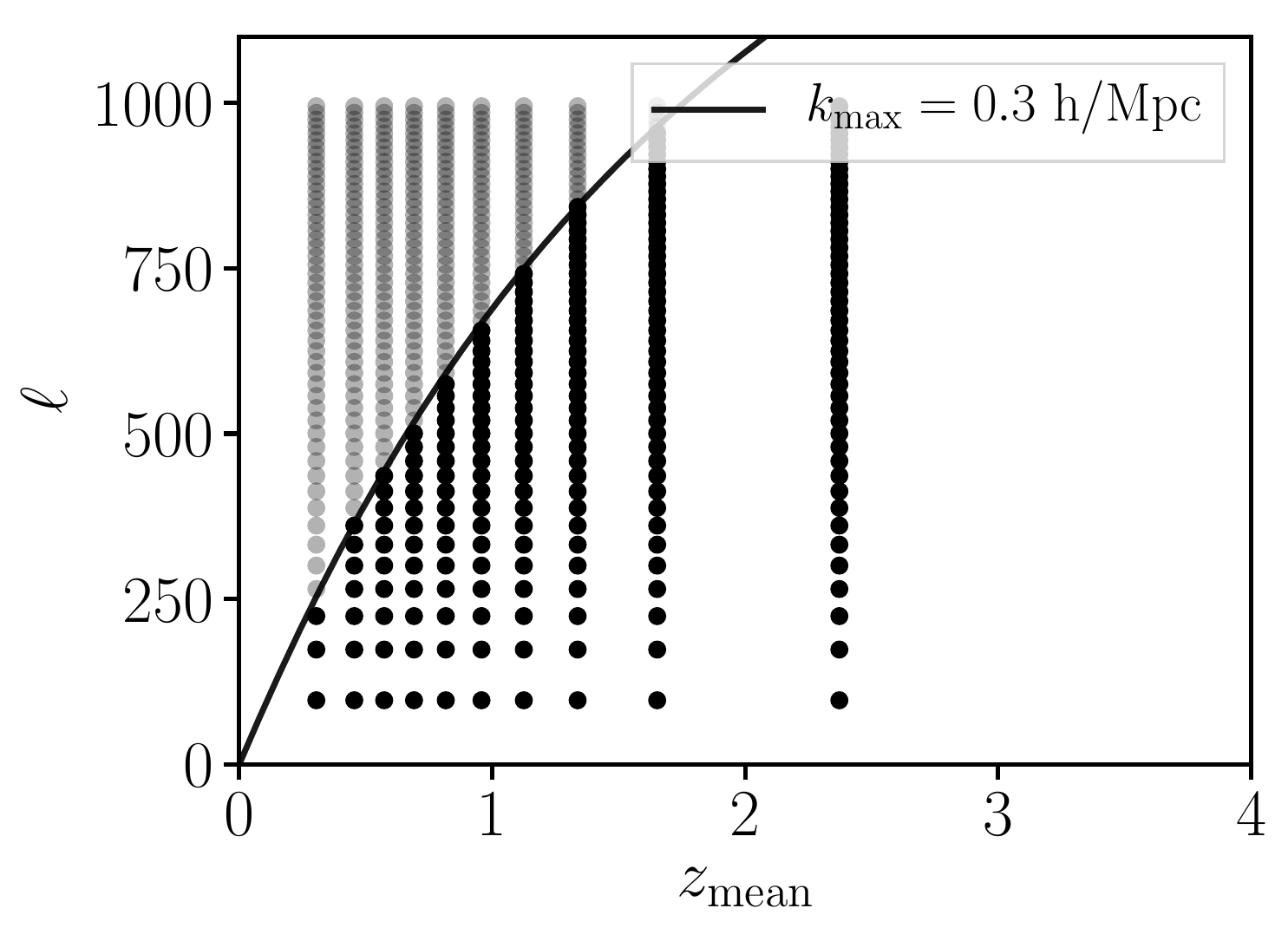}
\caption{
\textbf{Left:} True redshift distribution of the tomographic galaxy samples (red-yellow) used as sources, lenses and tracers, including a fiducial outlier fraction of $10\%$.
\textbf{Right:} Fourier multipoles included in the analysis.
For all power spectra, angular scales smaller than $\ell=1000$ are discarded. For clustering and galaxy - lensing correlations, angular modes with $\ell \geq 0.3 (\text{h/Mpc}) \chi_{(\langle z_i \rangle)} - 1/2$, corresponding roughly to comoving scales $k\geq 0.3$ $h$/Mpc, are further discarded.}
\label{fig:dndz}
\end{figure}

Furthermore, we also include photo-z outliers through a mixing or leakage matrix $c_{ij}$ \cite{Huterer2006}.
Another promising approach is the parameterization of \cite{2020arXiv200714989H}.
The number $c_{ij}$ is the fraction of galaxies in the true redshift bin $i$ that were (mistakenly if $i\neq j$) placed in the photometric redshift bin $j$.
In particular, $\sum_{j} c_{ij} = 1$, leading to 90 free parameters for the 10 tomographic bins.
The resulting redshift distribution for the tomographic sample $i$, including Gaussian errors and outliers, is thus given by
\beq
\frac{dn_i}{dz}
=
\frac{dn_i^G}{dz} \left( 1 - \sum_{j\neq i} c_{ij} \right)
+
\sum_{j\neq i}  \frac{dn_j^G}{dz} c_{ji}.
\eeq
For the fiducial values of these mixing coefficients, we assume $c_{ij} = 0.1 / \left( N_\text{bins} - 1 \right) \simeq 0.01$ for $i\neq j$, corresponding to $10\%$ of galaxies within in each bin being attributed a photometric redshift outside of the bin, as in \cite{LSSTScienceBook}.
The corresponding true redshift distributions of the tomographic samples are shown in Fig.~\ref{fig:dndz}.
In the analysis, we quantify the effect of marginalizing not only over the 20 Gaussian photo-z parameters, but also over the 90 coefficients $c_{ij}$ of the mixing matrix.

Having enough flexibility in the photo-z parameterization is crucial:
too little flexibility could lead to biases in the inferred cosmology, whereas too much flexibility should only lead to inefficiencies in the analysis (sampling too many, potentially poorly constrained and correlated, parameters).
Our parameterization includes 11 free parameters per tomographic bin, i.e. much more than the usual photo-z shift and scatter.
We show in App.~\ref{app:dndz_pca} a method to ensure that a given parametrization is sufficiently flexible.

\subsubsection{Observables in the 3$\times$2 analysis}

The galaxy number overdensity $g$ and the lensing convergence $\kappa_\text{g}$ from galaxies are related to the 3d matter overdensity field $\delta$ via a line-of-sight projection:
\beq
A(\vec{n}) = \int d\chi \, W_A(\chi) \, \delta (\chi \vec{n}, z(\chi)),
\eeq
with:
\beq
\left\{
\bal
W_{\kappa} (\chi, \chi_S) &= \frac{3}{2} \left( \frac{H_0}{c} \right)^2 \frac{\Omega_m^0}{a} \; \chi \left( 1 - \frac{\chi}{\chi_S} \right), \\
W_{\kappa_g} (\chi) &= \int dz_S \; \frac{1}{n}\frac{dn}{dz_S} \; W_\kappa (\chi, \chi(z_S)),\\
W_g(\chi) &= b(z) \; \frac{1}{n}\frac{dn}{dz} \; \frac{dz}{d\chi}
\; +\;  2(\alpha_g - 1) W_{\kappa_g} (\chi).\\
\eal
\right.
\eeq
In the last line, the second term is the magnification bias.
For a flux-limited galaxy catalog,
it is determined by the luminosity function as $\alpha_g = - d\ln N(>S_\text{cut}) / d\ln S_\text{cut}$, where $N(>S_\text{cut})$ is the total number of galaxies whose flux is above the cut $S_\text{cut}$.
In the flat sky and Limber approximation, the power spectrum of observables A and B is then simply:
\beq
C_\ell^{AB}
=
\int d\chi \;
\frac{W_A(\chi) W_B(\chi)}{\chi^2}
P_m\left( k=\frac{\ell+1/2}{\chi}, z(\chi)\right),
\eeq
where $P_m$ is the nonlinear matter power spectrum, computed using the \texttt{classylss}\footnote{\url{https://classylss.readthedocs.io/en/stable/\#}} 
Python wrapper to \texttt{CLASS} \cite{2011JCAP...07..034B}.
As is standard in forecasting, we assume that the expected level of shot noise and shape noise have been subtracted from the observed two-point functions, leaving the residual shot/shape noise as a noise only. In other words, the shot and shape noises are included in the covariance matrix.
In practice, because these are not exactly predicted by the $\propto 1 / n_\text{gal}$ scaling, a realistic analysis will likely need to introduce free parameters to fit for them.

\begin{figure}[h!!!]
\centering
\includegraphics[width=0.3\textwidth]{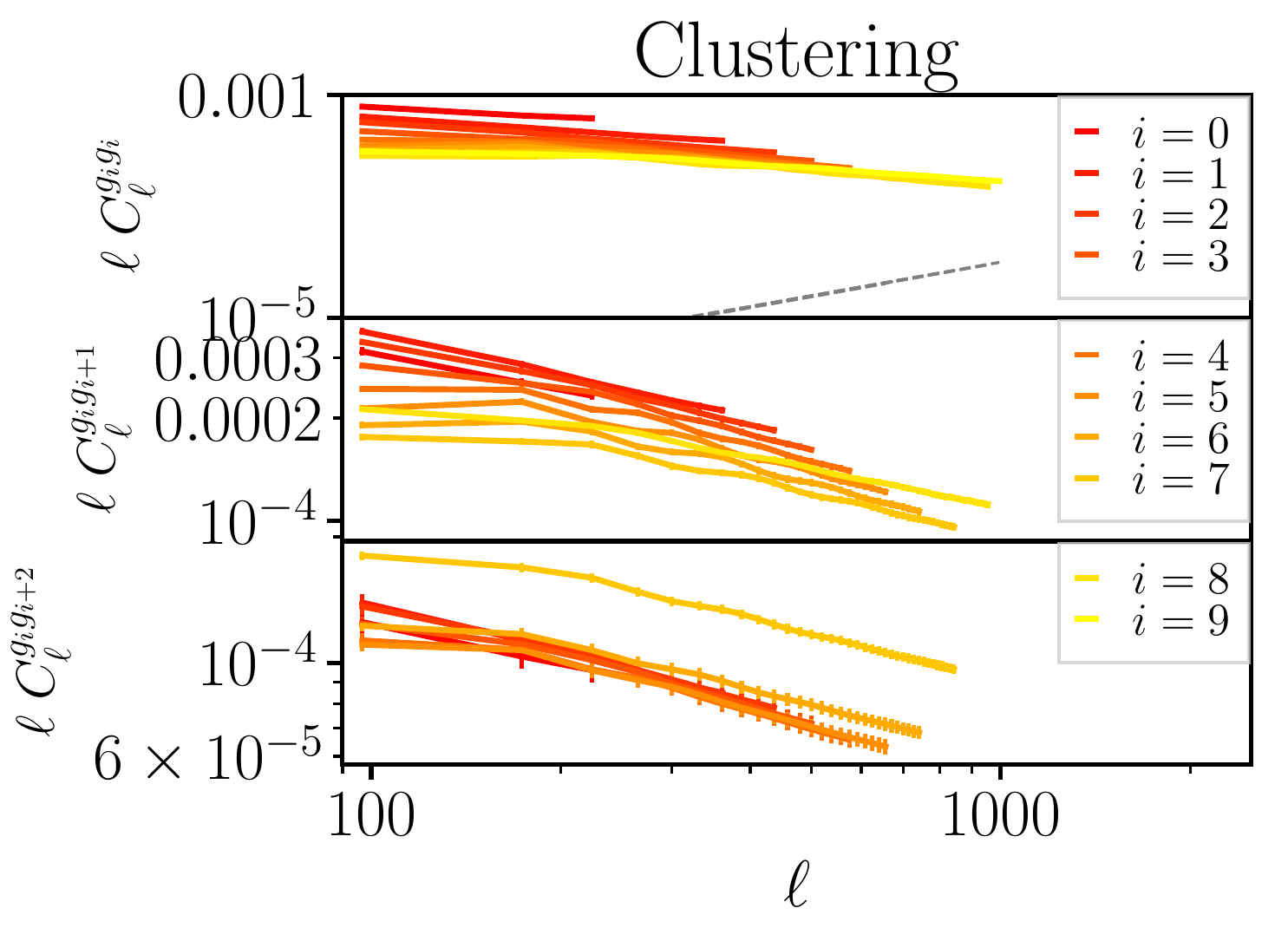}
\includegraphics[width=0.3\textwidth]{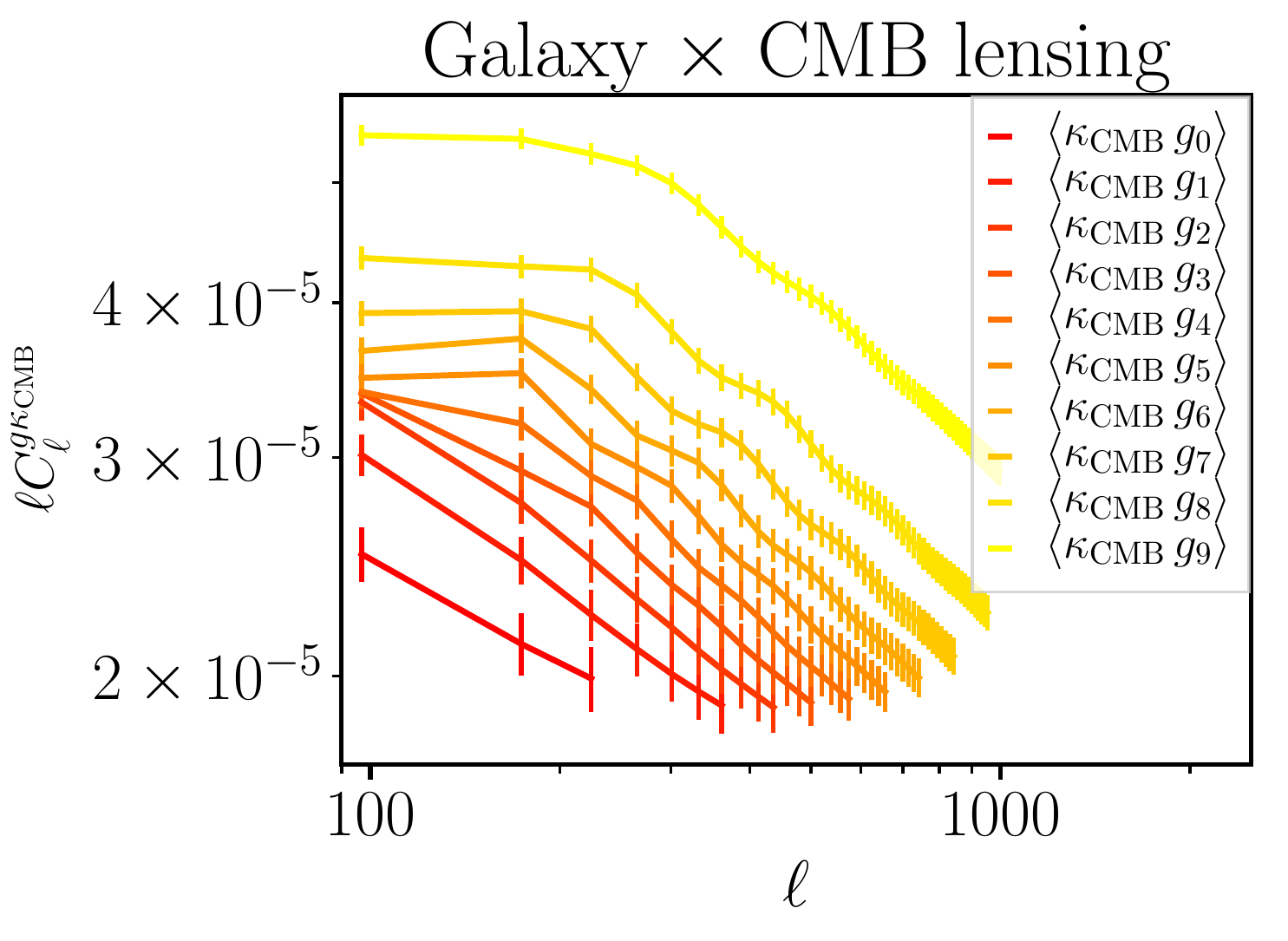}
\includegraphics[width=0.3\textwidth]{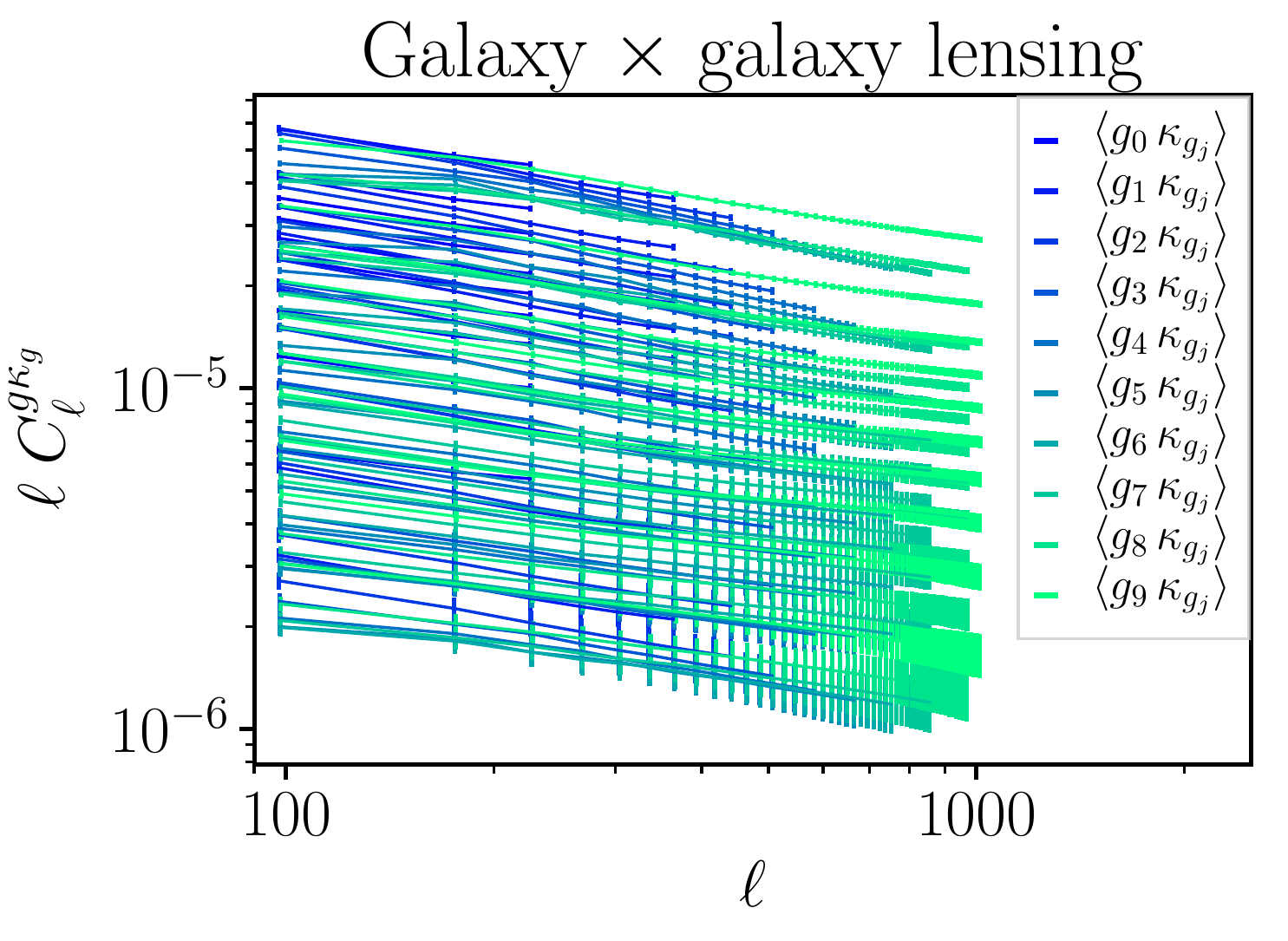}
\includegraphics[width=0.3\textwidth]{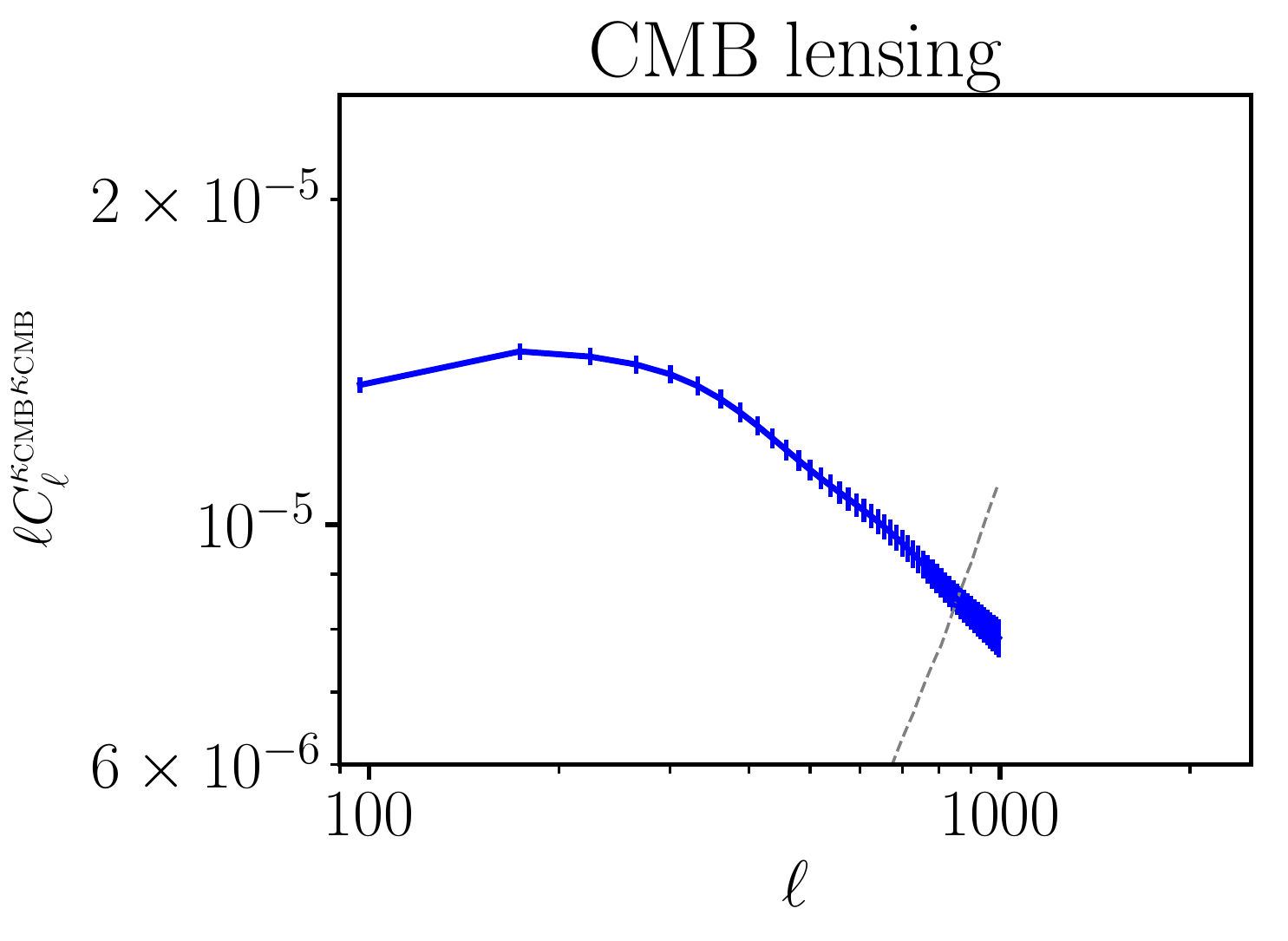}
\includegraphics[width=0.3\textwidth]{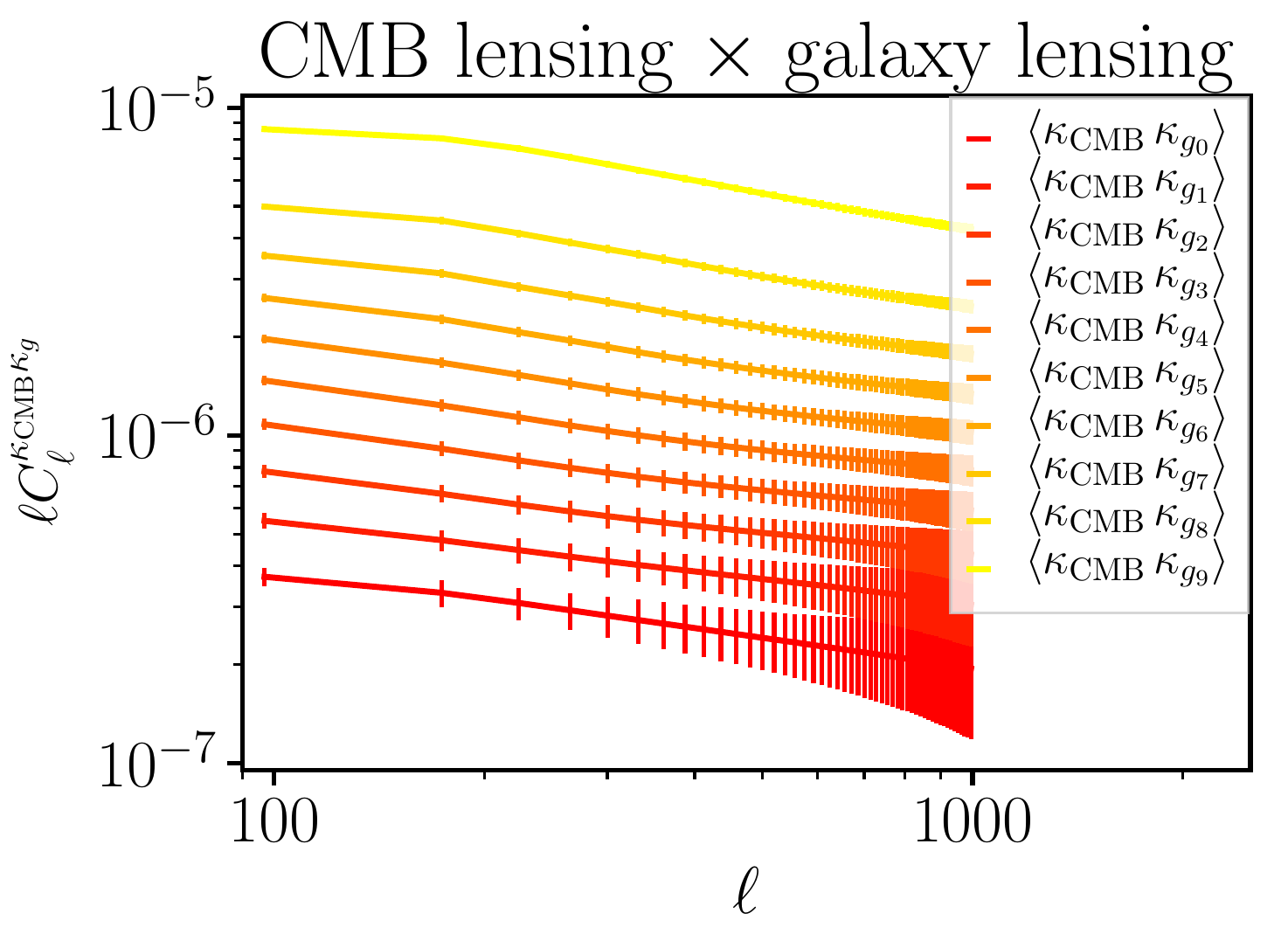}
\includegraphics[width=0.3\textwidth]{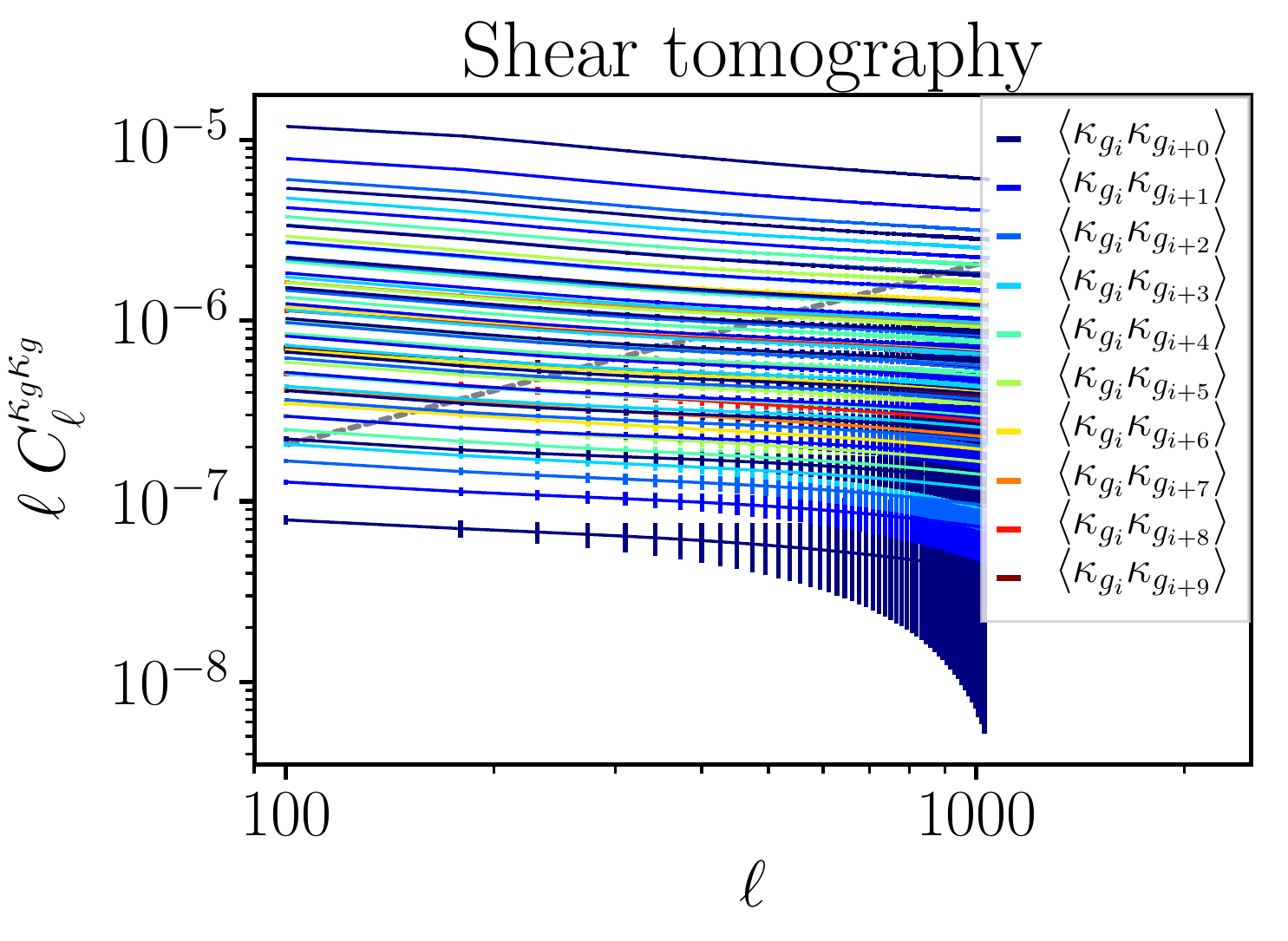}
\caption{
Auto and cross-power spectra included in the forecast (colored solid lines) compared to the galaxy shot noise, shape noise and CMB lensing noise (grey dashed lines).
The clustering auto-spectra and the cross-spectra of nearest-neighboring and next-to-nearest-neighboring tomographic bins are shown (top left).
The more distant cross-correlations, not shown here, are also included in the analysis.
These would be null in the absence of photo-z outliers, and are thus a useful diagnostic.
The clustering auto-spectra are cosmic variance limited on all scales and at all redshifts.
CMB lensing (bottom left) is almost cosmic variance limited on all the scales included,
and cosmic shear (bottom right) auto-spectra are cosmic variance limited on all scales in several of the higher redshift tomographic bins.
The galaxy - galaxy lensing spectra (top right) also include the ``null'' configurations where the source bin is at lower redshift than the lens bin, in order to detect photo-z outliers.
Finally, the galaxy - CMB lensing (top center) and CMB lensing - galaxy lensing correlations are also included.
}
\label{fig:data_vector}
\end{figure}

In the 3$\times$2 analysis, we include the photometric clustering $C_\ell^{g_i g_j}$,
the cosmic shear $C_\ell^{\kappa_{g_i} \kappa_{g_j}}$
and galaxy-galaxy lensing power spectra $C_\ell^{g_i \kappa_{g_j}}$
for all tomographic bins $i$ and $j$.
All cross- and auto-correlations are included, 
even the clustering cross-spectra of distant tomographic bins, 
or the galaxy-galaxy lensing in configurations where the lens bins is at higher redshift than the source bin.
These power spectra are shown in Fig.~\ref{fig:data_vector}.

\subsubsection{Scale cuts and systematics}

We limit the scales included in our forecast to $\ell_\text{max}=1000$ for all power spectra.
This scale cut is much more conservative than in most forecasts (e.g. $\ell_\text{max}=3000$ in \cite{Alonso2018}).
We further discard small scales such that 
$\ell \geq k_\text{max} \chi_{(\langle z_i \rangle)} - 1/2$
in the lens and tracer sample,
roughly corresponding to scales smaller than $k_\text{max}=0.3$ $h$/Mpc, as in \cite{Alonso2018}.
The corresponding $\ell$-modes included in the analysis are shown in Fig.~\ref{fig:dndz}.
We chose these stringent scale cuts to avoid modeling or marginalizing over intrinsic alignments, baryonic effects on the matter power spectrum and nonlinear biasing in the relationship between galaxy number density and the underlying dark matter density.
This is important, as some features of the small-scale nonlinear matter power spectrum may otherwise act as spurious standard rulers, leading to overestimate the performance of the photo-z self-calibration \cite{2010ApJ...720.1351H}.
In practice, nonlinear bias and intrinsic alignment may not be negligible at all redshifts for our highest multipoles. We leave a thorough exploration of these systematic effects to future work.
As the redshift gets higher, and the detected objects are brighter, nonlinear bias becomes more and more concerning \cite{2019JCAP...10..015W}. 
As Fig.~\ref{fig:dndz} shows, it would be easy to adopt a more stringent scale cut in the higher redshift bins without sacrificing too many Fourier modes.

We assume $\ell_\text{min}=20$, since the larger scales may be difficult to access observationally due to potential large angle systematics, and in order to keep the Limber approximation valid.
We bin the multipoles into 50 $\ell$-bins, defined to have the same number of 2D Fourier modes in each bin.
They are shown in Fig.~\ref{fig:dndz}.
This keeps the uncertainty (cosmic variance plus noise) roughly constant across $\ell$-bins, which improves the condition number of the covariance matrix.

We also assume that CMB lensing is free of any systematic.
One common concern in CMB lensing is extragalactic foregrounds such as the cosmic infrared background (CIB), the thermal and kinematic Sunyaev-Zel'dovich effects (tSZ \& kSZ) and radio point sources.
These extragalactic foregrounds are non-Gaussian and correlated with the lensing convergence, causing complex biases to CMB lensing from temperature
\cite{2014ApJ...786...13V, 2014JCAP...03..024O, 2018PhRvD..97b3512F, 2019PhRvL.122r1301S}.
However, at the high sensitivity of CMB S4, most of the lensing information comes from polarization, rather than temperature. In this regime, only polarized radio and IR sources contribute.
The measured percent-level polarization fractions of these sources \cite{2019MNRAS.490.5712G, 2019MNRAS.486.5239D, 2018ApJ...858...85P, 2019arXiv191109466L} imply that they are too small to cause significant bias to CMB lensing \cite{2009AIPC.1141..121S, 2018JCAP...04..018C}.
Polarized Galactic dust is significant on large scales in microwave maps, relevant to the search for primordial B-modes, but not on the small scales where CMB lensing comes from, and is therefore likely not a major concern.
Other instrumental systematics may affect CMB lensing, such as improperly characterized pointing, anisotropies in the instrumental noise or the beam ellipticity.
Quantifying these biases is beyond the scope of this paper, and we shall proceed assuming that they are under control.

\subsubsection{Gaussian likelihood and covariance matrix}
\label{sec:cov}

We assume the Likelihood to be Gaussian in the observed power spectra:
\beq
\ln \mathcal{L} (D|\theta)
=
-\frac{1}{2}
\left( D - M(\theta) \right)^t
\Sigma^{-1}
\left( D - M(\theta) \right),
\eeq
where $D$ is the data vector made of the observed power spectra, the mean $M(\theta)$ is their predicted value for cosmological and nuisance parameters $\theta$, and $\Sigma$ is the covariance matrix.

For simplicity, the covariance matrix includes only the Gaussian cosmic variance, galaxy shot noise, shape noise and CMB lensing noise. 
We do not include the trispectrum and supersample variance contributions to the covariance.
In \cite{Alonso2018}, a comparison of \texttt{CosmoLike} and \texttt{GoFish} demonstrated that the non-Gaussian covariances only changed the dark energy figure of merit by $2-5\%$, even when including much smaller scales in lensing ($\ell_\text{max}=3000$ in \cite{Alonso2018} instead of $\ell_\text{max} = 1000$ here).
We therefore expect the Gaussian covariance to produce cosmological uncertainties accurate to percent level here.
Ref.~\cite{Barreira2018} also showed that the connected non-Gaussian (i.e. trispectrum) term in the covariance is negligible for cosmic shear, and their Fig.~3 suggests that the impact of the supersample covariance is also small.

We include the cross-covariances between all pairs of power spectra but neglect the non-Gaussian covariances everywhere (trispectrum and supersample variance).
In other words,
\beq
\text{Cov} \left[ C_\ell^{ab}, C_{\ell '}^{cd} \right]_\text{Gaussian}
=
\frac{\delta^K_{\ell, \ell '}}{N_\text{modes}(\ell)}
\left[
C_\ell^{ac} C_\ell^{bd}
+
C_\ell^{ad} C_\ell^{bc}
\right]
\quad\text{with}\quad
N_\text{modes}(\ell)
\equiv
\frac{f_\text{sky}}{\pi}
\int_{\vl \in \ell} d^2\vl
.
\eeq
On the r.h.s., the clustering auto-spectra $C_\ell^{g_i g_i}$ receive a contribution from the galaxy shot noise $1/n_{\text{gal } i}$, 
the cosmic shear auto-spectra $C_\ell^{\kappa_{g_i}\kappa_{g_i}}$ receive a contribution from the galaxy shape noise $\sigma_e^2/n_{\text{gal } i}$ with $\sigma_e = 0.26$,
and the CMB lensing auto-spectrum receives a contribution from the CMB lensing reconstruction noise.
As shown in Fig.~\ref{fig:data_vector}, the clustering auto-spectra are cosmic variance limited on all scales and at all redshifts.
The cosmic shear auto-spectra are cosmic variance limited on all scales only at the higher redshifts, where the lensing signal is largest.

We include no shot/shape noise for cross-spectra. In practice, because galaxies from two tomographic bins may overlap in redshift and belong to the same halos, some level of shot/shape noise should be present. 
However, this should be smaller than the shot/shape noise on the auto-spectra, and an accurate modeling of the cross shot/shape noise from partially overlapping galaxy samples is delicate.
We thus differ its modeling to future work.

\begin{figure}[h!!!]
\centering
\includegraphics[width=0.3\textwidth]{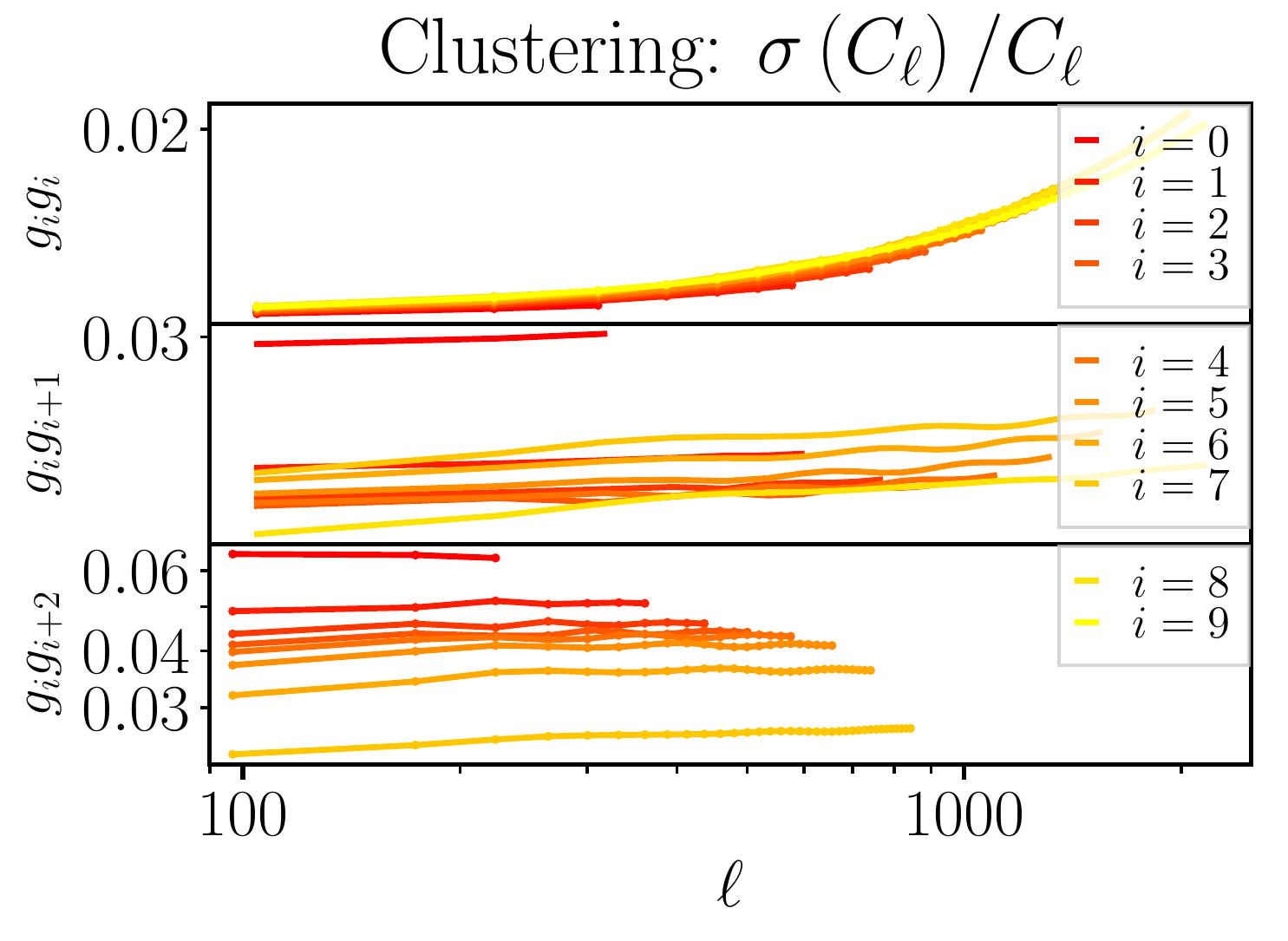}
\includegraphics[width=0.3\textwidth]{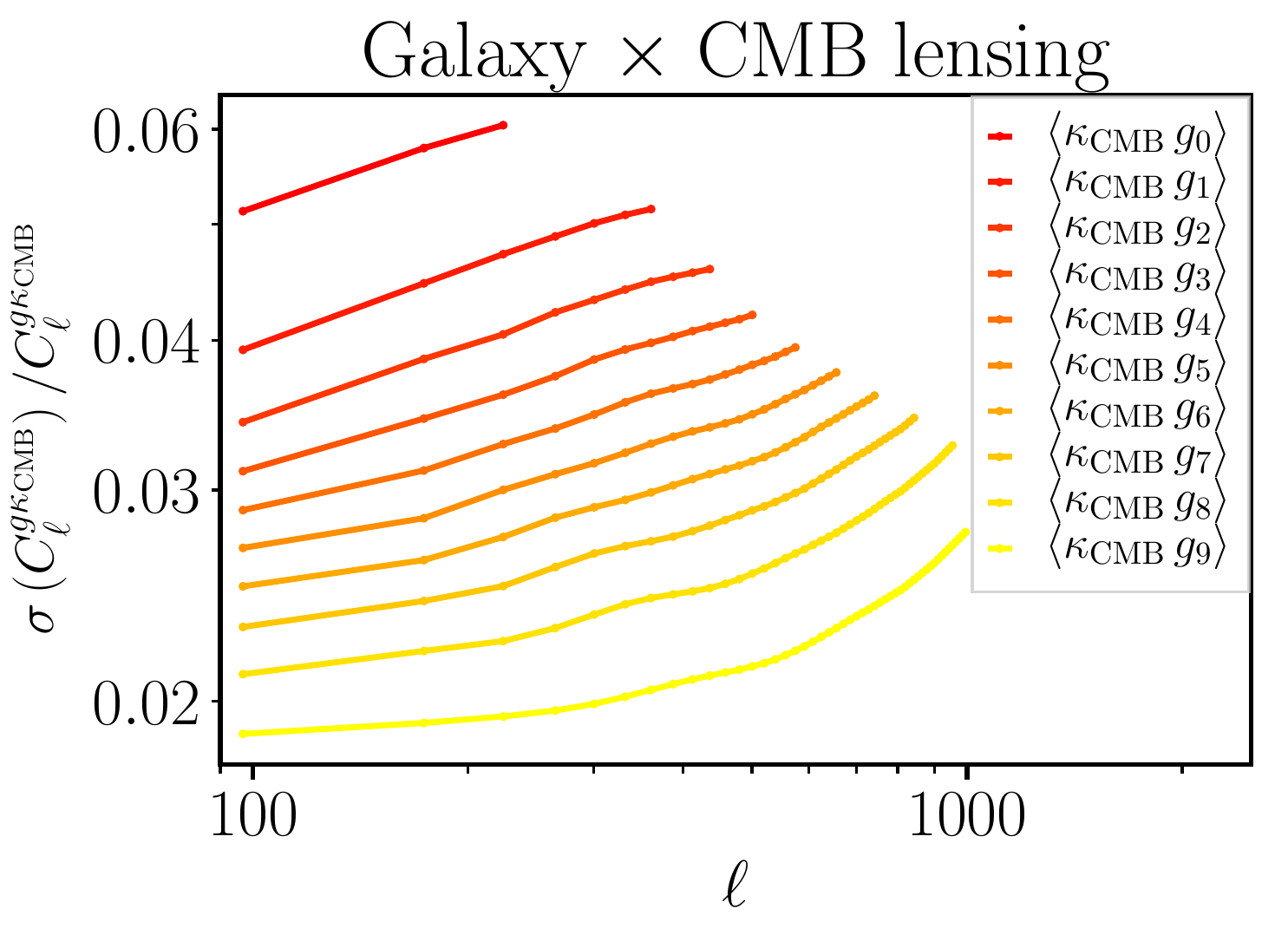}
\includegraphics[width=0.3\textwidth]{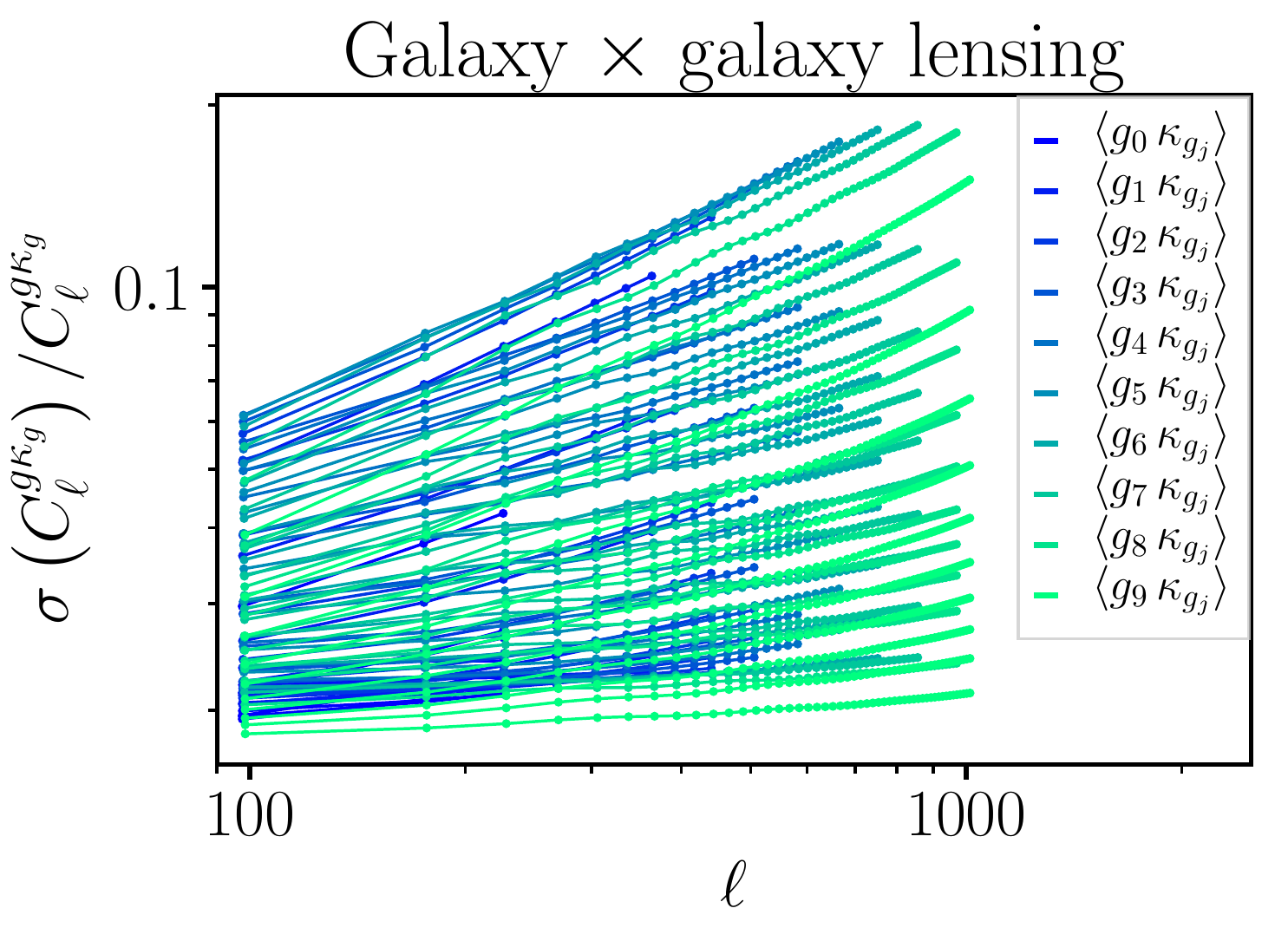}
\includegraphics[width=0.3\textwidth]{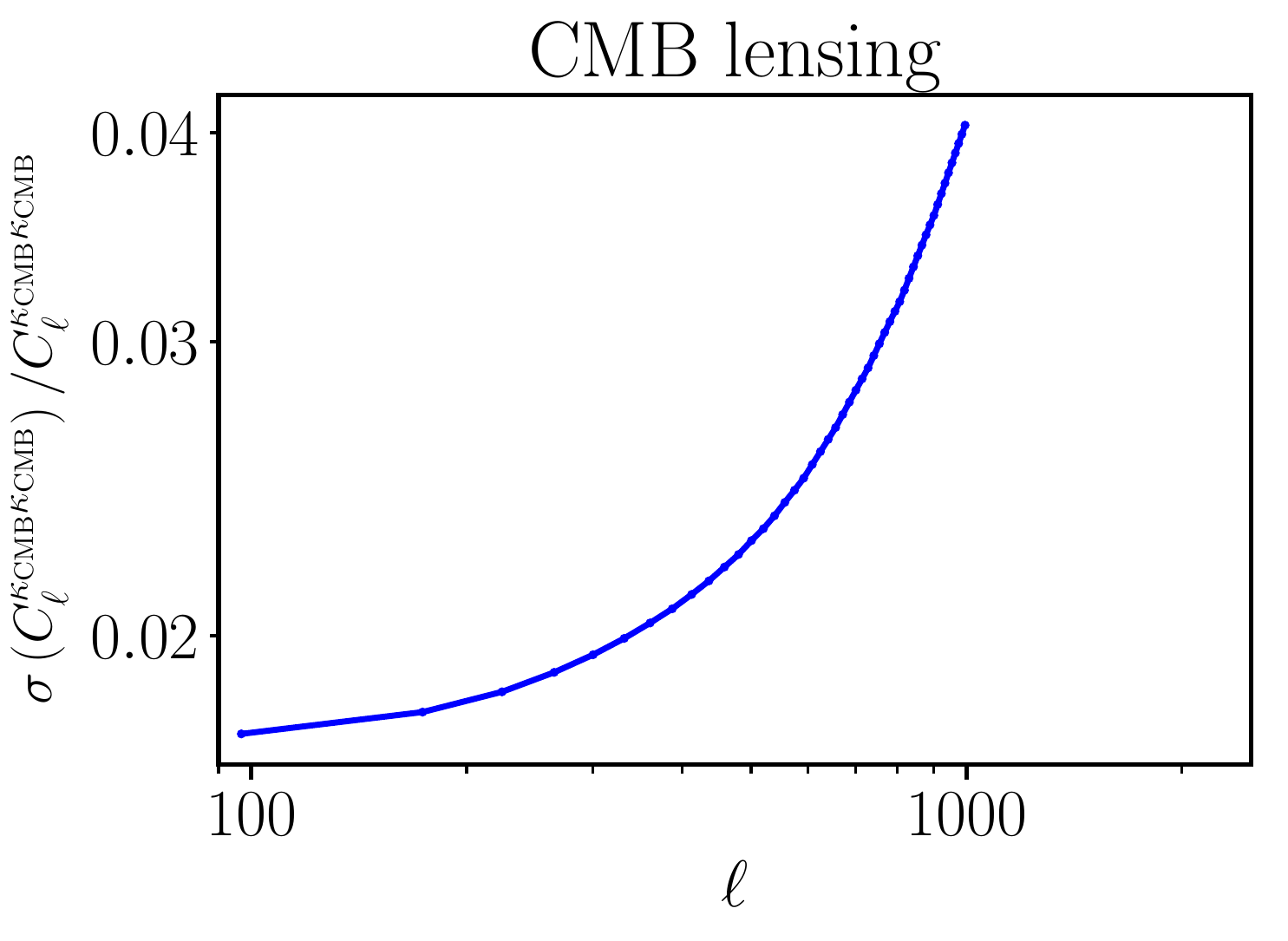}
\includegraphics[width=0.3\textwidth]{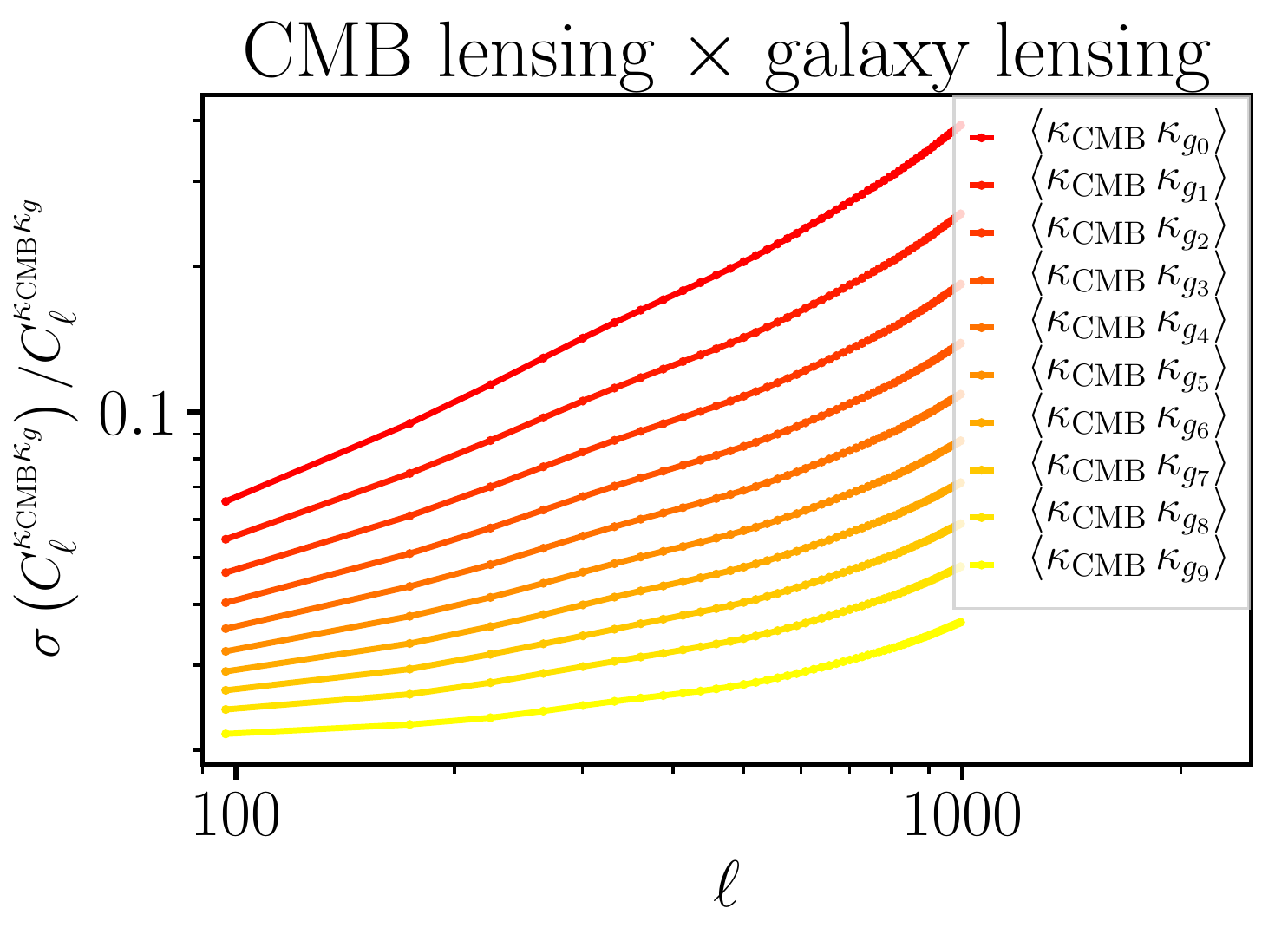}
\includegraphics[width=0.3\textwidth]{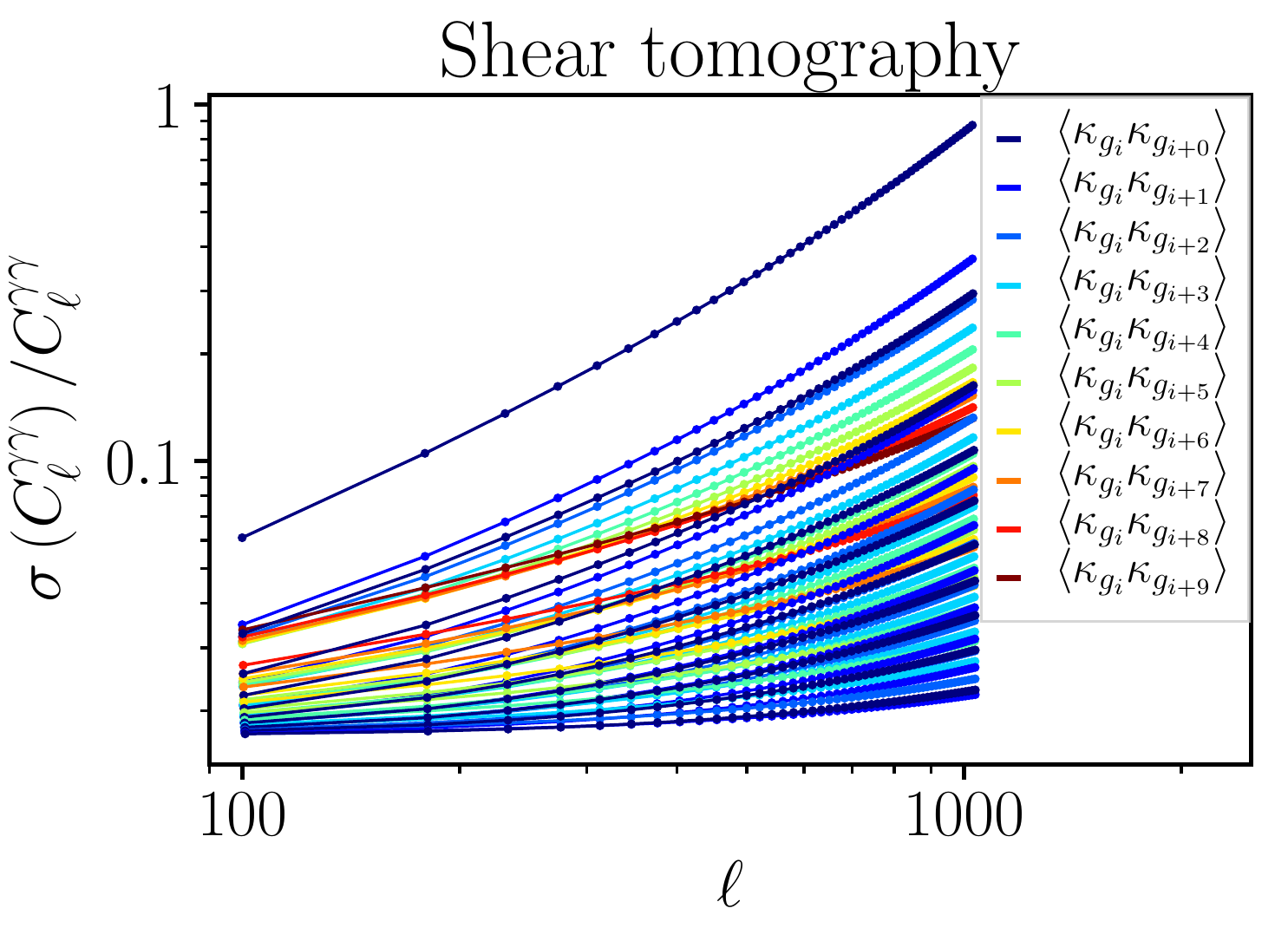}
\caption{
Relative uncertainty on the various power spectra included in the analysis, including cosmic variance and shot/shape noise.
Each $\ell$-bin of each two-point function is measured with a precision of a few percent or tens of percent.
Because the number of Fourier modes is identical in all $\ell$-bins, the relative uncertainty on $C_\ell^{g_i g_i}$ and $C_\ell^{\kappa_{g_i} \kappa_{g_i}}$ is independent of $\ell$ when the probe is cosmic variance limited, and grows with $\ell$ in the shot/shape noise dominated regime.
}
\label{fig:sp2d}
\end{figure}

The diagonal elements of the covariance matrix are shown in Fig.~\ref{fig:sp2d}.
Each $\ell$-bin of each two-point function is measured with a precision of a few percent to a few tens of percent.
This suggests that one may use even finer $\ell$-bins or tomographic bins in the analysis.
To simplify the numerical evaluation, and in particular inverting the large covariance matrix, we do not explore this possibility.
Because the number of Fourier modes is identical in all $\ell$-bins, the relative uncertainty on $C_\ell^{g_i g_i}$ and $C_\ell^{\kappa_{g_i} \kappa_{g_i}}$ is independent of $\ell$ when the probe is cosmic variance limited, and grows with $\ell$ in the shot/shape noise dominated regime.

From our Gaussian covariance, we can compute the total signal-to-noise ratios (SNR) for the various probes, including cosmic variance, shot noise, shape noise and CMB lensing noise.
These are shown in Tab.~\ref{tab:snr_lsst}.
They give an idea of the relative weight of each probe in the combined analyses. 
CMB lensing has a smaller SNR than galaxy lensing, but comparable.
It provides a source plane for lensing with a higher and well-known redshift, and is assumed to be free of multiplicative bias.
We also show the loss in SNR by discarding the null cross-correlations, i.e. the ones that would be zero in the absence of photo-z uncertainties. These contribute 25\% of the SNR in clustering, but only 3\% in galaxy-galaxy lensing and less than 10\% in the combined analysis. 
However, as we show below, these low SNR null cross-correlations contain precious information about photo-z errors, and in particular outliers.
\begin{table}[H]
\centering
\begin{tabular*}{0.95\textwidth}{@{\extracolsep{\fill}}| l l |}
\hline
LSST \& CMB S4 lensing & SNR \\  
\hline
\hline
\multicolumn{2}{|c|}{\textbf{Individual probes}} \\
\hline
clustering all: $g_ig_j$ & 793 \\
clustering auto (``no null''): $g_ig_i$ & 597 (25\% loss) \\
\hline
galaxy-galaxy lensing all: $g_i \kappa_{g_j}$ & 431 \\
galaxy-galaxy lensing ``no null'': $g_i \kappa_{g_{j \geq i}}$ & 416 (3\% loss) \\
\hline
shear tomography: $\kappa_{g_i} \kappa_{g_j}$ & 428 \\
\hline
CMB lensing auto: $\kappa_\text{CMB} \kappa_\text{CMB}$ & 282 \\
\hline
galaxy-CMB lensing: $g \kappa_\text{CMB}$ & 322 \\
\hline
galaxy lensing-CMB lensing: $\kappa_\text{gal} \kappa_\text{CMB}$ & 255 \\
\hline
\hline
\multicolumn{2}{|c|}{\textbf{Combinations}} \\
\hline
LSST: $gg, g\kappa_g, \kappa_{g}\kappa_g$ & 851 \\
LSST: $gg, g\kappa_g, \kappa_{g}\kappa_g$, ``no null'' & 787 (8\% loss) \\
\hline
LSST+CMB lensing: $g, \kappa_g, \kappa_\text{CMB}$ & 868 \\
LSST+CMB lensing: $g, \kappa_g, \kappa_\text{CMB}$, ``no null'' & 816 (6\% loss)\\
\hline
\end{tabular*}
\caption{
Individual and combined signal-to-noise ratios (SNR) for LSST and CMB S4 lensing,
including the Gaussian cosmic variance as well as shot noise, shape noise and CMB lensing noise.
They provide insight as to the relative weight of each probe in the combined analyses. 
We also show the loss in SNR when discarding the ``null'' cross-correlations, i.e. the ones that would be zero in the absence of photo-z uncertainties. These contribute 25\% of the SNR in clustering, but only 3\% in galaxy-galaxy lensing and less than 10\% in the combined analysis. 
However, as we show below, these low SNR ``null'' cross-correlations contain precious information about photo-z errors, and in particular outliers.
}
\label{tab:snr_lsst}
\end{table}

Comparing with our previous forecast \cite{2017PhRvD..95l3512S}, several assumptions have changed.
The assumed number density of source galaxies is identical, but we reduced the maximum scale included in lensing from $\ell_\text{max}=5000$ to $\ell_\text{max}=1000$. This produced a reduction of $25\%$ in the cosmic shear SNR, and of $40\%$ in the CMB lensing SNR, thus reducing the relative weight of CMB lensing compared to galaxy lensing in the joint analysis.
The maximum scale included in clustering is similar in both forecasts, amounting to $k_\text{max}=0.3$ $h$/Mpc. 
However, while \cite{2017PhRvD..95l3512S} assumed an extremely conservative, RedMaGiC-like clustering sample with only $0.25\ \text{galaxies/arcmin}^2$, we now use the same sample as sources, lenses and tracers, leading to $27\  \text{galaxies/arcmin}^2$.
This large increase in sample size makes the clustering cosmic variance limited on all scales and at all redshifts, as shown above, and leads to a factor two improvement in clustering SNR.

Finally, we do not vary the covariance matrix with cosmological parameters, and simply evaluate it at the fiducial cosmology. 
While varying the covariance matrix with cosmology may improve the constraints \cite{2009A&A...502..721E}, it is not clear whether varying it or not gives a better approximation to the true non-Gaussian likelihood \cite{2013A&A...551A..88C}.
Finally, we perform several forecasts, for different data and parameter combinations, by applying the Fisher formalism.
For our parameter-independent covariance matrix, the Fisher matrix takes the simple form:
\beq
F_{\theta_i, \theta_j}
=
\frac{\partial M^t}{\partial \theta_i}
\Sigma^{-1}
\frac{\partial M}{\partial \theta_j},
\eeq
and we will assume that
$\text{cov}\left[ \theta_i, \theta_j \right] = \left( F^{-1} \right)_{\theta_i, \theta_j}$ as is routinely done in Fisher forecasts.
Given the large parameter space in this analysis, an accurate inversion of the Fisher matrix requires very accurate Fisher matrix elements.
In particular, the convergence of the numerical derivatives is key.
In App.~\ref{app:full_fisher_plots}, we carefully explore the derivative step sizes, and show that our derivatives are converged to percent accuracy, and that the marginalized uncertainties (from the inverse Fisher matrix) are accurate to better than 10\%.

\subsubsection{Cosmological and nuisance parameters}

The parameters we vary are summarized in Tab.~\ref{tab:params_fisher}.
The cosmological parameters include extensions to $\Lambda$CDM, namely massive neutrinos, varying dark energy equation of state and curvature:
$\Omega_\text{cdm}, \Omega_b, A_s, n_s, h, \tau_\text{reio}, M_\nu, w_0, w_a, \Omega_k$.
Their definition and implementation follows those in \texttt{CLASS} \cite{2011JCAP...07..034B}.
In particular, for each value $M_\nu$ of the sum of the neutrino masses, we use as input to \texttt{CLASS} the three distinct neutrino masses in the normal hierarchy, with sum $M_\nu$ and solar and atmospheric splittings 
$\Delta m^2_\text{atm} = 2.5\times 10^{-3} \text{eV}^2$
and
$\Delta m^2_\text{solar} = 7.6\times 10^{-5} \text{eV}^2$.
Our nuisance parameters are 
the galaxy biases for each tomographic bin $b_{g0}, ..., b_{g9}$ (10 parameters), 
the shear biases for each tomographic bin $m_{0}, ..., m_{9}$ (10 parameters),
the Gaussian photo-z shifts and scatters $\delta_{z_0}, \sigma_{z_0}, ..., \delta_{z_9}, \sigma_{z_9}$ (20 parameters),
and the photo-z outlier parameters in the form of the mixing matrix $c_{ij}$ for $i\neq j$ (90 parameters).

The fiducial values and priors for all the parameters are shown in Tab.~\ref{tab:params_fisher}.
The fiducial cosmological parameters follow the Planck 2018 TT+TE+TT+lowE+CMB lensing+BAO in Tab.~2 of \cite{PlanckParam2018},
and we assume a CMB-only Planck-like prior on the cosmological parameters, from a CMB-only Fisher forecast which reproduces the 2018 parameter results.
To include the off-diagonal parameter correlations in the prior, we calculate a Fisher matrix including all 6 Planck frequency channels with their respective sensitivities and we assume Gaussian beams for simplicity.  We use both temperature and polarization information, with $\ell_{\rm max} = 2500$ for both, while noting that the results are insensitive to $\ell_{\rm max}$, since the beam provides for an effective cutoff on larger scales. We additionally set $\ell_{\rm min} = 6$ in polarization only, in order to reproduce the constraint $\sigma(\tau) \approx 0.007$. We find good agreement with the Planck 2018 parameters for the published diagonal constraints.

One interesting extension of this work would be to
explore the photo-z self-calibration in the absence of CMB priors on cosmology.
While the constraints on e.g., neutrino masses would be degraded, the relative improvement in photo-z self-calibration when including the null correlations may be greater.
This could also be useful for consistency checks between low and high redshift probes, for instance for the Hubble parameter.
\begin{table}[H]
\centering
\begin{tabular*}{0.95\textwidth}{@{\extracolsep{\fill}}| c c l |}
\hline
Parameter & Fiducial value & Prior \\  
\hline
\hline
\multicolumn{3}{|c|}{$\Lambda$\textbf{CDM cosmology}} \\
$\Omega_m$ & 0.26 & $\mathcal{N}(0.26, 0.003)$  \\ 
$\Omega_b$ &  0.049 &  $\mathcal{N}(0.049, 0.00004)$ \\
$A_S$ & $2.105 \times 10^{-9}$ &  $\mathcal{N}(2.105  \times 10^{-9}, 3.010\times 10^{-11})$ \\ 
$n_s$ & 0.9665 & $\mathcal{N}(0.9665, 0.0038)$  \\
$h_0$  & 0.6766 &  $\mathcal{N}(0.6766, 0.638)$ \\
$\tau$  & 0.0561 & $\mathcal{N}(0.0561, 0.0070)$ \\
\hline
\multicolumn{3}{|c|}{$\Lambda$\textbf{CDM extensions}} \\
$w_0$ &  -1 &  $\mathcal{N}(-1, 2.559)$  \\
$w_a$ &  0 & $\mathcal{N}(0, 14.57)$  \\
$M_\nu$ &  0.1 eV & $\mathcal{N}(0.1, 0.92)$  \\
$\Omega_k$ &  0 & $\mathcal{N}(0, 0.0368)$  \\
\hline
\multicolumn{3}{|c|}{\textbf{Galaxy bias parameter}} \\
$b_0 ... b_9$ & 1  & Uniform \\
\hline
\multicolumn{3}{|c|}{\textbf{Shear calibration}} \\
$m_i $ & 0 & Uniform or $\mathcal{N}(0, 0.005)$\\
\hline
\multicolumn{3}{|c|}{\textbf{Photo-z}} \\
$\delta z _i $ & 0 & Uniform or $\mathcal{N}(0, 0.002)$ \\
$\sigma_{z_i} / (1+\langle z\rangle_i) $ & 0.05 & Uniform or $\mathcal{N}(0.05, 0.003)$ \\
$c_{ij, i\neq j} $ & $0.1 / \left( N_\text{bins}-1 \right)$ & Uniform or $\mathcal{N}(0.011, 0.00556)$ \\
\hline
\end{tabular*}
\caption{
We constrain the $\Lambda$CDM model (6 parameters) and several extensions (up to 10 parameters).
We also marginalize over 130 nuisance parameters: 10 galaxy biases, 10 shear biases and 110 photo-z parameters.
The fiducial cosmological parameters follow the Planck 2018 TT+TE+TT+lowE+CMB lensing+BAO \cite{PlanckParam2018}.
The bias parameters $b_i$ with fiducial values of 1 are scaling parameters for the galaxy bias such that $b(z) = b_i \times 0.95/D(z)$ where $D(z)$ is the linear growth factor. 
Our ``Planck prior'' is a Gaussian prior from a CMB-only Fisher forecast, whose marginalized errors match the table 3 in the Planck 2015 parameters \cite{2015arXiv150201589P} (see main text for more details). It includes the non-zero covariance between the cosmological parameters.
}
\label{tab:params_fisher}
\end{table}

\subsubsection{Summary: comparison with the DESC SRD}
Our assumptions differ from those in the DESC SRD in several points.
Our source sample is split into 10 tomographic bins instead of 5 in the DESC SRD. 
The use of only 5 bins was justified in the DESC SRD by numerical convergence issues when inverting the covariance matrix.
The procedure described in App.~\ref{app:cov_conditiong_number} allows us to improve the conditioning number of the covariance matrix by many orders of magnitude, and avoid this issue.
Another justification for using 5 bins instead of 10 in the DESC SRD was to reduce the sensitivity to photo-z outliers, which were not modeled. 
In this study, we specifically investigate this question by marginalizing over outlier parameters.

Contrary to the DESC SRD, we assume the same galaxy sample for sources and lenses, to help with photo-z self-calibration.
In practice, this means that our lens and tracer sample is much sparser than in the DESC SRD, with only $27$ galaxies per squared arcminute, instead of $47$ for the gold sample.
We expect this pessimistic assumption not to impact our constraints too negatively, since we find the clustering measurements to be cosmic variance limited and not shot noise dominated over most scales.
Furthermore, our lens and tracer sample has the same photo-z scatter as our source sample, i.e. $\sigma_z/(1+z) = 0.05$, compared to $0.03$ in the DES SRD.
On the other hand, our approach divides by two the number of photo-z parameters to constrain, which are the most numerous nuisance parameters in this analysis.
Another difference is that our tracer and lens sample extends to higher redshift than assumed in the DESC SRD, where it was cut off at $z=1.2$. 
However, the SRD mentions that future versions of the analysis will extend this to higher redshift.

Our scale cuts are more conservative in lensing, with $\ell_\text{max}=1000$ here compared to 3000 in the SRD.

Finally, contrary to the DESC SRD, we include all the ``null'' cross-correlations, i.e. the clustering cross-correlations, as well as the galaxy-galaxy lensing spectra where the lens bin is at higher redshift than the source bin.

\subsection{Results}

\subsubsection{Baseline constraints}

\begin{figure}[H]
\centering
\includegraphics[width=\columnwidth]{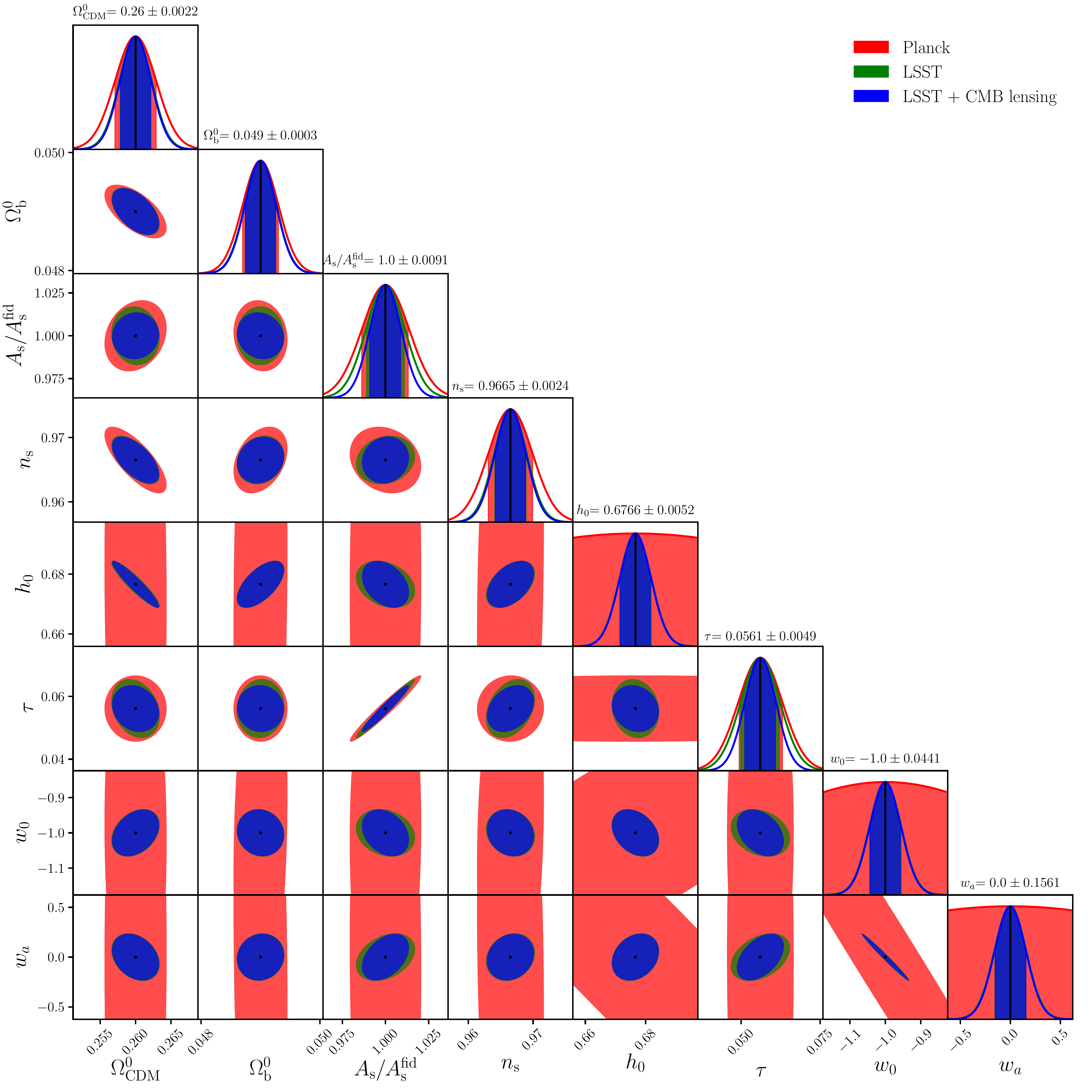}
\caption{
Baseline cosmological constraints for $w_0w_a$CDM,
including priors on the Gaussian and outlier photo-z errors consistent with the LSST requirements, as well as Planck priors on cosmology.
The parameter 1$\sigma$ uncertainties listed along the diagonal are for LSST+CMB S4 lensing, very similar to LSST alone in this analysis.
  }
\label{fig:baseline_cosmo_contours_lcdmw0wa}
\end{figure}

The baseline cosmology constraints, for LSST with Planck priors and for $w_0w_a$CDM, are shown in Fig.~\ref{fig:baseline_cosmo_contours_lcdmw0wa}.
These posterior contours marginalize over all the nuisance parameters (galaxy biases, shear biases, Gaussian and outlier photo-z) with their fiducial priors.
The dark energy equation of state parameters are constrained with a $0.04, 0.16$ statistical uncertainty on $w_0, w_a$ respectively.
The corresponding constraints for $\Lambda$CDM, $w_0$CDM, $\Lambda$CDM + $M_\nu$ and $\Lambda$CDM + curvature are shown in App.~\ref{app:other_cosmo_params}.
The neutrino masses are measured with an uncertainty of $0.019$ eV, and the uncertainty on curvature $\Omega_k$ is $0.0007$.
The Hubble parameter $h_0$ is not improved compared to the Planck prior in the $\Lambda$CDM case, but much improved in the $\Lambda$CDM extensions, due to degeneracy breaking.
Our results are comparable to \cite{2018PhRvD..97l3544M} and \cite{2018JCAP...01..022B}, although slightly more optimistic in terms of the neutrino masses.

Fig.~\ref{fig:baseline_cosmo_contours_lcdmw0wa} also shows that CMB lensing only provides a mild improvement on the $w_0w_a$CDM cosmological parameters. 
This is most likely due to the high SNR in clustering and lensing in this analysis, the large number (10) of tomographic bins, and the stringent scale cuts, which make CMB lensing appear less important.

\subsubsection{Importance of marginalizing over photo-z outliers}

\begin{figure}[H]
\includegraphics[width=0.45\columnwidth]{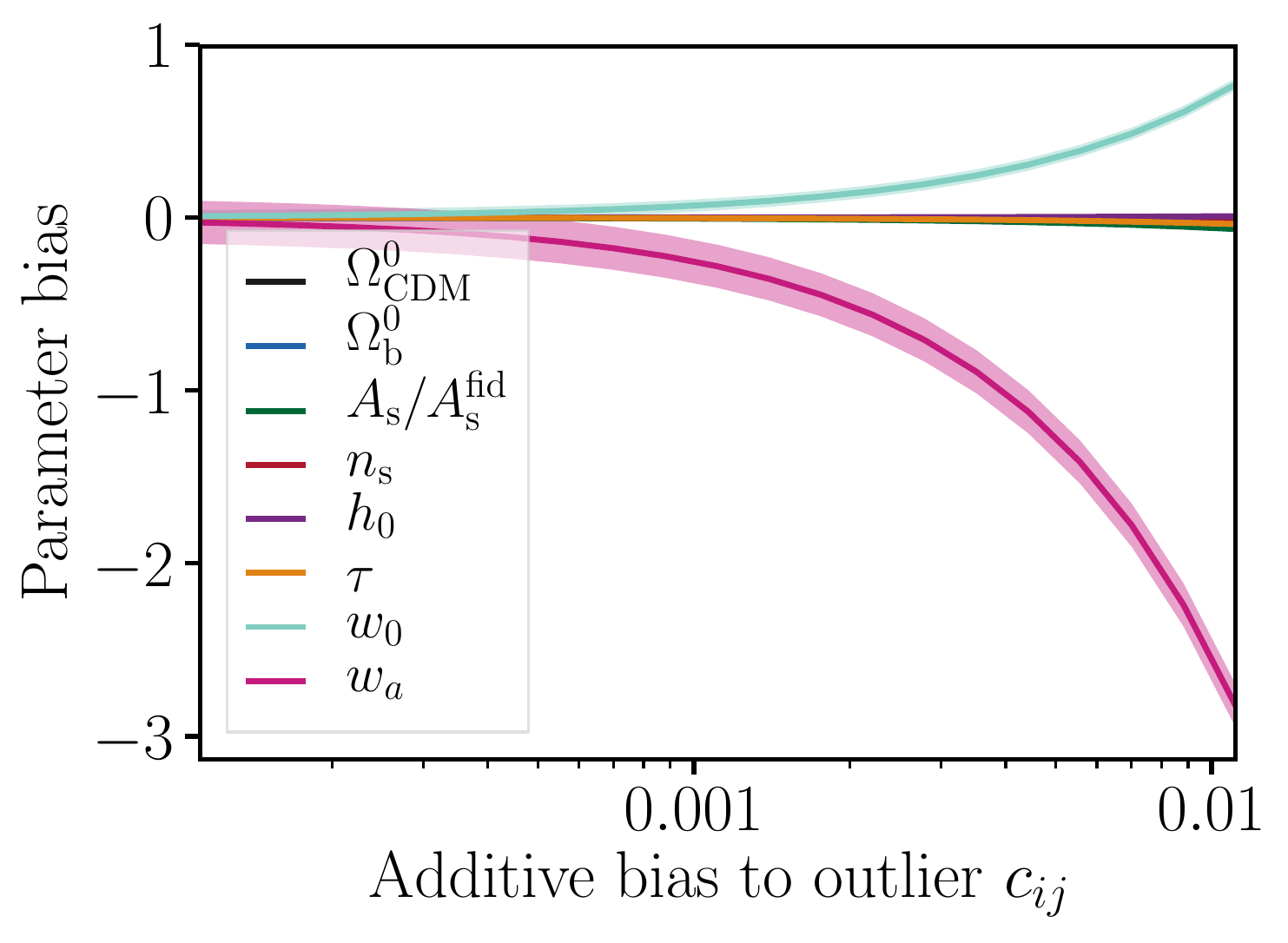}
\includegraphics[width=0.45\columnwidth]{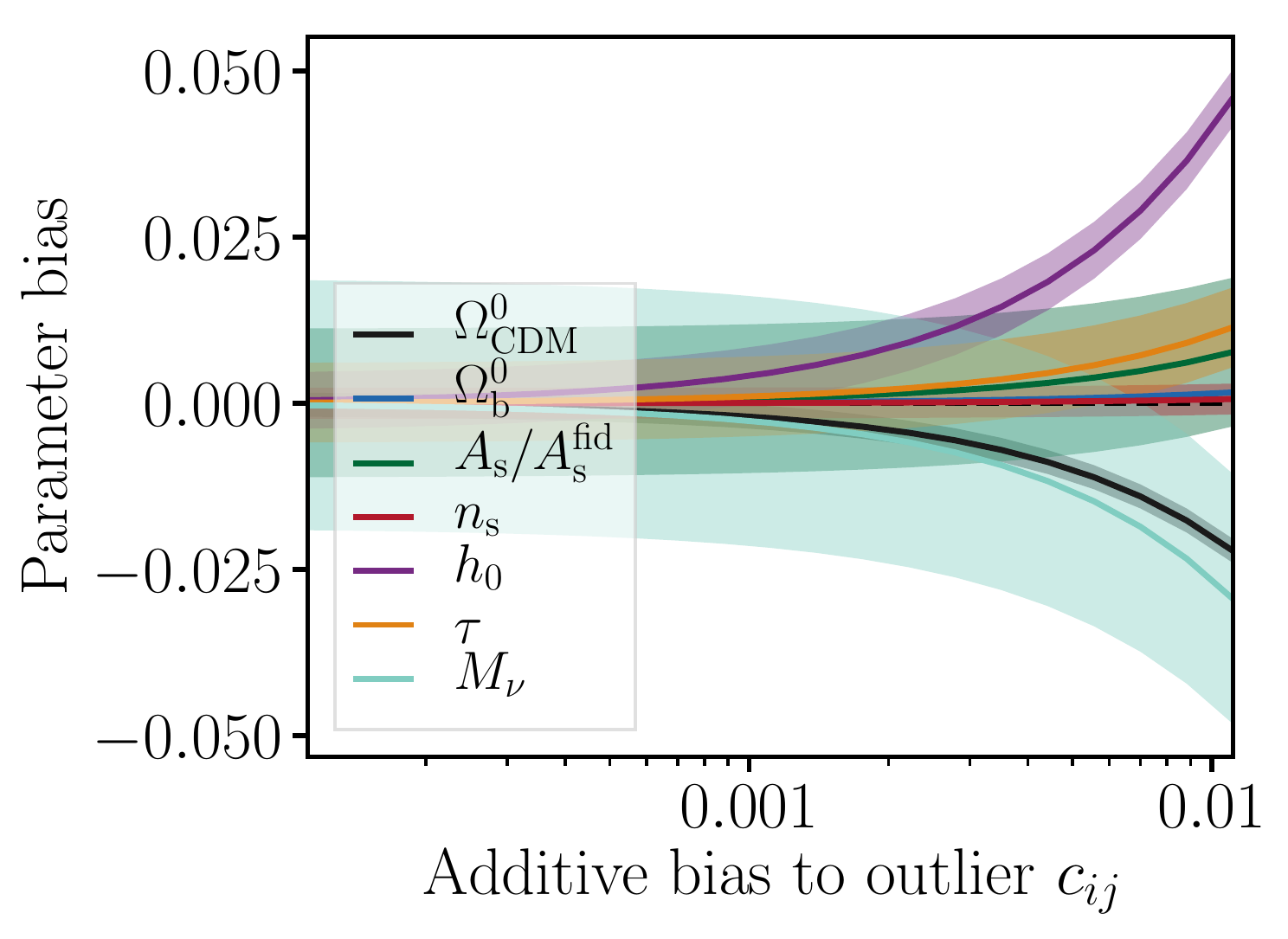}
\caption{
When the photo-z outlier fractions $c_{ij}$ are fixed instead of marginalized over, a bias in the assumed outlier fraction results in a bias in cosmological parameters.
The additive absolute bias to each parameter is shown by the colored lines.
The bands indicate the $1\sigma$ statistical error bar on each cosmological parameter, when the photo-z outliers are not marginalized.
\textbf{Left:} an additive error on the outlier fractions $c_{ij}$ of 0.0005, i.e. only 5\% of the fiducial value assumed here ($c_{ij}^\text{fiducial}=0.011$) is enough to bias the dark energy equation of state parameters $w_0$ and $w_a$ at the $1\sigma$ level.
\textbf{Right:} an additive error on assumed $c_{ij}$ of 0.007 causes a $1\sigma$ bias to the neutrino masses, and a more significant bias in the Hubble parameter.
}
\label{fig:bias_photozoutliers_cosmo}
\end{figure}

If the photo-z outliers fractions $c_{ij}$ are known perfectly, they should be fixed and not marginalized over.
How well then do they need to be known in order to avoid biasing the cosmology?
To answer this question, we use the Fisher formalism to propagate a bias in some fixed parameters (the photo-z outliers) into the resulting bias on cosmological parameters (see \cite{2007MNRAS.374.1377T} and references therein).
Indeed, if the full parameter vector $\theta = \{ \{p_i\}, \{\psi_\alpha\} \}$ is split into the varied parameters $\{p_i\}$ and the fixed parameters $\{ \psi_\alpha \}$, the bias $\delta p_i$ in the inferred parameters is related to the bias $\delta \psi_\alpha$ in the fixed parameters via:
\beq
\delta p_i
=
- \left( F_p^{-1} \right)_{i j}
\left(F_\theta\right)_{j \alpha}
\delta \psi_\alpha,
\eeq
where $F_p$ is the Fisher matrix for the $\{p_i\}$ parameters only and $F_\theta$ is the Fisher matrix for the full parameter set $\theta$.
Fig.~\ref{fig:bias_photozoutliers_cosmo} shows the results in the case of $w_0w_a$CDM and $\Lambda$CDM + $M_\nu$.

The dark energy equation of state parameters acquire a $1 \sigma$ bias, i.e. a bias as large as the statistical error bar, as soon as the assumed fixed outlier fractions $c_{ij}$ are incorrect by 0.0005.
For our choice of $c_{ij}^\text{fiducial}=0.011$, this means that the outlier fractions would have to be known to better than 5\% of their values.
Achieving such a knowledge of the photo-z outliers \textit{a priori} seems difficult, which motivates us to marginalize over them.

The neutrino masses are much less sensitive to the bias in the outlier fraction, acquiring a 1$\sigma$ bias only when the outlier fractions are wrong by 0.007.
In our analysis, this is a 60\% relative uncertainty on the $c_{ij}$.
This would more feasible, however the Hubble parameter would then by biased by more than 1$\sigma$.
This is still a very stringent requirement for the knowledge of the $c_{ij}$, which again motivates us to marginalize over them.

\subsubsection{Gaussian and outlier photo-z self-calibration}

Can the Gaussian and outlier photo-z parameters be self-calibrated from the data alone?
We answer this question by examining the photo-z posterior constraints as we vary their priors. 
We also assess the degradation in cosmology, as the photo-z priors are loosened.

\begin{figure}[H]
\centering
\includegraphics[width=0.45\columnwidth]{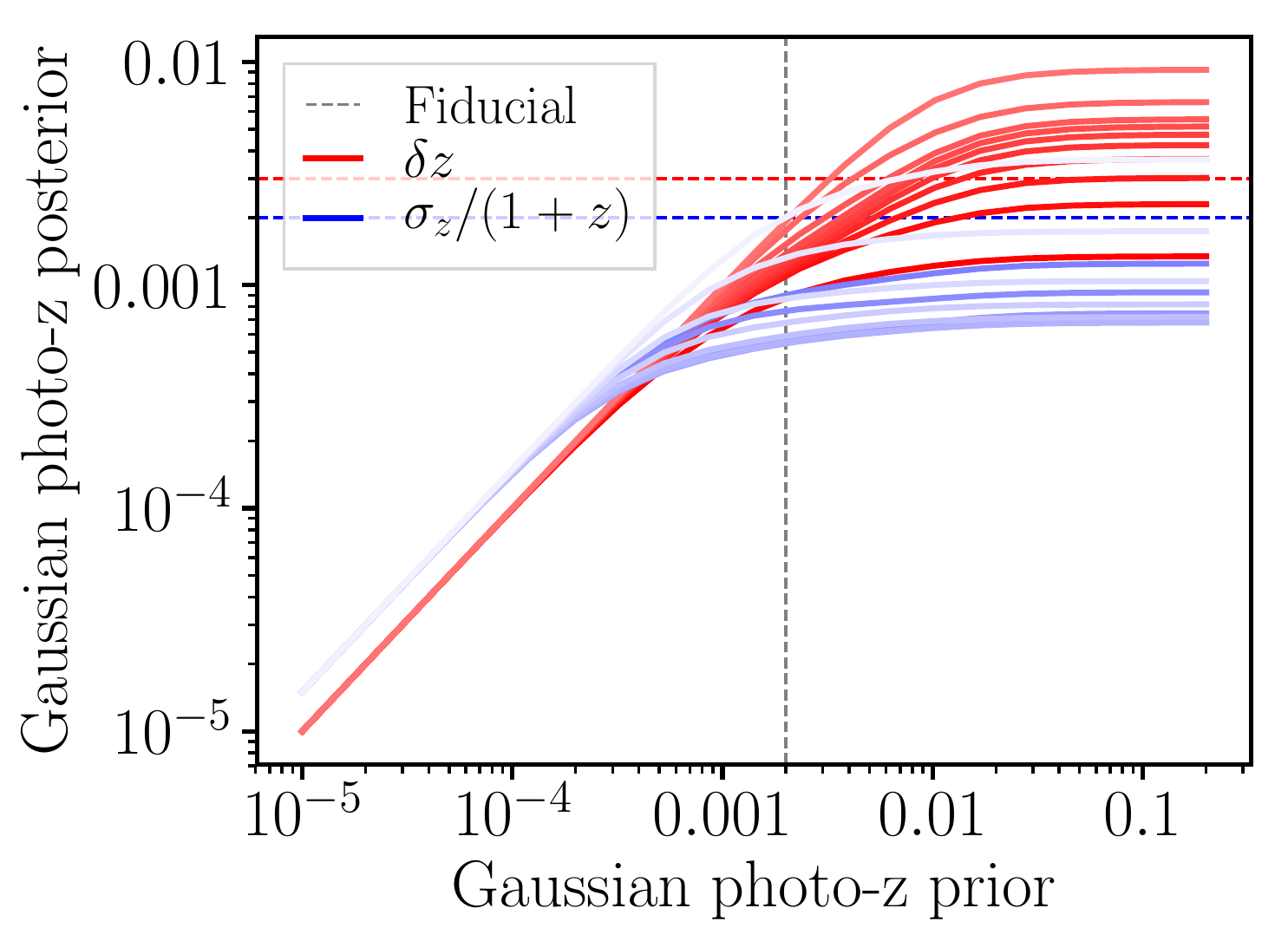}
\includegraphics[width=0.45\columnwidth]{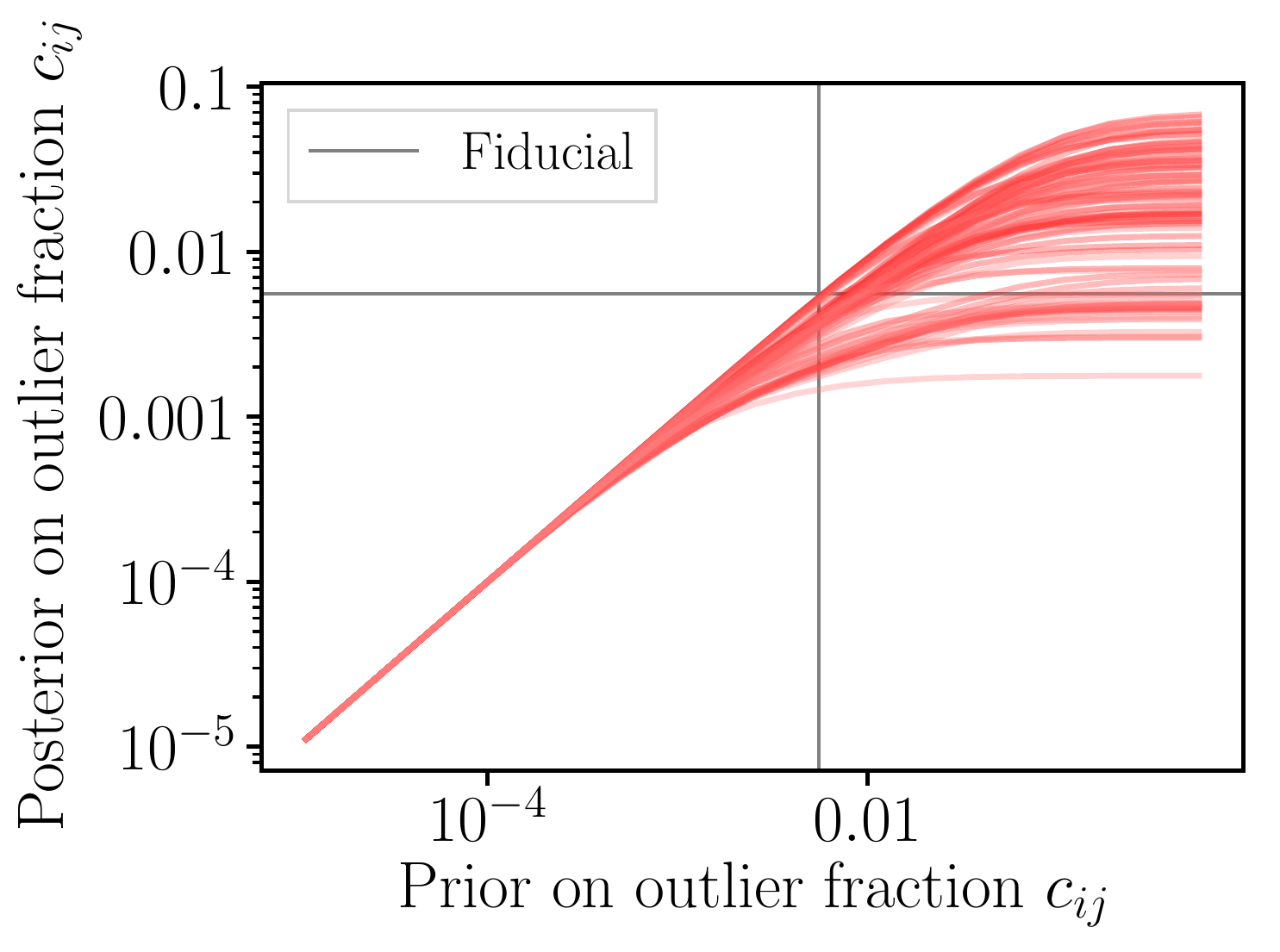}
\caption{
Self-calibration of Gaussian (left) and outlier (right) photo-z parameters.
In both cases, the cosmological parameter set is $w_0w_a$CDM, but the results are almost identical for $\Lambda$CDM + $M_\nu$ and $\Lambda$CDM + curvature.
\textbf{Left:}
The priors on $\delta_{z_i}$ and $\sigma_{z_i}$ are simultaneously varied, with a value equal to the x-axis and 1.5 times the x-axis, respectively.
Here the photo-z outlier fractions are marginalized over, with a fixed (fiducial) prior.
Lighter red lines correspond to higher redshift bins.
For priors larger than a few percent, the degradation in Gaussian width and shift parameters saturates, indicating that the self-calibration regime is reached.
The self-calibration level for the scatter (resp. shift) parameter is comparable to (resp. a factor of a few worse than) the DESC SRD requirements.
\textbf{Right:}
Self-calibration of the photo-z outlier fractions $c_{ij}$.
The Gaussian photo-z nuisance parameters are marginalized over, with a prior equal to the standard requirements (DESC SRD).
For priors of order unity, the degradation saturates, indicating self-calibration.
The self-calibration values for the outlier fractions vary by a large amount depending on the pair of tomographic bins considered.
}
\label{fig:photoz_self_calibration}
\end{figure}

\begin{figure}[H]
\centering
\includegraphics[width=0.45\columnwidth]{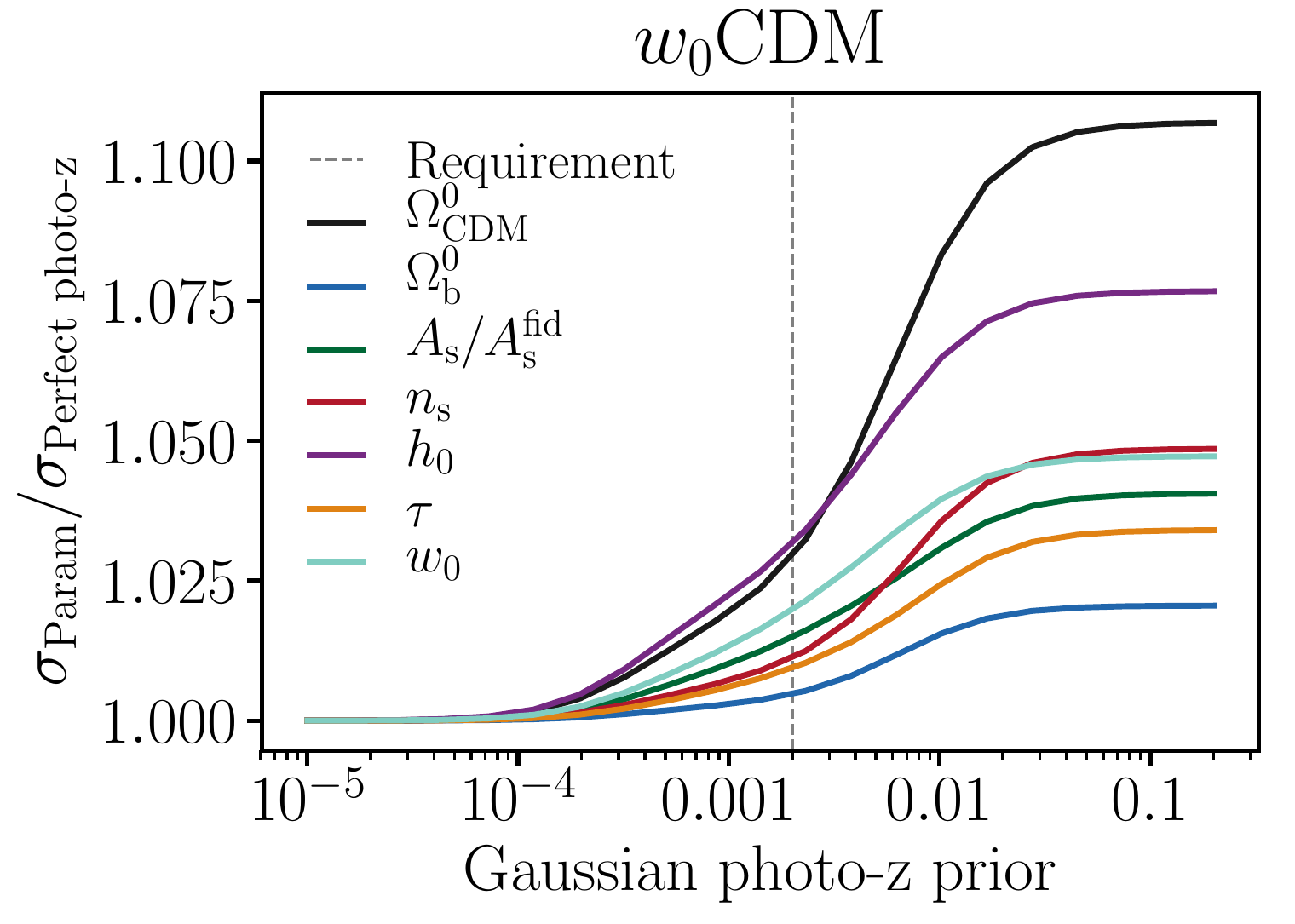}
\includegraphics[width=0.45\columnwidth]{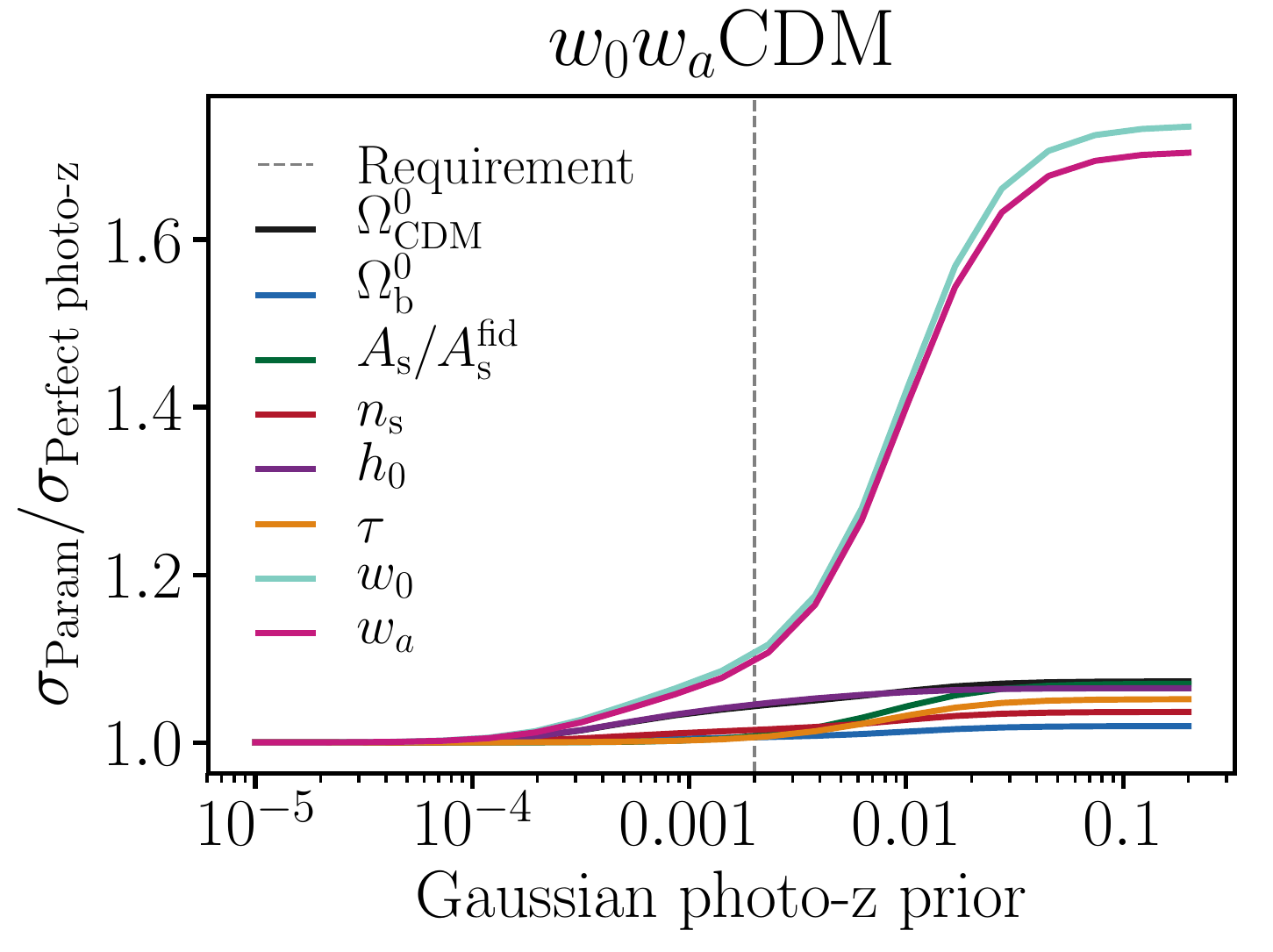}
\includegraphics[width=0.45\columnwidth]{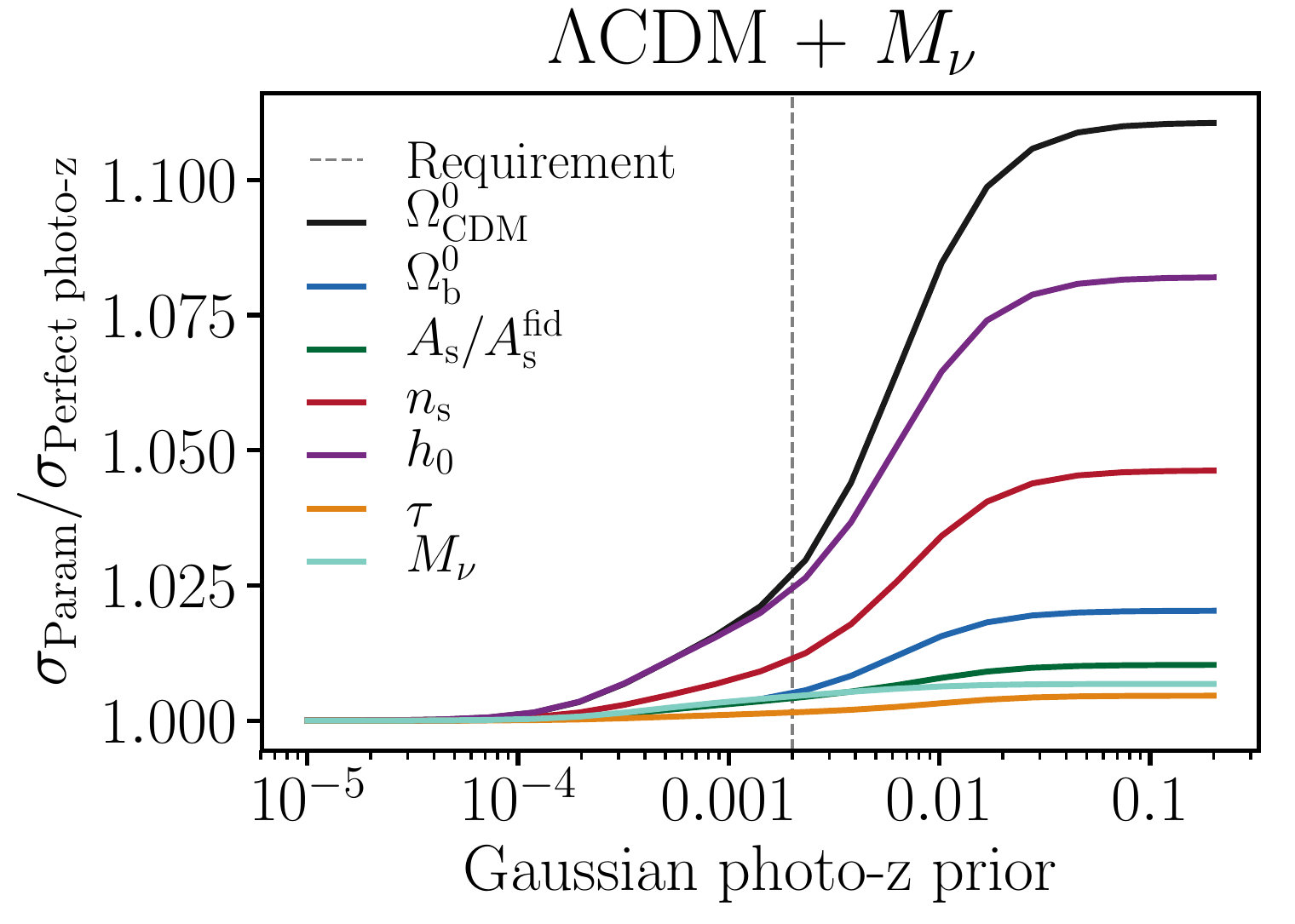}
\includegraphics[width=0.45\columnwidth]{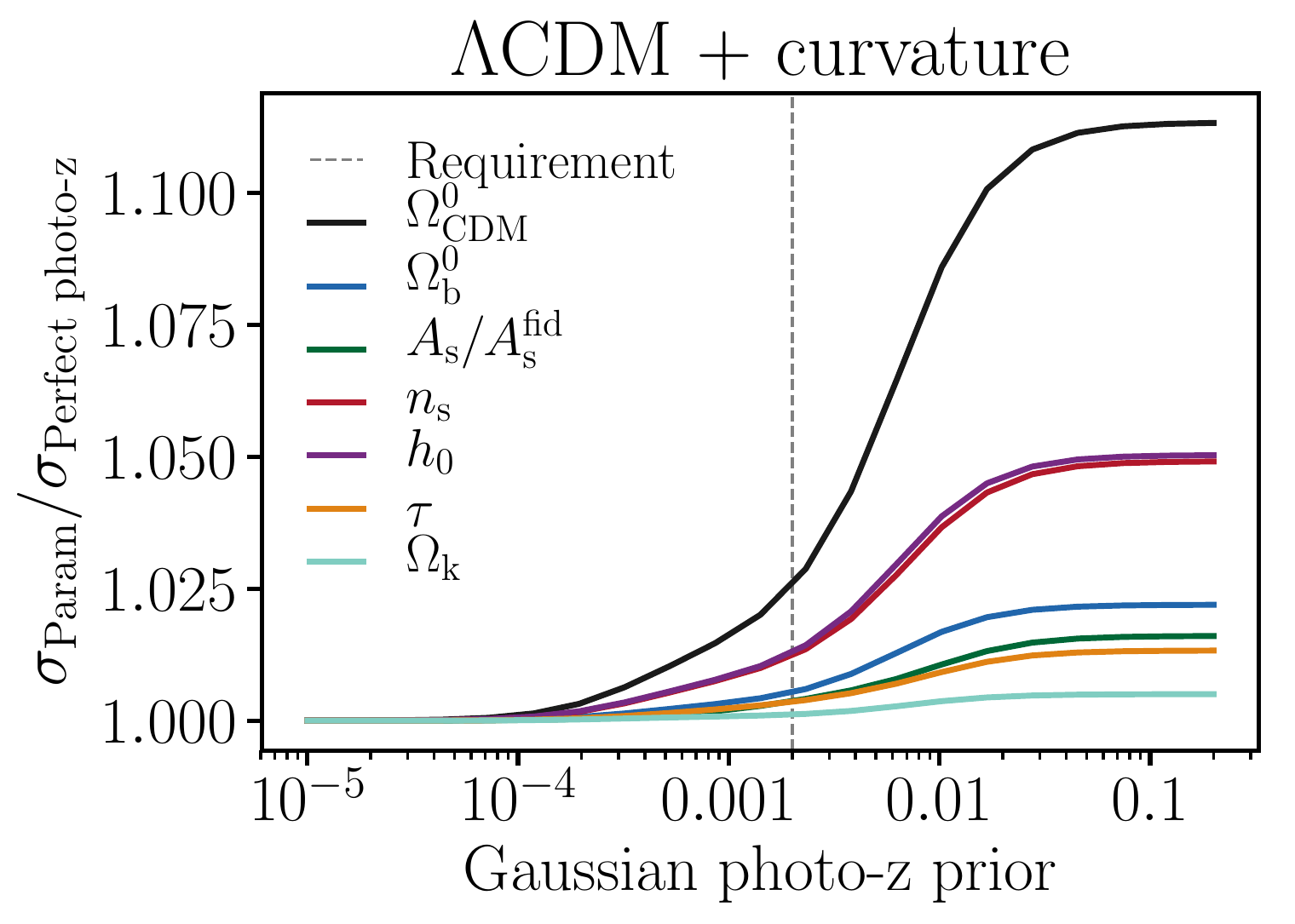}
\caption{
Degradation in cosmological parameters when the Gaussian photo-z priors are varied as in Fig.~\ref{fig:photoz_self_calibration}.
The priors on $\delta_{z_i}$ and $\sigma_{z_i}$ are simultaneously varied, with a value equal to the x-axis and 1.5 times the x-axis, respectively. The photo-z outlier fractions are also marginalized over, with a fixed (fiducial) prior.
The different panels correspond to the various cosmological parameter sets: $w_0$CDM, $w_0w_a$CDM, $\Lambda$CDM + $M_\nu$ and $\Lambda$CDM + curvature.
When the Gaussian photo-z priors are removed, the degradation in cosmology is at most $10\%$ for all cosmological parameters, except when $w_0$ and $w_a$ are varied together ($65\%$ degradation).
}
\label{fig:gphotozreq_deg_cosmo}
\end{figure}

In Fig.~\ref{fig:photoz_self_calibration} (left panel) and Fig.~\ref{fig:gphotozreq_deg_cosmo}, we vary the Gaussian photo-z priors while keeping the priors on the outlier fractions $c_{ij}$ fixed at their fiducial values.
Specifically, we rescale the priors on the Gaussian widths and shifts of all the tomographic bins by the same factor, which we increase progressively.
This way, the priors on widths are always 1.5 times wider than on the shifts, as in the fiducial case.
As the Gaussian photo-z priors are loosened, the corresponding posteriors saturate to a self-calibration level.
This level is mostly independent of the choice of cosmological parameters ($w_0$CDM, $w_0w_a$CDM, $\Lambda$CDM + $M_\nu$ or $\Lambda$CDM + curvature).
For the Gaussian scatter parameters, this level is very similar to the LSST requirements.
For the Gaussian shift parameters, the level of self-calibration is worse than the LSST requirements by a factor of a few.
Considering the cosmological constraints (Fig.~\ref{fig:gphotozreq_deg_cosmo}), completely removing the Gaussian photo-z priors only causes a $10\%$ degradation only for most parameters,
e.g., neutrino masses, curvature, $w_0$.
However, if both $w_0$ and $w_a$ are varied, then the degradation is $65\%$.
This is again consistent with our finding in the previous subsection.
In summary, the self-calibration of Gaussian photo-z parameters is possible, at a minimum cost for most cosmological parameters, except for the dark energy equation of state parameters $w_0, w_a$.

\begin{figure}[H]
\centering
\includegraphics[width=0.45\columnwidth]{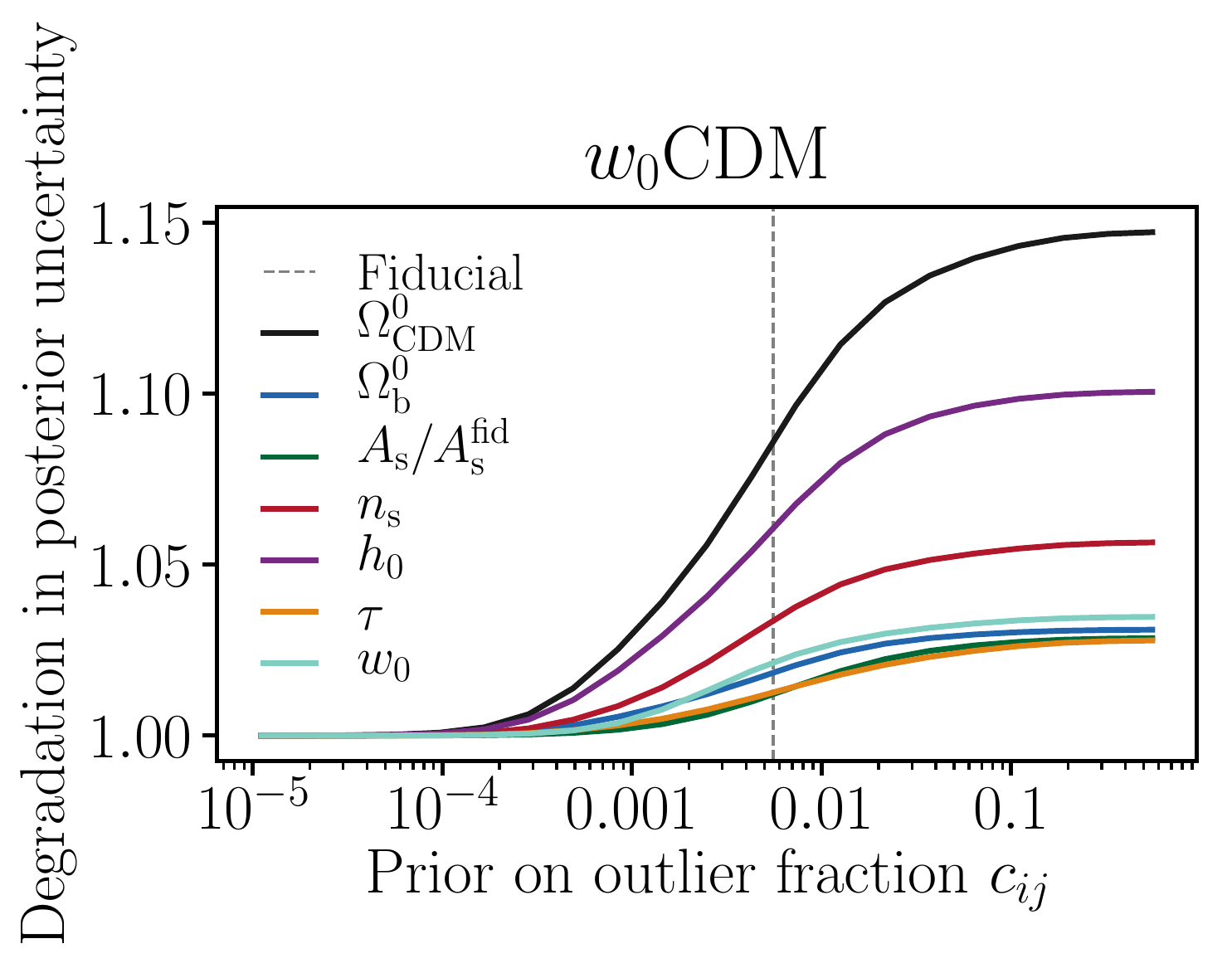}
\includegraphics[width=0.45\columnwidth]{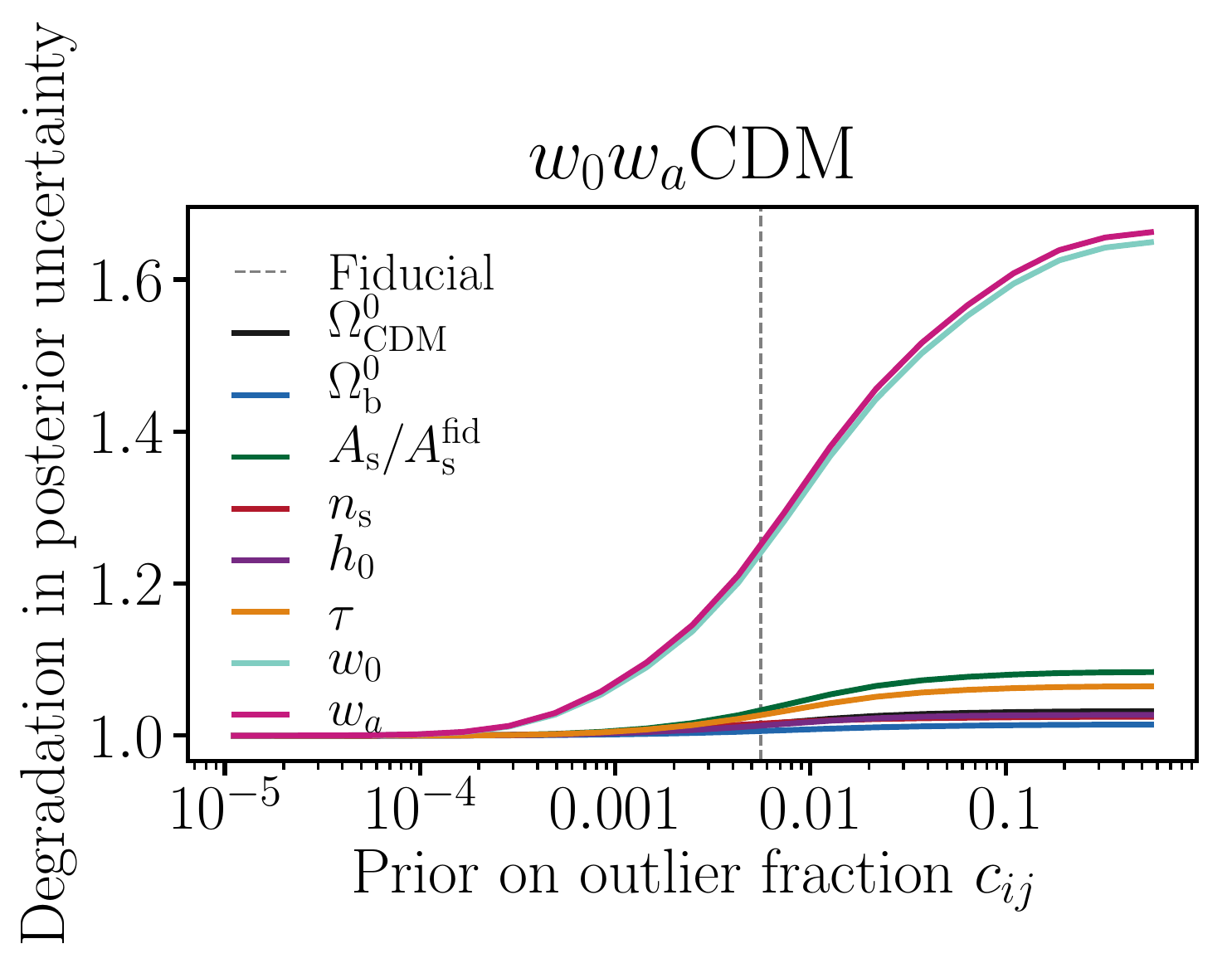}
\includegraphics[width=0.45\columnwidth]{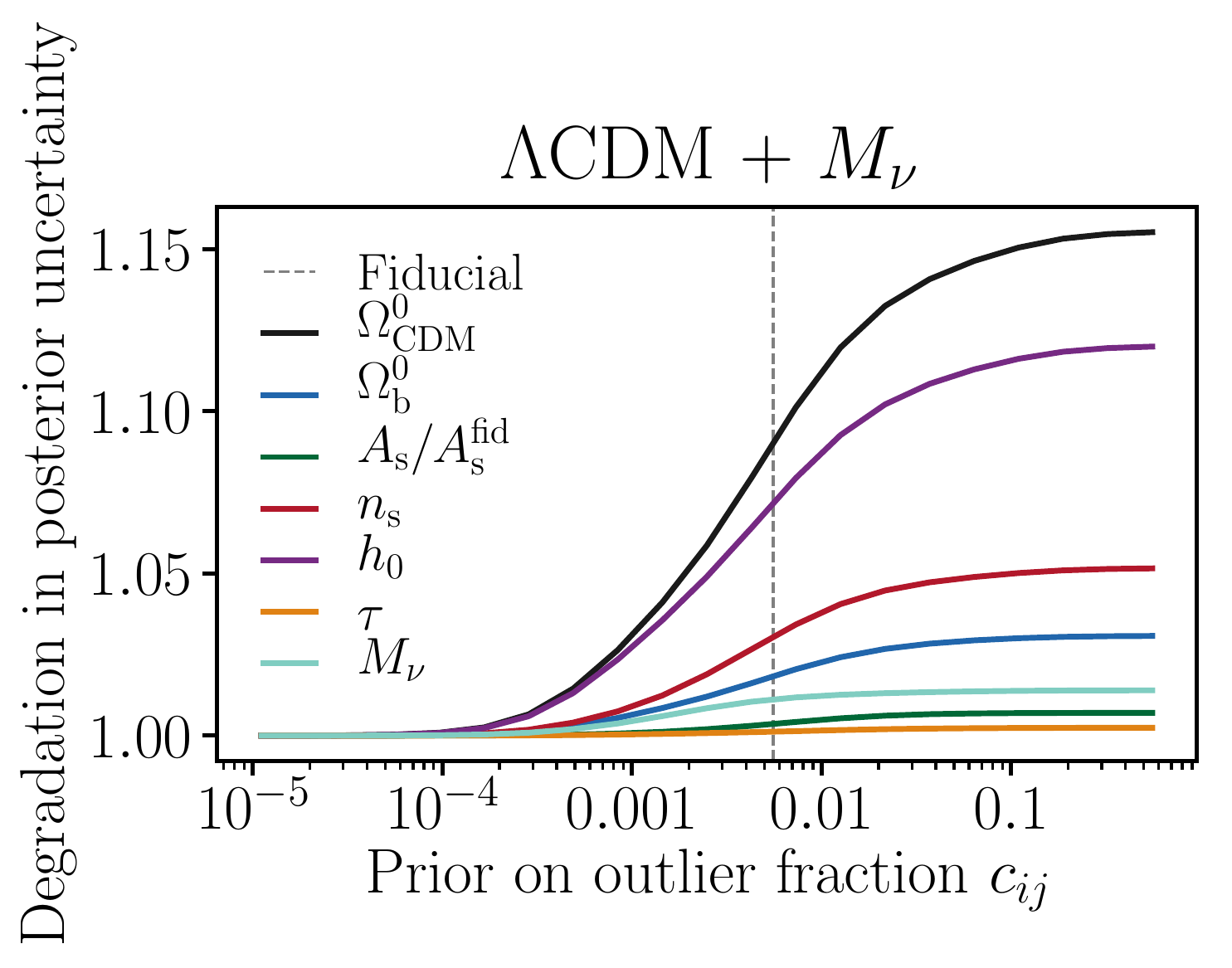}
\includegraphics[width=0.45\columnwidth]{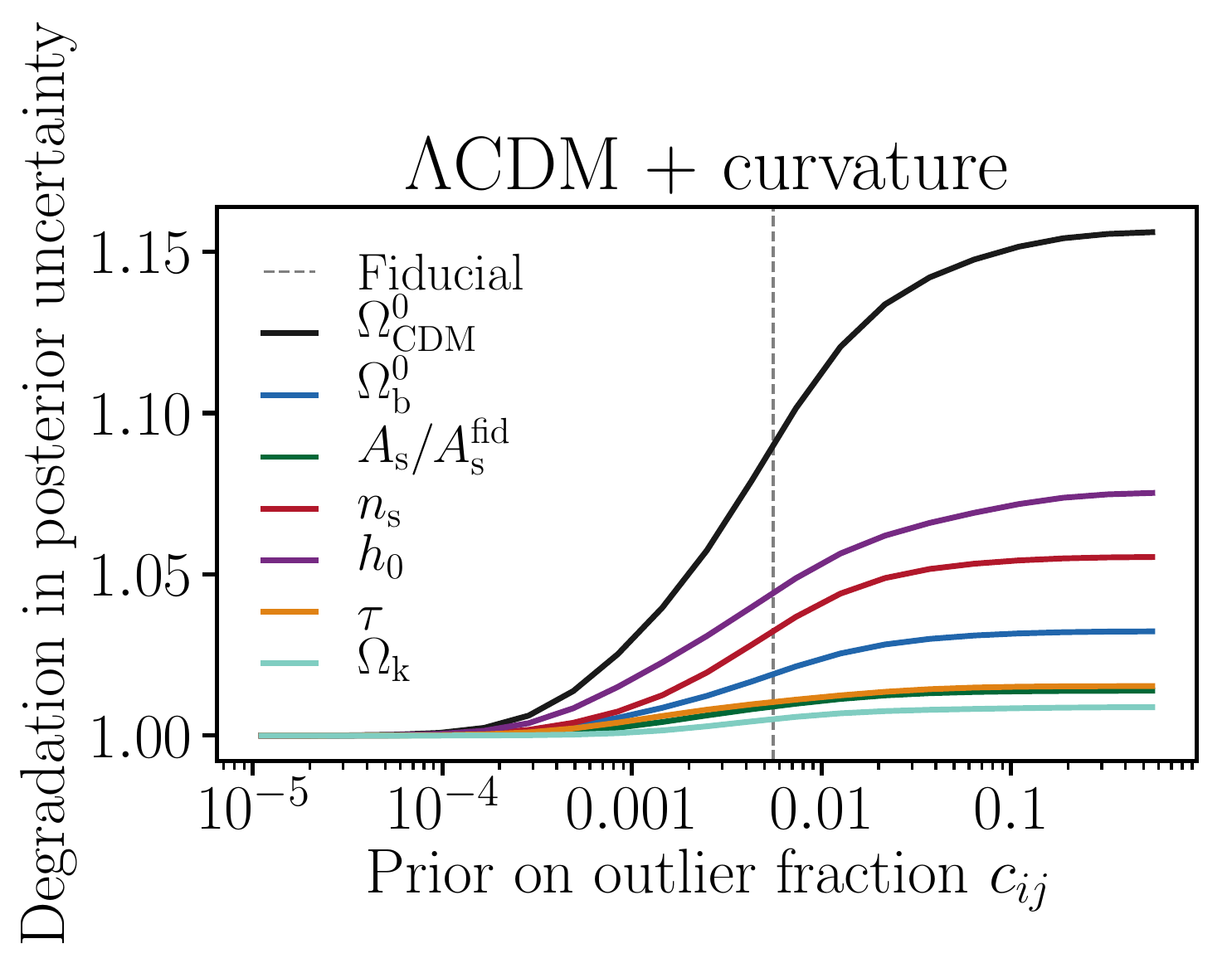}
\caption{
Same as Fig.~\ref{fig:gphotozreq_deg_cosmo}, but varying the outlier priors rather than the Gaussian photo-z priors.
Again, when the outlier priors are removed, the degradation in cosmology is small for all cosmological parameters, except when $w_0$ and $w_a$ are varied together ($65\%$ degradation).
}
\label{fig:outlierphotozreq_deg_cosmo}
\end{figure}

Fig.~\ref{fig:photoz_self_calibration} (right panel) and Fig.~\ref{fig:outlierphotozreq_deg_cosmo}
instead show the self-calibration of photo-z outliers.
Here the Gaussian photo-z parameters are still marginalized over, with their fiducial prior.
As the prior on the outlier fractions $c_{ij}$ are simultaneously loosened by a multiplicative factor, 
their posteriors saturate to a self-calibration level.
This level is again mostly independent of the choice of cosmological parameters ($w_0$CDM, $w_0w_a$CDM, $\Lambda$CDM + $M_\nu$ or $\Lambda$CDM + curvature).
Depending on the pair of tomographic bins considered, the self-calibration level can be a few times better than the fiducial prior, or up to ten times worse.
In the absence of prior on the photo-z outliers, the constraints on $w_0$ and $w_a$ are degraded by $65\%$, but the neutrino mass uncertainties are worsened by only $3\%$ and curvature by only 1\%.

\subsubsection{Importance of the null cross-correlations}

In the absence of photo-z uncertainties (and magnification bias), many cross-correlations included in the analysis would be null.
This is the case of the clustering cross-correlations and the galaxy - galaxy lensing when the source bin is at lower redshift than the lens bin.
With Gaussian photo-z uncertainties, this changes: the clustering cross-correlations of adjacent bins $g_i g_{i+1}$ become detectable, as well as some of the inverted galaxy - galaxy lensing correlations.
With our fiducial level of photo-z outliers, all the clustering cross-correlations $g_i g_j$ become detectable.
This suggests that these null cross-correlations contain useful information about photo-z errors: the adjacent clustering correlations tell us about Gaussian photo-z errors, and the distant correlations about photo-z outliers.
To quantify this, we compare the posterior constraints on photo-z and cosmology with and without the null cross-correlations.

\begin{figure}[H]
\includegraphics[width=0.49\columnwidth]{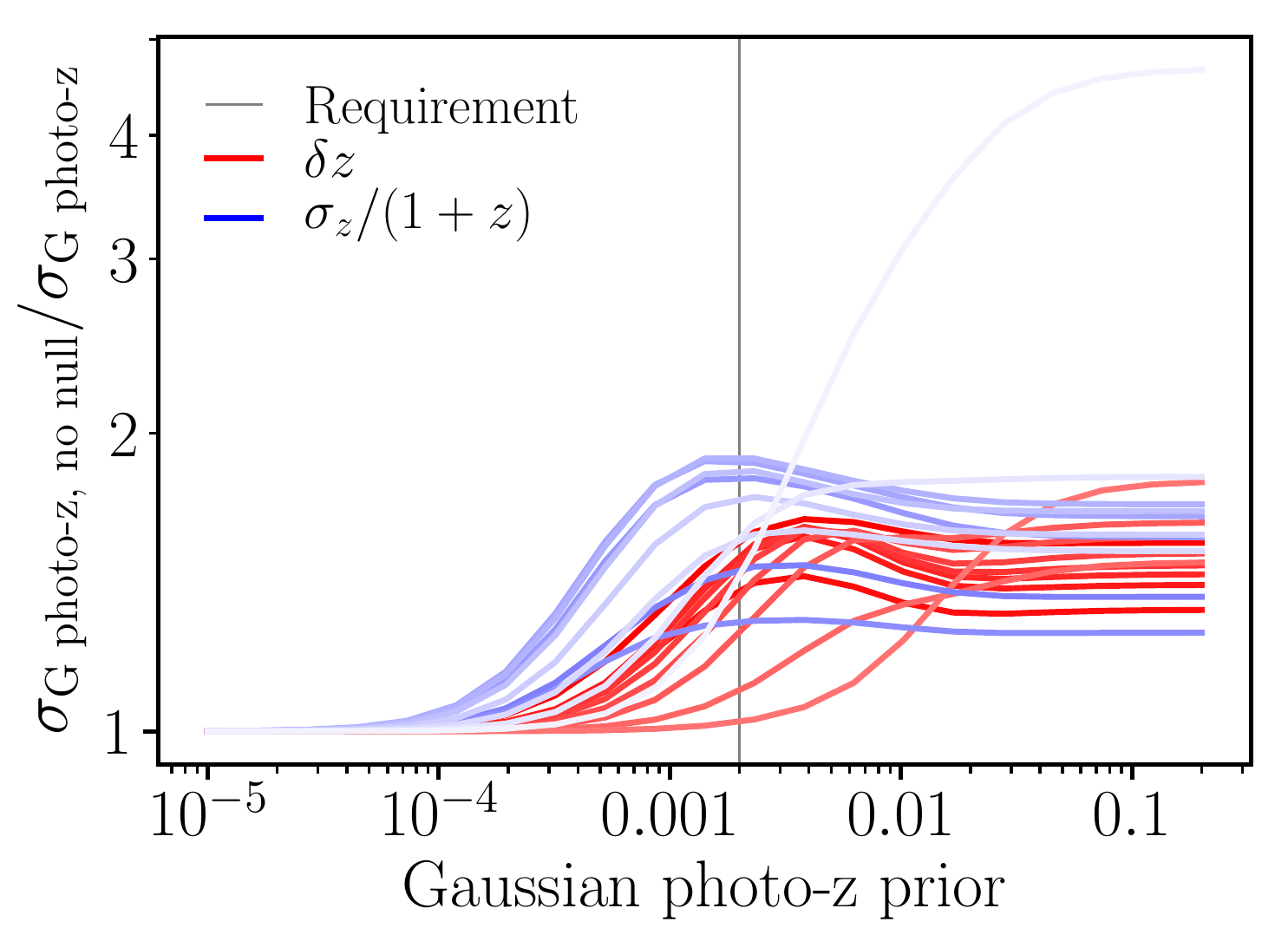}
\includegraphics[width=0.49\columnwidth]{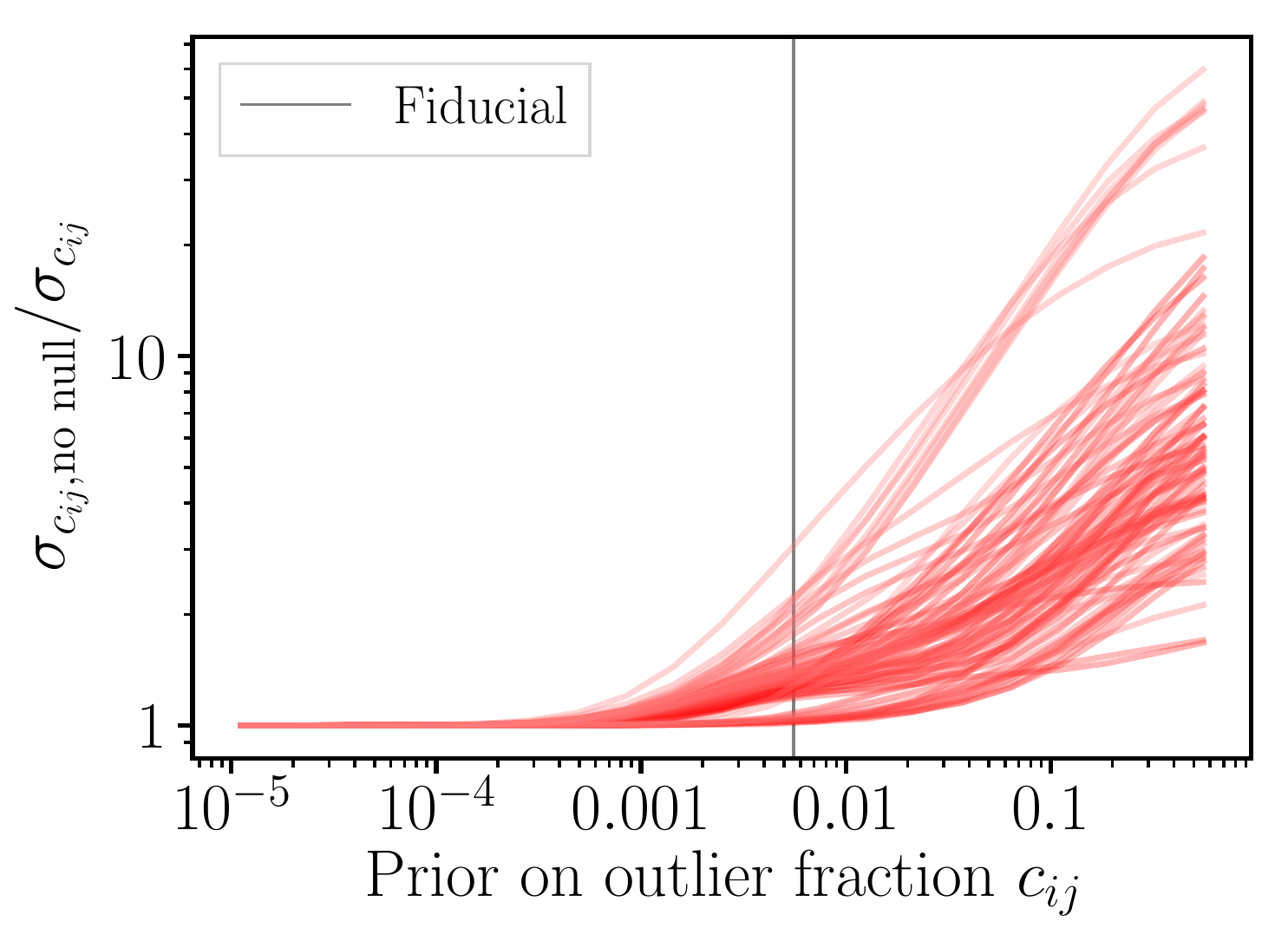}
\caption{
Discarding the null correlations causes a large degradation in Gaussian photo-z (left) and photo-z outlier (right) uncertainties.
\textbf{Left:} 
The priors on $\delta z_i$ and $\sigma_{z_i}$ are simultaneously varied, with a value equal to the x-axis and 1.5 times the x-axis, respectively. Here the photo-z outlier fractions are marginalized over, with a fixed (fiducial) prior.
In the absence of Gaussian photo-z priors, the degradation is about a factor of 2.
The vertical grey line indicates the standard photo-z requirement from the DESC SRD.
\textbf{Right:}
The Gaussian photo-z nuisance parameters are marginalized over, with a prior equal to the standard requirements (DESC SRD), while the priors on outlier fractions are varied.
In the absence of outlier priors, ignoring the null cross-correlations can degrade the photo-z posteriors by more than an order of magnitude.
}
\label{fig:photozreq_vs_nonull}
\end{figure}

Fig.~\ref{fig:photozreq_vs_nonull} shows the large degradation in Gaussian photo-z errors (factor 2) and outlier uncertainties (more than an order of magnitude) when the null correlations are discarded, in the absence of photo-z priors.
This degradation is still non-negligible when the fiducial photo-z priors are used.

\begin{figure}[H]
\centering
\includegraphics[width=0.45\columnwidth]{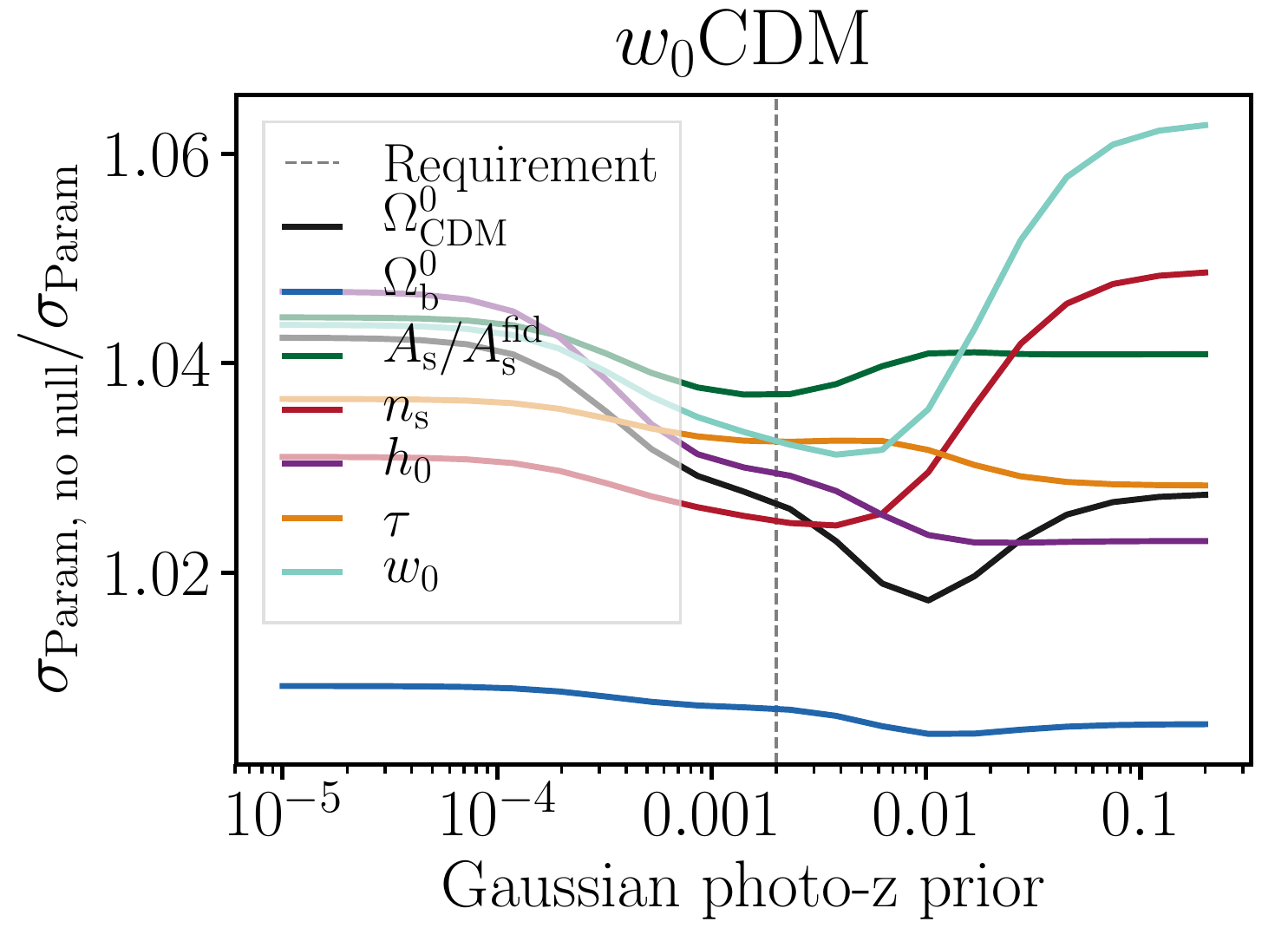}
\includegraphics[width=0.45\columnwidth]{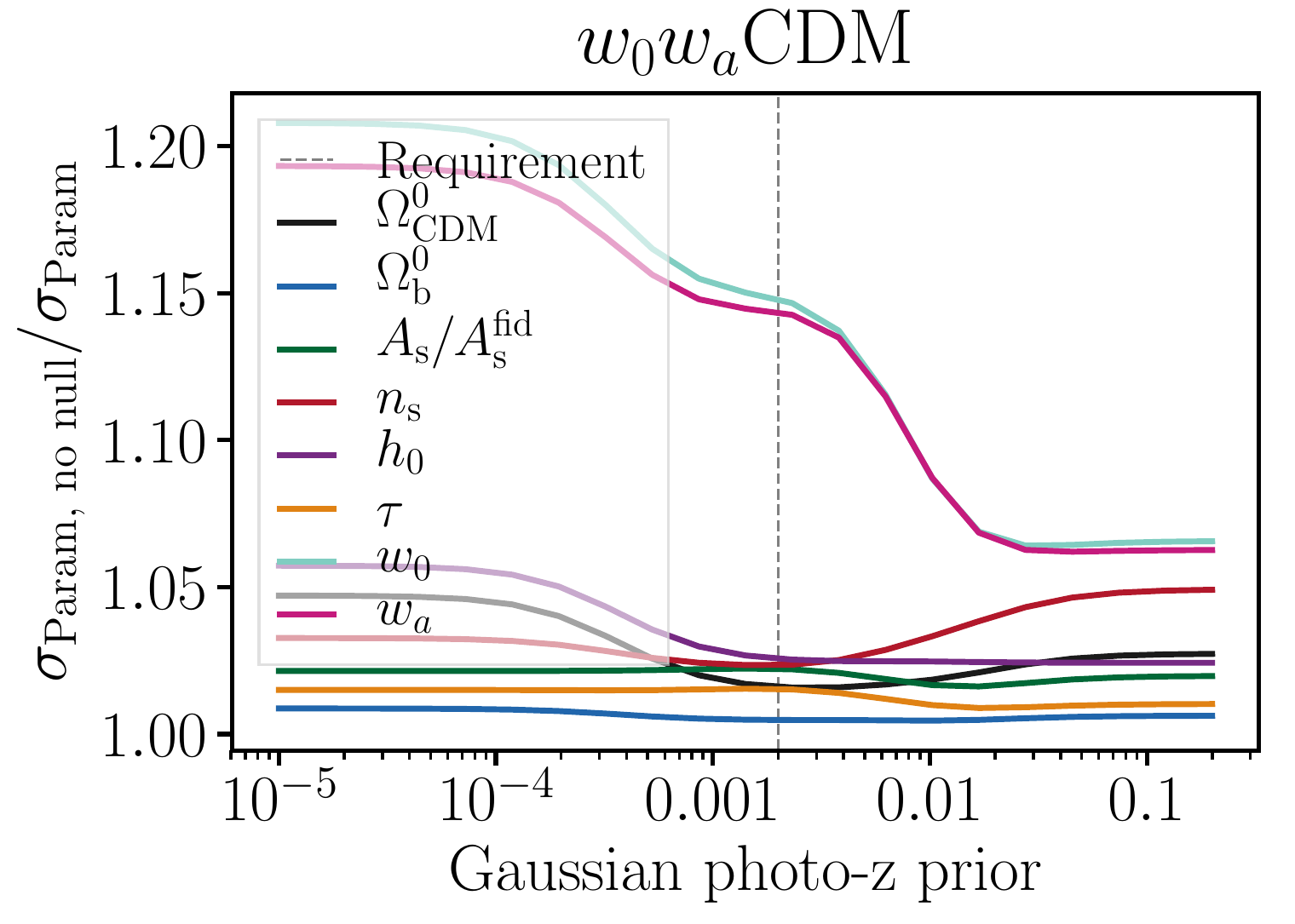}
\includegraphics[width=0.45\columnwidth]{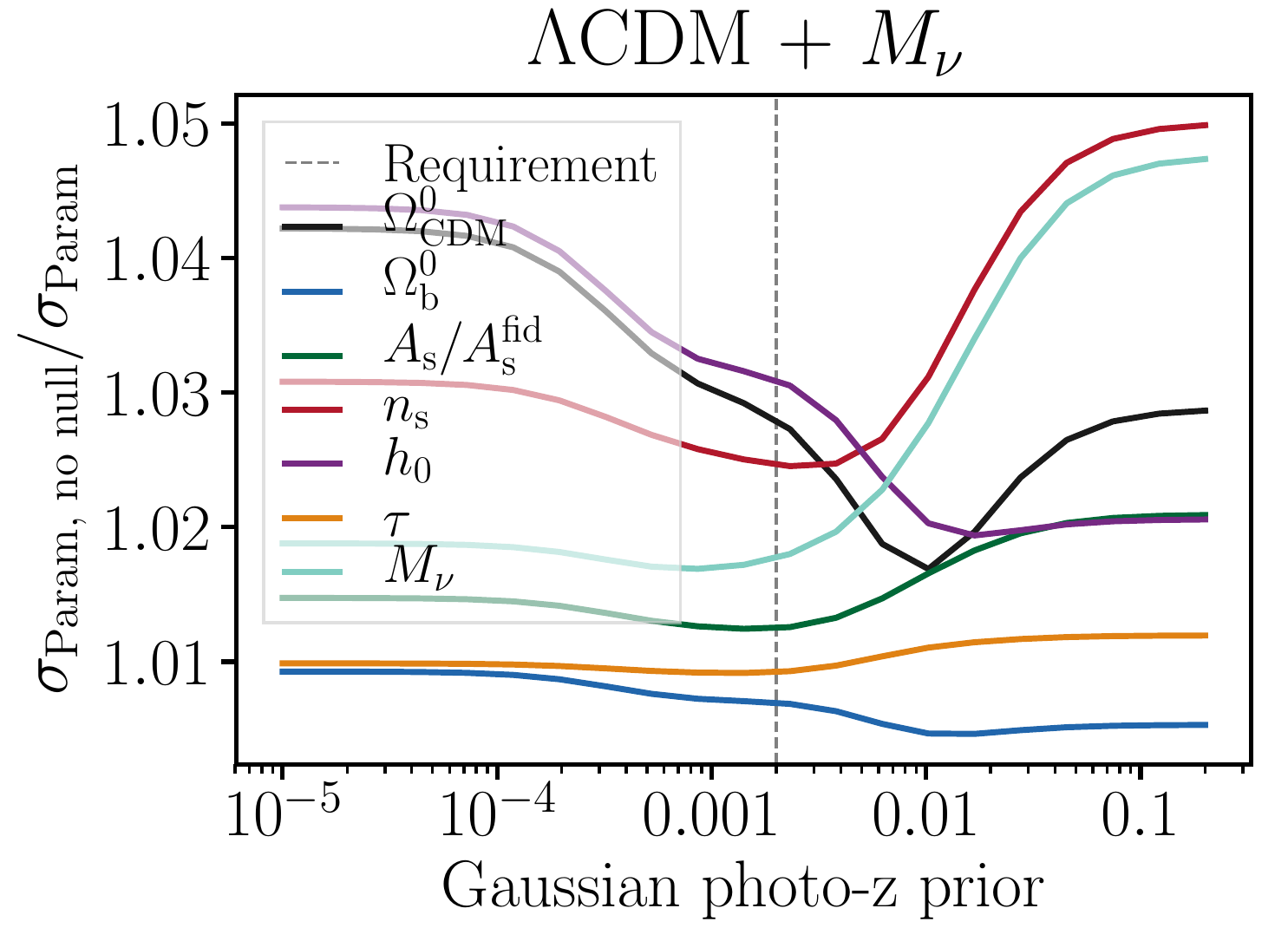}
\includegraphics[width=0.45\columnwidth]{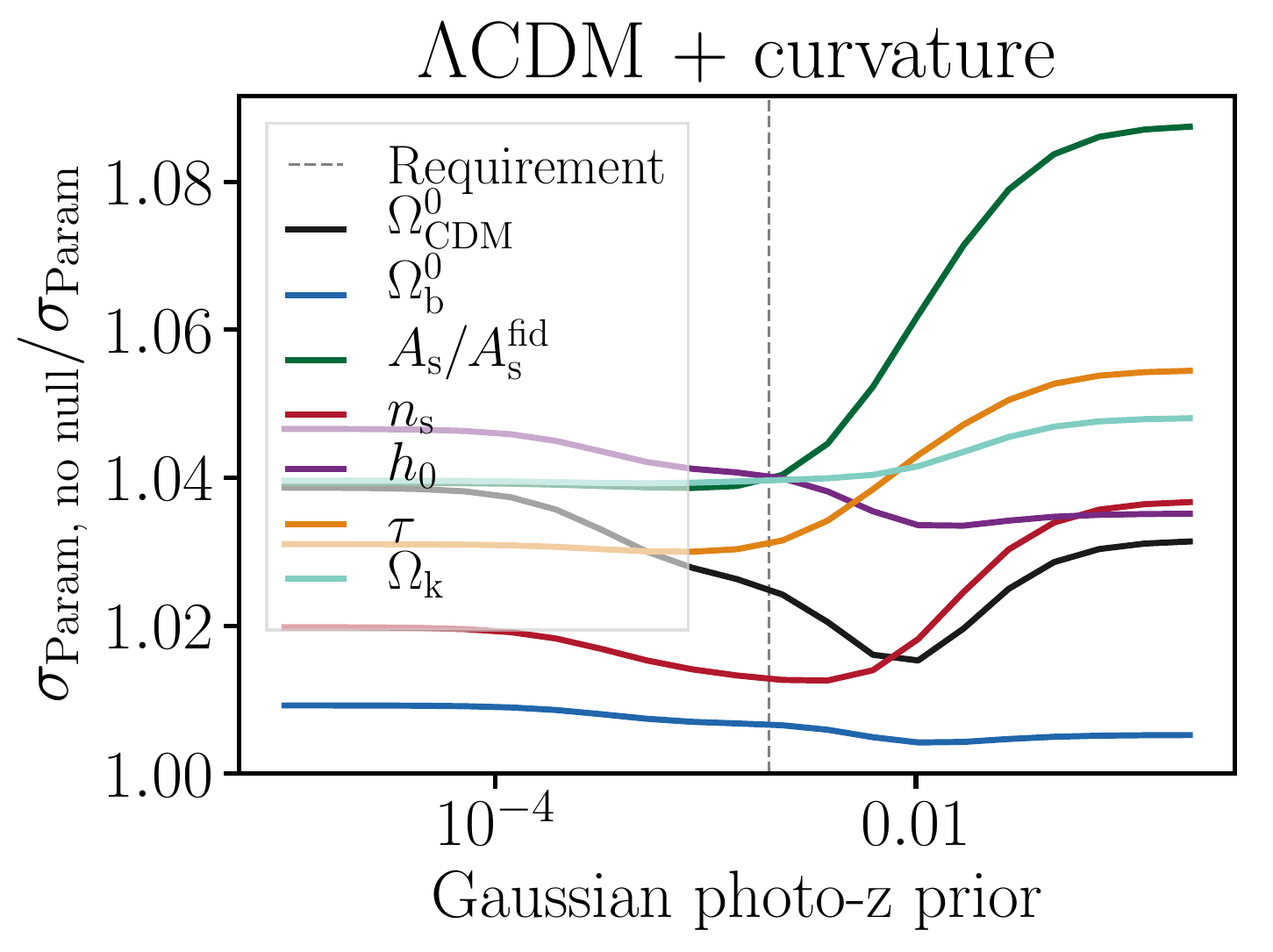}
\caption{
With the fiducial Gaussian prior (vertical dashed line), 
discarding the null correlations causes a minimal degradation in most cosmological parameters (e.g., $2\%$ degradation only in $M_\nu$), but a significant degradation in the dark energy equation of state parameters (15\%).
This is further motivation to include these cross-correlations.
Because this degradation persists with perfect Gaussian photo-z priors, it is likely due to the added redshift resolution from the cross-correlations, whose effective redshifts are in-between the ten tomographic bins.
The priors on $\delta z_i$ and $\sigma z_i$ are simultaneously varied, with a value equal to the x-axis and 1.5 times the x-axis, respectively. The photo-z outlier fractions are also marginalized over, with a fixed (fiducial) prior.
}
\label{fig:gphotozreq_cosmo_vs_nonull}
\end{figure}

\begin{figure}[H]
\centering
\includegraphics[width=0.45\columnwidth]{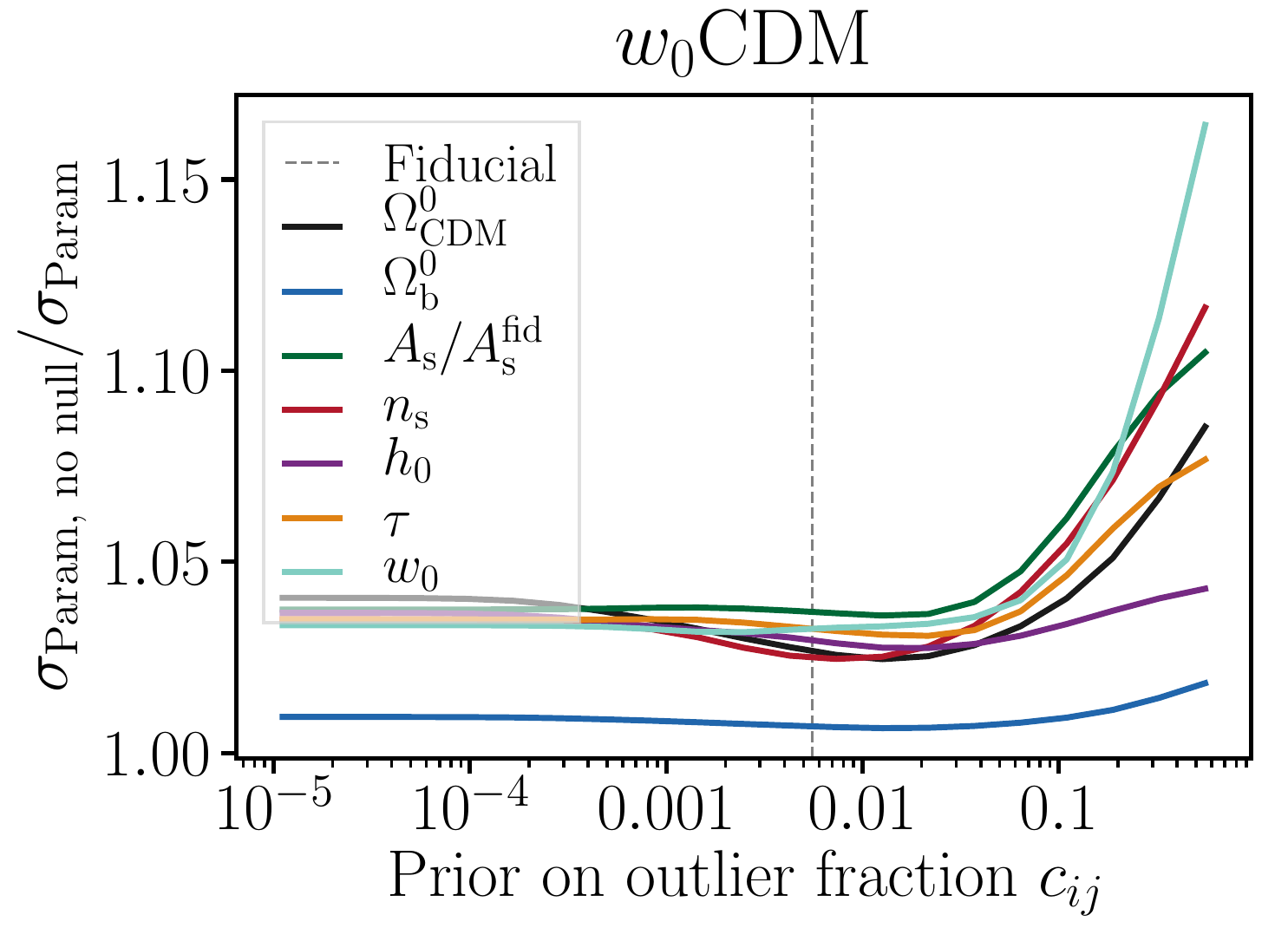}
\includegraphics[width=0.45\columnwidth]{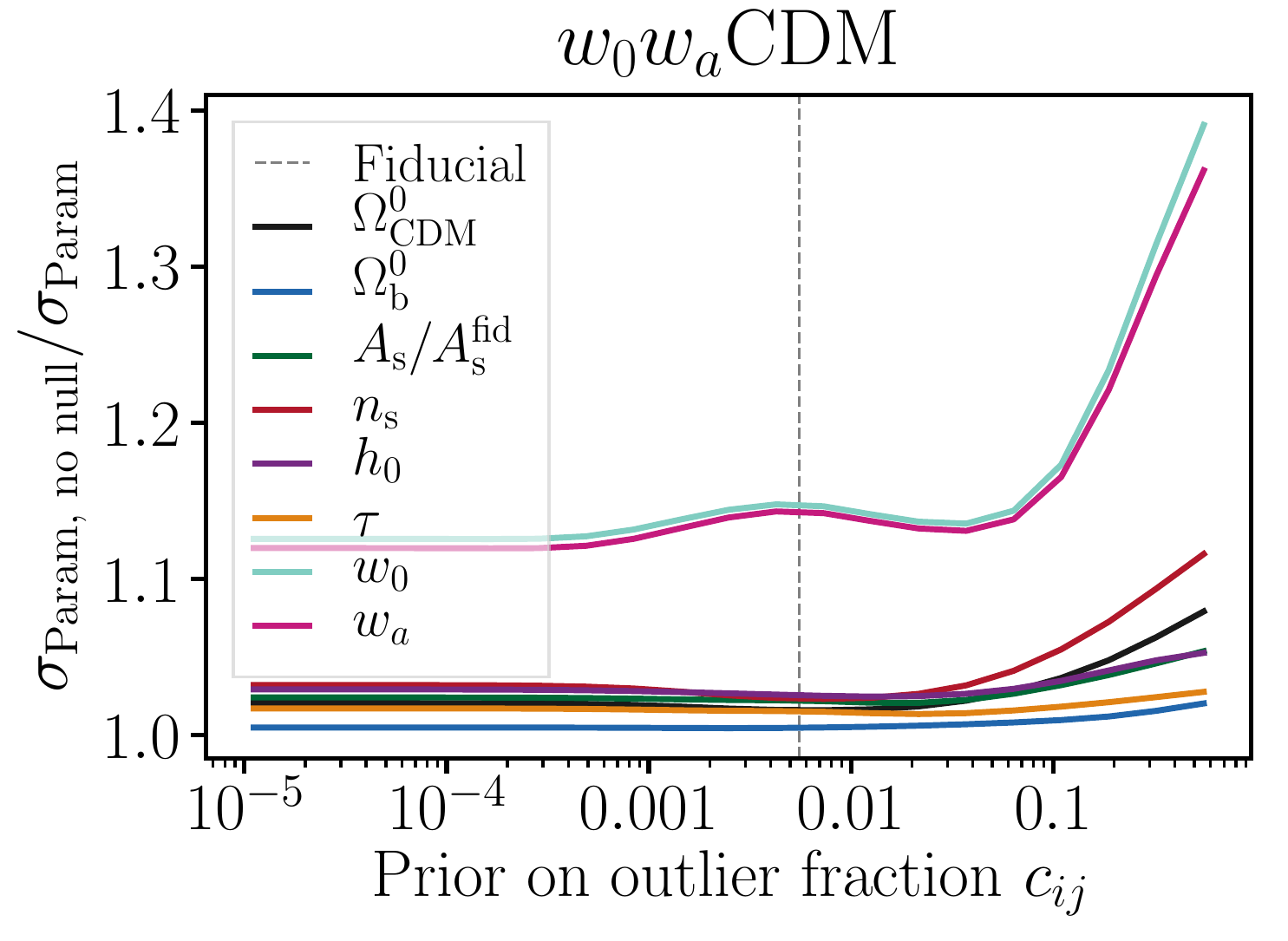}
\includegraphics[width=0.45\columnwidth]{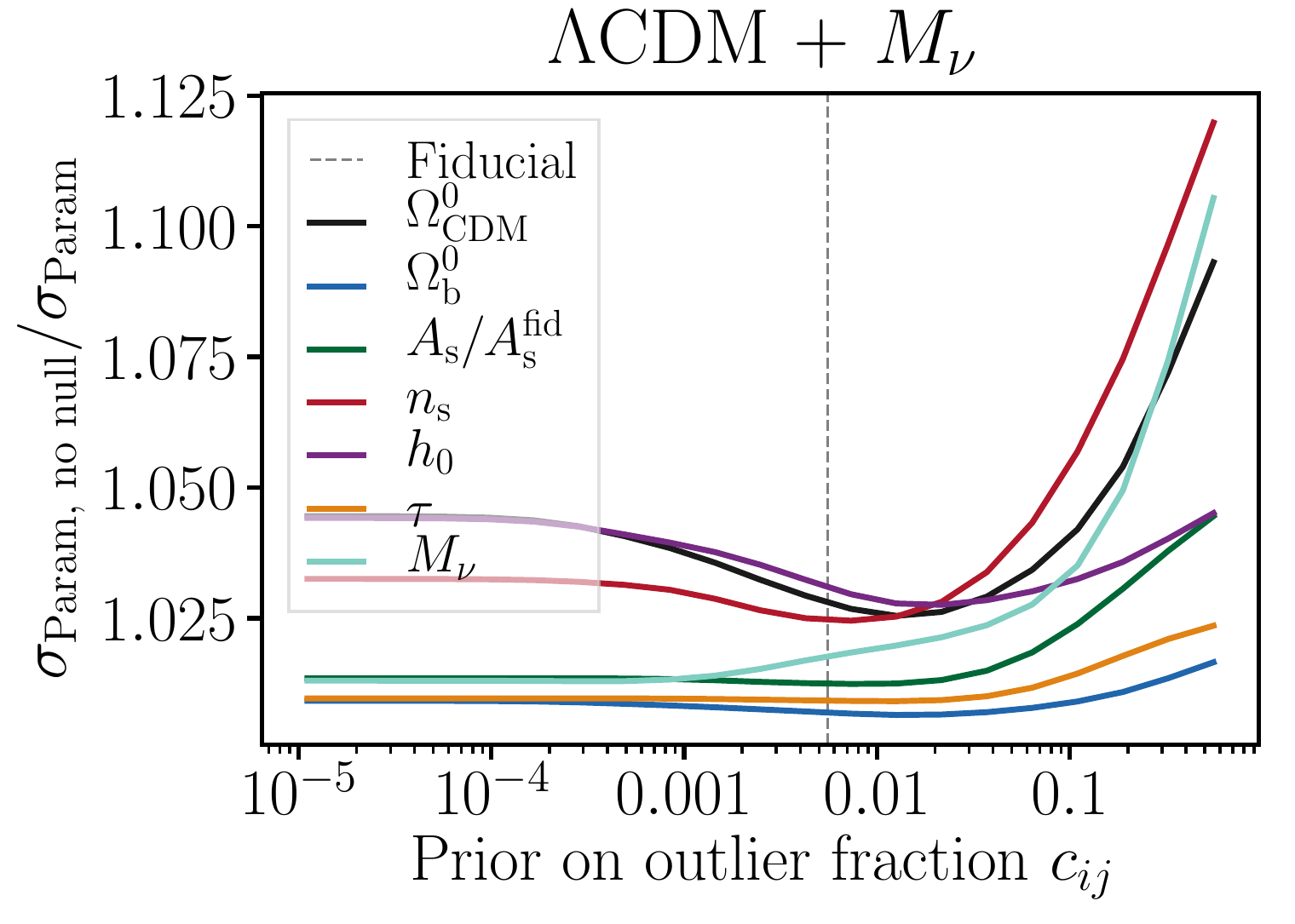}
\includegraphics[width=0.45\columnwidth]{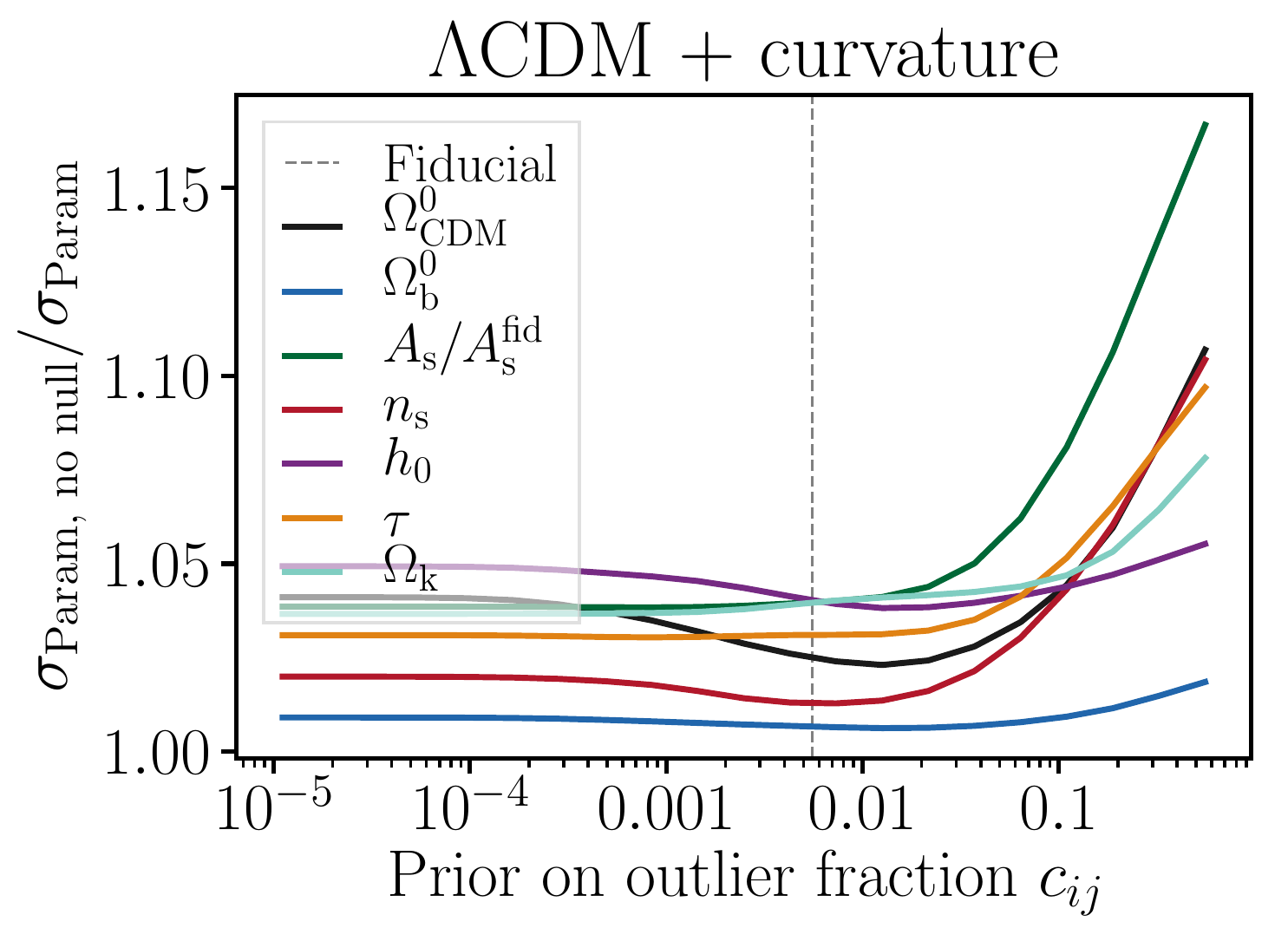}
\caption{
Same as Fig.~\ref{fig:gphotozreq_cosmo_vs_nonull}, but varying the photo-z outlier priors.
The conclusion is similar: discarding the null correlations causes a minimal degradation in most cosmological parameters (e.g., $2\%$ degradation only in $M_\nu$), but a 15\% degradation in the dark energy equation of state parameters $w_0, w_a$, when using the fiducial prior.
In the absence of outlier prior, the null correlations are even more important.
}
\label{fig:outlierphotozreq_cosmo_vs_nonull}
\end{figure}

Fig.~\ref{fig:gphotozreq_cosmo_vs_nonull} and \ref{fig:outlierphotozreq_cosmo_vs_nonull} show the corresponding degradation in cosmology when the null correlations are discarded, as a function of the Gaussian and outlier priors.
The large degradation in photo-z only causes a small degradation in cosmology (a few percent degradation only in $M_\nu$ and curvature), except for the dark energy equation of state parameters, when they are both varied.
The constraints on $w_0$ and $w_a$ are degraded by 15\%, a non-negligible amount.

\begin{figure}[H]
\includegraphics[width=0.45\columnwidth]{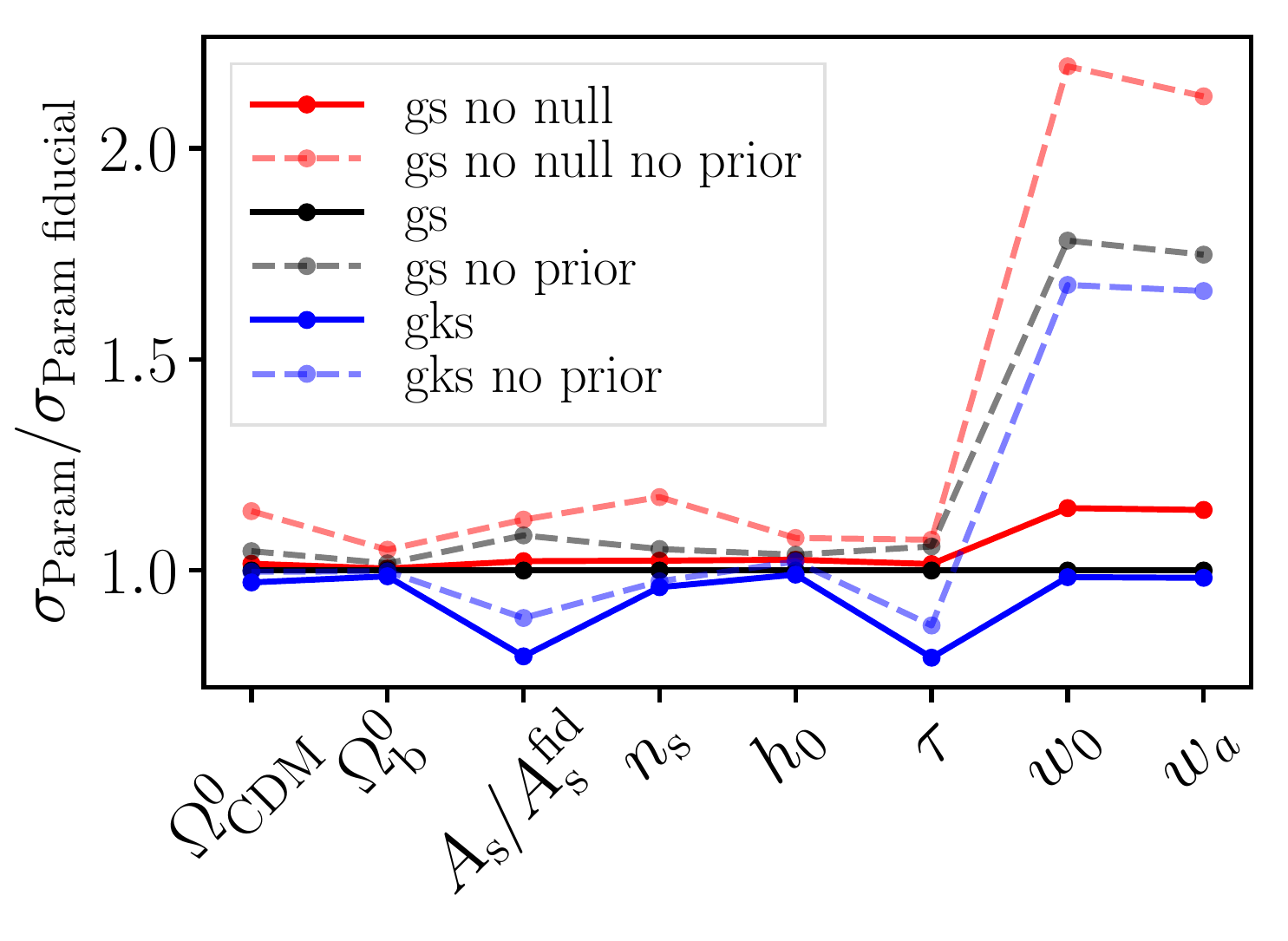}
\includegraphics[width=0.45\columnwidth]{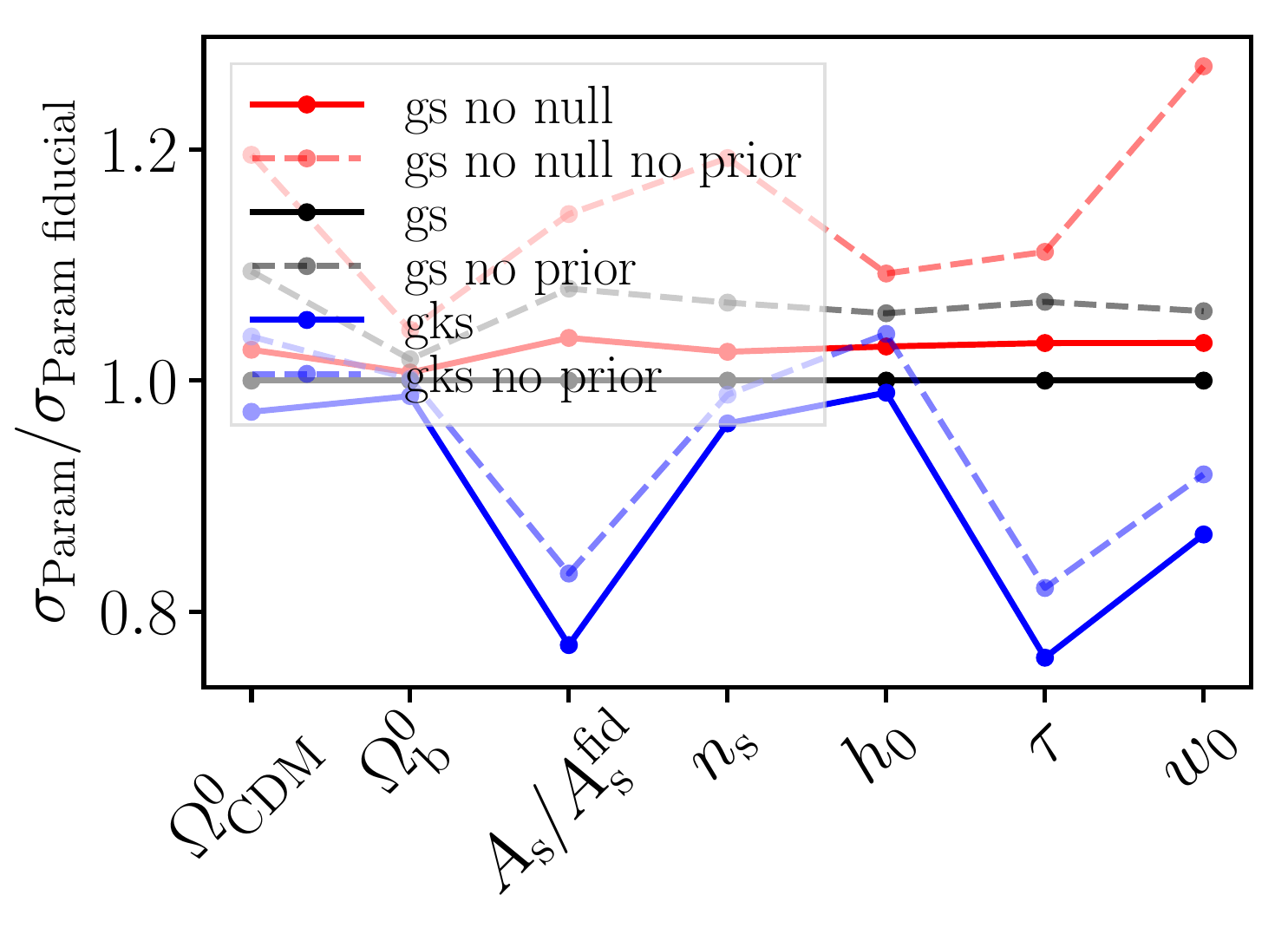}
\caption{
Comparison of the $w_0w_a$CDM cosmological constraints from the various datasets considered in this analysis.
Black lines correspond to the fiducial dataset, including galaxy number density and shear (``gs'').
The red lines discard the null correlations from this dataset (``gs no null''), whereas the blue lines add CMB lensing to the fiducial dataset (``gks'').
\textbf{Left:} The null cross-correlations in the $3\times 2$ analysis, which only contribute $10\%$ of the total signal-to-noise ratio, and would be null in the absence of photo-z errors, actually produce a dramatic improvement in the dark energy equation of state parameters $w_0$ and $w_a$.
In the absence of null correlations, removing the photo-z priors also leads to a much larger degradation in the dark energy equation of state.
\textbf{Right:}
If $w_a$ is fixed and only $w_0$ is varied, then the null correlations no longer bring such a large improvement on $w_0$. This shows that the null correlations really help measuring $w_a$ and breaking its degeneracy with $w_0$.
}
\label{fig:comparison_cosmo_nonull}
\end{figure}

Interestingly, this degradation in $w_0$ and $w_a$ remains or is even enhanced for tighter photo-z priors.
This suggests that the null correlations do not simply help $w_0$ and $w_a$ by improving the photo-z errors, but instead by providing additional redshift resolution in the tomographic analysis.
Indeed, $w_a$ determines the subtle evolution of $w$ across the LSST tomographic bins, and the null cross-correlations provide measurements of growth and expansion at intermediate effective redshifts, in-between those of the ten tomographic bins.
Indeed, Fig.~\ref{fig:comparison_cosmo_nonull} shows that the null correlations are no longer as crucial if $w_0$ is varied with $w_a$ fixed (right panel). 
This suggests that the null correlations really help pin down $w_a$, and break its degeneracy with $w_0$, thus reducing the marginalized uncertainty on both.

\subsubsection{Benefits of using the same galaxy sample for sources and lenses}

\begin{figure}[H]
\centering
\includegraphics[width=0.45\columnwidth]{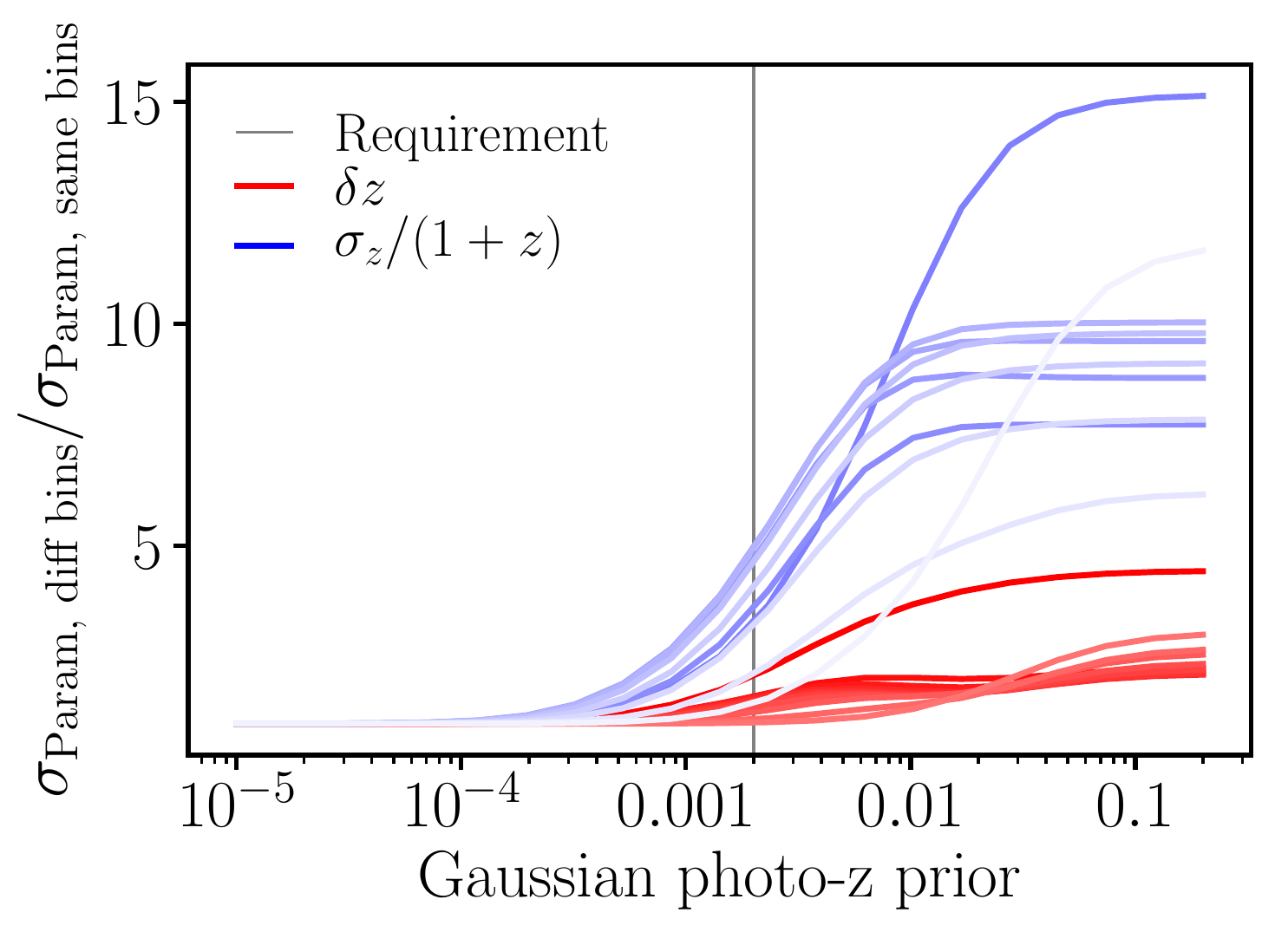}
\includegraphics[width=0.45\columnwidth]{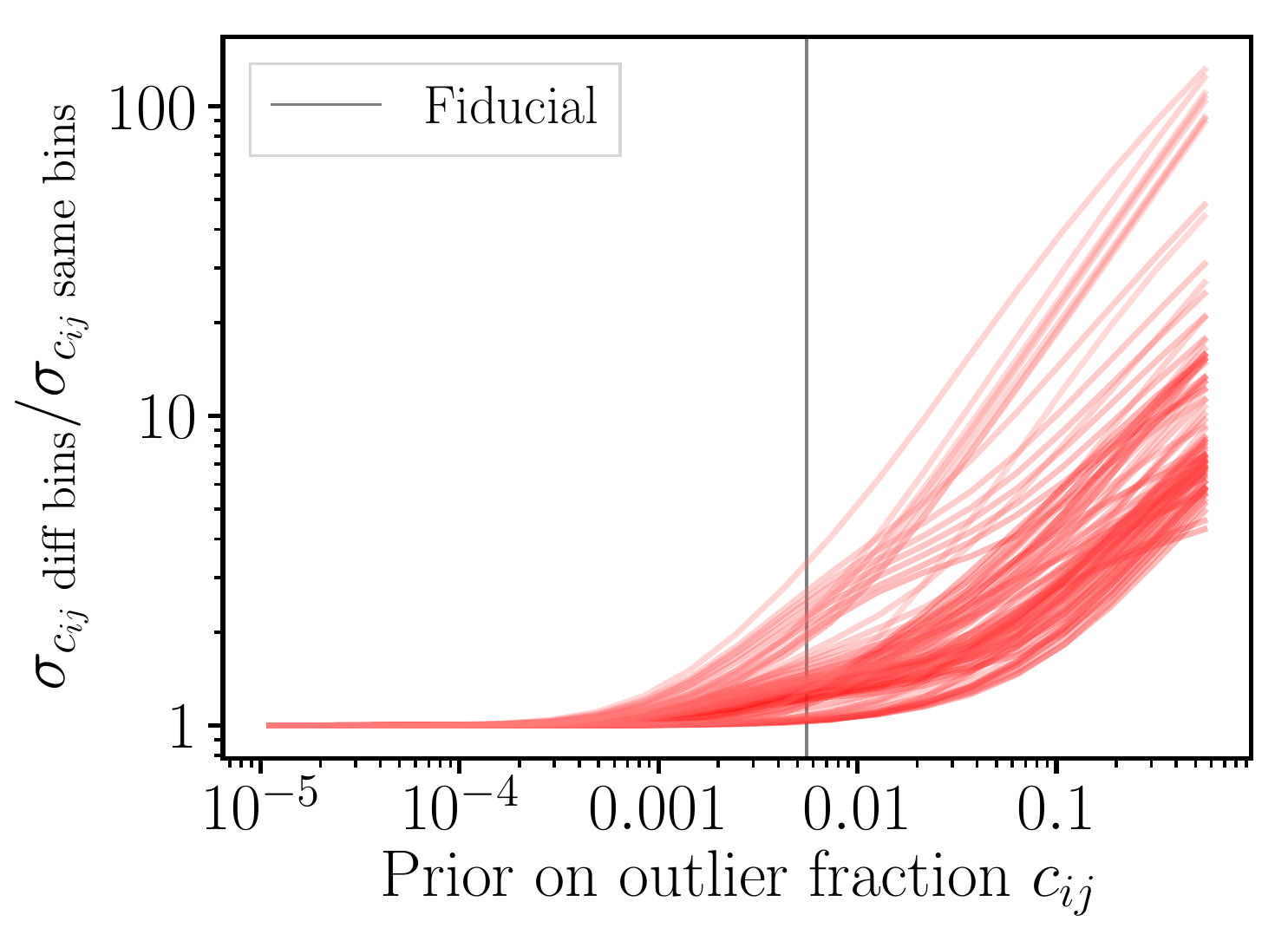}
\caption{
Comparing the self-calibration of Gaussian photo-z (left) and photo-z outliers (right) when the source and lens galaxies are the same or different. In the absence of photo-z priors, using different bins results in a large degradation in the photo-z posteriors.
}
\label{fig:photoz_vs_diffgs}
\end{figure}

In this analysis, we used the same galaxy sample as lenses and sources.
We would like to know whether this was a crucial choice or not.
To answer this question, we compare the fiducial forecast with one where the tomographic bins for lenses and sources have the same redshift distributions as the fiducial case, but have independent photo-z nuisance parameters.
In practice, this multiplies by two the number of photo-z nuisance parameters, from 110 to 220.
Furthermore, the clustering cross-correlations, which are helpful in constraining photo-z errors, now only constrain the lens photo-z, and contain no information about the source photo-z.
One would thus naively expect a large degradation from having different samples as lenses and sources.
As expected, Fig.~\ref{fig:photoz_vs_diffgs} shows this very large degradation in Gaussian and photo-z outlier posteriors.

\begin{figure}[H]
\centering
\includegraphics[width=0.45\columnwidth]{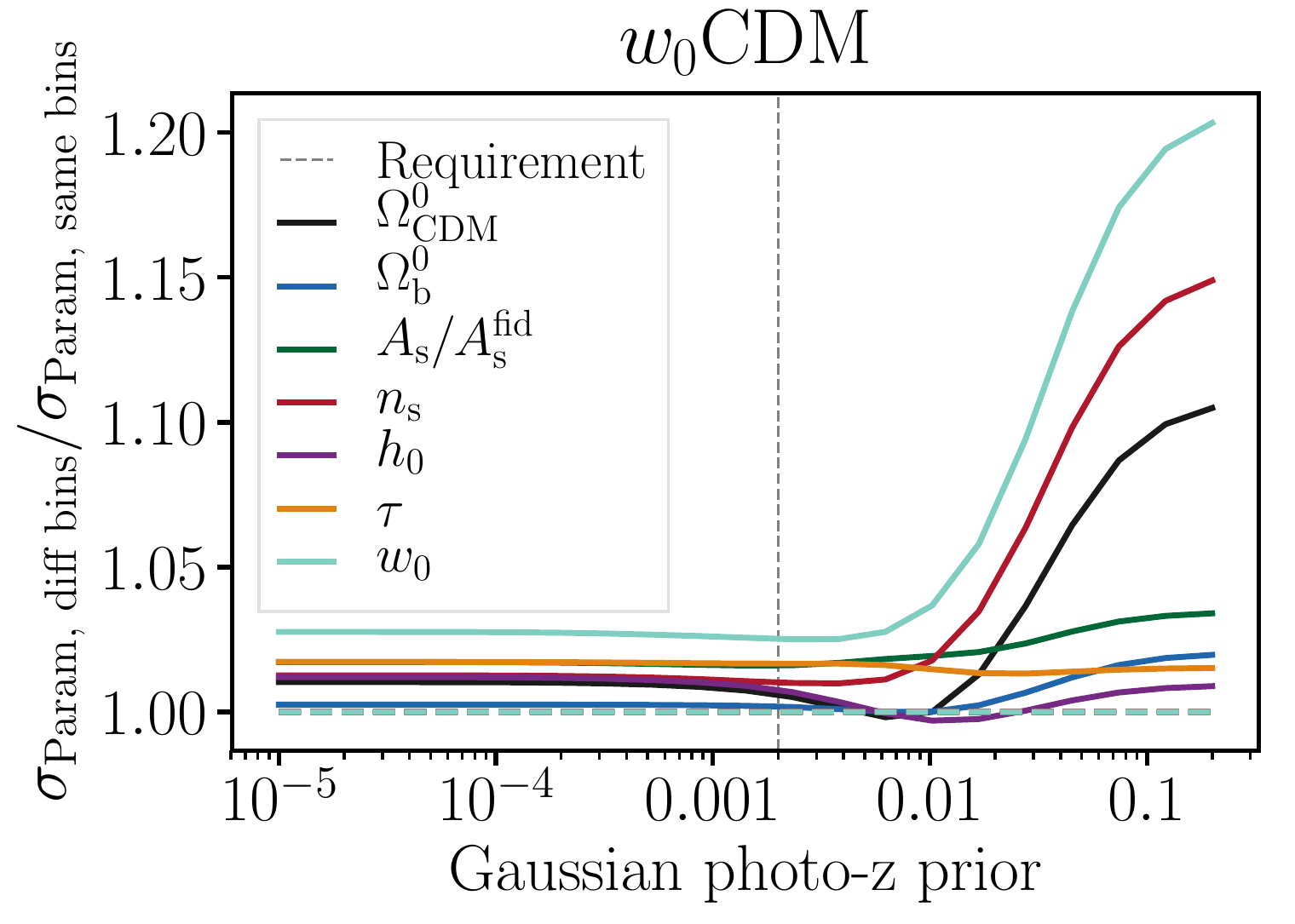}
\includegraphics[width=0.45\columnwidth]{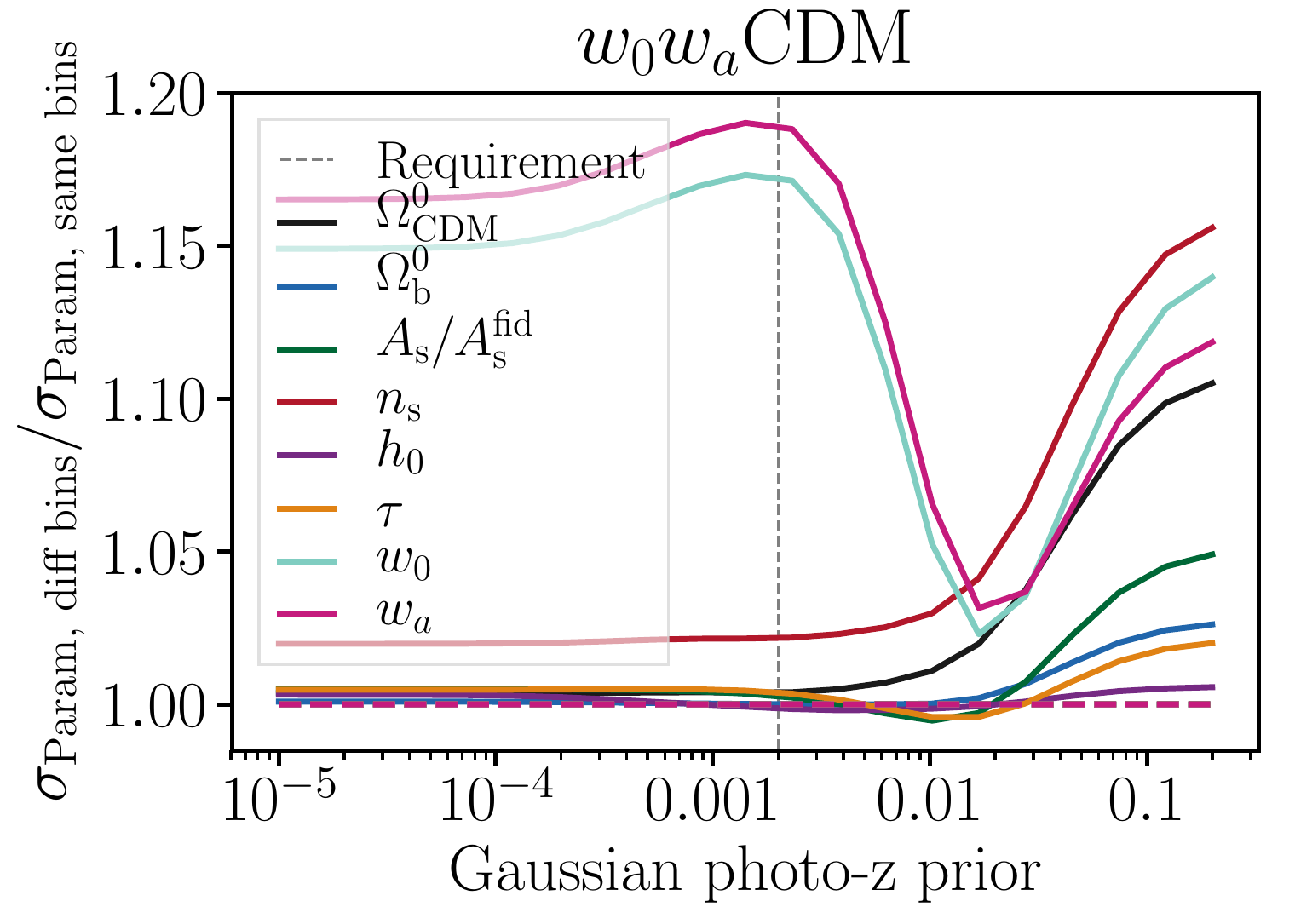}
\includegraphics[width=0.45\columnwidth]{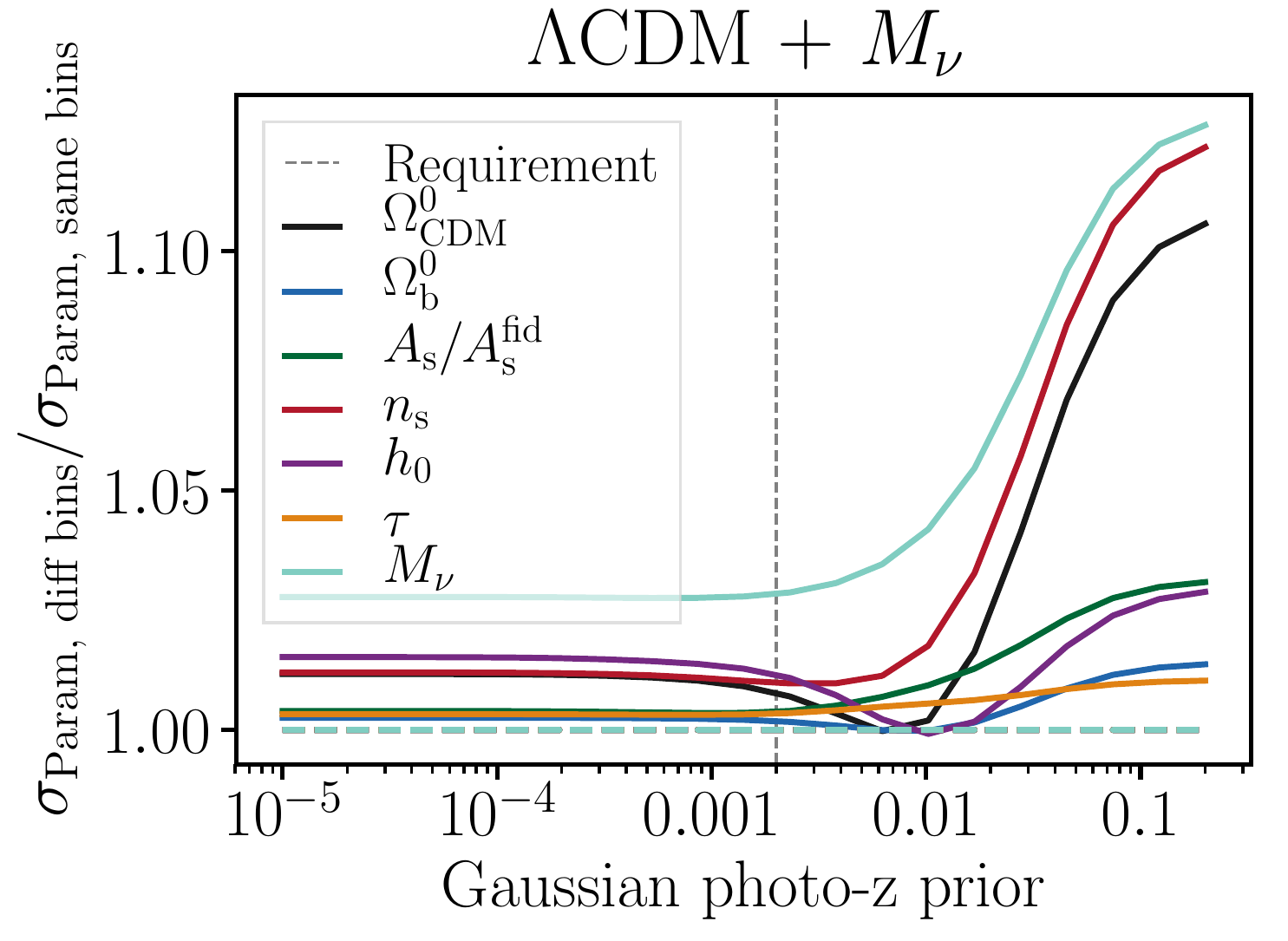}
\includegraphics[width=0.45\columnwidth]{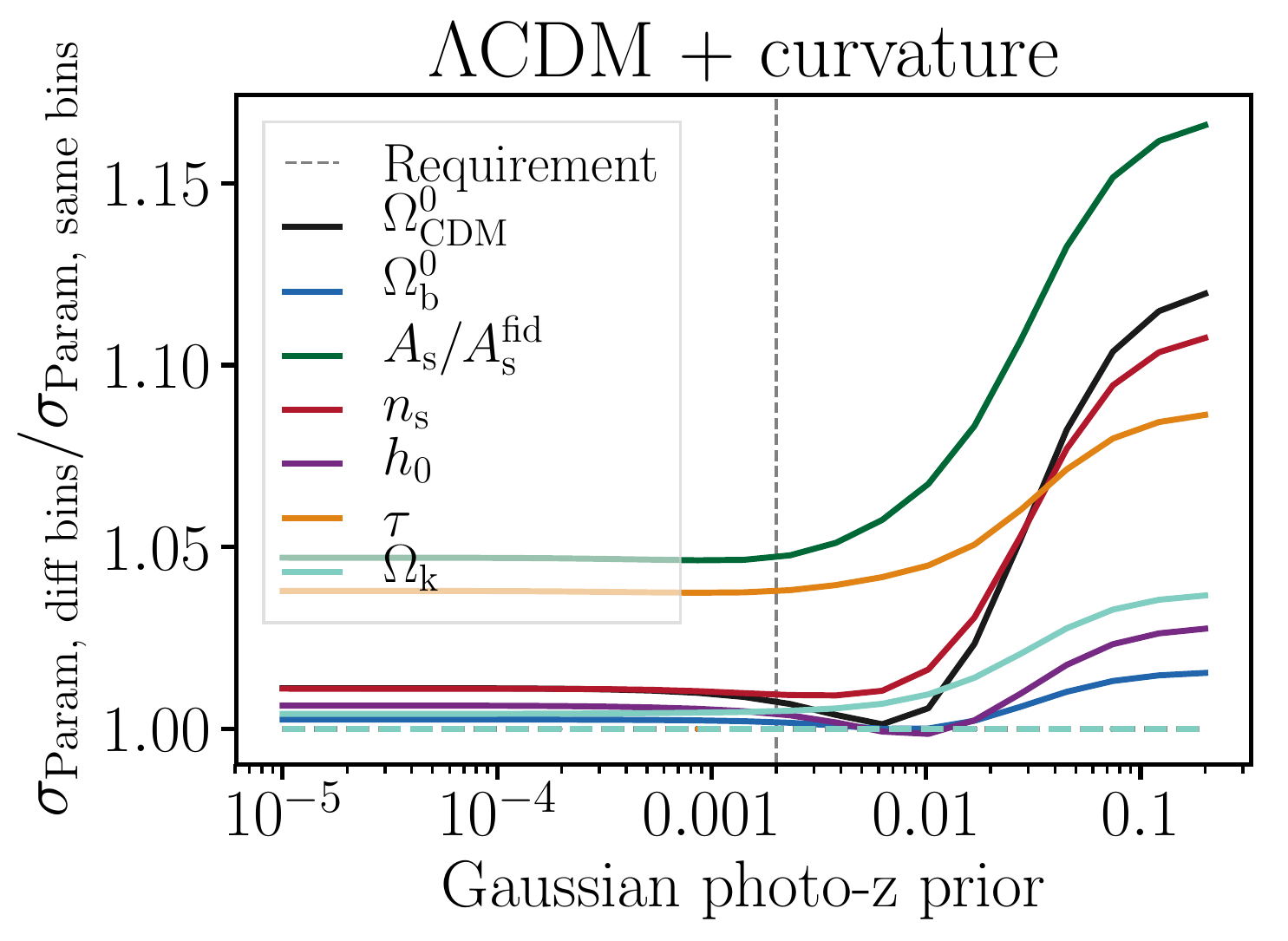}
\caption{
Comparing the cosmological parameter posteriors when the source and lens galaxies are the same or different, as a function of the Gaussian photo-z prior. 
For the fiducial priors, using different lens and source samples degrade cosmology by up to 18\% for $w_0, w_a$, which motivates using the same galaxy samples.
}
\label{fig:gphotoz_cosmo_vs_diffgs}
\end{figure}

However, surprisingly, the cosmological parameters are minimally degraded by using different lens and source bins.
In the absence of Gaussian photo-z prior, the degradation on the dark energy equation of state parameters is at most $20\%$, and the degradation in neutrino masses at most $15\%$, as shown in Fig.~\ref{fig:gphotoz_cosmo_vs_diffgs}.
For the fiducial photo-z priors, the degradation is only about $1\%$ for $w_a$ and $3\%$ for $M_\nu$.

\begin{figure}[H]
\centering
\includegraphics[width=0.45\columnwidth]{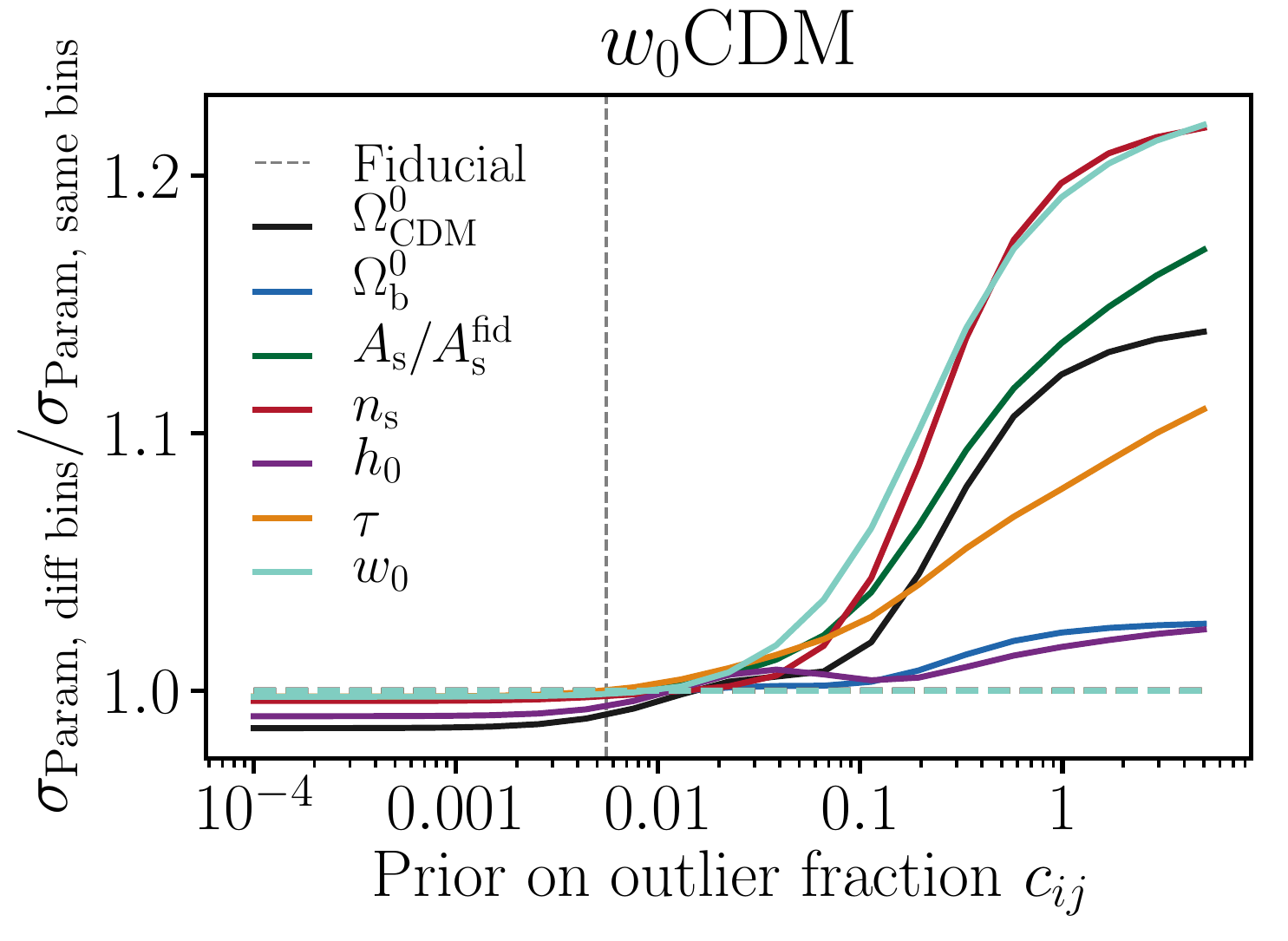}
\includegraphics[width=0.45\columnwidth]{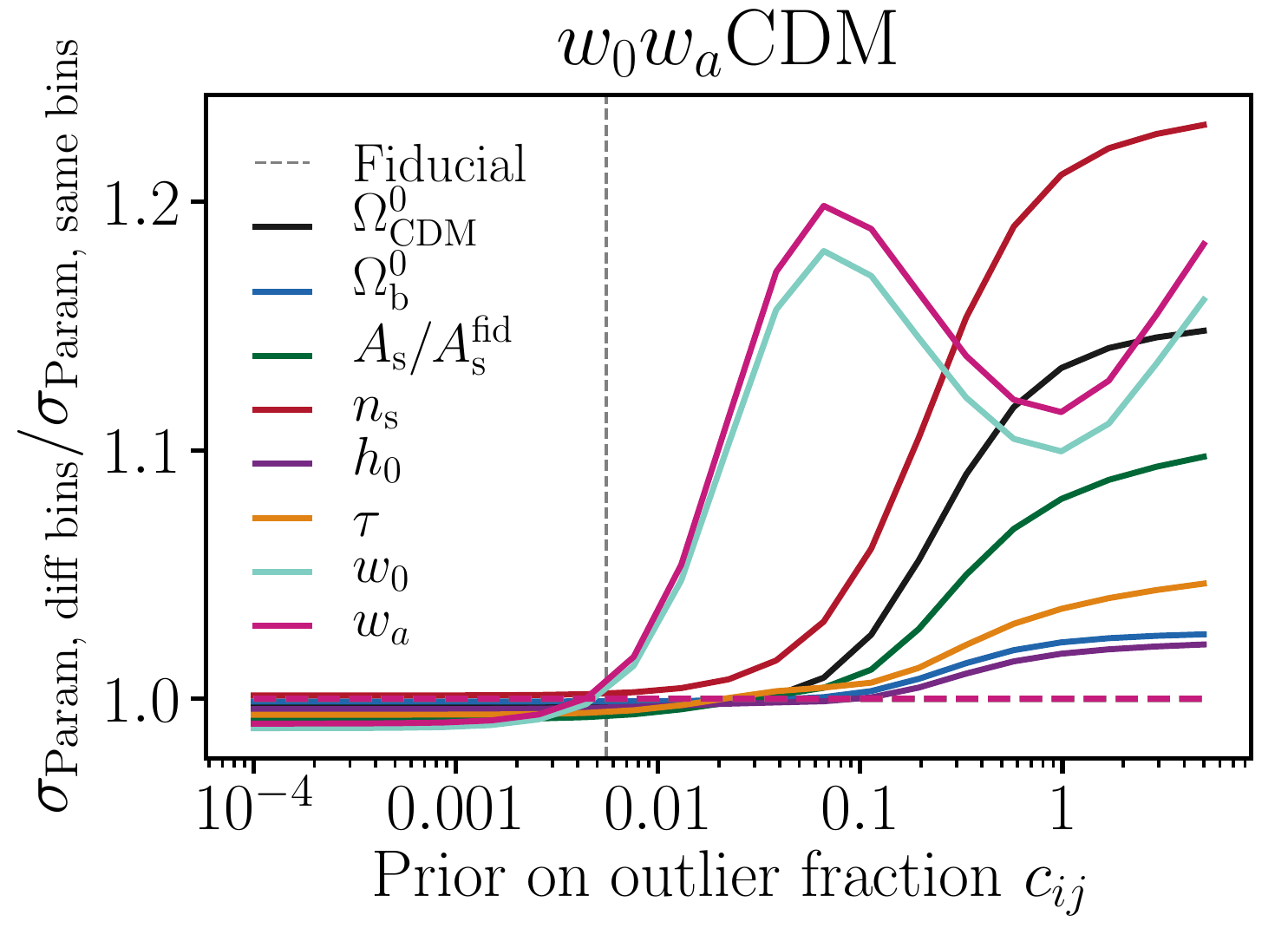}
\includegraphics[width=0.45\columnwidth]{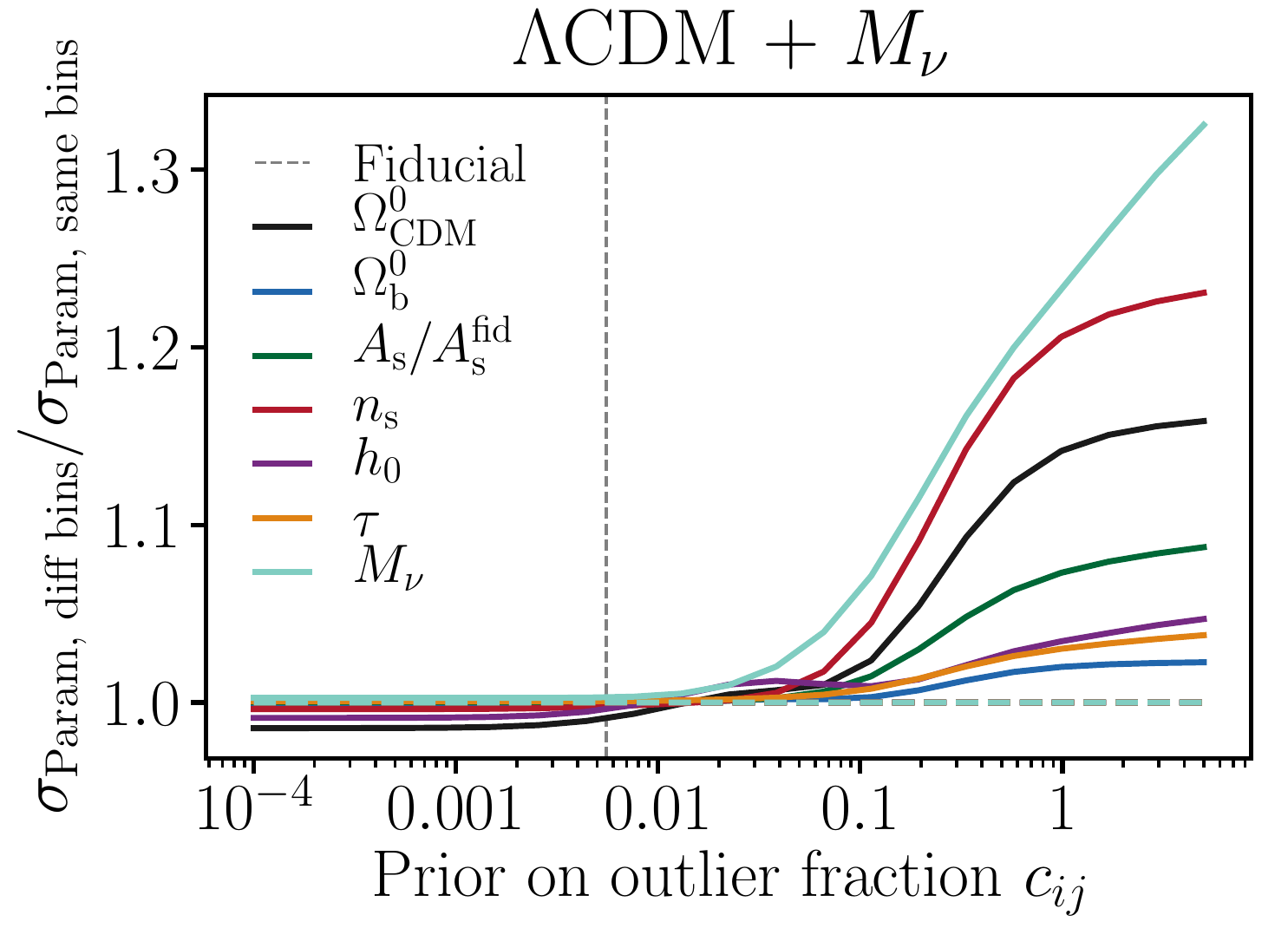}
\includegraphics[width=0.45\columnwidth]{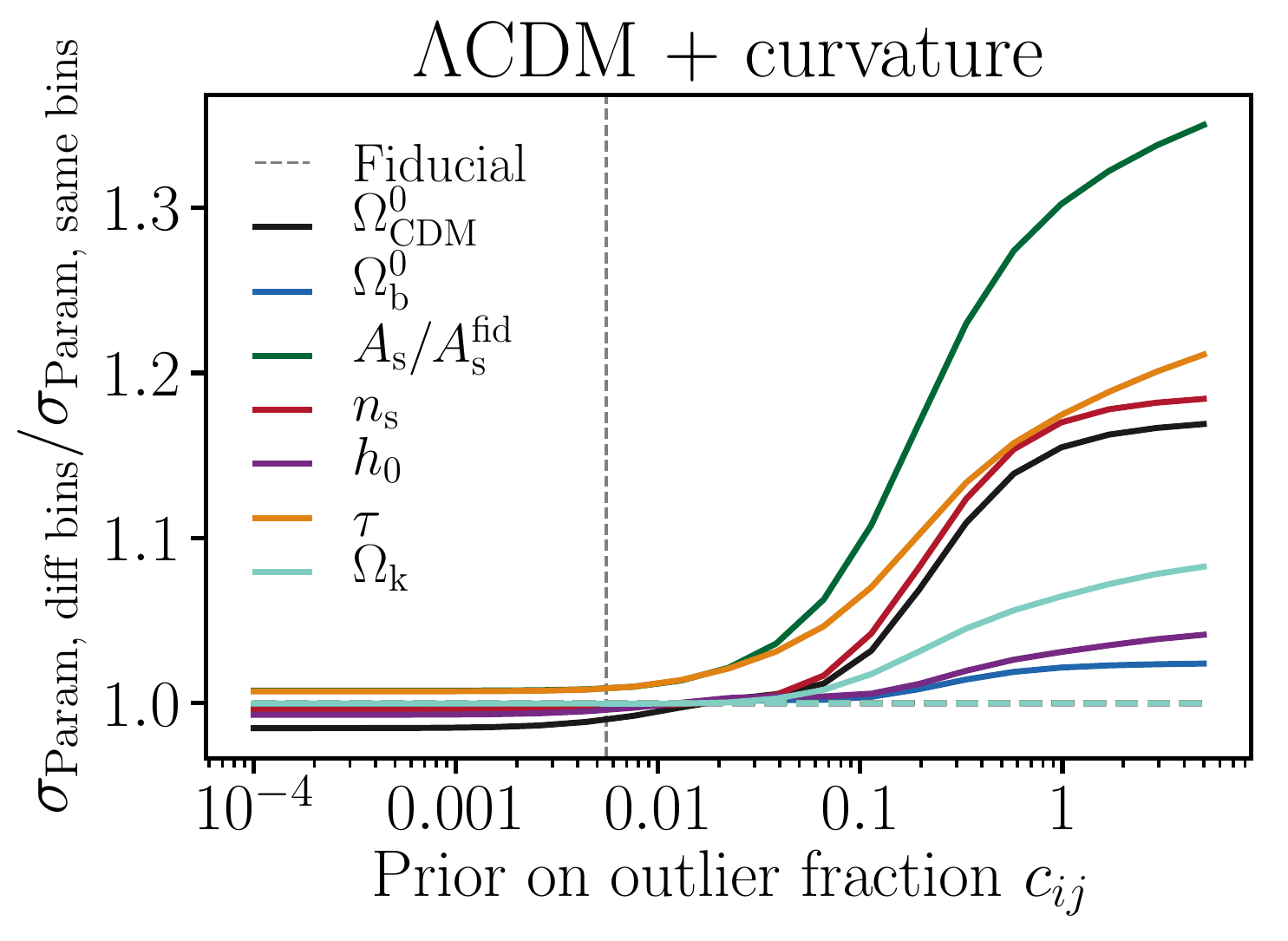}
\caption{
Comparing the cosmological parameter posteriors when the source and lens galaxies are the same or different, as a function of the photo-z outlier prior. 
For the fiducial outlier prior, there is no degradation. However, in the absence of outlier prior, using different source and lens bins can degrade cosmology by up to 30\%.
}
\label{fig:outlierphotoz_cosmo_vs_diffgs}
\end{figure}

Fig.~\ref{fig:outlierphotoz_cosmo_vs_diffgs} shows a similarly small degradation in cosmological parameters from using two different galaxy samples, as a function of the photo-z outlier prior\footnote{The observant reader may even notice that for very tight photo-z priors, the cosmological constraints can actually be better (although by a negligible amount) when using two different galaxy samples for sources and lens.
This can likely be understood as follows. In the case with two different galaxy samples, we have twice as many photo-z priors. If the data is good enough to calibrate the difference between the redshifts of the corresponding source and lens bins, then the two priors effectively get combined into one single prior, tighter by a factor $\sqrt{2}$.
In practice though, this improvement is sub-percent, and thus negligible.}.

\subsubsection{Improvements with CMB lensing}

\begin{figure}[H]
\includegraphics[width=0.45\columnwidth]{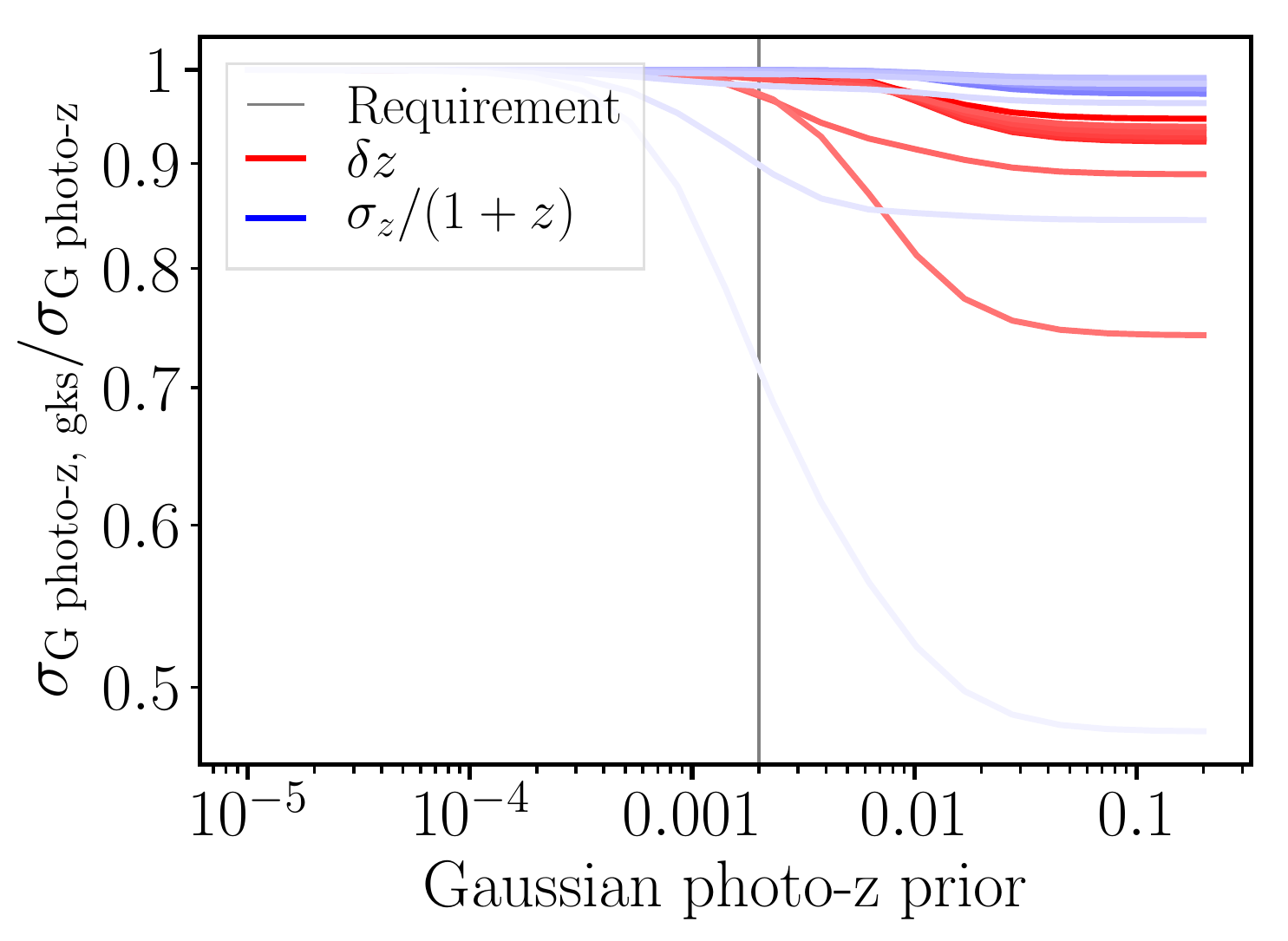}
\includegraphics[width=0.45\columnwidth]{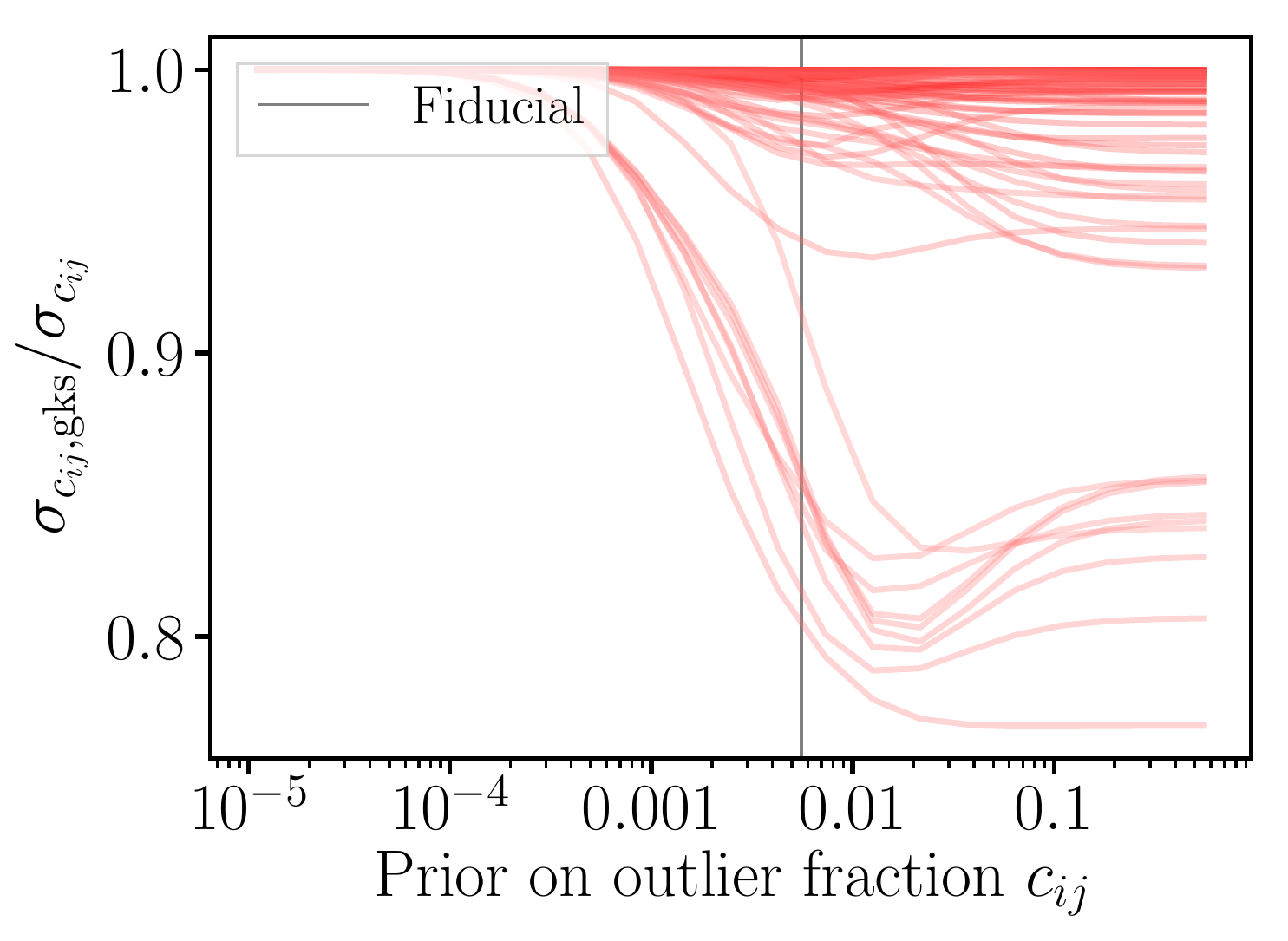}
\caption{
Adding CMB lensing improves the Gaussian photo-z (left) and outlier photo-z (right) posteriors by several tens of percent, even when the fiducial photo-z priors are used.
}
\label{fig:photozreq_vs_gks_lcdmw0wa}
\end{figure}

Does CMB lensing help constrain the photo-z parameters?
CMB lensing has been shown to help with shear calibration \cite{2013arXiv1311.2338D, 2017PhRvD..95l3512S, 2012ApJ...759...32V}, intrinsic alignments \cite{2014PhRvD..89f3528T, 2014MNRAS.443L.119H}, as well as photo-z errors \cite{Cawthon2018a}.
We compare the fiducial $3\times 2$ LSST analysis with the $6 \times 2$ analysis of LSST + CMB lensing from a CMB S4-like experiment.
We thus add the CMB lensing auto-spectrum and the correlations of CMB lensing with galaxy number density and shear from each of the ten tomographic bins.
Fig.~\ref{fig:photozreq_vs_gks_lcdmw0wa} shows an improvement in Gaussian and outlier photo-z uncertainties by several tens of percent when adding CMB lensing.

However, the cosmological parameters are only improved at the $10\%$ level.
This is likely a consequence of the lower SNR in CMB lensing (SNR = 282) compared to galaxy lensing (SNR = 428) in our forecast, due to our stringent scale cuts.
In CMB lensing auto-spectrum, one may hope to increase the maximum multipole beyond $\ell_\text{max}=1000$, since CMB lensing comes from higher redshift than galaxy lensing and clustering, where nonlinearities and baryonic effects are less important.
However, \cite{2020PhRvD.101f3534C} shows that baryonic effects on the matter power spectrum may already be a percent bias at this scale.
In summary, the improvement in photo-z and cosmology from including CMB lensing cannot compensate for discarding the null correlations.
\begin{figure}[H]
\centering
\includegraphics[width=0.45\columnwidth]{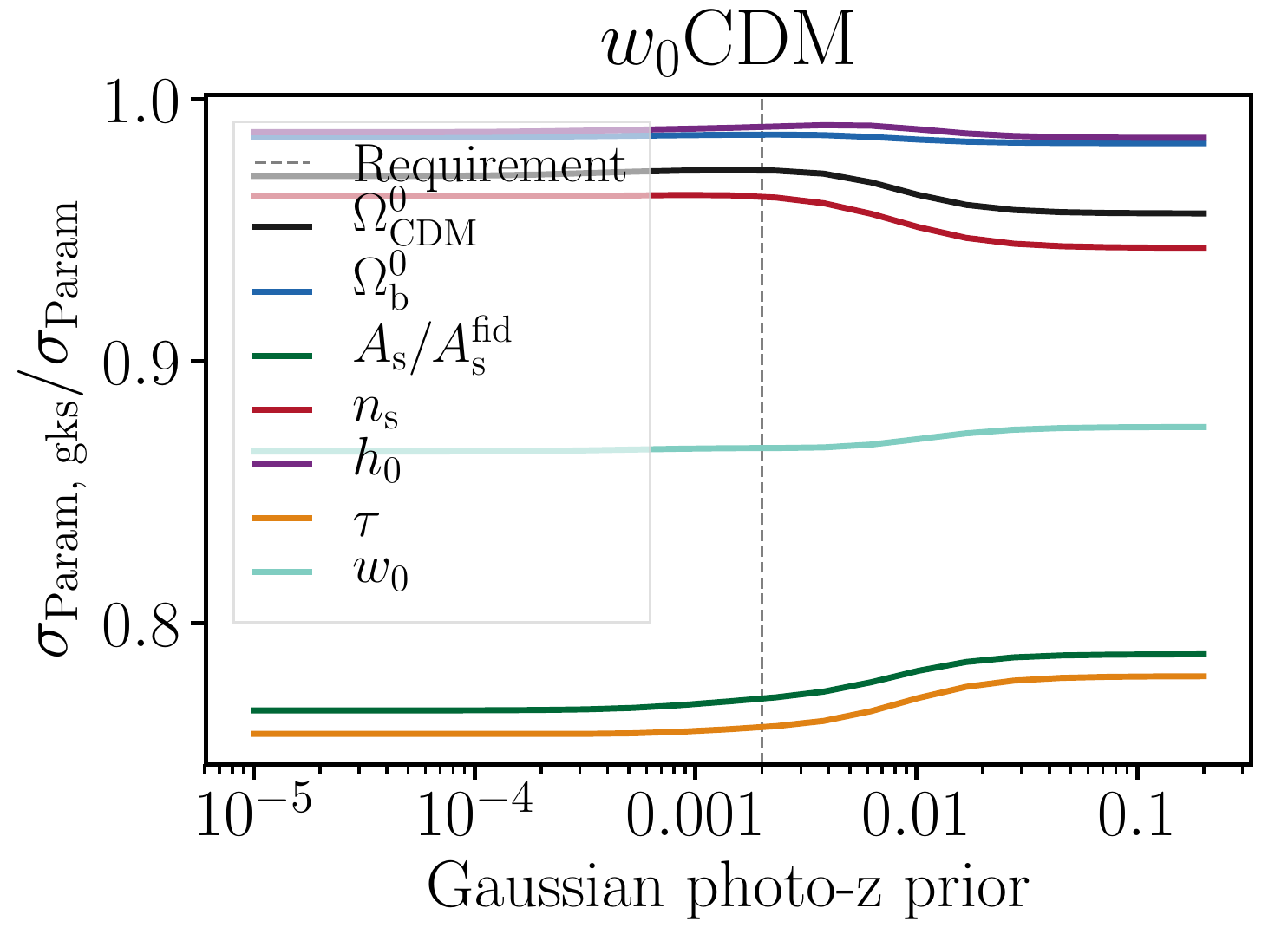}
\includegraphics[width=0.45\columnwidth]{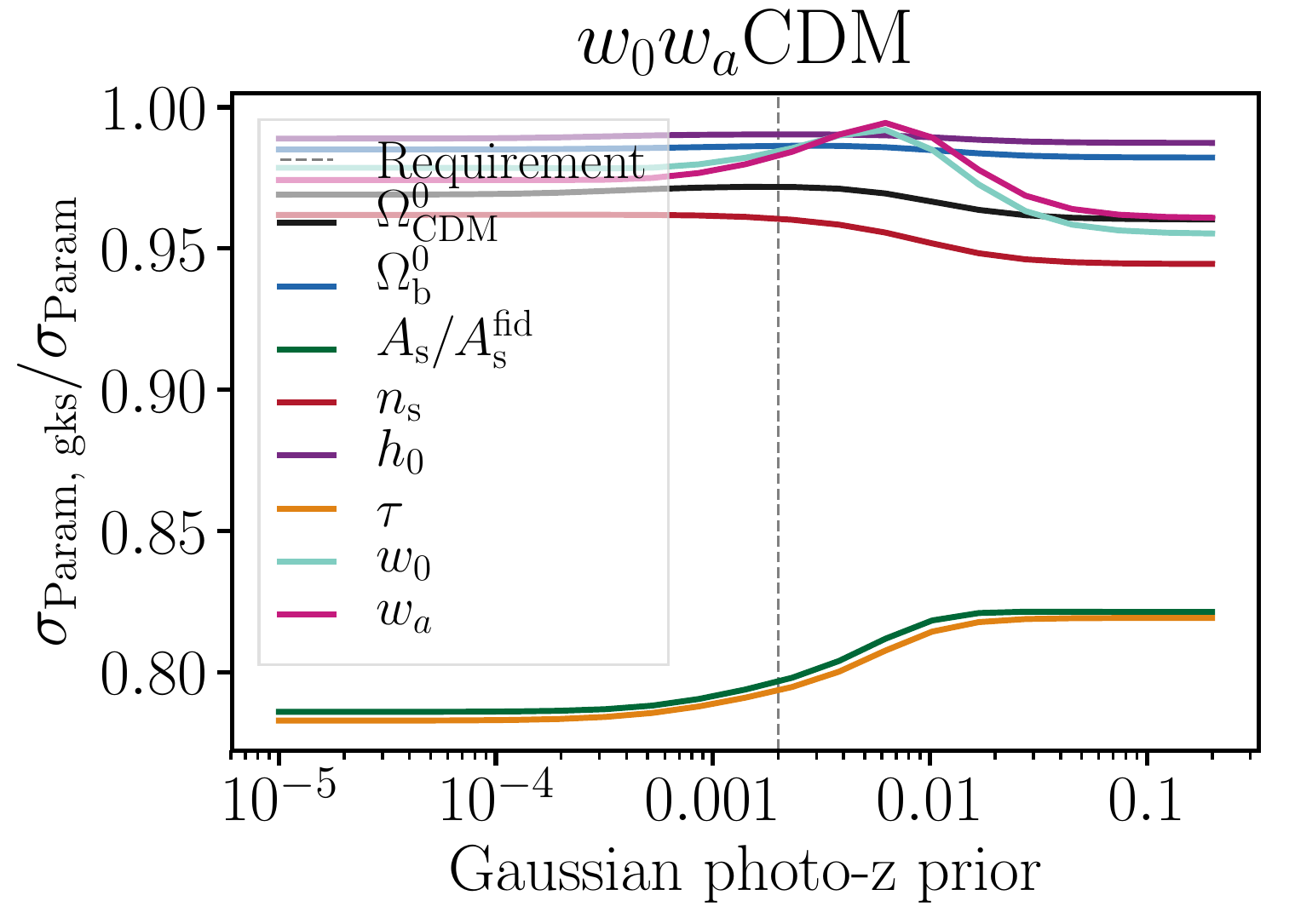}
\includegraphics[width=0.45\columnwidth]{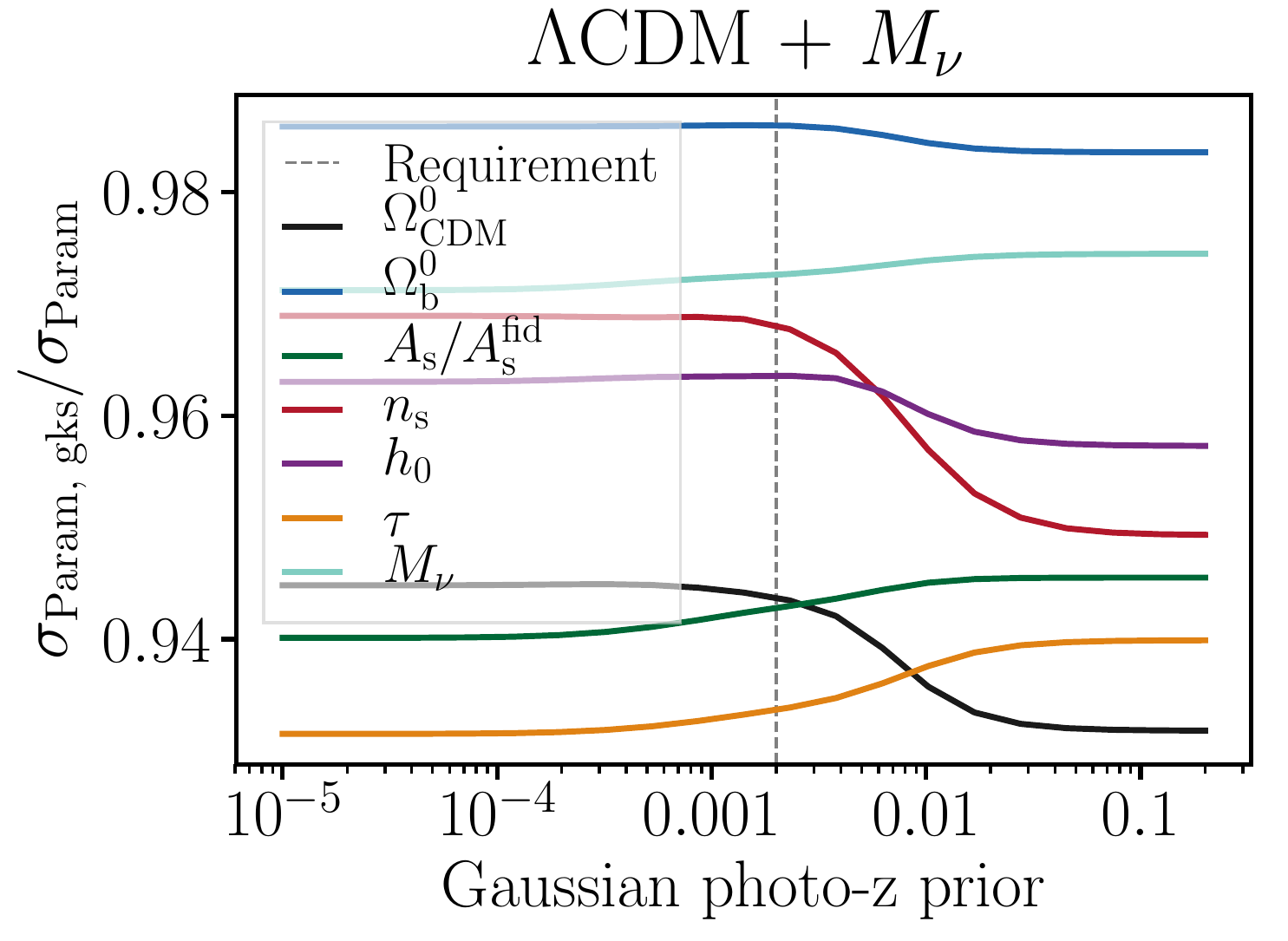}
\includegraphics[width=0.45\columnwidth]{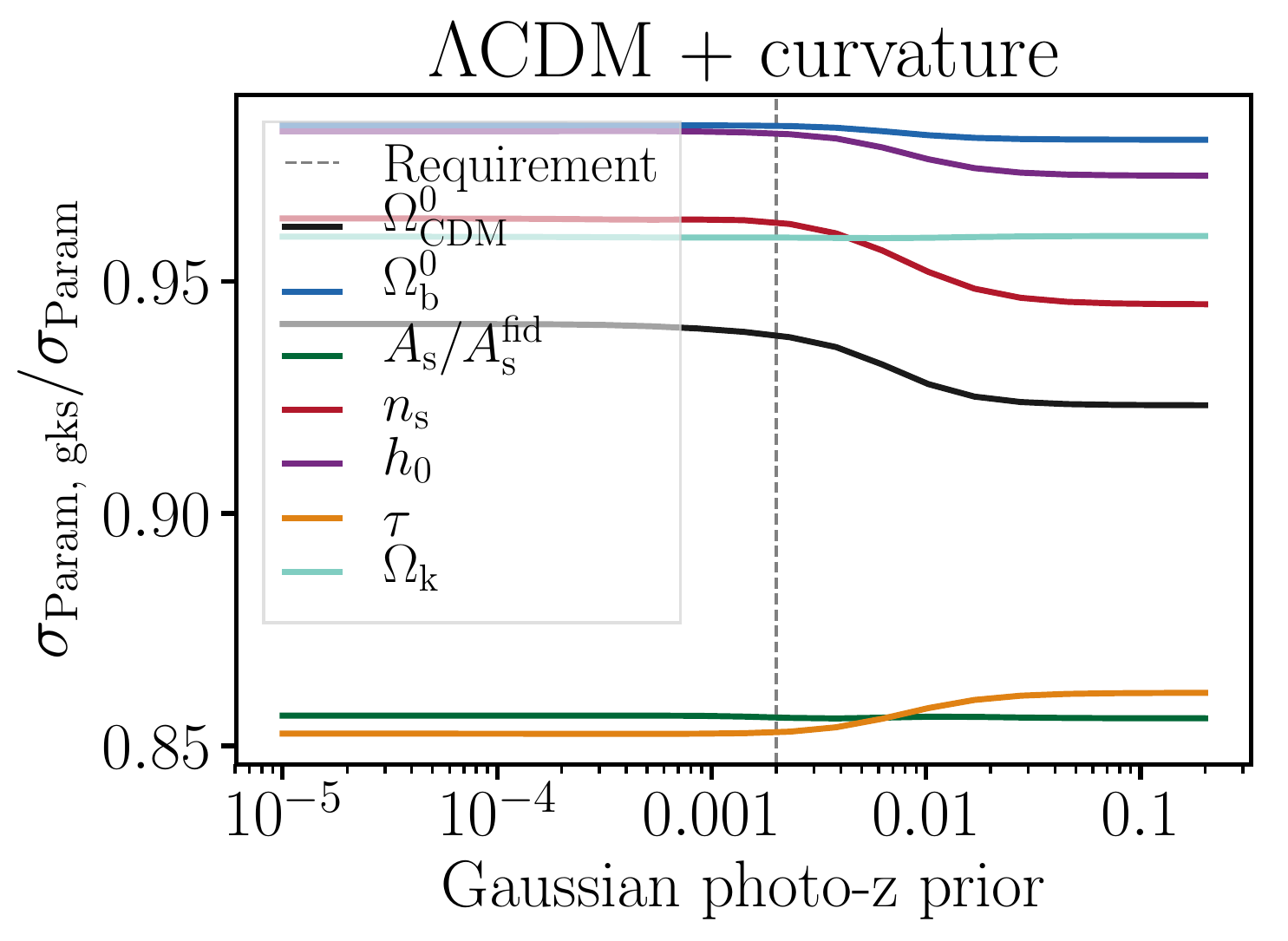}
\caption{
Adding CMB lensing improves cosmological constraints by about $10\%$ only, likely due to our stringent scale cuts which reduce the relative importance of CMB lensing compared to galaxy lensing.
}
\label{fig:gphotozreq_vs_gks}
\end{figure}

\begin{figure}[H]
\centering
\includegraphics[width=0.45\columnwidth]{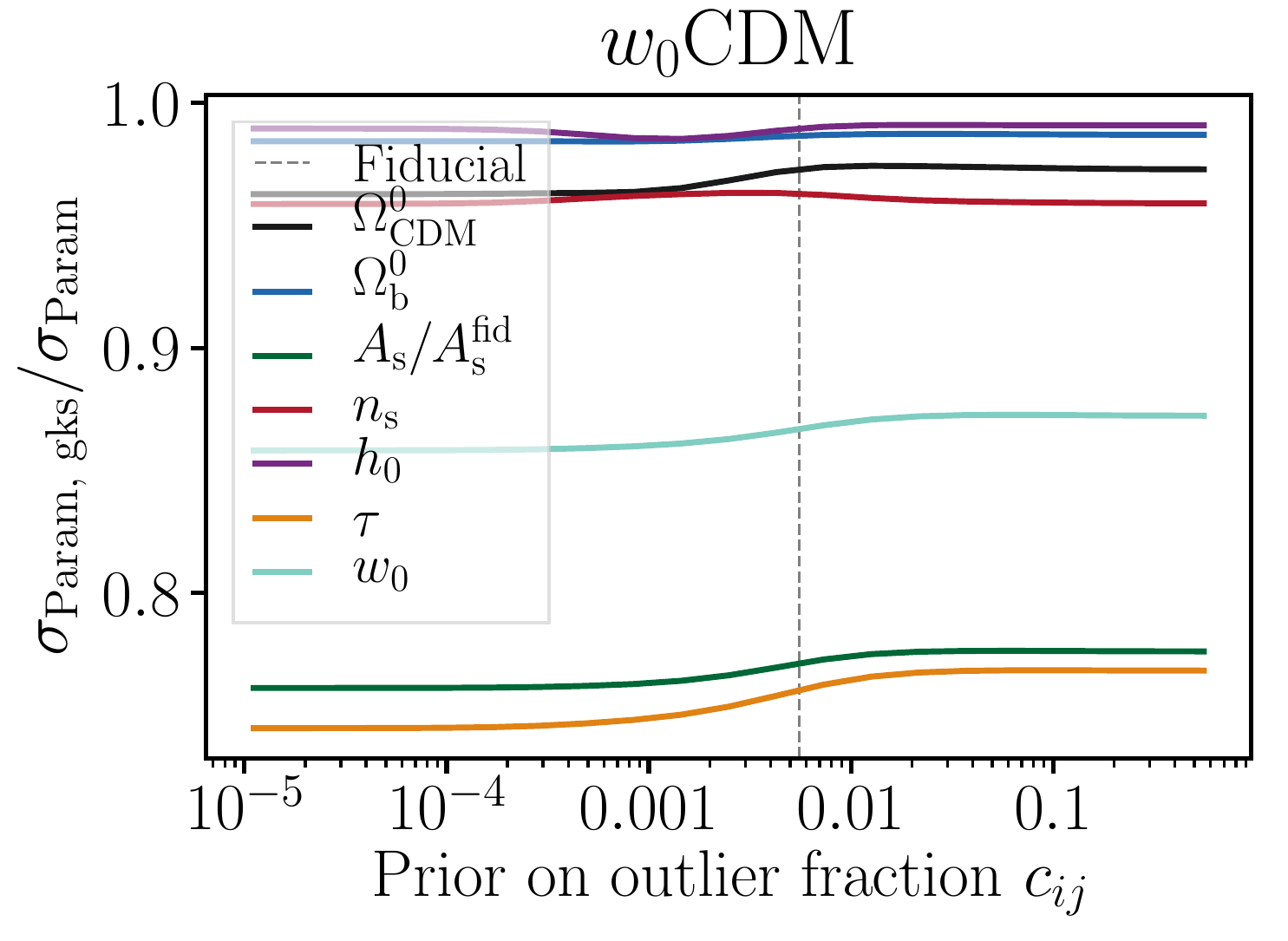}
\includegraphics[width=0.45\columnwidth]{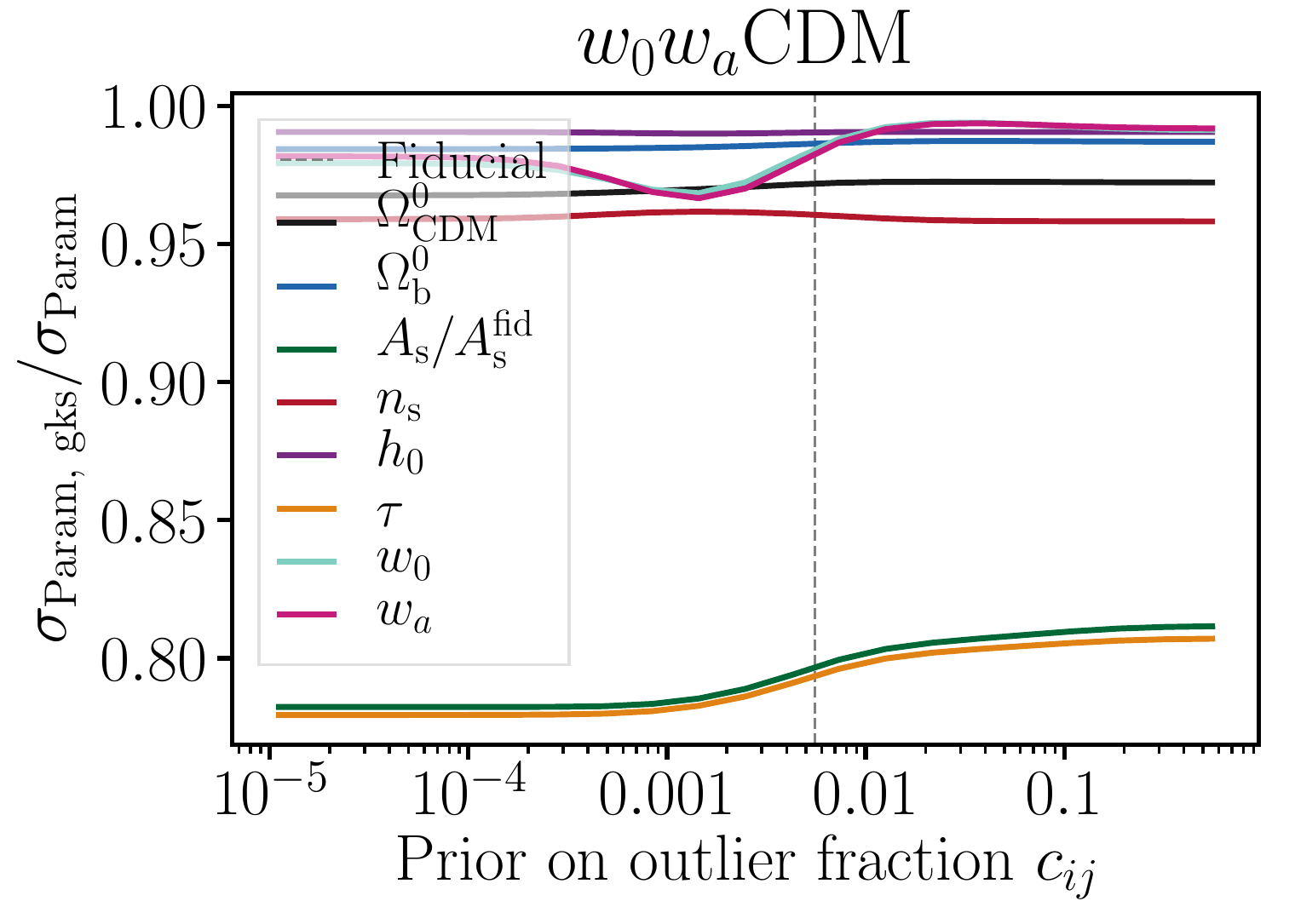}
\includegraphics[width=0.45\columnwidth]{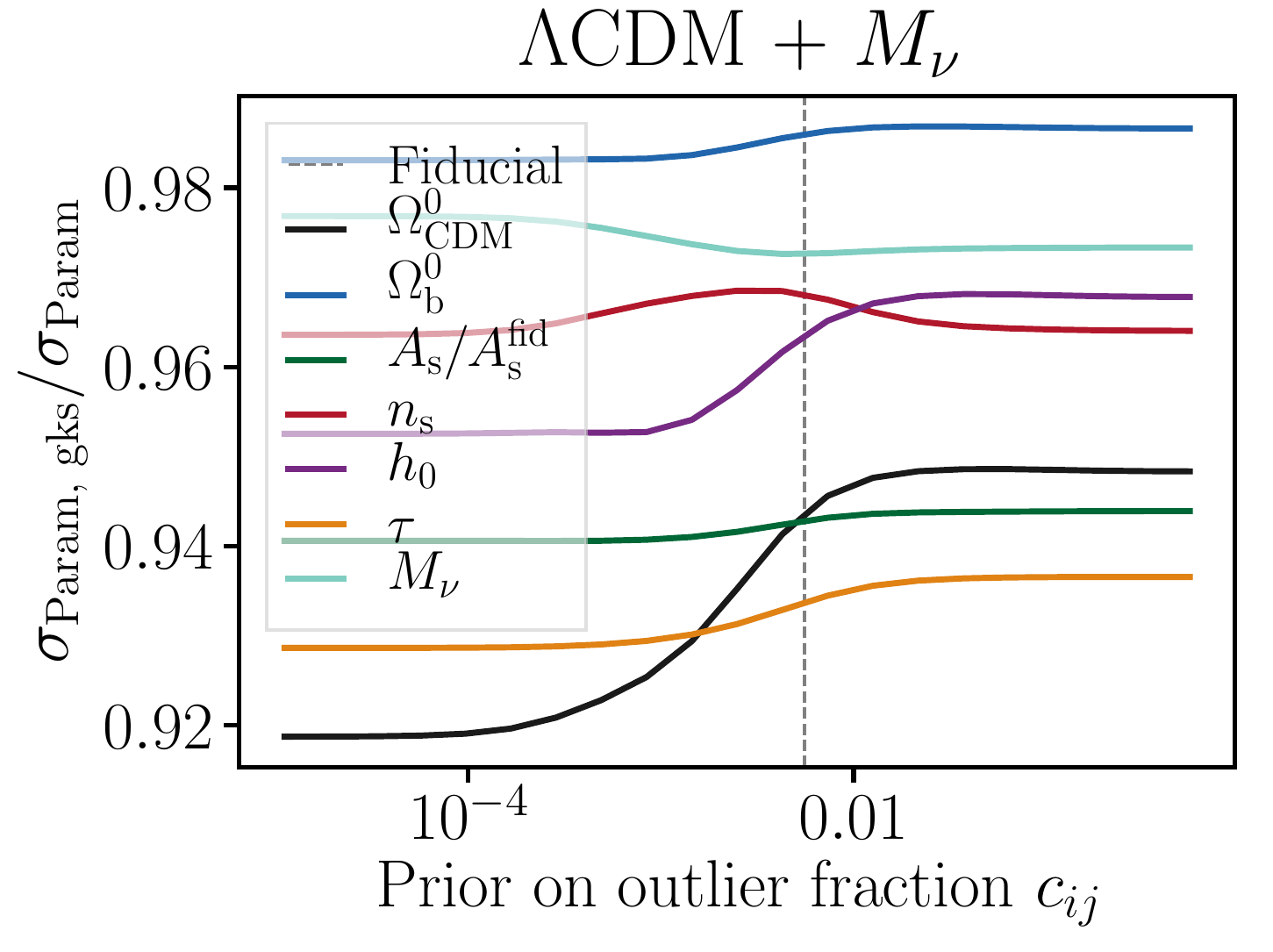}
\includegraphics[width=0.45\columnwidth]{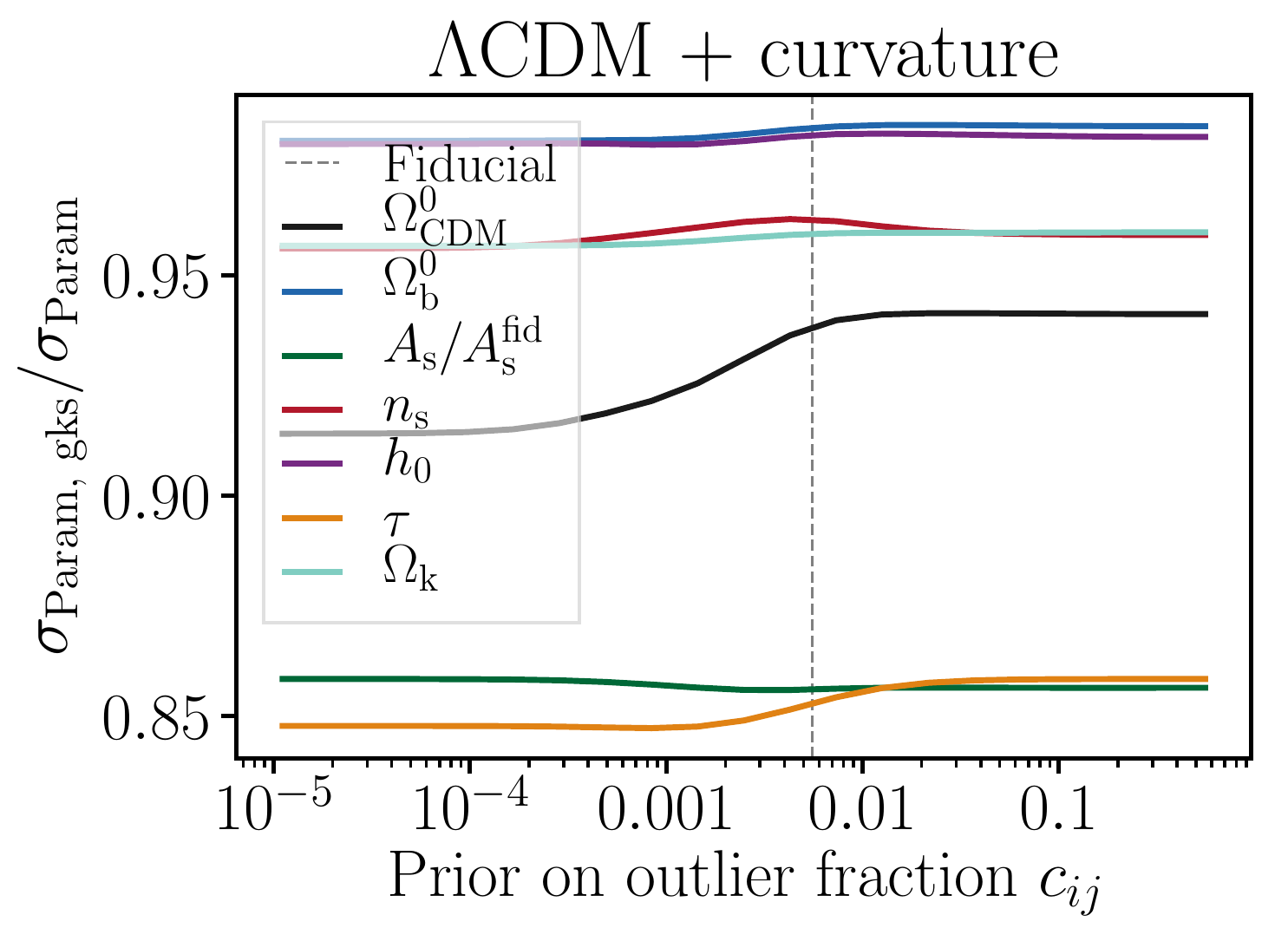}
\caption{
Same as Fig.~\ref{fig:gphotozreq_vs_gks}, but as a function of the outlier photo-z priors.}
\label{fig:outlierphotozreq_vs_gks}
\end{figure}

\section{Deconstructing Fisher forecasts: disentangling the sources of information}

The possibility of self-calibrating photo-z internally, without relying on datasets external to the photometric survey, has been explored several times in the literature.
Ref.~\cite{Huterer2006} shows that this is possible using cosmic shear alone, although at the cost of a factor two degradation in cosmological parameters.
Refs.~\cite{2013MNRAS.433.2857M,Elvin-Poole2018,Schneider2006,Benjamin2010,Zhang2010,Erben2009,Hildebrandt2009,Hildebrandt2009a,Zhang2017} explore the possibility of using clustering cross-correlations of the photometric samples to constrain their redshift distributions.
In this section, we pose the problem in a slightly different way, and extend these results.
Three potential sources of information enable the self-calibration:
\begin{itemize}
\item{\textbf{Algebra: projection.}}
For a given tomographic bin, the observed galaxy number density and lensing convergence fields are two independent linear combinations of the underlying matter density fields at various redshifts, weighted by the bin's redshift distribution.
In some cases, the redundancy between galaxy density and shear alone allows to invert this algebraic relation
and solve for the true matter density fields and the redshift distributions of the tomographic galaxy samples.
We explore this in Sec.~\ref{sec:algebra}.
\item{\textbf{Statistics: decorrelation.}}
In the absence of photo-z errors, and within the Limber approximation, the non-overlapping bins should have independent clustering.
Detecting a non-zero correlation between the clustering of two tomographic bins indicates that they must contain galaxies at the same redshift.
Other effects, such as magnification bias or common contamination (e.g., by stars) or modulation (e.g., by Galactic extinction, airmass, etc.) may mimic this effect. We discuss this in Sec.~\ref{sec:algebra_statistics}.
\item{\textbf{Model rigidity: $\Lambda$CDM.}}
When varying freely the $\Lambda$CDM cosmological parameters (and extensions thereof), not all shapes or redshift evolutions are allowed for the matter power spectrum.
As a result, the $\ell$-dependence and redshift evolution of the observed power spectra contains information on the redshift distributions of the tomographic samples.
We explore this in Sec.~\ref{sec:model_rigidity}.
\end{itemize}
We note that all three sources of information about the redshift distribution are automatically and properly included in the realistic Fisher forecast above (Sec.~\ref{sec:full_fisher}). 
Our goal here is simply to look inside the Fisher black box and gain intuition.

\subsection{Algebraic information: Projection}
\label{sec:algebra}

In this section, we give intuition on the role of the `algebraic' source of information, by considering a toy model where nothing is assumed about the statistics of the matter density field nor the shape of the matter power spectrum.
In other words we ask the following question:
\textbf{given observed galaxy number density fields and convergence fields for some number of tomographic bins, can we recover the redshift distribution of the tomographic bins and the underlying matter density field, without assuming anything about the statistics of the matter density field nor the shape of the matter power spectrum?}
In this subsection, we give a positive answer to this question.

We develop the basic idea using a toy version of the problem that highlights all the basic elements. 
Assume that we have defined $N_S$ photometric samples ($S$ for ``samples'') of galaxies. 
These correspond for example to the $N_S = 10$ tomographic bins used in the LSST $3\times 2$ analysis, split according to their photometric redshifts.
Here, we assume no knowledge of the quality of the photo-z algorithm, and consider the true redshift distribution of each tomographic bin completely unknown.

\begin{figure}[H]
\centering
\includegraphics[width=0.7\textwidth]{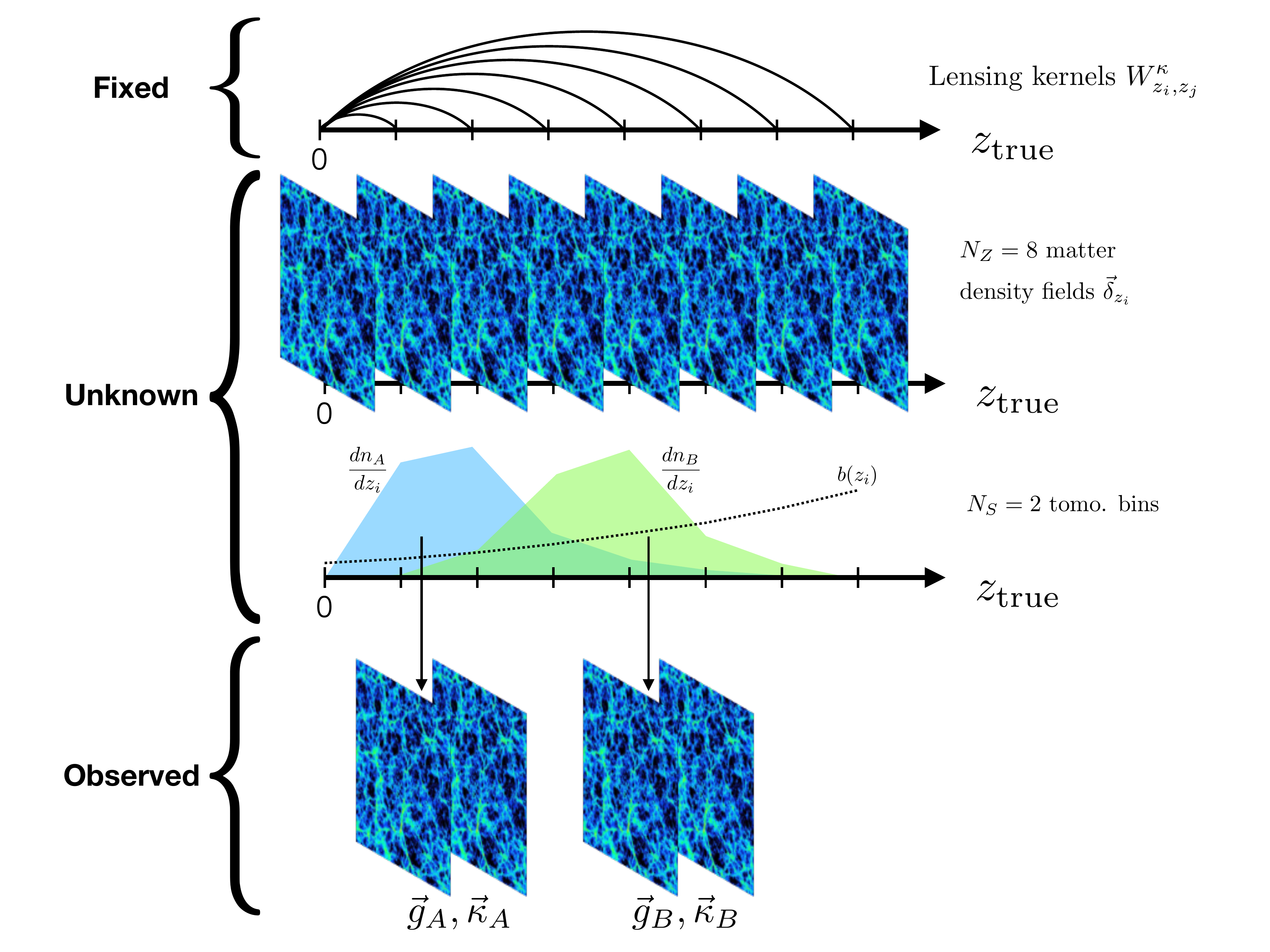}
\caption{
Illustration of the toy model.
The observed fields are the galaxy density fields $\vec{g}_A$ and lensing convergence $\vec{\kappa}_A$ for each of the tomographic bins $A, B, ...$.
These tomographic bins are linear combinations of the galaxy sample in $N_Z$ true-redshift bins, with unknown $dn/dz$. At each true redshift, there is an unknown matter density field.
In the text, we show under which assumptions the true matter density fields at each redshift and the redshift distributions $dn/dz$ can be solved for, given only the observed galaxy number density and convergence fields  $\vec{g}_A$ and $\vec{\kappa}_A$.
}
\label{fig:schematic_toy_model}
\end{figure}

For tomographic galaxy sample $A$, we assume that both the galaxy number density field $g_A$ and the convergence field $\kappa_A$ are observed, and similarly for the other galaxy samples $B,C$, etc.
Effectively, this assumes that the lens and source tracer samples are the same, as we did in the realistic Fisher forecast.
In this toy model, we assume these measurements to be noiseless.
The galaxy density $g_A$ and the convergence $\kappa_A$ are fields, or maps, observed at many pixel locations or Fourier space multipoles, and we will therefore denote them as vectors:
$\vec{g}_A = \left( g_A(\vl_1), ..., g_A(\vl_{N_\ell}) \right)$
and
$\vec{\kappa}_A = \left( \kappa_A(\vl_1), ..., \kappa_A(\vl_{N_\ell}) \right)$, where $N_\ell$ is the number of Fourier multipoles (or pixels) observed.
We then relate these observed fields to the underlying matter density field and to the redshift distribution of the tomographic samples.
To do so, we assume that the Universe is made of $N_z$ true redshift bins, such that the matter density field in true redshift bin $z_i$ is 
$\vec{\delta}_{z_i} = \left( \delta_{z_i}(\vl_1), ..., \delta_{z_i}(\vl_{N_\ell}) \right)$.
The number $N_z$ can be made arbitrarily large in principle, making this assumption arbitrarily accurate.
In our toy model, we therefore have:
\beq
\left\{
\bal
&\footnotesize{\text{\textbf{Gal. density:} }}
&&
\small{\vec{g}_A = \sum_i  \frac{dn_A}{dz_i} \left[  b_{A,z_i}\vec{\delta}_{z_i} + 2 \left( \alpha_{A, z_i} - 1 \right)  \sum_j W^\kappa_{z_i ,z_j} \vec{\delta}_{z_j}\right]}
&&&\footnotesize{\left( N_S \times N_\ell \text{ eq.} \right)}\\
&\footnotesize{\text{\textbf{Gal. lensing:} }} 
&&\vec{\kappa}_A = \sum_i \left(1+m_{A,z_i}\right) \frac{dn_A}{dz_i} \sum_j W^\kappa_{z_i ,z_j} \vec{\delta}_{z_j}
&&&\footnotesize{\left( N_S \times N_\ell \text{ eq.} \right)}\\
&\footnotesize{\text{\textbf{CMB lensing:} }} 
&&\vec{\kappa}_\text{CMB} = \sum_i  W^{\kappa_\text{CMB}}_{z_i} \vec{\delta}_{z_i} 
+ W^{\kappa_\text{CMB}}_\text{high z} \vec{\delta}_\text{high z}
&&&\footnotesize{\left( N_\ell \text{ eq.} \right)}\\
\eal
\right.
\label{eq:toy_model_system}
\eeq
where we assume that the lensing kernels $W^\kappa_{z_i ,z_j}$ and the magnification biases $\alpha_{A, z_i} \equiv \frac{d \ln N_{A, z_i}}{d\ln S}$ are known, but the galaxy biases $b_{A, z_i}$ and shear multiplicative biases $m_{A, z_i}$ are unknown nuisance parameters.
Here $ \vec{\delta}_\text{high z}$ represents the matter density field at redshifts higher than any of the LSST galaxies, which contribute to CMB lensing but to none of the galaxy observables.
This term can be thought of as a noise term.
In this toy model the data are the galaxy number density fields $\vec{g}_A$ and the convergence fields $\vec{\kappa}_A$.
The unknowns we wish to infer are the redshift distributions $\frac{dn_A}{dz_i}$ of the tomographic bins and the underlying matter density fields $\vec{\delta}_{z_i}$.
This is illustrated in Fig.~\ref{fig:schematic_toy_model}.
We choose to express these relations at the field or map level, rather than in terms of the auto- and cross-spectra, to emphasize that we are simply solving an algebraic system of equations, with no assumption about the statistics (decorrelation, Gaussianity) of the fields or about the shape of their power spectra.

We note that the relationship between observed fields ($\vec{g}_A$, $\vec{\kappa}_A$) and the true underlying matter density fields ($\vec{\delta}_{z_i}$) is linear, since it represents a projection from 3D to 2D. In other words, if the redshift distributions, galaxy, shear and magnification biases were known, solving the system for the true matter density fields $\vec{\delta}_{z_i} $ would be as simple as inverting a matrix. 
\footnote{
The matrix in question is typically invertible, because it is diagonally dominant}: the redshift distribution for a given tomographic bin typically peaks in the corresponding true redshift bin, even in the presence of photo-z errors.

In what follows, we make the simplifying assumption that the galaxy bias and shear bias depend only on true redshift, and not on photo-z:
$b_{A, z_i} \equiv b_{z_i}$ and $m_{A, z_i} \equiv m_{z_i}$.
These assumptions are routinely made in $3\times 2$ analyses, and we have adopted them throughout this paper.
However, we keep in mind that they are likely wrong in detail:
galaxies at the same true redshift but with different photo-z are likely of different types (e.g. red vs blue), 
so they likely have different clustering bias and shear multiplicative bias.
Furthermore, we will assume that $N_\ell \gg N_S, N_z$, i.e. that the number of observed Fourier multipoles is much larger than the number of tomographic bins and true redshift bins.
At this stage, since we assumed nothing about the shape of the matter power spectra, it does not matter whether the various multipoles $\ell$ included have the same modulus or different moduli.

We wish to conclude whether the matter density field and redshift distributions can be inferred by solving the system Eq.~\eqref{eq:toy_model_system},
without assuming anything about the statistics of the matter density field or the shape of its power spectrum.
To do so, we simply count the effective number of equations and unknowns.
We also use a more rigorous approach in App.~\ref{app:algebra_fisher}, searching for potential continuous degeneracies in these equations using the Fisher formalism, and find the simple counting of equations and unknowns to be valid.
To properly count unknowns, we note that the observed galaxy density and convergence fields $\vec{g}_A$ and $\vec{\kappa}_A$ are linear combinations of the underlying matter density fields $\vec{\delta}_{z_i}$. As a result, they live in the vector space spanned by the $(\vec{\delta}_{z_i})_{i\in [\![ 1, N_z ]\!]}$. This vector space thus has dimension $N_z$ (at most), and so each vector $\vec{\delta}_{z_i}$, despite having $N_\ell \gg N_z$ components, should really count as $N_z$ scalar unknowns.
Another consequence is that one can at most generate $N_z$ independent linear combinations of the $(\vec{\delta}_{z_i})_{i\in [\![ 1, N_z ]\!]}$. In other words, one can at most construct $N_S=N_z$ linearly independent tomographic samples. Any additional tomographic sample would simply be a linear combination of the others, and should therefore not change whether the system can be solved or not.
In short, when counting unknowns, we replace $N_\ell$ and $N_S$ with $N_z$.

To summarize, we have $N_z^2$ equations from the galaxy density fields $\vec{g}_A$ and $N_z^2$ equations from the convergence fields $\vec{\kappa}_A$.
On the other hand, we have $N_z^2$ unknowns from the matter density fields $\vec{\delta}_{z_i}$, $N_z\left( N_z-1 \right)$ unknowns from the redshift distributions $dn_A/dz_i$ (normalized to integrate to unity), $N_z$ for the galaxy biases $b_{z_i}$ and $N_z$ for the shear multiplicative biases $m_{z_i}$.
We now compare the number of equations and unknowns in several data combinations.
Again, this counting is validated more rigorously in App.~\ref{app:algebra_fisher}.

\paragraph*{\textbf{Clustering-only: the system is underconstrained, and degeneracies exist}}

If only the galaxy density fields are observed, we have only $N_z^2$ equations (galaxy density fields alone) 
for $N_z^2 + N_z\left( N_z-1 \right) + N_z = 2 N_z^2$ unknowns (matter density fields, redshift distributions and galaxy biases). 
In particular, we expect $N_z^2$ unconstrained parameter combinations.
Indeed, for each true redshift $z_i$, rescaling the galaxy bias by any factor $\alpha$ and the matter density field by $1/\alpha$ leaves the observed galaxy density field unchanged. 
These constitute $N_z$ rescaling degeneracies.
The $N_z(N_z-1)$ remaining degeneracies are somewhat less obvious, and correspond to rotations of each vector $\vec{\delta}_{z_j}$ in the vector space spanned by $(\vec{\delta}_{z_i})_{i\in [1, N_z]}$, accompanied by a corresponding operation applied to the matrix $dn_A/dz_i$.
These degeneracies are related to the case of clustering redshifts, where one infers $b(z)*dn/dz$ and not $b(z)$ and $dn/dz$ separately.
Mathematically, there is an additional reordering degeneracy: any permutation of the true redshifts $z_i$ is allowed.
In other words, the system alone does not constrain the ordering in redshift of the tomographic bins. 
In practice, this last degeneracy is irrelevant, since the observer always knows how to rank the tomographic samples in order of growing redshift.

\paragraph*{\textbf{Clustering and shear: the system is perfectly constrained if the shear biases are known, and underconstrained otherwise.}}

If galaxy density and convergence are observed and the shear multiplicative biases are known, 
we have $2N_z^2$ equations (galaxy density and convergence fields)
and $2N_z^2$ unknowns (matter density fields, redshift distributions and galaxy biases).
The system Eq.~\eqref{eq:toy_model_system} has no degeneracy, and we can solve simultaneously for the redshift distributions, the galaxy biases and the underlying matter density fields.
Since we made no assumption about the statistics (e.g. the power spectrum) of the matter density field, the redshift distributions and galaxy biases are recovered without any cosmic variance.
\textbf{This is therefore a form of sample variance cancellation, and the precision of the inferred redshift distribution is only limited by instrumental noise, galaxy shot noise and shape noise.}

Since no modeling of the matter power spectrum was required, the small scales where nonlinearities and baryonic effects are important can be included without incurring the usual model bias.
However, the validity of the linear bias model assumed limits the scales that can be included.
Including magnification with a known magnification bias does not change the conclusion.
On the other hand, if the magnification bias is not known, the system becomes underconstrained.
If the shear multiplicative biases are unknown, they constitute $N_z$ additional unknowns, and the system becomes underconstrained again.
For the system to admit a unique solution, the shear multiplicative biases must therefore be constrained either by priors, or by information not included here, e.g. their different effect on the shape of the power spectra compared to other cosmological and nuisance parameters.

\paragraph*{\textbf{Clustering, shear and CMB lensing: In the absence of any prior on the statistics of the underlying matter density field, adding CMB lensing does not help}}

Indeed, while we add a new measured field $\vec{\kappa}_\text{CMB}$,
we also add a new unknown field $\vec{\delta}_\text{high z}$.
However, this conclusion changes as soon as we include a decorrelation prior for the matter density fields at different redshifts, as we now examine.

\subsection{Algebra \& Statistics: Decorrelation prior}
\label{sec:algebra_statistics}

We expect that the matter density fields $\vec{\delta}_{z_i}$ at different true redshifts should be uncorrelated, as long as the true redshifts are far enough from each other.
This decorrelation prior is effectively included in any Fisher forecast. 
This is particularly clear when using the Limber approximation: in the absence of photo-z errors, the predicted clustering cross-spectra of non-overlapping tomographic bins vanish.
Na\"ively, this decorrelation prior could thus be implemented as follows
\cite{Zhang2010}:
\beq
\langle \vec{\delta}_{z_i} \cdot \vec{\delta}_{z_j}^*  \rangle
=
\sum_{\vl \in \ell \text{ bin}} \langle \delta_{z_i} (\vl) \delta_{z_j}^* (\vl) \rangle
= 0 \text{ if } i\neq j
\hspace{1cm}
\left( N_z\left( N_z- 1\right)/2 \text{ equations} \right)
\label{eq:toy_model_decorrelation}
\eeq
This decorrelation prior is helpful because it adds $N_z\left( N_z- 1\right)/2$ new equations to those in Eq.~\eqref{eq:toy_model_system}, thus making it more constrained.

However, if only the galaxy density fields $\vec{g}_A$ are observed, this prior does not add enough equations to allow the system to be solved.
One would need to add two such decorrelation constraints, for two different $\ell$-bins, to get enough equations.
Indeed, by splitting the various multipoles into two $\ell$-bins where the decorrelation prior is imposed, we effectively duplicate the system Eq.~\eqref{eq:toy_model_system}, thus multiplying by two the number of equations. 
However, the number of unknowns is not multiplied by two, since galaxy bias and redshift distribution are independent of multipole.
As a result, in this clustering-only case with two $\ell$-bins, we get $5N_z^2-N_z$ equations for only $3N_z^2$ unknowns, which is now even over-constrained.
Ref.~\cite{Zhang2010} then concludes that one can solve for the redshift distributions, in a way that is not limited by sample variance.
However, the decorrelation prior of Eq.~\eqref{eq:toy_model_decorrelation} is actually only approximate, valid only within cosmic variance:
random fluctuations will cause chance correlations between the matter density fields in the different redshift bins.
In particular, if the decorrelation prior is needed for the system to admit a unique solution, this means that no sample variance cancellation occurs, and the accuracy of the inferred solution will be limited by cosmic variance\footnote{Na\"ively, one can increase the number of decorrelation constraints by increasing the number of $|\vl|$ bins. However, each decorrelation constraint then becomes looser, because the sum over $\vl$ has fewer terms and is therefore farther from zero due to cosmic variance.}.
This subtlety is automatically included in the standard Fisher forecasts.

In fact, the correct prior, which encodes the decorrelation of the matter density fields at different redshifts, within cosmic variance, is the following:
\beq
\text{Proba}\left[ \vec{\delta}_{z_1}, ..., \vec{\delta}_{z_{N_z}} \right]
\propto
e^{- \sum_i   \frac{|\vec{\delta}_{z_i}|^2}{2 \sigma_i^2}}
\eeq
where $\sigma_i^2$ is effectively the power spectrum of $\delta_{z_i}$. 
In App.~\ref{app:algebra_fisher}, we add this prior to the clustering-only case, and show that this only allows to constrain the underlying matter density field to the level of the prior, and does not improve the constraints on the redshift distribution beyond the prior.

In the case of CMB lensing, this decorrelation prior implies that the $\vec{\delta}_\text{high z}$ is uncorrelated with all the observed $\vec{g}_A$ and $\vec{\kappa}_A$. It can therefore be projected out directly, and its contribution removed from $\vec{\kappa}_\text{CMB}$, leaving only a noise residual. This effectively removes $\vec{\delta}_\text{high z}$ from the count of unknowns.
As a result, including CMB lensing adds $N_z$ observables and no extra unknowns.
These $N_z$ new observables can be used to solve for $N_z$ additional unknowns, such as the shear multiplicative biases. 
Thus, \textbf{when the statistical decorrelation prior is included, adding CMB lensing to galaxy density and shear now allows to solve for the shear multiplicative bias}.

We performed the analysis above at the field level, rather than the power spectrum, to show that no prior was assumed on the matter density power spectrum.
However, for the purpose of determining whether the system admits a unique solution or not, i.e. for counting equations and unknowns, these two approaches are equivalent.
This can be understood because the system Eq.~\eqref{eq:toy_model_system} is linear in the observed fields and the matter density fields.
It is therefore equivalent to solve it in terms of vectors (the various fields) or in terms of their coefficients in a given base (their cross and auto-spectra).
In particular, our counting of equations and unknowns at the field level agrees with Ref.~\cite{Zhang2010} at the power spectrum level.

\subsection{Power spectrum shape information}
\label{sec:model_rigidity}

In this section we highlight the information on photo-z outliers carried by the shape of the clustering and lensing power spectra. To build intuition, we will discuss a toy example, referring the reader to the treatment in the main body of the paper above for the full and more realistic case.

In particular, let's assume that a fraction $c_{21}$ of galaxies, thought to belong to a photometric bin at redshift $z_1$, are actually outliers at true redshift $z_2$. For simplicity, here we will consider top-hat photometric bins of width $\Delta z = 0.5$.
Then the redshift kernels get modified as follows:
\beq
W_g(z) = b(z) \frac{dn}{dz} \longrightarrow W_g(z, z_1, z_2, c_{21}) = (1-c_{21}) b_{z_1} \frac{dn_1}{dz} + c_{21} b_{z_2} \frac{dn_2}{dz}
\eeq
where $\frac{dn_i}{dz} = 0$ if $z$ is outside a bin of (full) width $\Delta z$ around $z_i$ and is normalized such that $\int dz \frac{dn}{dz} = 1$. A similar expression holds for lensing, and we shall assume that each redshift bin is independent.

\begin{figure}[H]
\centering
\includegraphics[width=0.49\columnwidth]{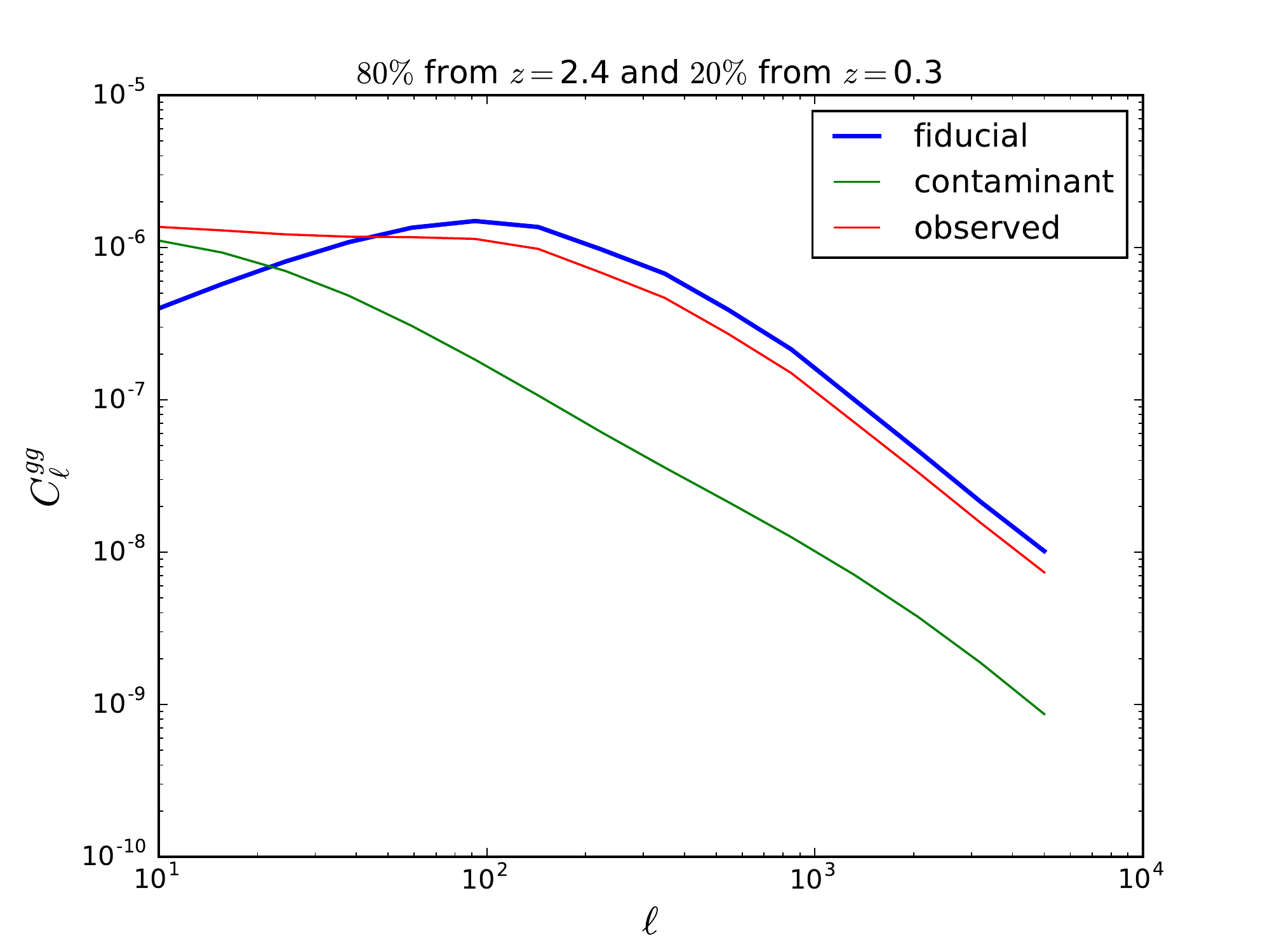}
\includegraphics[width=0.49\columnwidth]{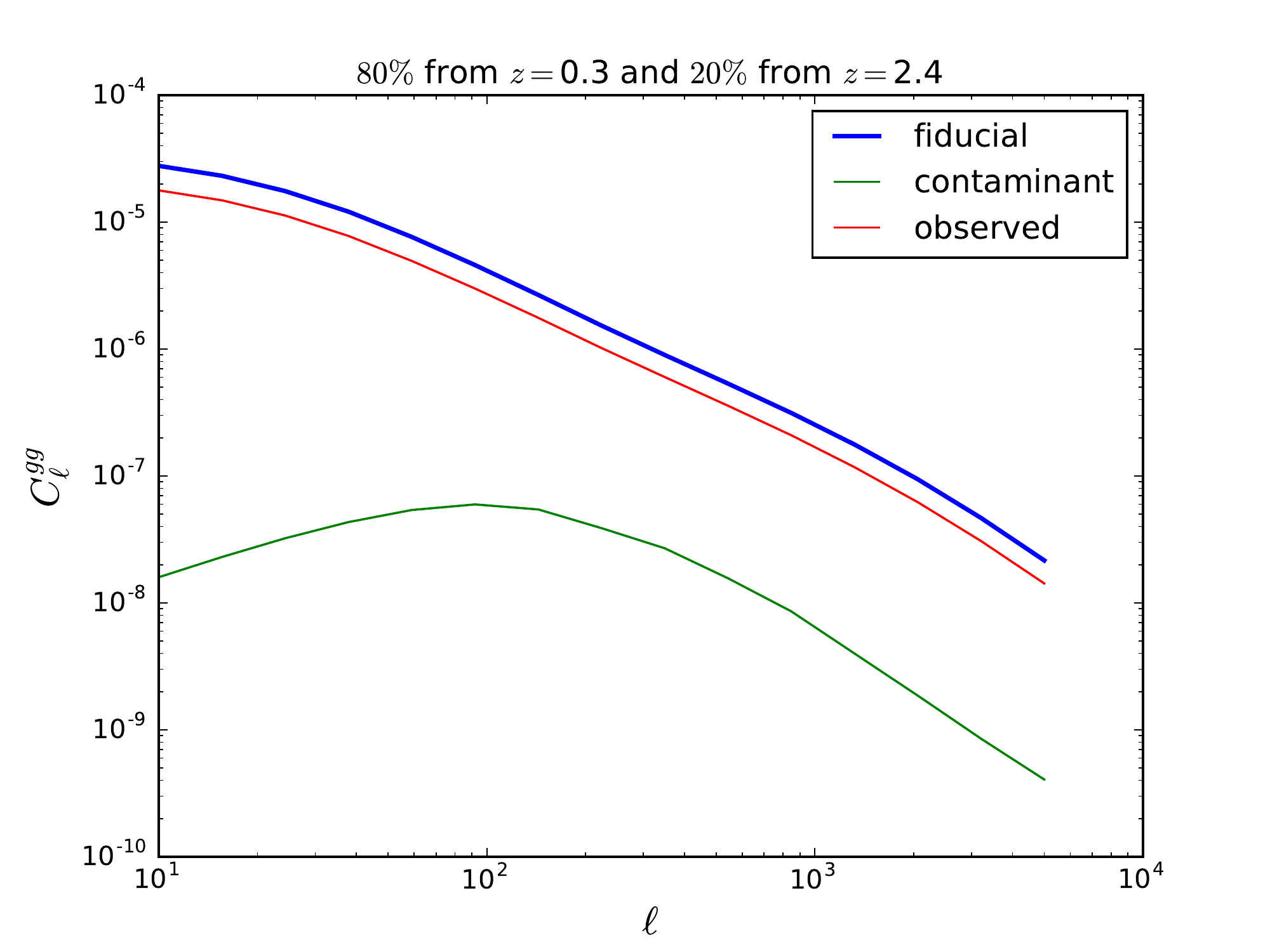}
\includegraphics[width=0.49\columnwidth]{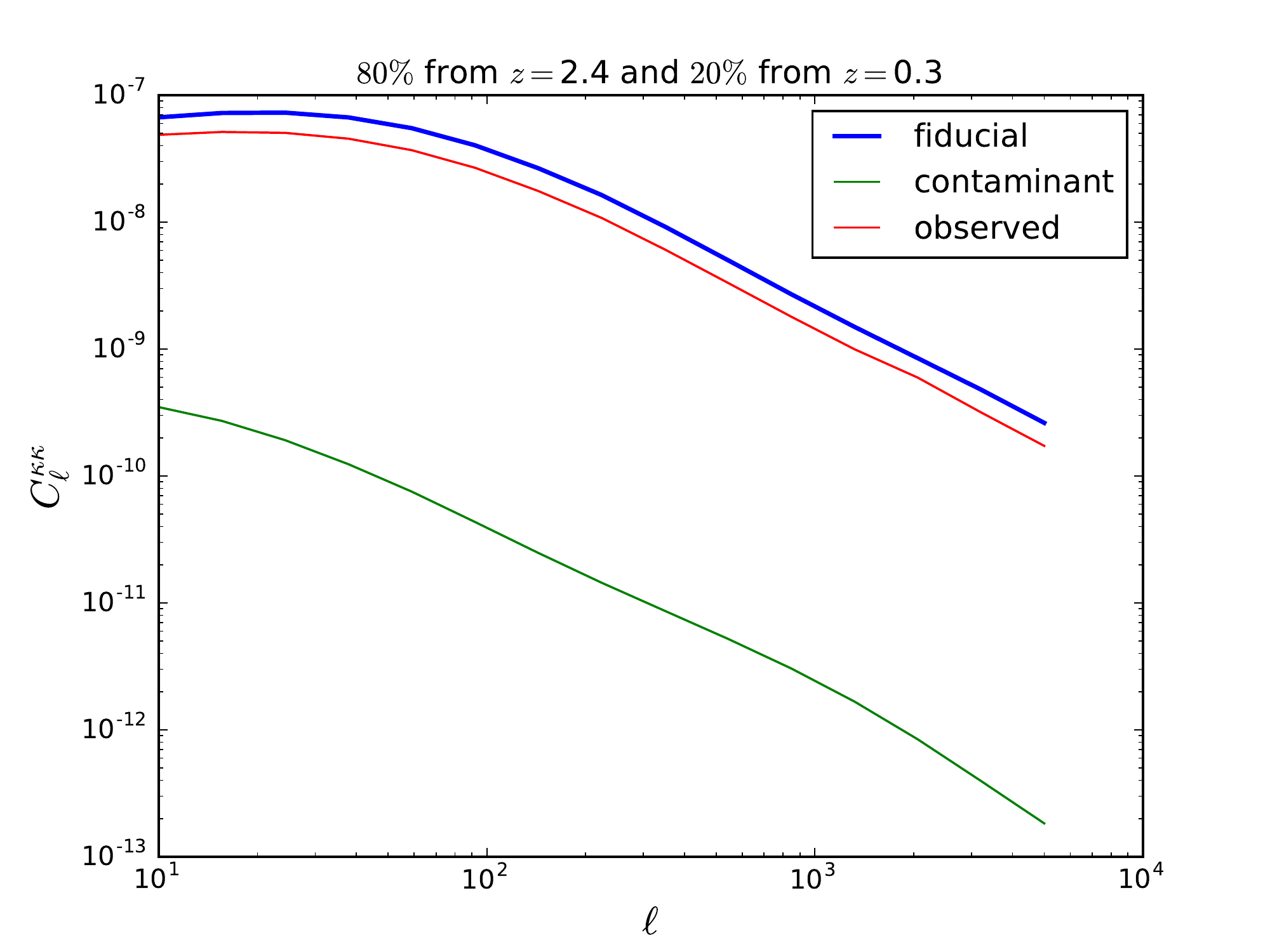}
\includegraphics[width=0.49\columnwidth]{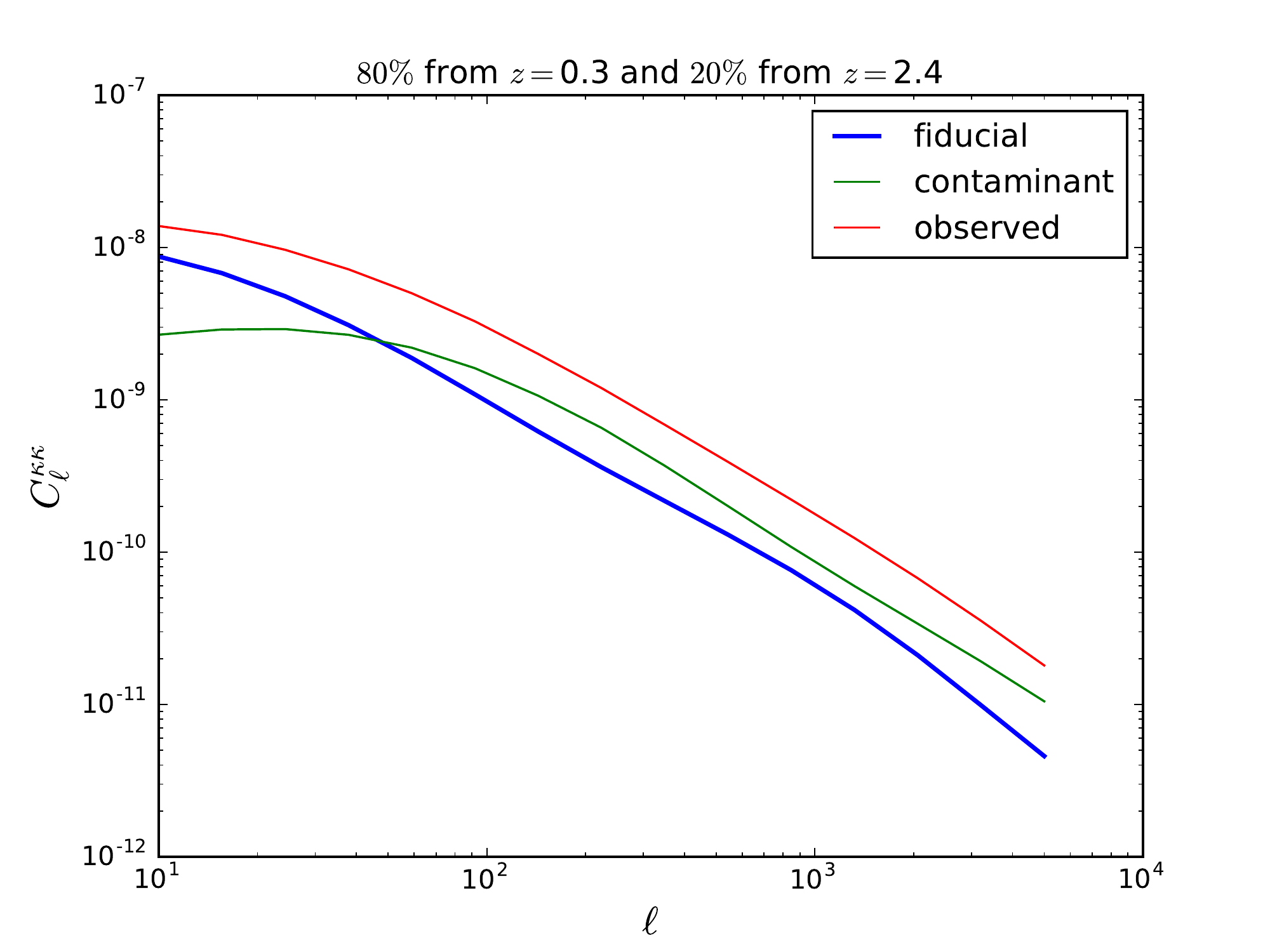}
\caption{
Example of shape dependence, for a 20\% outlier fraction ($c_{21} = 0.2$), between the lowest and highest redshift bins in our setup. We show clustering (top) and lensing auto-power spectrum (bottom).
}
\label{fig:shapeinfo}
\end{figure}

In Fig.~\ref{fig:shapeinfo}, we show the effect of scattering from low $z$ to high $z$ and vice-versa for both clustering and lensing.
From this figure, we conclude the following:
\begin{itemize}
    \item 
    For low-redshift clustering, contamination by galaxies at high redshift decreases the power by a factor $\approx 2c_{21}(1-c_{21})$, if the clustering power is approximately independent on redshift.
    \item 
    For high-redshift clustering, contamination by low-redshift galaxies can significantly increase the measured power on large scales, while suppressing the smaller scales by a factor of $\approx 2c_{21}(1-c_{21})$.
    This is because the peak in the power spectrum is moved to lower $\ell$ for the lower redshift sample, therefore increasing the power there. We note that this not only can affect the determination of cosmological parameters, but if unmitigated, it's likely to bias attempts to measure local $f_{\rm NL}$ through scale-dependent bias \cite{Dalal2008}.
    \item For lensing the effect is reversed: 
    high-redshift contaminants in a low redshift bin enhance the lensing power spectrum, potentially by a large factor.
    This is because the amplitude of lensing is larger when a fraction of the sample is at higher redshift.
    \item Conversely, 
    low redshift contaminants in a high-redshift tomographic bin suppress the lensing power, by a fraction approximately $c_{21}(2-c_{21})^2$.
\end{itemize}

Furthermore, photometric redshift errors dramatically modify cross-correlations.
Indeed, the clustering cross-correlations, which should be null for non-overlapping tomographic bins, are now non-zero.
In summary, the different dependence on outliers of the galaxy clustering and lensing spectra, together with the large effects on cross-correlations, provide valuable information on the unknown photo-z errors.
This information is automatically included in the full Fisher forecast of Sec.~\ref{sec:full_fisher}.

\section{Conclusions}

We performed a realistic Fisher forecast for the Rubin Observatory LSST $3\times 2$ analysis, which consists of the auto and cross-correlations of galaxy number density and shear, and for the LSST + CMB lensing $6\times 2$ analysis, which additionally includes all the auto and cross-correlations with CMB lensing from a CMB S4-like experiment.
We focused on the posterior constraints on Gaussian photo-z errors (shifts and scatters for each tomographic bin) and on photo-z outliers (a leakage fraction $c_{ij}$ for each pair of tomographic bins), producing a very flexible description of the photo-z uncertainties.
We considered the cosmological constraints for $w_0w_a$CDM, $\Lambda$CDM + $M_\nu$ and $\Lambda$CDM + curvature.
We simultaneously marginalized over galaxy bias and shear bias in each tomographic bin.

We showed that fixing the photo-z outlier fractions $c_{ij}$ at values off by only an additive 0.0005 can bias the dark energy equation of state and neutrino masses by more than their statistical uncertainties.
We thus concluded that photo-z outliers need to be marginalized over in a realistic analysis.

We demonstrated that both the Gaussian photo-z and outlier parameters can be self-calibrated, at a level close to the LSST requirements.
While the cost is negligible (percent increase in uncertainties) for neutrino masses, curvature of $w_0$, it is significant when both $w_0$ and $w_a$ are varied (65\% degradation).

We included the ``null'' cross-correlations usually discarded in similar forecasts and analyses. Despite contributing only $10\%$ of the total SNR, these correlations improve the photo-z constraints by large factors, up to an order of magnitude for some of the outlier parameters.
This also leads to a percent improvement in neutrino masses, and a 15\% improvement in dark energy.
This non-negligible improvement in $w_0$ and $w_a$ is likely not the result of the improved photo-z constraints though, but instead of the additional redshift resolution provided by these correlations, whose effective redshifts are in-between the ten tomographic bins.

We showed that using the same galaxy sample as lens and source improves the photo-z posteriors significantly, but only produces a minor $\sim 10\%$ improvement in cosmological parameters.

Because of the large galaxy lensing SNR in this analysis (428) compared to CMB lensing (282), adding CMB lensing only produces limited improvements, of order $10\%$, on photo-z and cosmology posteriors uncertainties.
This is likely the result of our stringent scale cuts ($\ell_\text{max}=1000$ instead of the usually assumed 3000), which degrades the SNR in CMB lensing more than shear.

Because our analysis does not include any multipoles higher than $1000$ in lensing and any scales smaller than $k=0.3$ $h$/Mpc in clustering, it should be robust to nonlinear bias, and insensitive to the uncertain nonlinearities and baryonic effects on the matter power spectrum.
We further included magnification bias, but with a known magnification bias in each tomographic bin.

To gain intuition on these results, we deconstructed the Fisher forecast and highlighted that the internal information on the photo-z errors comes from three sources: the algebraic constraint that the 2d observables are all related to the same 3d matter density field, the decorrelation of the matter density field for differing redshift bins, and the rigidity in the power spectrum shape for $\Lambda$CDM and the extensions considered here.

\section{Future directions}

While we have explored the photo-z self-calibration problem with an increased level of realism compared to previous work, several questions remain to be answered.
Here we discuss some of these possible extensions.
\begin{itemize}
    \item Photometric outliers can also significantly bias measurements of local $\fnl$ from scale-dependent bias. The self-calibration technique discussed in this paper can potentially provide a powerful mitigation strategy. The power spectrum on small scales can be used to solve for the outlier fraction, while the large scales can provide a robust measurement of $\fnl$.
    \item In this work, we have not used any prior on the relative size of the various $c_{ij}$, and we have not explored whether some outlier fractions $c_{ij}$ are more crucial to constrain than others. In practice, $z_\text{photo} - z_\text{true}$ plots show which outlier fractions are expected to be larger. One could use this as a prior on the $c_{ij}$. This would reduce the effective number of free parameters, and would probably improve the self-calibration.
	\item Common systematics can affect multiple redshift bins, such as selection effects, variations in photometric calibration and reddening or extinction. These are expected to affect the clustering and lensing signals differently, so their auto and cross-correlations should provide partial mitigation of these systematics \cite{2019JCAP...04..023M}. It is also often possible to obtain spatial templates for many of these (i.e. maps of the systematics), and deproject the affected modes with little effect on the statistical uncertainties.
	\item Since neighboring tomographic bins overlap in redshift, due to the photo-z errors, there could be a non-zero shot noise term in the cross-correlation between neighboring bins. This effect should be modeled by fitting for an arbitrary additive constant for each pair of bins, potentially with a tight prior on its size. 
	\item We have assumed that the linear galaxy bias is only a function of the true galaxy redshift and not of the photometric redshift. If galaxies with catastrophic photo-z failures are physically different from the other galaxies at the same redshift (i.e. different colors, host halo masses, etc), they should have a different clustering bias. So the bias likely depends on both true redshift and photometric redshift.
	\item Following the DESC SRD \cite{Alonso2018}, we modeled the redshift dependence of the galaxy bias as $b(z)\propto 1/D(z)$, corresponding to a passively evolving galaxy population. Future work should quantify the uncertainty on this \cite{2018MNRAS.477.1664G} and propagate it to parameter uncertainties. Note that this is not specific to our photo-z self-calibration method, and is instead a general concern for future 3x2pt analyses.
	\item As in the DESC SRD and most current forecasts, we assumed linear biasing at all redshifts and on all the scales included in the analysis ($k \leq 0.3h/$Mpc). Exploring the validity of this assumption would be very valuable.
	\item We did not propagate the uncertainty in the faint-end slope of the galaxy luminosity function, which determines the amplitude of magnification bias, although we have included a fiducial value for this slope and calculated the effect of magnification self-consistently in all auto and cross correlations.
	\item We have neglected the non-Gaussian contributions to the covariance matrix, the effect of intrinsic alignments and the baryonic effects on the lensing power spectrum. 
	As we discuss in Sec.~\ref{sec:cov}, Ref.~\cite{Alonso2018, Barreira2018} show that the supersample and trispectrum terms in the covariance should be negligible in this analysis, especially given our conservative $\ell_{\rm max} = 1000$ choice.
	While we believe they can be modeled and shouldn't affect the validity of the method presented here, they should be included in a real analysis and may degrade the cosmological constraints. Moreover, our conservative choice of $\ell_{\rm max} = 1000$ for lensing ensures that the contributions due to these effects are subleading.
	\item We have not included the effect of source blending, since in our formalism, the total number of galaxies is conserved. It should be possible to extend the analysis by allowing the total number of galaxies to change (within some prior region), and therefore capturing the effect of blends.
	\item Further work is required to determine the best parametrization for the deviation from a fiducial redshift distribution due to photo-z errors (e.g., quantiles \cite{2018AJ....156...35M}). If too few parameters are used, the resulting constraints could be biased because the photo-z model would not be flexible enough to encompass the truth.
	\item Using the same galaxy sample for lenses and sources poses a data analysis challenge, as these samples are generally selected in different ways, with different optimization schemes and weights.
	Exploring how to optimally select a single sample for sources and lenses would be very interesting.
	\item Since the clustering and many of the cosmic shear power spectra are cosmic variance limited rather than shot/shape noise limited, it should be possible to split the galaxy samples (based on color, Sersic index, etc) without losing much SNR. This would enable consistency checks on the redshift distributions and cosmology.
\end{itemize}

\section*{Acknowledgments}

We thank David Alonso, Alex Amon, Enea di Dio, Cyrille Doux, Patrick McDonald, Chihway Chang, Tim Eifler, Daniel Gruen, Benjamin Joachimi, Elisabeth Krause, Alex Krolewski, Jeffrey Newman, Andrina Nicola, Kimmy Wu, Wayne Hu, Bhuvnesh Jain, Markus Rau, Sukhdeep Singh, David Spergel and Martin White for useful discussions.
E.S. is supported by the Chamberlain fellowship at Lawrence Berkeley National Laboratory. S.F. is supported by the Physics Division of Lawrence Berkeley National Laboratory.
This material is based upon work supported by the National Science Foundation under Grant Numbers 1814370 and NSF 1839217, and by NASA under Grant Number 80NSSC18K1274.


\bibliographystyle{JHEP}
\bibliography{references}

\appendix

\section{Flexibility of the photo-z parametrization}
\label{app:dndz_pca}

In this appendix, we present a general method to determine which modes of the redshift distribution of a tomographic bin need to be marginalized over. 
Intuitively, these are the modes which are small-enough perturbations of $dn/dz$ to be undetected by the usual photo-z calibration, yet cause large-enough changes to the observed power spectra to bias the cosmological inference.
This method relies on principal component analysis, and automatically provides a well-motivated parametrization of redshift errors, in the form of the principal components.
We implement it in the case of a single tomographic bin, and leave the natural generalization to multiple tomographic bins to future work.
This method also enables assessing whether a given parametrization of the redshift errors (e.g. shift, width and outlier fractions like in our main analysis) is sufficiently flexible to avoid biasing the cosmological inference.

\begin{figure}[h!!!!]
\centering
\includegraphics[width=0.49\columnwidth]{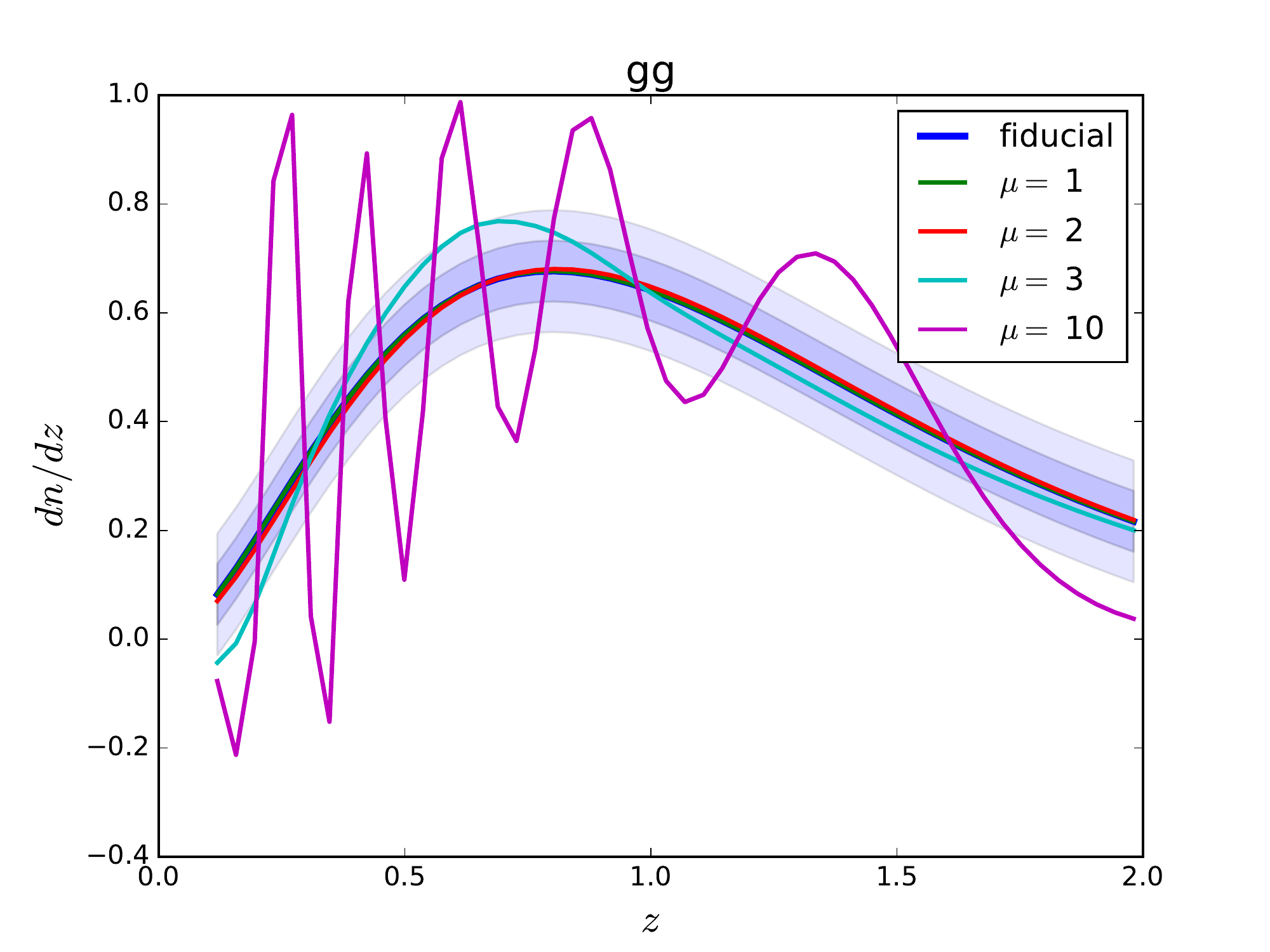}
\includegraphics[width=0.49\columnwidth]{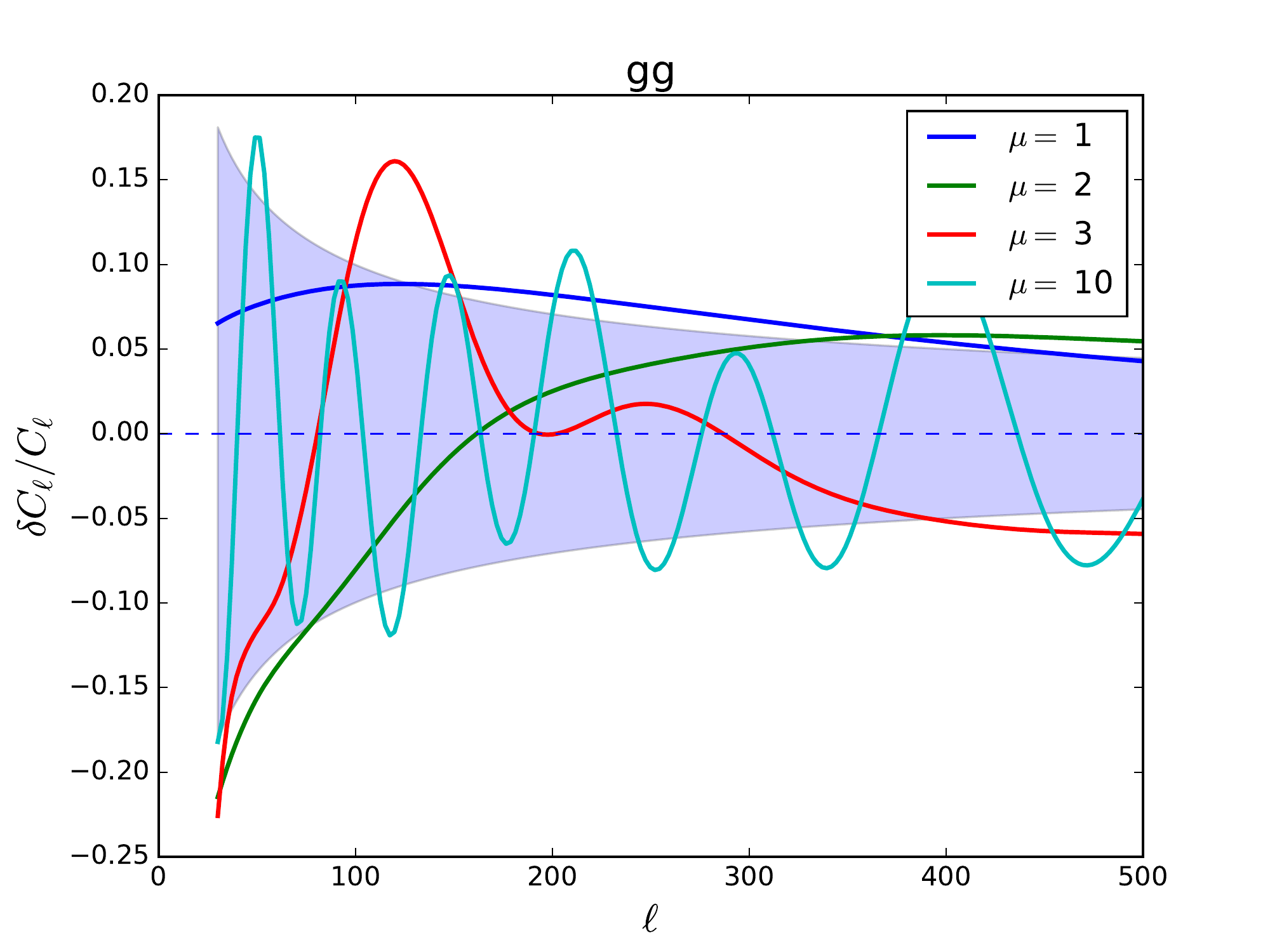}
\caption{
{\bf Left panel:} Effect of the first 3 principal components (PCs) $S_\mu(z)$ (ones with largest eigenvalue $\sigma^{-2}_{\mu}$) on $dn/dz$, as well as the 10th. The amplitude is set by $ \sigma_\mu$. The shaded bands correspond to a 10\% and 20\% uncertainty on $dn/dz$. This highlights that the first 3 PCs are unlikely to be constrained by photo-z alone, while the 10th is. 
{\bf Right panel:} Effect of the $dn/dz$ shown in the left panel on the galaxy power spectrum.  While the changes in $dn/dz$ vary significantly between each PC, the overall effect on the measurement of $C^{gg}_{\ell}$ is the same.  The shaded band represents the cosmic variance error.
}
\label{fig:pca}
\end{figure}

Consider a binned redshift distribution such that $dn/dz$ at redshift $z$ is given by
\begin{equation}
    \frac{dn}{dz} (z) = (1 + A(z)) \left(\frac{dn}{dz}\right)^{\rm fid} (z)
\end{equation}
and normalized such that at all redshifts the fiducial amplitude $A^{\rm fid}(z) = 0$.
Given tracers $X, Y \in \{g_i, \gamma_i, \kappa_{\rm CMB}\}$, we can define the response of their correlation $C^{XY}_\ell$ to a perturbation in a bin at redshift $z_i$ as
\begin{equation}
    \mathcal{R}_{\ell, i}^{XY} = \frac{\partial C^{XY}_\ell} {\partial A(z_i)}.
\end{equation}
The small perturbation $A(z)$ in the redshift distribution thus causes the following change in the observed power spectrum, to first order:
\begin{equation}
    \delta C^{XY}_\ell = \sum_{i = 1}^{N_z} \mathcal{R}_{\ell, i}^{XY} \delta A(z_i) 
    .
\end{equation}
One way to assess how much the redshift perturbation $\delta A(z_i)$ affects the observed power spectrum, within the measurement uncertainty,
is to ask how well the former is constrained by the latter.
We answer this question by forecasting the covariance matrix of the $\delta A(z_i)$:
\begin{equation}
    \langle \delta A(z_i) \delta A(z_j) \rangle = \left( F^{-1} \right)_{ij},
\end{equation}
where $F$ is the Fisher matrix for the $\delta A(z_i)$ given the observed power spectra. It is given by:
\begin{equation}
    F_{ij} = \frac{f_{\rm sky}}{2} \sum_{\ell} \frac{2\ell+1}{\left(C^{XY}_\ell + N^{XY}_\ell \right)^2} \mathcal{R}_{\ell, i}^{XY} \mathcal{R}_{\ell, j}^{XY} 
\end{equation}
Diagonalizing the Fisher matrix, we obtain the principal components $S_\mu(z_i)$:
\begin{equation}
    F_{ij} = \sum_{\mu = 1}^{N_z} S_\mu(z_i) \sigma_\mu^{-2} S_\mu(z_j)
\end{equation}
The redshift perturbation $A(z)$ can be expressed in terms of the principal components as
\begin{equation}
    A(z) = A^{\rm fid}(z) + \sum_{\mu = 1}^{N_z} m_\mu S_\mu(z)
\end{equation}
and the variance of the $\mu$-th principal component is given by the $\mu$-th eigenvalue: 
\begin{equation}
    \langle m_\mu m_\nu \rangle = \sigma^2_\mu \delta_{\mu \nu}
\end{equation}
We compute and display in Fig.~\ref{fig:pca} these principal components in the case of a single tomographic bin, given the galaxy clustering power spectrum.
We show the effect of each mode on the redshift distribution (left panel) of the tomographic bin and on the clustering power spectrum (right panel).
Each mode is shown with an amplitude $\sigma_\mu$, corresponding to the $1\sigma$ uncertainty on the mode from the observed power spectrum.
For each mode, we then compare the amplitude of the redshift perturbation (left panel in Fig.~\ref{fig:pca}) to the colored bands, corresponding to a $10\%$ and $20\%$ uncertainty on the redshift distribution. These numbers are chosen as the typical accuracy to which the redshift distribution of a tomographic bin is known \textit{a priori}, after the usual photo-z training and calibration.
The criterion is then simple: any mode within the colored bands should be marginalized over, while modes which go outside the band can be ignored.

For example, in Fig.~\ref{fig:pca}, the first three modes $\mu=1,2,3$ cause a change in $dn/dz$ smaller than $20\%$. This means that for these modes, the change in $dn/dz$ is small enough that it will not be ruled out by the photo-z priors (assumed to be about $20\%$ in $dn/dz$), yet these modes can cause a $1\sigma$ bias in the prediction of the power spectrum. These are therefore unconstrained by the data, and should be marginalized over.
On the other hand, the tenth mode $\mu=10$, causes changes in $dn/dz$ much larger than the $20\%$ band. This means that for this mode to cause a $1\sigma$ bias in the observed power spectrum, the mode would have to change the redshift distribution by a large amount, ruled out by the photo-z calibration. This mode can therefore be safely ignored.

This procedure identifies a finite number of modes which are small-enough perturbations of the $dn/dz$ that they are not ruled out by the photo-z calibration, yet cause large-enough changes to the measured power spectra to bias the inference.
These modes need to be marginalized over in the analysis.
In practice, any parametrization of the redshift errors is acceptable, as long as it has enough freedom to span the principal modes.
In this simple example, only three modes matter. They look like a shift, a change of width, and a change of skewness.
In the full analysis with ten tomographic bins, the modes will be more complex, involving relative changes of different tomographic bins (e.g., the centers of two tomographic bins moving towards or away from each other).
As the constraining power of the data increases, e.g., when including galaxy number density, lensing and CMB lensing for each of these bins, the number of modes that needs to be included will also increase.
We leave answering this question in detail to future work, but note that the number of relevant modes is unlikely to exceed the 110 free parameters included in the present analysis.

\section{Numerical derivatives and step sizes}
\label{app:full_fisher_plots}

\begin{figure}[H]
\centering
\includegraphics[width=0.45\columnwidth]{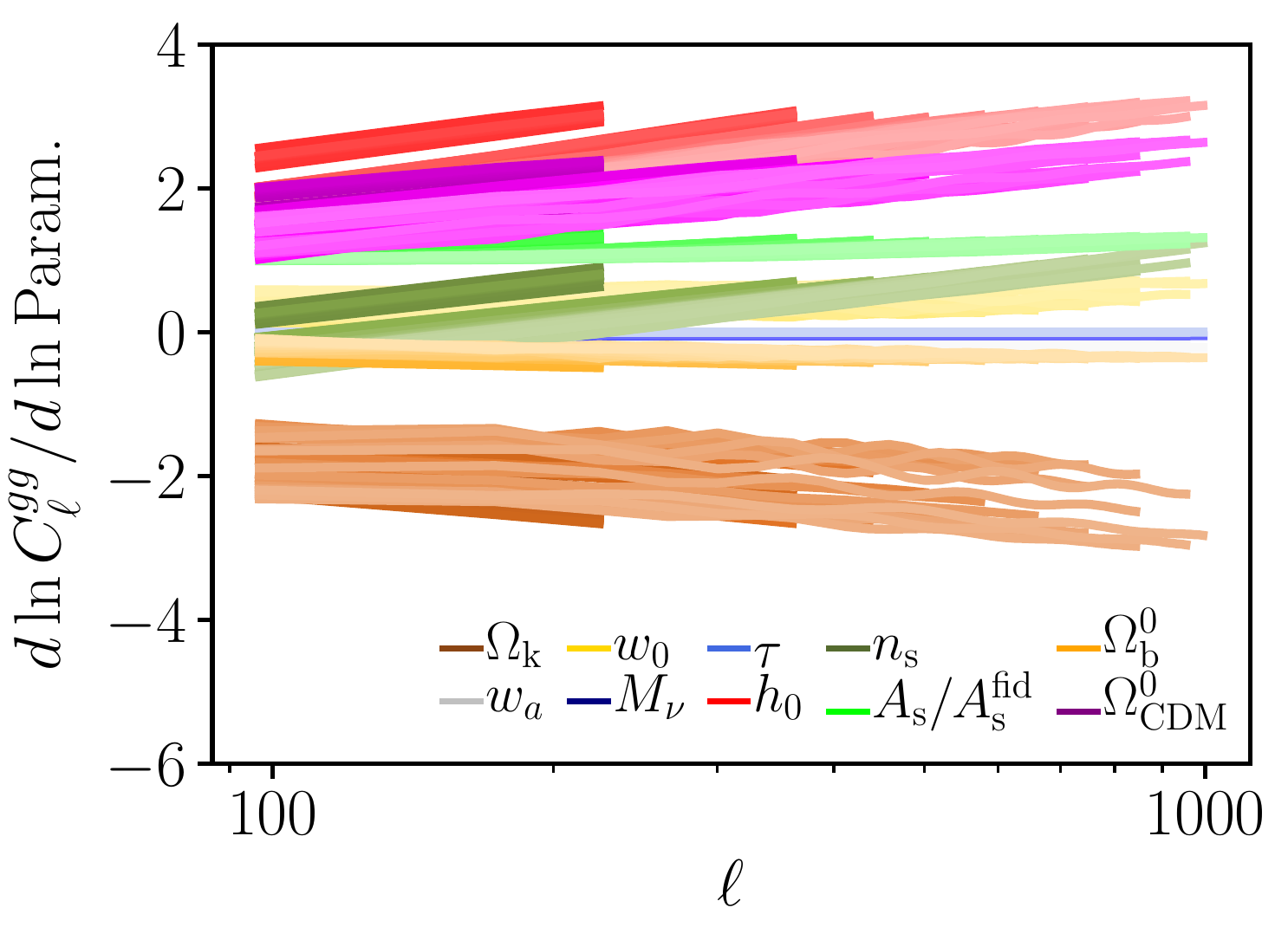}
\includegraphics[width=0.45\columnwidth]{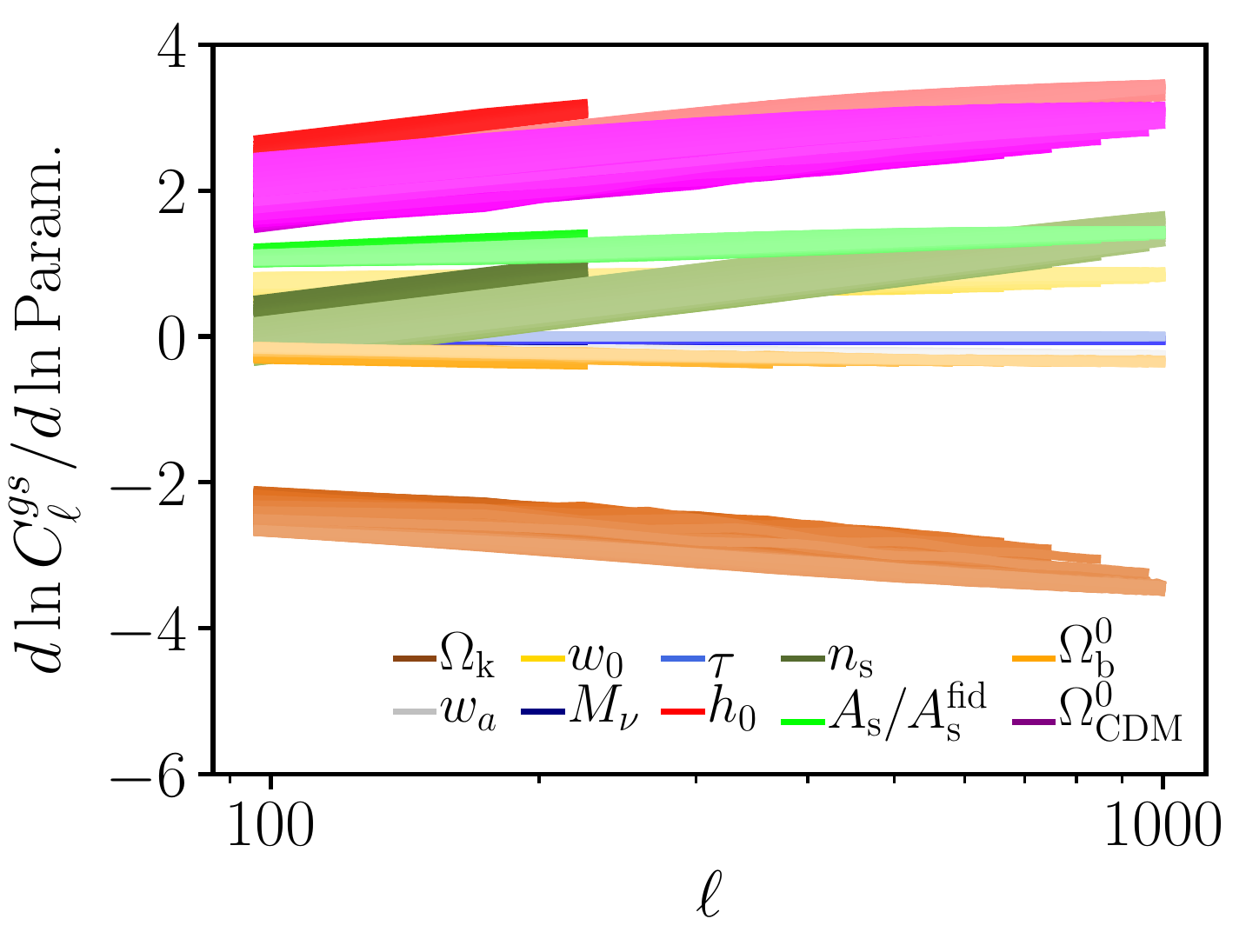}
\includegraphics[width=0.45\columnwidth]{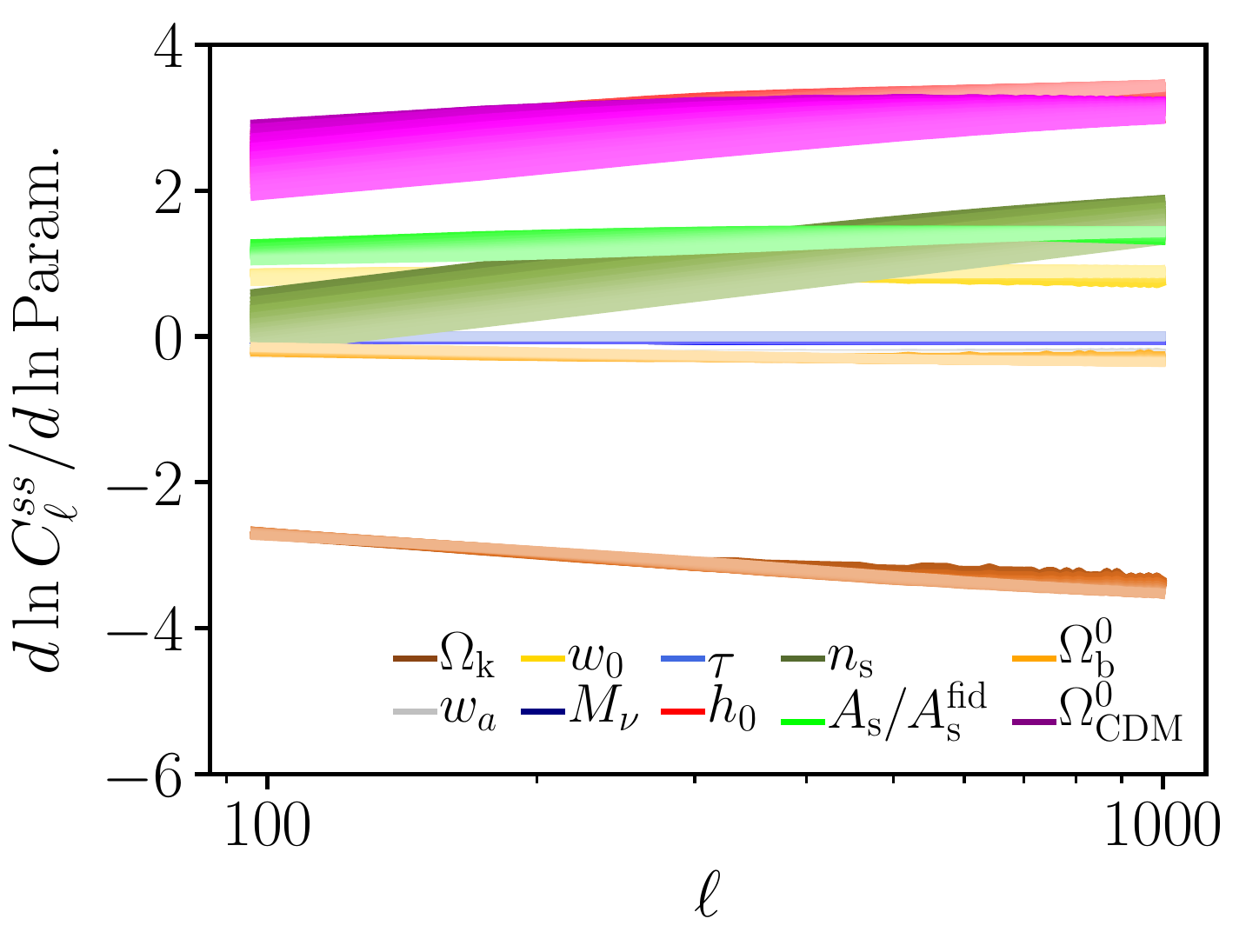}
\includegraphics[width=0.45\columnwidth]{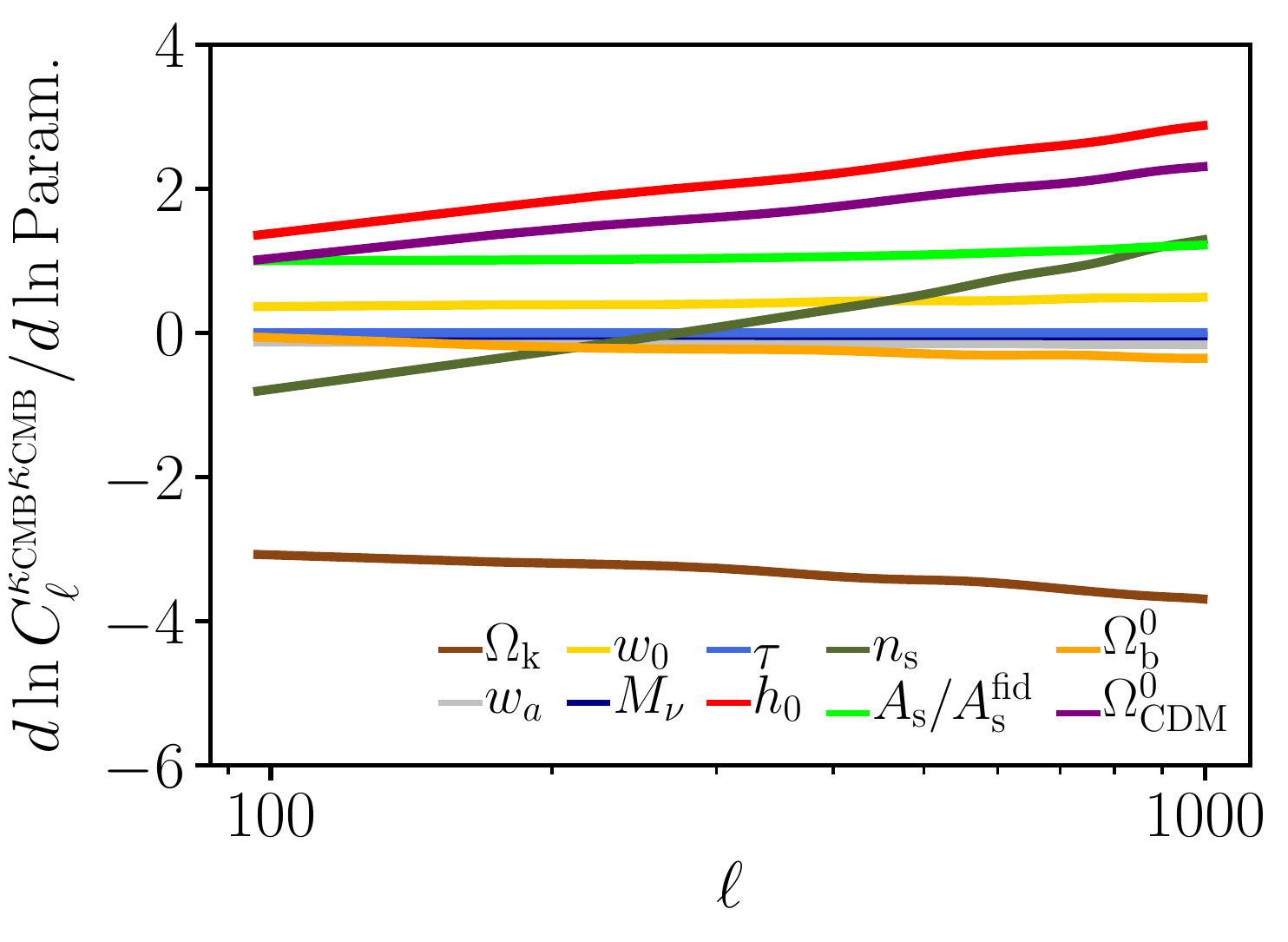}
\includegraphics[width=0.45\columnwidth]{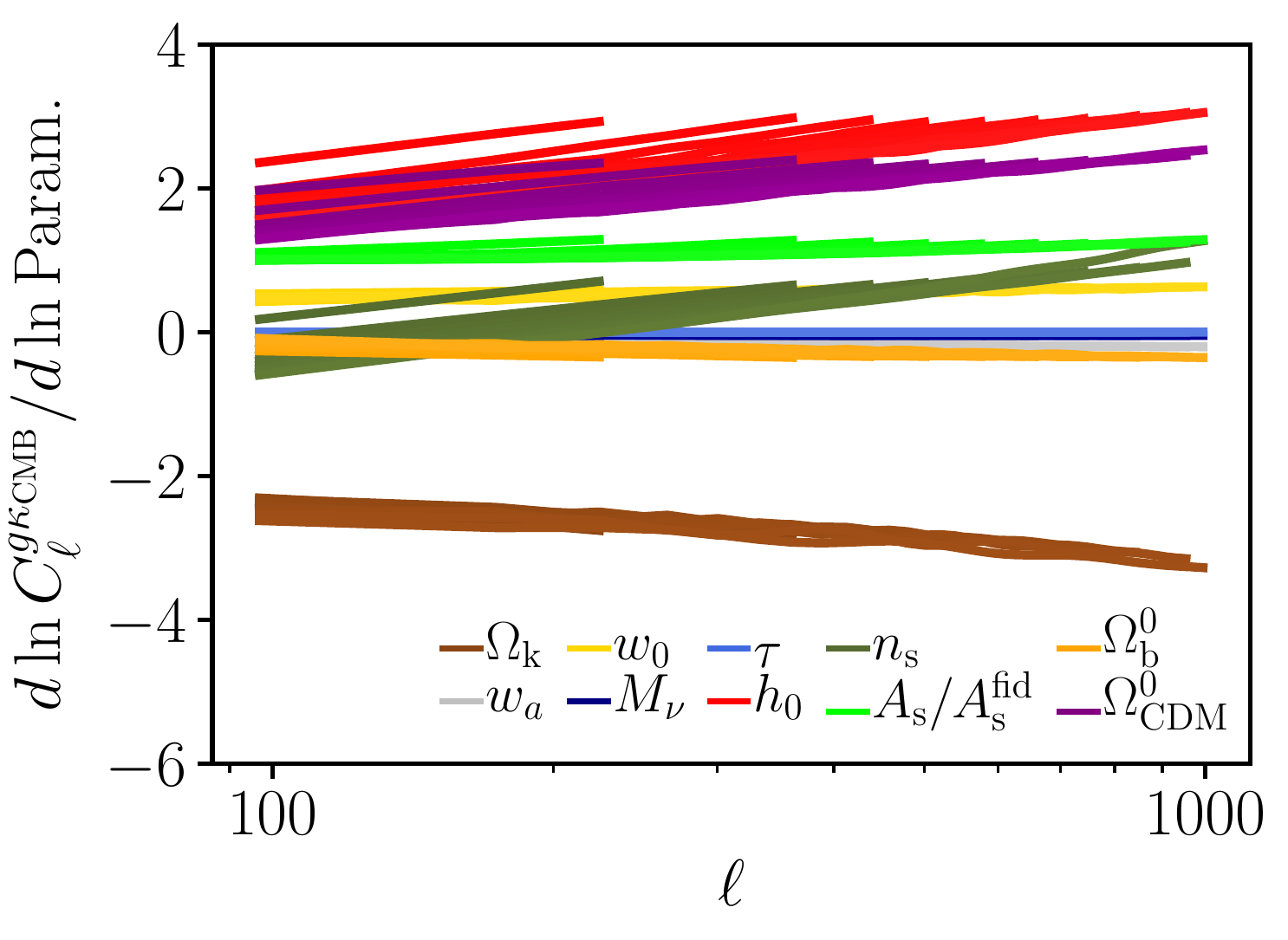}
\includegraphics[width=0.45\columnwidth]{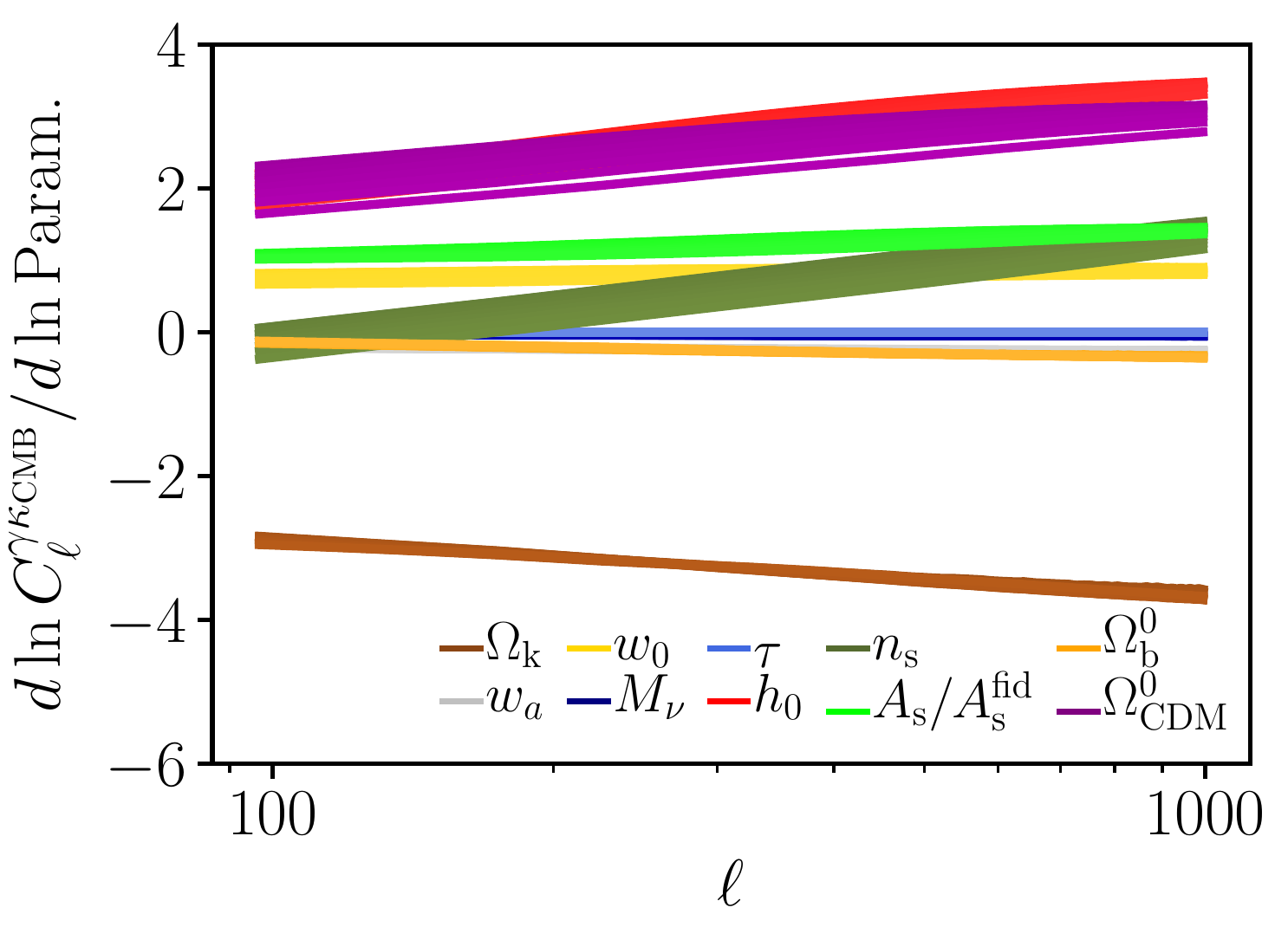}
\caption{
Derivatives of the data vector with respect to cosmological parameters.
Many features of the parameter dependence can be read off this plot.
A 1\% change in a given cosmological parameter causes a $y$\% change in the power spectra, where y is the value read on the plot.
A strong parameter dependence corresponds to a high absolute value of the derivative.
A positive value of the derivative corresponds to an observable growing with the parameter.
A horizontal curve indicates a multiplicative factor, and a slanted curve corresponds to a tilt in the observable, when the cosmological parameter is varied.
If two curves differ only by a multiplicative factor, then the corresponding cosmological parameters are perfectly degenerate.
}
\label{fig:dp2d_dcosmo}
\end{figure}

We show the derivative of the data vector with respect to the cosmological parameters in Fig.~\ref{fig:dp2d_dcosmo}.

A key check for any Fisher forecast is the numerical convergence of the derivatives.
In the absence of automatic differentiation in the Boltzmann codes, the derivatives are typically estimated with finite differences.
Here, we use the two-sided three-point estimate\footnote{This may seem like a two-point estimate, since only $f(x-h)$ and $f(x+h)$ are used. However, this is really a three-point estimate, in that it is the best linear combination of $f(x-h), f(x)$ and $f(x+h)$: the best coefficient for $f(x)$ is simply zero.}:
\beq
f'(x)
=
\frac{f(x+h) - f(x-h)}{2h}
- h^2 \frac{f^{(3)}(\xi)}{6}
\quad
\text{for some }\xi \in [x-h, x+h].
\label{eq:deriv_finite_diferences}
\eeq
We use the two-sided derivative for all cosmological parameters, including $M_\nu$ (with fiducial value 66 meV), $w_0$ and $w_a$.
If the step size is too large, the estimate becomes inaccurate because of the $h^2 \frac{f^{(3)}(\xi)}{6}$ and higher order terms. 
If the step size is too small, the estimate  is affected by the numerical accuracy of the Boltzmann code.
Our step sizes are inspired from \cite{2015PhRvD..92l3535A, 2018arXiv180902120Y} and shown in Table~\ref{tab:deriv_step_sizes}.
\begin{table}[H]
\centering
\begin{tabular*}{0.95\textwidth}{@{\extracolsep{\fill}}| c c l |}
\hline
Parameter & Fiducial value & Step size $h$ \\  
\hline
\hline
\multicolumn{3}{|c|}{$\Lambda$\textbf{CDM cosmology}} \\
$\Omega_m$ & 0.26 & 0.0066  \\ 
$\Omega_b$ &  0.049 & 0.0018 \\
$A_S$ & $2.105 \times 10^{-9}$ & $10^{-10}$ \\ 
$n_s$ & 0.9665 & 0.01  \\
$h_0$  & 0.6766 & 0.1 \\
$\tau$  & 0.0561 & 0.02 \\
\hline
\multicolumn{3}{|c|}{$\Lambda$\textbf{CDM extensions}} \\
$w_0$ &  -1 & 0.06 \\
$w_a$ &  0 & 0.15 \\
$M_\nu$ &  0.1 eV & 0.02 eV \\
$\Omega_k$ & 0 & 0.02  \\
\hline
\multicolumn{3}{|c|}{\textbf{Galaxy bias parameter}} \\
$b_0 ... b_9$ & 1  & 0.05 \\
\hline
\multicolumn{3}{|c|}{\textbf{Shear calibration}} \\
$m_i $ & 0 & 0.05 \\
\hline
\multicolumn{3}{|c|}{\textbf{Photo-z}} \\
$\delta z _i $ & 0 & 0.002 \\
$\sigma_{z_i} / (1+\langle z\rangle_i) $ & 0.05 & 0.003 \\
$c_{ij, i\neq j} $ & $0.1 / \left( N_\text{bins}-1 \right)$ & 0.05 \\
\hline
\end{tabular*}
\caption{
Step sizes $h$ used to estimate the derivatives of the data vector via finite differences (Eq.~\eqref{eq:deriv_finite_diferences}).
These steps are shown below to give percent accurate derivatives.
}
\label{tab:deriv_step_sizes}
\end{table}

\begin{figure}[H]
\centering
\includegraphics[width=0.8\columnwidth]{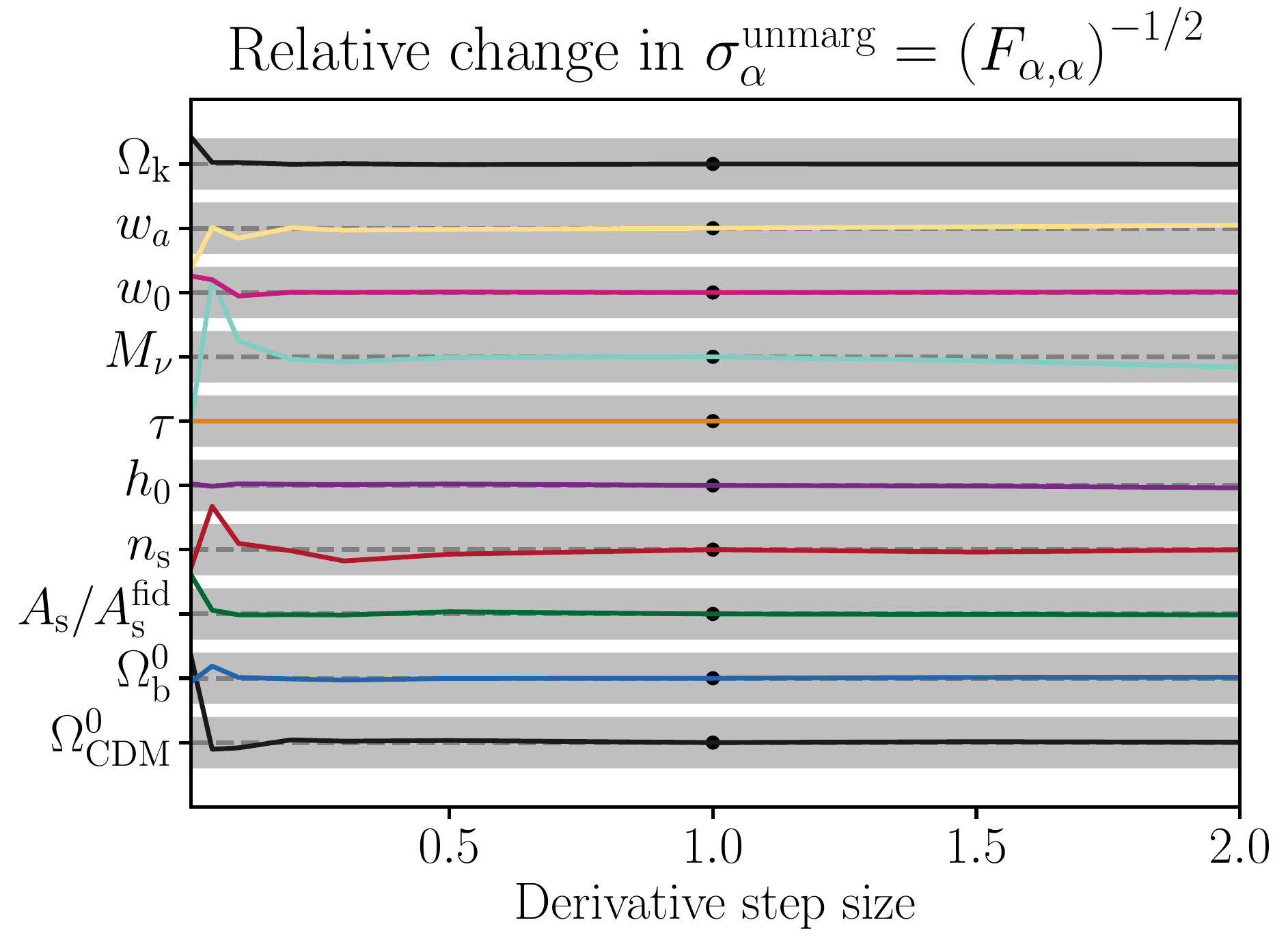}
\includegraphics[width=0.32\columnwidth]{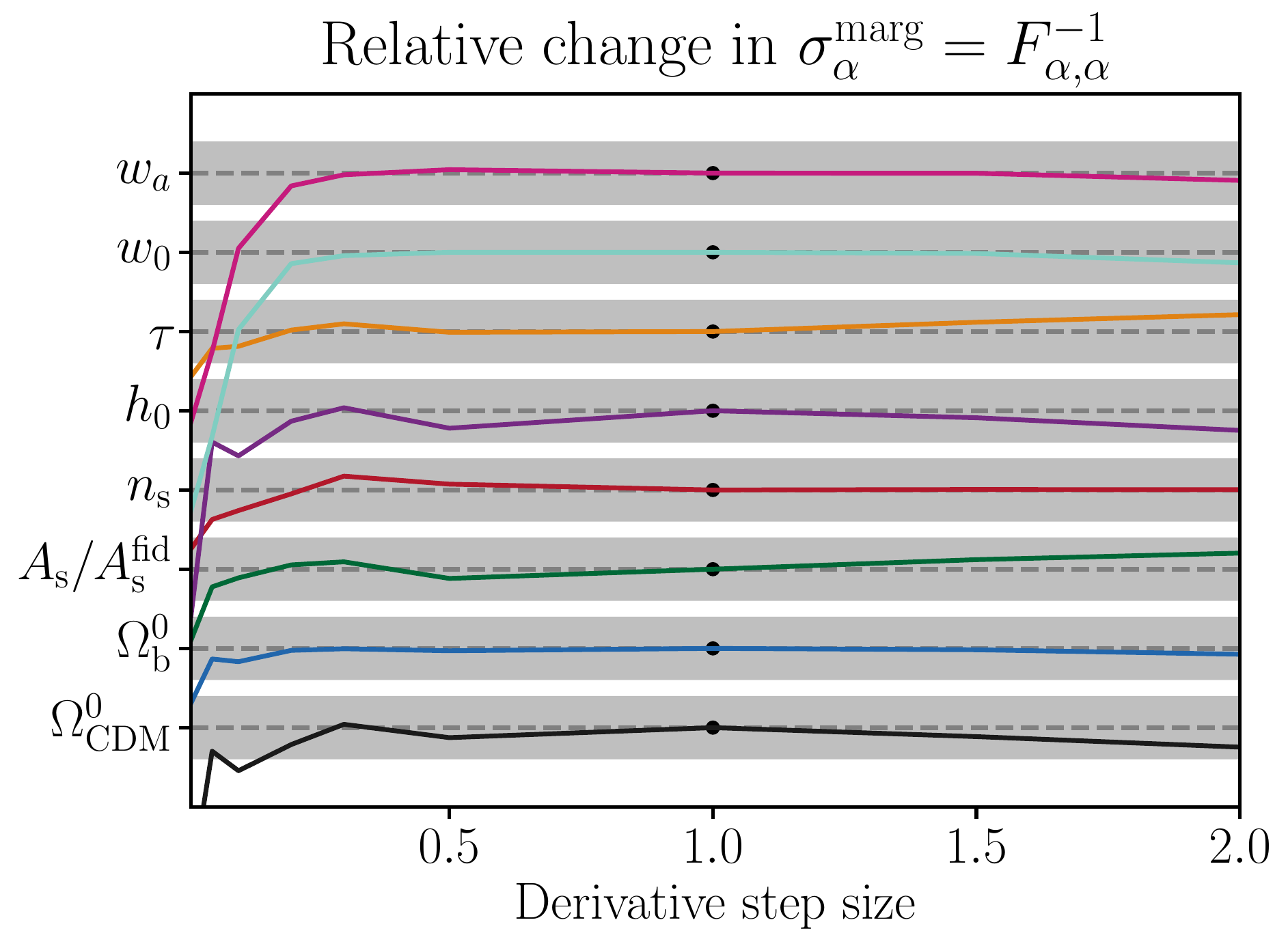}
\includegraphics[width=0.32\columnwidth]{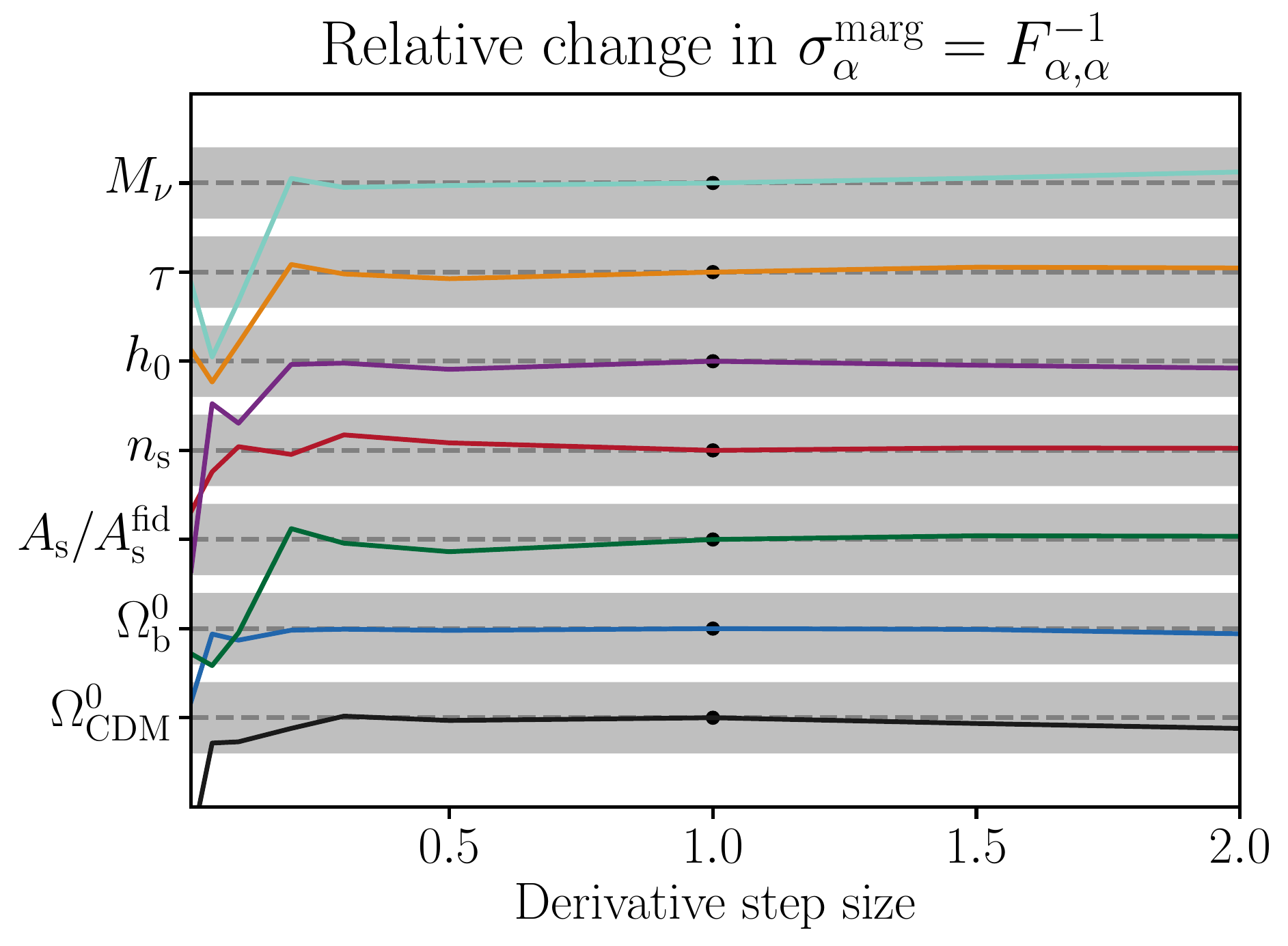}
\includegraphics[width=0.32\columnwidth]{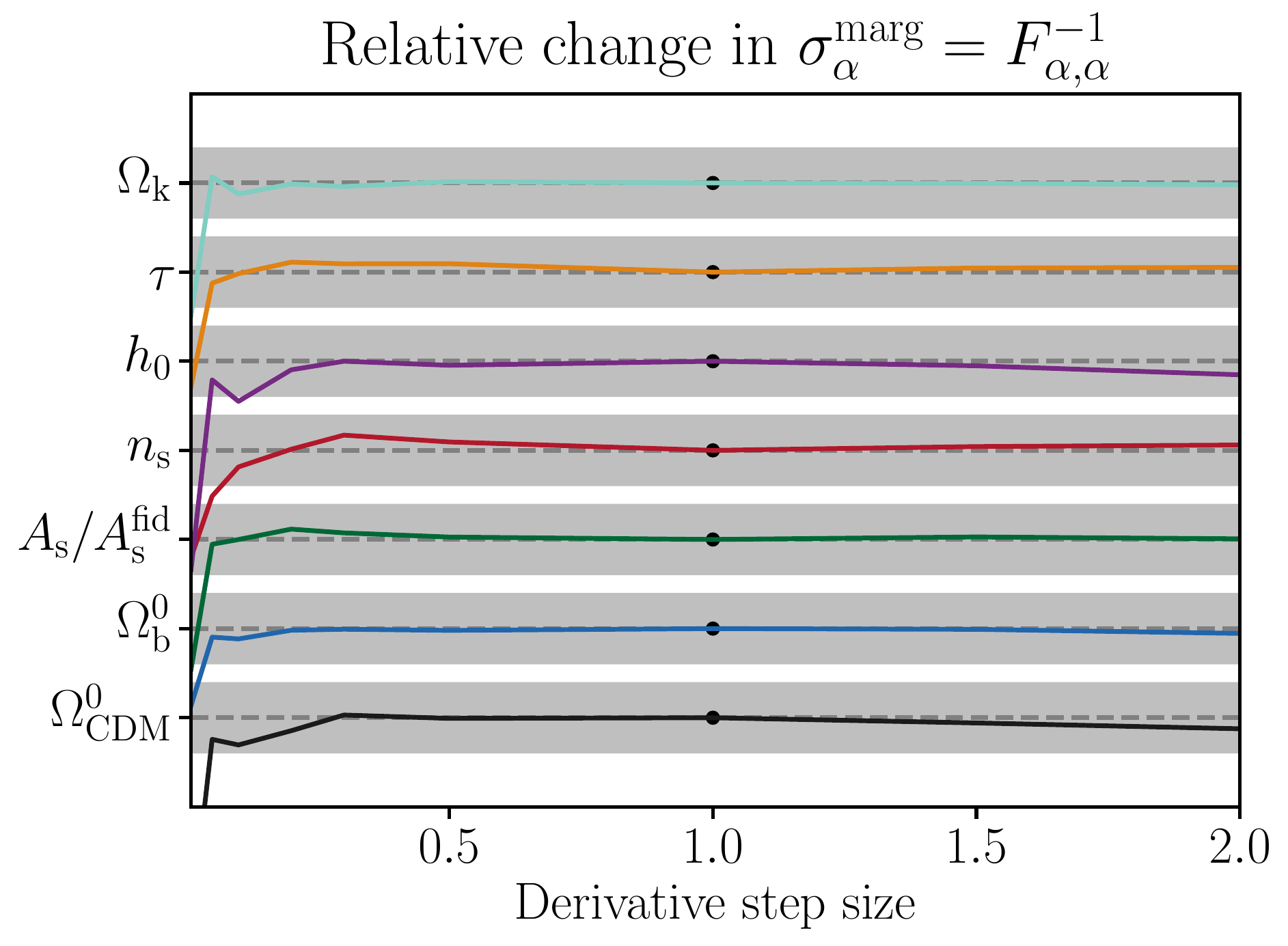}
\caption{
Convergence of the numerical derivatives.
In both panels, the grey band for each parameter corresponds to a $\pm 10\%$ relative variation of the derivative, compared to the fiducial case (solid points).
The x-axis is a scaling factor by which the fiducial step size of each parameter is multiplied.
For small derivative steps, inaccuracies come from the Boltzmann code. For high derivative steps, inaccuracies come from the higher order terms in the finite difference approximation.
For every parameter, the fiducial step size is indeed found to be appropriate.
\textbf{Top:} variation of the unmarginalized posterior constraint with step size. 
This is the relevant compression of the data vector into one number for each cosmological parameter, and without mixing the derivatives with respect to different parameters.
It shows that the individual derivatives are converged to about 1\%.
\textbf{Bottom:} variation of the marginalized posterior constraints with step size. This combines all the data vector elements and mixes all the derivatives with respect to all the parameters.
These are the key quantities whose accuracy we want to check.
It shows that the posterior uncertainties are converged to about 10\%.
}
\label{fig:convergence_dp2d_dcosmo}
\end{figure}

We therefore vary the step size for each cosmological parameter, and observe the convergence of the derivatives in Fig.~\ref{fig:convergence_dp2d_dcosmo}.

For a given cosmological parameter $\alpha$, the derivative vector $\partial \vec{D} / \partial \alpha$ has 11550 elements. We compress them into one number.
This number is the appropriately inverse-variance weighted combination, the diagonal Fisher matrix element, i.e.:
\beq
\sigma_\alpha^\text{unmarginalized}
=\left(F_{\alpha \alpha}\right)^{-1/2}
=
\left( \frac{\partial D^t}{\partial \alpha } \Sigma^{-1} \frac{\partial D}{\partial \alpha } \right)^{-1/2}.
\eeq
This quantity is meaningful, as it is the unmarginalized posterior constraint, and has the advantage of not mixing the derivatives with respect to different parameters. 
This is therefore a useful diagnostic to identify if one cosmological parameter step size is inappropriate. It is shown in the top panel of Fig.~\ref{fig:convergence_dp2d_dcosmo}.

Eventually, the quantity we care most about is the marginalized posterior uncertainty $\sigma_\alpha^\text{marginalized} = F^{-1}_{\alpha \alpha}$ for each cosmological parameter. 
We show them in the bottom panels of Fig.~\ref{fig:convergence_dp2d_dcosmo}, when marginalizing over all other parameters (cosmology and nuisance parameters).
This test is more stringent, in that it takes into account the loss of accuracy from inverting the Fisher matrix, and it includes the mixing of the derivatives with respect to every cosmological parameter.

Finally, we do not explore the convergence of the derivatives with respect to the nuisance parameters.
Indeed, the data is either linear or quadratic in the galaxy bias, shear bias, and photo-z parameters, making our finite difference estimate mathematically exact.

\section{Improving the covariance matrix condition number}
\label{app:cov_conditiong_number}

Performing the Fisher forecast for the $6\times 2$ analysis of LSST and CMB lensing requires inverting the data covariance matrix, which is more than $11,000 \times 11,000$.
At face value, the condition number of this matrix can be so large that inverting it numerically is impossible, even with 16 digit floating point numbers.
This issue can be avoided with the following precautions.
First, we choose an arbitrary ``galaxy convergence unit'' and ``CMB convergence unit'', i.e. we multiply these fields by an arbitrary number (here 10 for galaxy lensing and 3 for CMB lensing) so that the power spectra involving shear and galaxy  number densities now have very similar amplitudes. We rescale the covariance matrix elements correspondingly, such that the Fisher matrix is mathematically unchanged.
We also substitute $\ell C_\ell$ to $C_\ell$ in the data vector, and change the covariance matrix correspondingly. This allows all of the data vector entries to have very similar sizes.
Finally, we choose the binning in $\ell$ such that the number of Fourier modes per $\ell$-bin is the same for all $\ell$-bins. This makes all the diagonal covariance matrix entries similar.
Together, these procedures improve the conditioning number of the covariance and Fisher matrices by many orders of magnitude, making them well-conditioned.

\section{Other cosmological parameter sets}
\label{app:other_cosmo_params}

\begin{figure}[H]
\centering
\includegraphics[width=0.45\columnwidth]{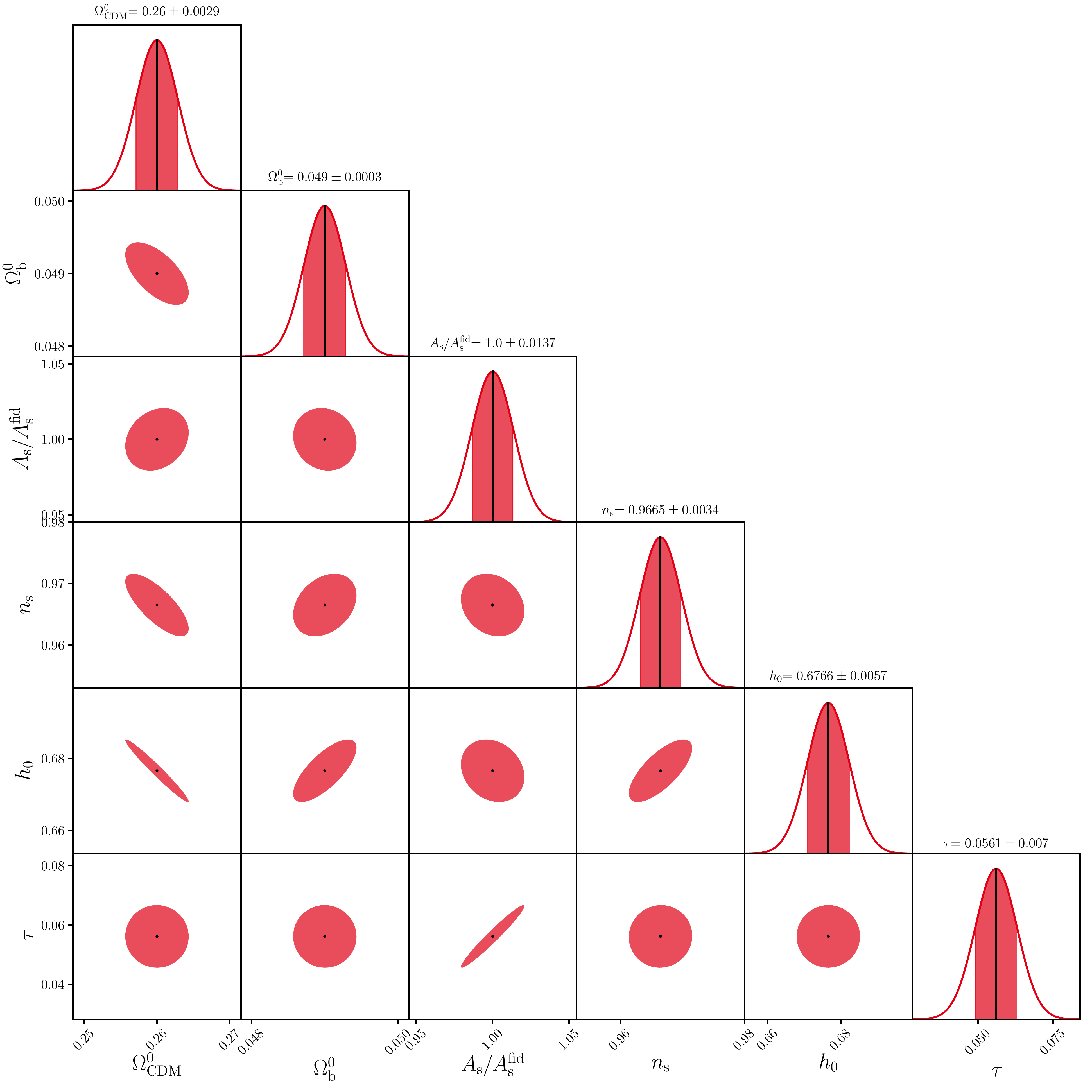}
\includegraphics[width=0.45\columnwidth]{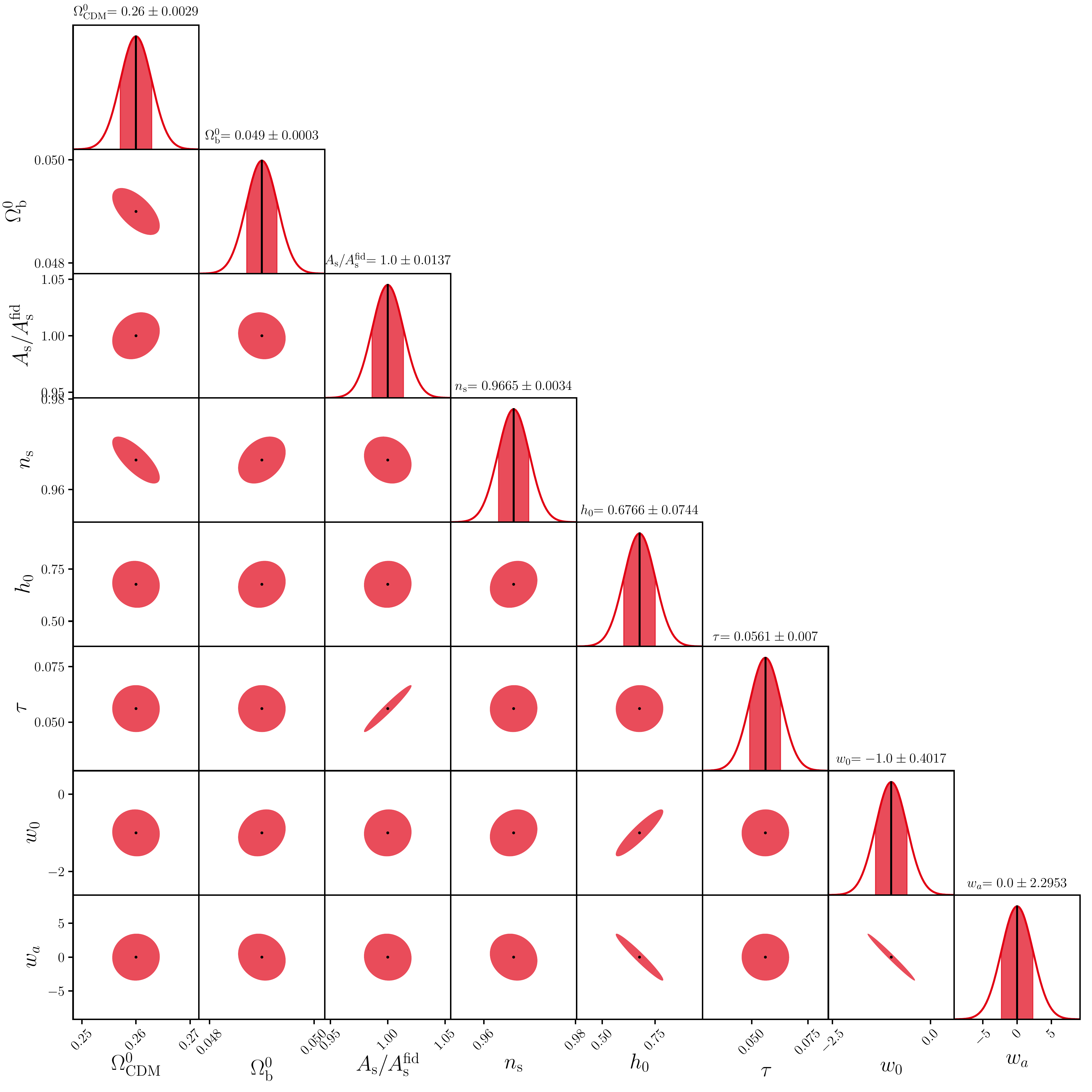}
\includegraphics[width=0.45\columnwidth]{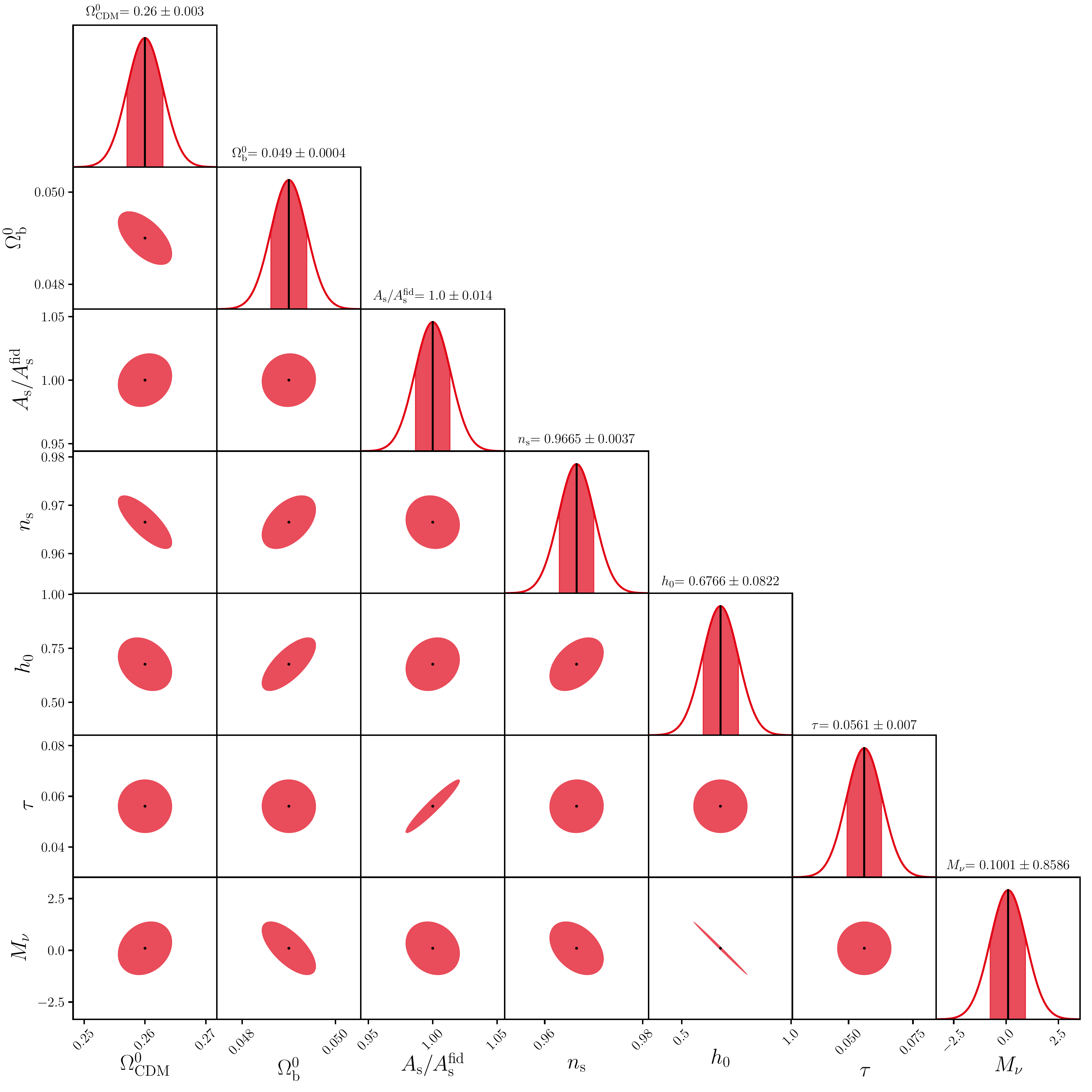}
\includegraphics[width=0.45\columnwidth]{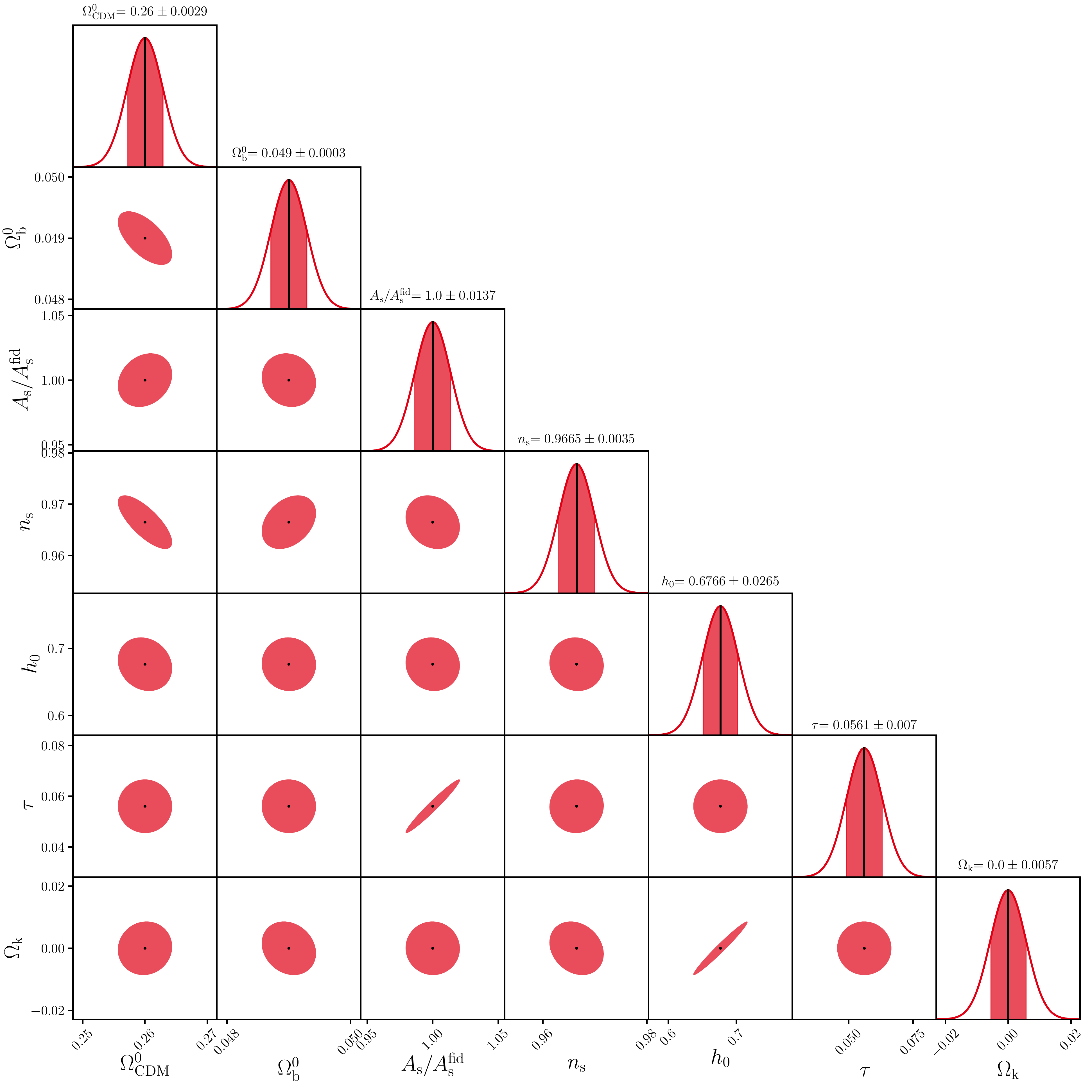}
\caption{
Planck-like prior used in this analysis, visualized for different parameter combinations: $\Lambda$CDM (top left), $w_0w_a$CDM (top right), $\Lambda$CDM + $M_\nu$ (bottom left) and $\Lambda$CDM + curvature (bottom right).
Comparing $\Lambda$CDM to each of the extensions, the constraints on the Hubble parameter are degraded by a large factor, due to a degeneracy with the dark energy equation of state, the masses of the neutrinos and curvature. The addition of the LSST $3\times 2$ data in this analysis fixes these degeneracies.
} 
\label{fig:planck_prior}
\end{figure}

\begin{figure}[H]
\centering
\includegraphics[width=0.95\columnwidth]{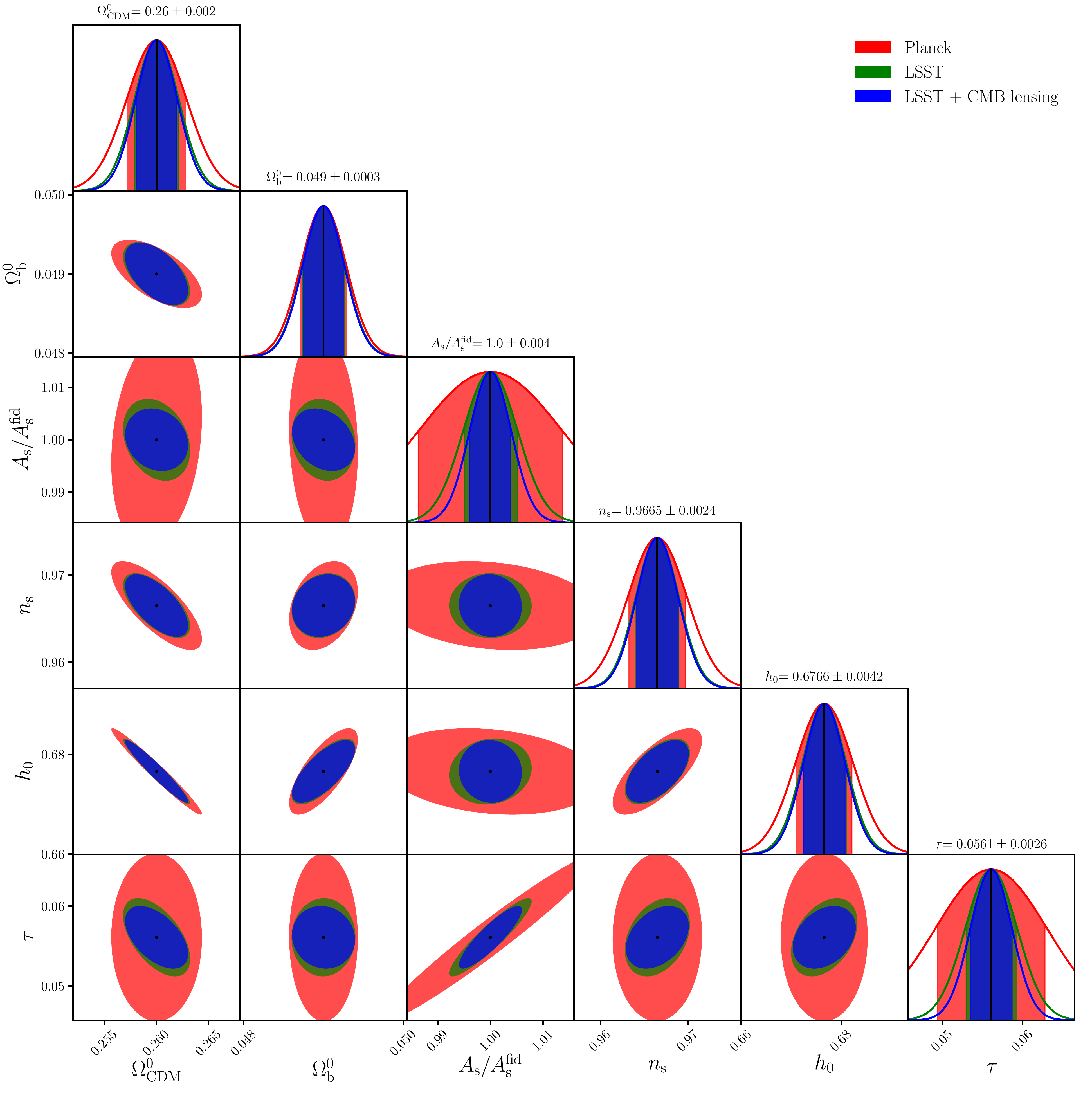}
\caption{
Baseline cosmological constraints for $\Lambda$CDM,
including priors on the Gaussian and outlier photo-z errors consistent with the LSST requirements, as well as Planck priors on cosmology.
The galaxy biases and shear biases are simultaneously marginalized over, with their fiducial priors.
} 
\label{fig:baseline_cosmo_contours_lcdm}
\end{figure}

\begin{figure}[H]
\centering
\includegraphics[width=0.95\columnwidth]{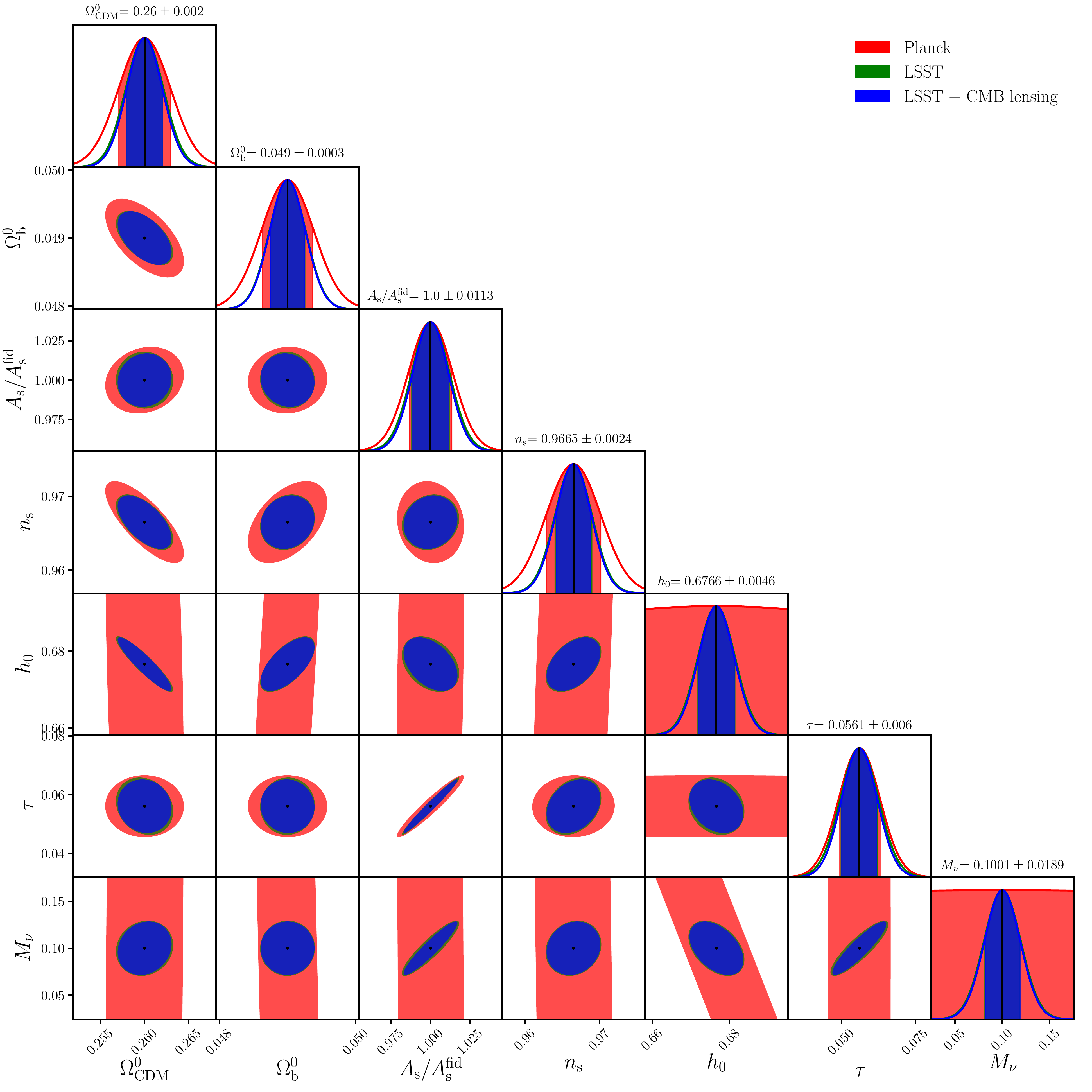}
\caption{
Baseline cosmological constraints for $\Lambda$CDM+$M_\nu$,
including priors on the Gaussian and outlier photo-z errors, consistent with the LSST requirements as well as Planck priors on cosmology.
The galaxy biases and shear biases are simultaneously marginalized over, with their fiducial priors.
} 
\label{fig:baseline_cosmo_contours_lcdmmnu}
\end{figure}

\begin{figure}[H]
\centering
\includegraphics[width=0.95\columnwidth]{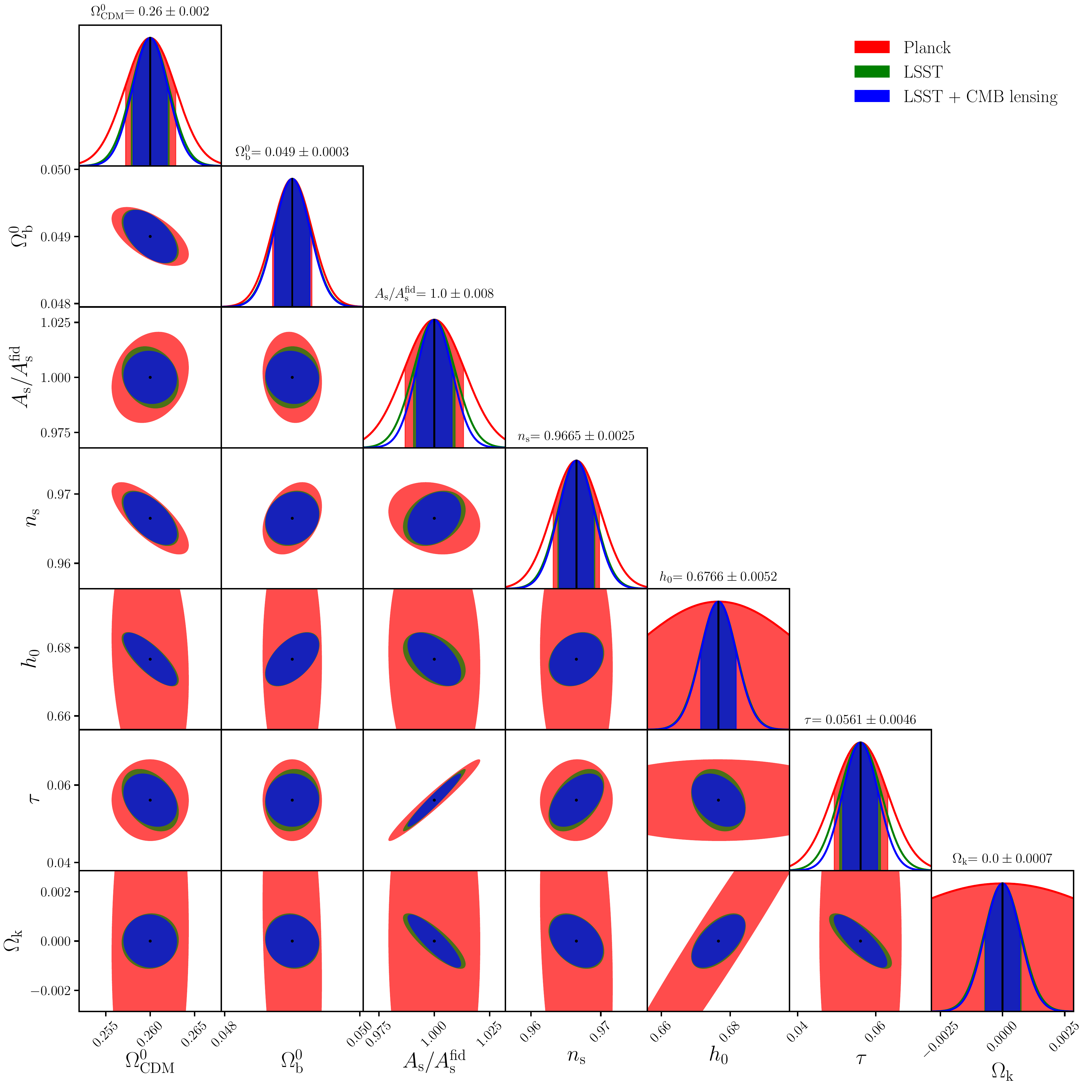}
\caption{
Baseline cosmological constraints for $\Lambda$CDM+curvature,
including priors on the Gaussian and outlier photo-z errors consistent with the LSST requirements, as well as Planck priors on cosmology.
The galaxy biases and shear biases are simultaneously marginalized over, with their fiducial priors.
}
\label{fig:baseline_cosmo_contours_lcdmcurv}
\end{figure}

\begin{figure}[H]
\includegraphics[width=0.45\columnwidth]{figures/fisher_lsst/comparison_cosmo_lcdmw0_vs_gs.pdf}
\includegraphics[width=0.45\columnwidth]{figures/fisher_lsst/comparison_cosmo_lcdmw0wa_vs_gs.pdf}
\includegraphics[width=0.45\columnwidth]{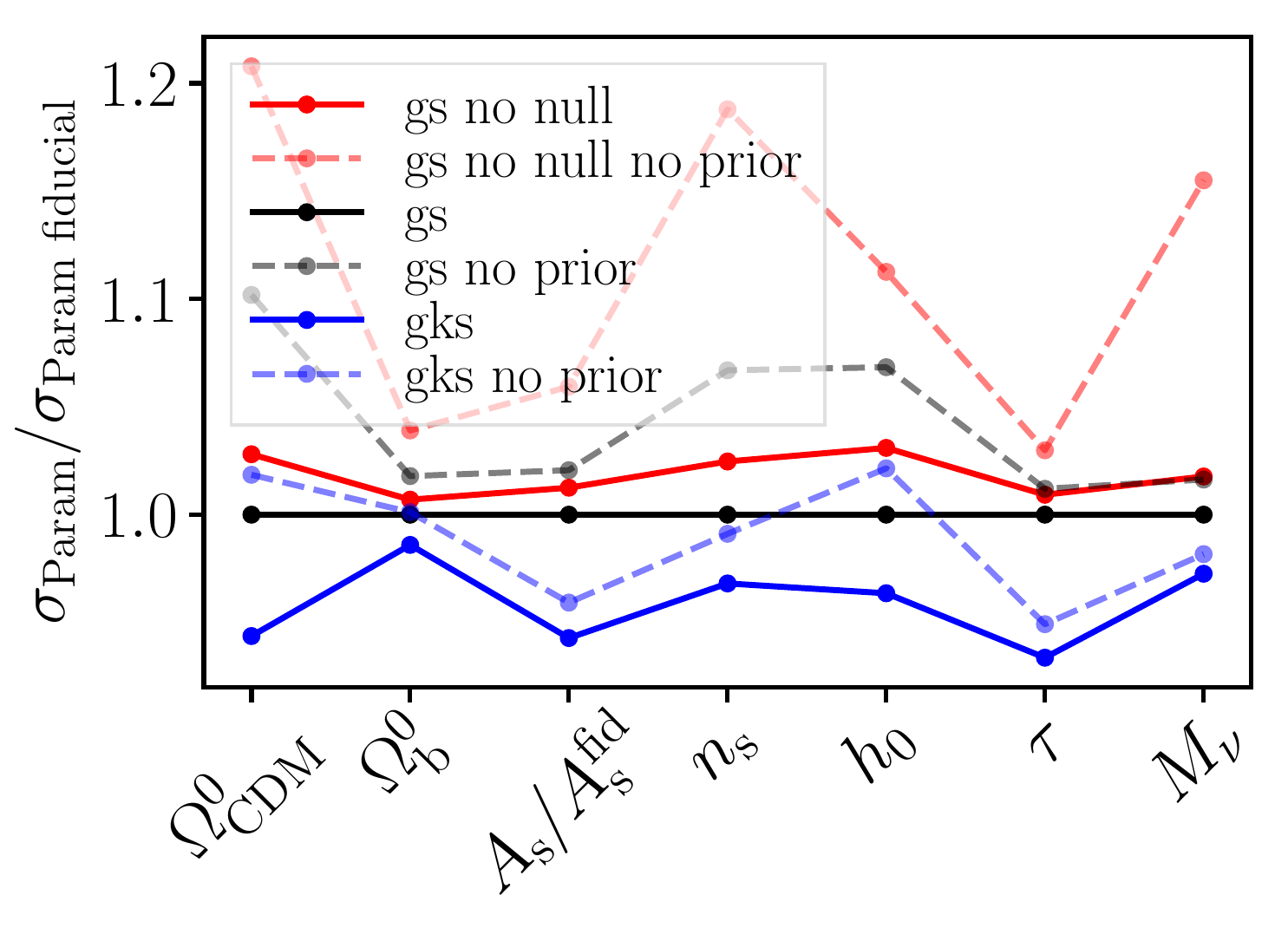}
\includegraphics[width=0.45\columnwidth]{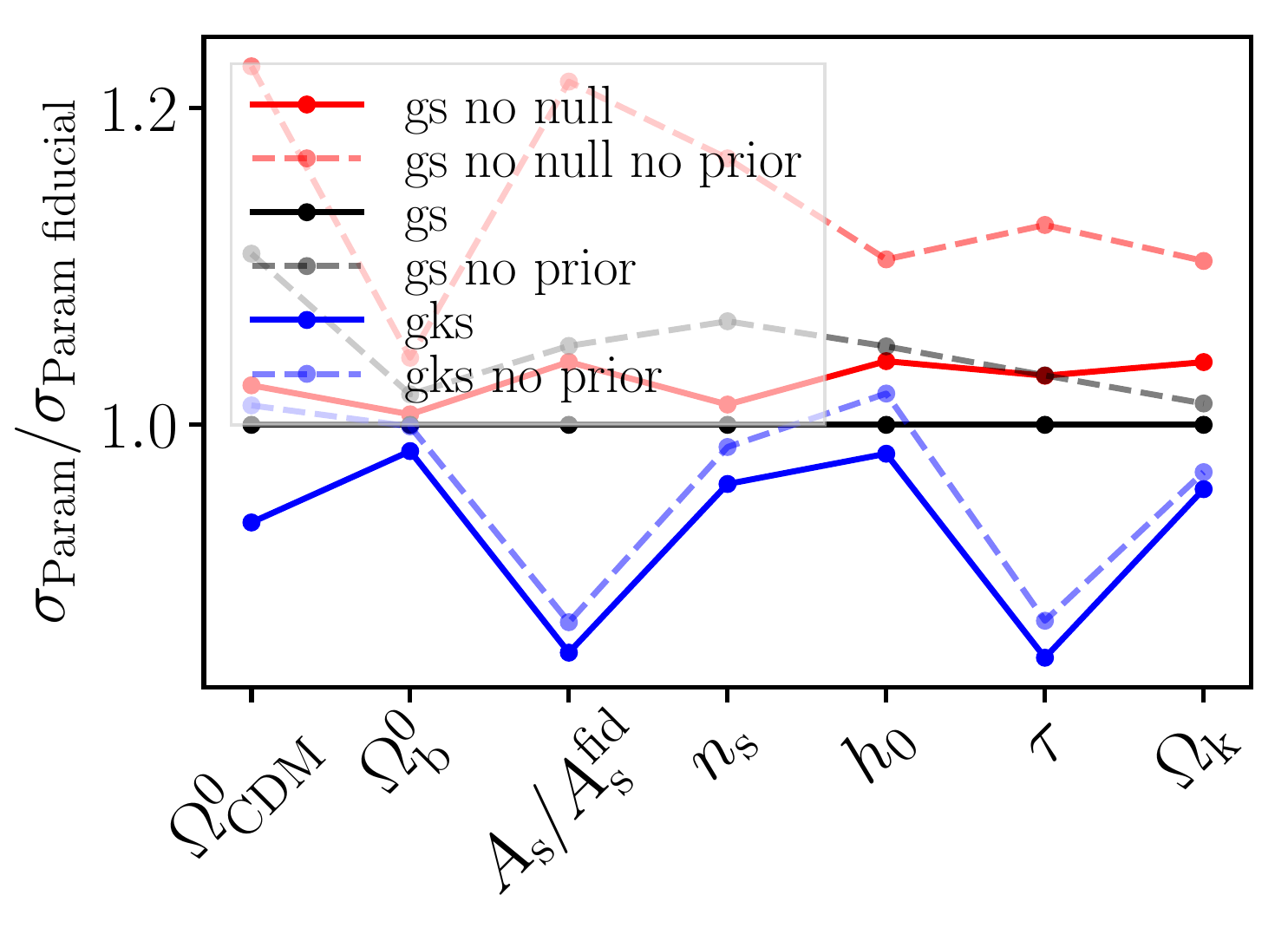}
\caption{
Comparison between the cosmological constraints from the various datasets in this analysis.
The different colors correspond to gs only (fiducial, black), adding CMB lensing (blue) and discarding the null correlations (red).
Top left: $w_0$CDM. 
Top right: $w_0 w_a$CDM.
Bottom left: $\Lambda$CDM + $M_\nu$.
Bottom right: $\Lambda$CDM + curvature.
}
\label{fig:comparison_cosmo_othercosmo}
\end{figure}

In this appendix, we visualize the Planck-like prior used in this analysis (Fig.~\ref{fig:planck_prior}) and the posterior constraints for the different cosmological parameter sets:
$\Lambda$CDM in Fig.~\ref{fig:baseline_cosmo_contours_lcdm},
$\Lambda$CDM + $M_\nu$ in Fig.~\ref{fig:baseline_cosmo_contours_lcdmmnu}
and
$\Lambda$CDM + curvature in Fig.~\ref{fig:baseline_cosmo_contours_lcdmcurv}.
In Fig.~\ref{fig:comparison_cosmo_othercosmo}, we compare the cosmological constraints from the various data included in this analysis and the various cosmological parameter sets.

\section{Algebraic information only: identifying degeneracies with the Fisher formalism}
\label{app:algebra_fisher}

In this appendix, we consider again Eq.~\eqref{eq:toy_model_system}:
\beq
\left\{
\bal
&\footnotesize{\text{\textbf{Gal. density:} }}
&&
\small{\vec{g}_A = \sum_i  \frac{dn_A}{dz_i} \left[  b_{A,z_i}\vec{\delta}_{z_i} + 2 \left( \alpha_{A, z_i} - 1 \right)  \sum_j W^\kappa_{z_i ,z_j} \vec{\delta}_{z_j}\right]}
&&&\footnotesize{\left( N_S \times N_\ell \text{ eq.} \right)}\\
&\footnotesize{\text{\textbf{Gal. lensing:} }} 
&&\vec{\kappa}_A = \sum_i \left(1+m_{A,z_i}\right) \frac{dn_A}{dz_i} \sum_j W^\kappa_{z_i ,z_j} \vec{\delta}_{z_j}
&&&\footnotesize{\left( N_S \times N_\ell \text{ eq.} \right)}\\
&\footnotesize{\text{\textbf{CMB lensing:} }} 
&&\vec{\kappa}_\text{CMB} = \sum_i  W^{\kappa_\text{CMB}}_{z_i} \vec{\delta}_{z_i} 
+ W^{\kappa_\text{CMB}}_\text{high z} \vec{\delta}_\text{high z}
&&&\footnotesize{\left( N_\ell \text{ eq.} \right)}\\
\eal
\right.
\eeq
Given the observed fields $\vec{\delta}_A$ and $\vec{\kappa}_A$ (and $\vec{\kappa}_\text{CMB}$), we wish to solve for the matter density fields $\vec{\delta}_{z_i}$, the redshift distributions $dn_A/dz_i$ and the galaxy biases and shear biases.
The existence of a solution is physically trivial, since the true matter density fields in the Universe, along with the actual redshift distributions of the tomographic bins and the true values of galaxy bias and shear biases must have produced the observed galaxy number density fields and shear fields (and CMB lensing field).
The question we wish to answer is that of unicity of the solution.
We wish to confirm that the counting of equations and unknowns we performed in Sec.~\ref{sec:algebra} correctly answers this question.

We use the Fisher formalism to assess the existence of continuous degeneracies around the true solution.
Indeed, \textbf{degenerate directions around the true parameters correspond to null eigenvectors of the Fisher matrix.}
This tells us about the existence of parameters, arbitrarily close to the truth, that produce exactly the same observables as the true parameters.
As a caveat, the Fisher formalism does not inform us on the existence of isolated islands in parameter space that may produce the same observables, such as the re-indexing of the bins or other discrete symmetries in the problem.

\textbf{Proof:}\\
Let $\mathbf{D}_{(\mathbf{p})}$ be the data vector, and $\mathbf{p}$ the parameter.
Let $\mathbf{D}_0$ be the actual observed data vector, and $\mathbf{p}_0$ be the corresponding true parameter.
Our question is: is there a parameter $\mathbf{p}$, arbitrarily close to $\mathbf{p}_0$, such that $\mathbf{D}_{(\mathbf{p})} = \mathbf{D}_0$.
The parameter being arbitrarily close to the truth means that the Taylor expansion 
$\mathbf{D}_{(\mathbf{p})} \simeq \mathbf{D}_0 + \frac{\partial \mathbf{D}}{\partial p_i} \delta p_i$
is arbitrarily accurate.
Here the sum over the repeated index $i$ is implicit.
Thus we are looking for parameter shifts such that 
$ \frac{\partial \mathbf{D}}{\partial p_i} \delta p_i = 0$.
This is equivalent to 
$\mathcal{Q} [\frac{\partial \mathbf{D}}{\partial p_i} \delta p_i] = 0$,
for any positive definite quadratic form $\mathcal{Q}$.
Let us choose
$\mathcal{Q}[\mathbf{X}] \equiv \mathbf{X}^t \; \mathbf{\Sigma}^{-1} \; \mathbf{X}$,
where $\mathbf{\Sigma}$ is any symmetric positive definite matrix, for example the covariance matrix for the data vector.
Then 
$\mathcal{Q} [\frac{\partial \mathbf{D}}{\partial p_i} \delta p_i] = 
\delta p_i \frac{\partial \mathbf{D}^t}{\partial p_i} 
\; \mathbf{\Sigma}^{-1} \;
\frac{\partial \mathbf{D}}{\partial p_j} \delta p_j
=
\mathbf{F}[\delta \mathbf{p}]$,
where 
$\mathbf{F}[\mathbf{X}] \equiv 
X_i^t \frac{\partial \mathbf{D}^t}{\partial p_i} 
\; \mathbf{\Sigma}^{-1} \;
\frac{\partial \mathbf{D}}{\partial p_j} X_j$
is the usual Fisher matrix.
In summary, 
a parameter $\mathbf{p} = \mathbf{p}_0 +\delta\mathbf{p}$ with $\delta\mathbf{p}$ infinitesimal satisfies $\mathbf{D}_{(\mathbf{p})} = \mathbf{D}_0$
if and only if 
$\mathbf{F}[\delta \mathbf{p}]=0$,
i.e. $\delta\mathbf{p}$ is a null eigenvector of the Fisher matrix.

\begin{figure}[h!!!!]
\centering
\includegraphics[width=0.6\columnwidth]{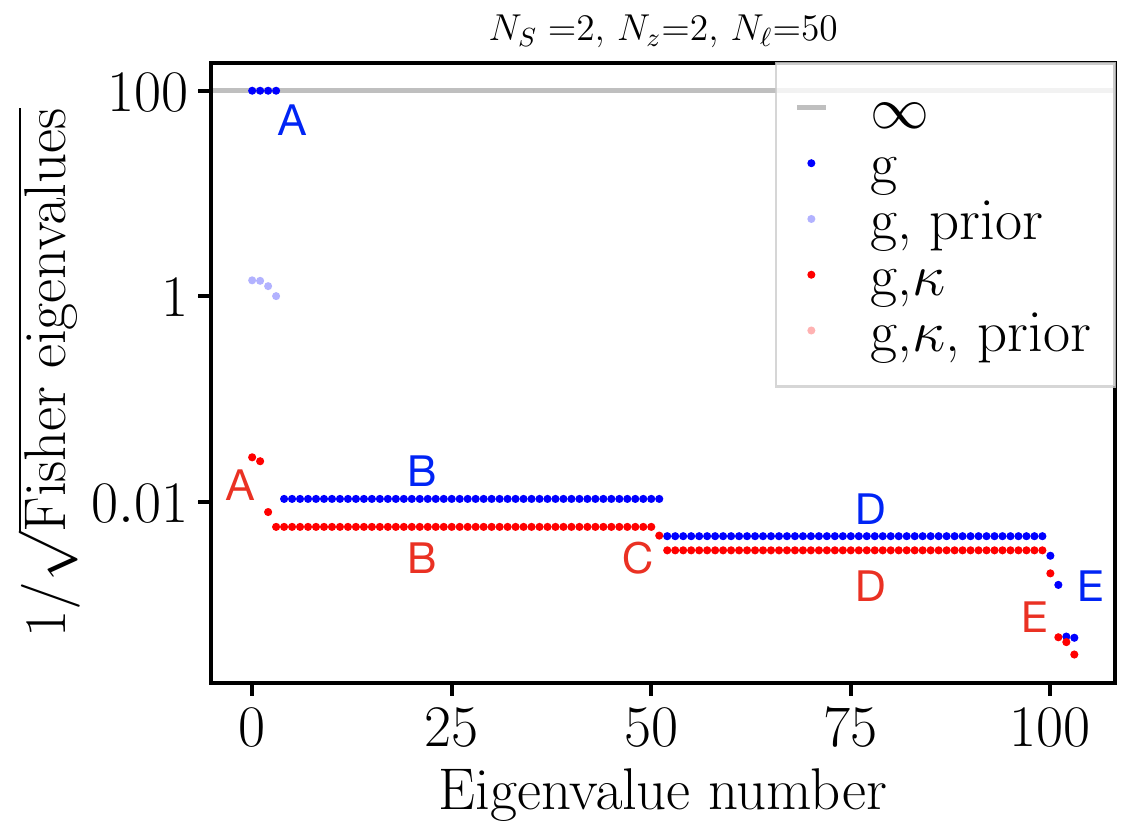}
\caption{
Eigenvalues of the Fisher matrix, expressed as posterior uncertainties, when only galaxy positions are observed (blue) or when both galaxy positions and shear and observed (red). The lighter color points to adding a Gaussian prior on the matter density.
With galaxy positions only (blue), and in the absence of prior, 4 ($=N_z^2$) null eigenvalues (labelled ``A'') exist as expected, corresponding to 4 combinations of the parameters that are perfectly degenerate. 
The prior makes these eigenvalues finite.
With both galaxy positions and shear (red), there are no unconstrained parameter combinations, whether or not a prior is included.
In both cases, the Fisher matrix only has at most 10 different eigenvalues, compared to the 100 matrix elements.
This is again as expected, and corresponds to the number of actual unknowns in our counting, which is the true dimensionality of the problem.
}
\label{fig:fisher_toy_model_eigenvalues}
\end{figure}

For this Fisher matrix, the data vector is 
$\mathbf{D} = (\delta_A(\vl))$ for clustering-only
or
$\mathbf{D} = (\delta_A(\vl), \kappa_A(\vl))$ for clustering and shear.
The parameter vector is $\mathbf{p} = (\delta_i(\vl), n_{Ai}, b_i)$.
When considering the effect of the decorrelation prior for the matter density fields in different true redshift bins, we add the priors that
$\delta_i = \mathcal{N}(0, \sigma_i)$, for arbitrarily chosen $\sigma_i$. 
The corresponding Fisher matrix eigenvalues are shown in Fig.~\ref{fig:fisher_toy_model_eigenvalues}, in the case $N_S=N_z=2$.

\paragraph{\textbf{Clustering-only}}

In the absence of prior on the $\delta_i$, Fig.~\ref{fig:fisher_toy_model_eigenvalues} shows that 4 parameter combinations are perfectly degenerate. 
Examining the eigenvectors gives further intuition on these parameter combinations.

\textbf{From worst to best constrained parameter combinations:}\\
A- 4 unconstrained modes corresponding to the rotation of $\delta_1$, rotation of $\delta_2$, scaling of $\delta_1$ and $b_1$, scaling of $\delta_1$ and $b_1$. These scaling degeneracies were already trivially visible from the equations: one can rescale each $\delta_{i_0}$ by some factor and divide the $b_{Ai_0}$ by the same factor, without changing the observables.\\
B- Only $\delta_1$ varies, all else being constant\\
C- Only $\delta_2$ varies, all else being constant\\
D- 4 best constrained parameter combinations, where galaxy bias and $dn/dz$ are varied while $\delta_1$ and $\delta_2$ are constant.\\

\paragraph{\textbf{Clustering \& shear}}

As seen in Fig.~\ref{fig:fisher_toy_model_eigenvalues}, all eigenvalues are finite, i.e. there is no exact degeneracy.
In other words, the system can be solved completely, and one can infer the true matter density field and the true redshift distribution of each tomographic bin.
This is true without priors on the $\delta_{z_i}$, which means that the ability to solve the system is not limited by cosmic variance.
Some parameter modes are better constrained than others. Again, examining the eigenvectors gives further intuition. 

\textbf{From worst to best constrained modes:}\\
A- 3 worst constrained modes, corresponding respectively to a rotation of $\delta_1$, a rotation of $\delta_2$, and a rotation of $\delta_1$ and $\delta_2$ keeping their angle constant.\\
B- only $\delta_1$ and $\delta_2$ vary, in opposite directions\\
C- only 1 mode, not very clear intuitively, where galaxy bias is fixed, $dn/dz$ and $\delta_1$ almost fixed, and $\delta_2$ varies.\\
D- only $\delta_1$ and $\delta_2$ vary, in the same direction\\
E- 4 best constrained modes, where galaxy bias and $dn/dz$ are varied while $\delta_1$ and $\delta_2$ are constant.\\

\subsubsection{Fewer samples than true redshift bins: $N_S < N_z$}

\begin{figure}[H]
\centering
\includegraphics[width=0.4\columnwidth]{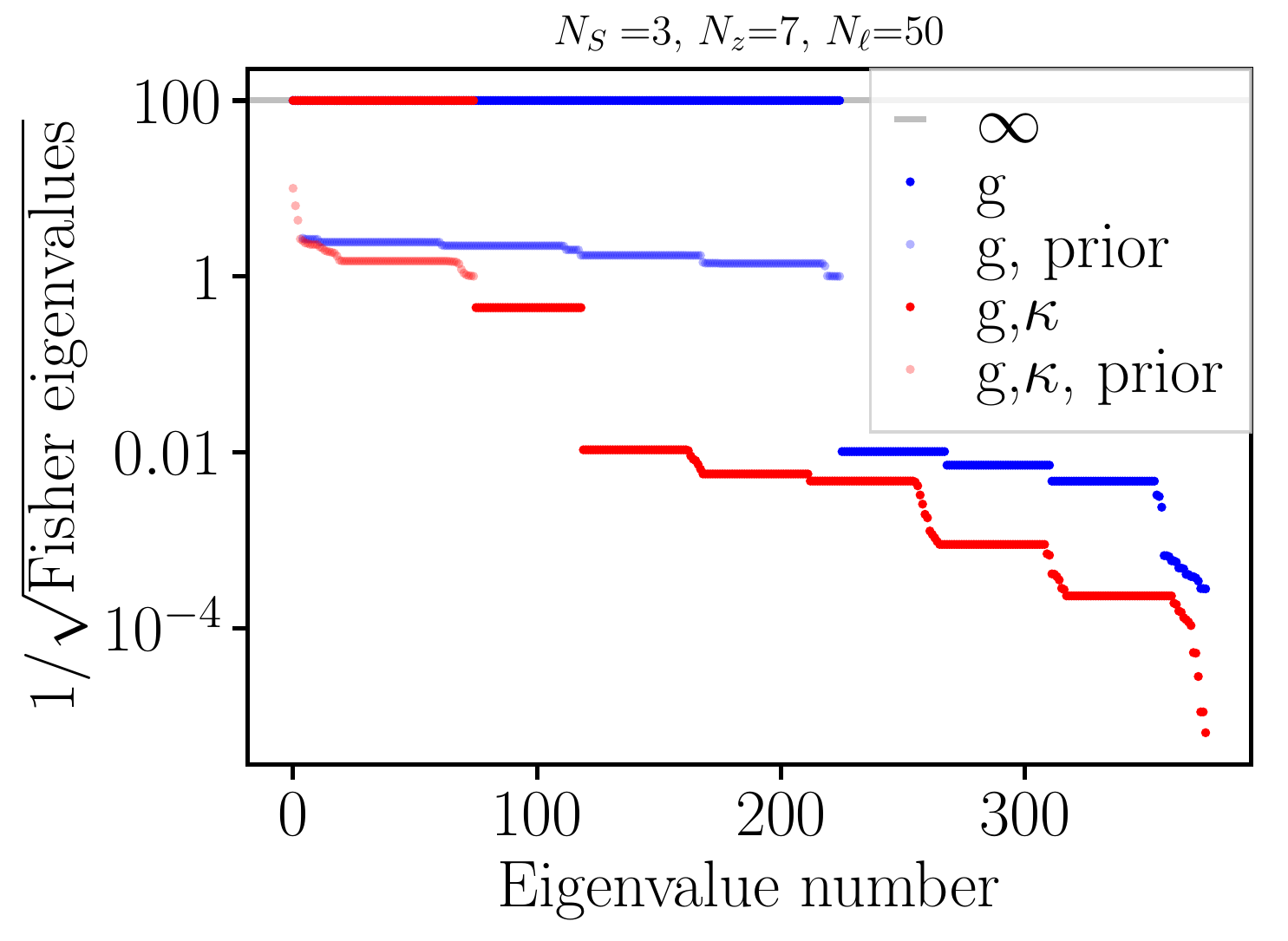}
\caption{
When the number of galaxy samples (here 3) is less than the number of true redshift bins (here 7), degeneracies appear, even when clustering and shear are included.
}
\label{fig:fisher_toy_model_eigenvalues_gs_ns3_nz7}
\end{figure}

As expected, degeneracies appear when $N_S<N_z$, even when clustering and shear are included.
This is shown in Fig.~\ref{fig:fisher_toy_model_eigenvalues_gs_ns3_nz7}.

\subsubsection{Larger values of $N_S=N_z$}

\begin{figure}[H]
\centering
\includegraphics[width=0.4\columnwidth]{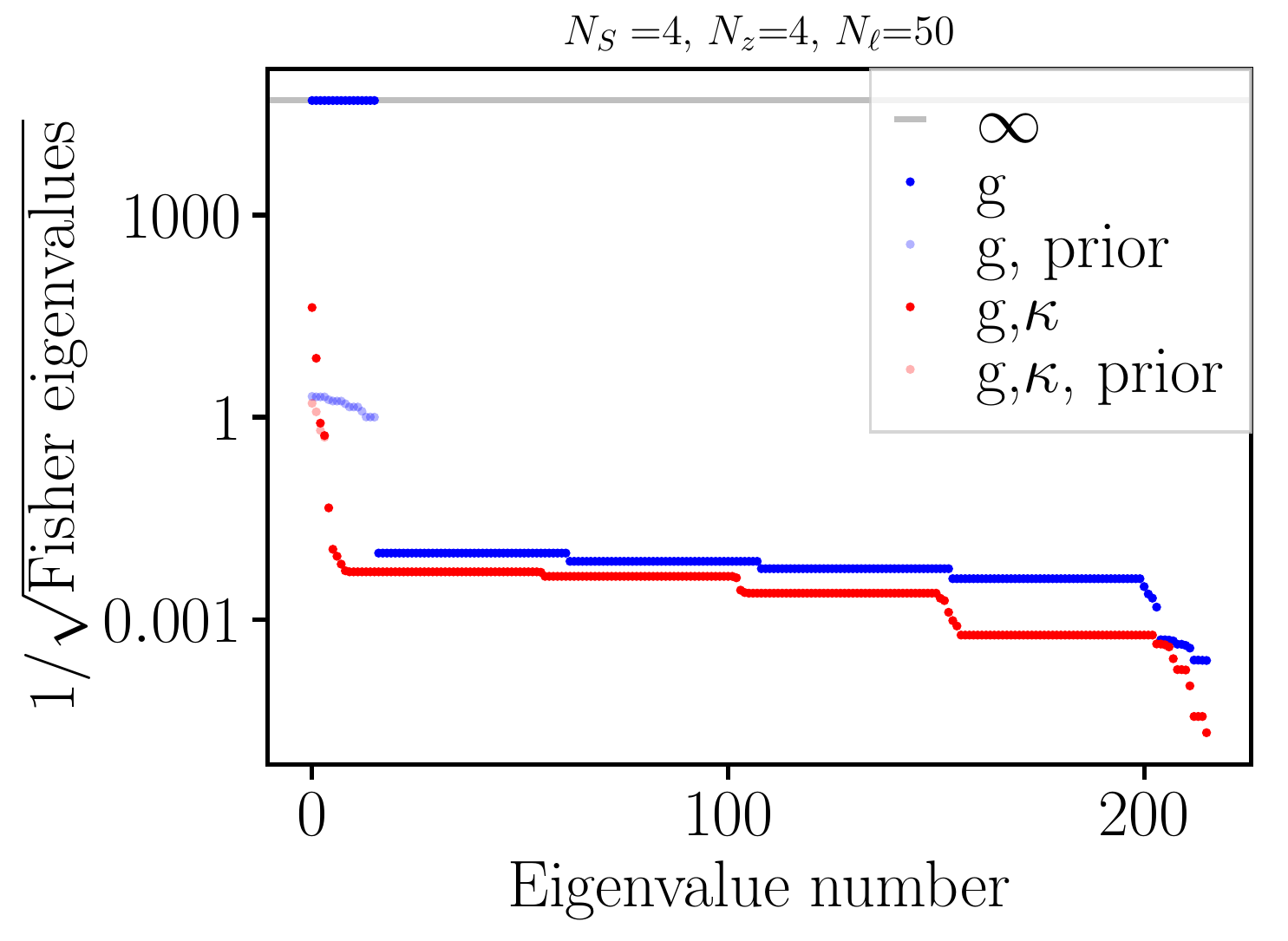}
\caption{
Clustering \& Shear: posterior uncertainties of the various parameter combinations, for $N_S= N_Z=4$.
As $N_S= N_Z$ increases, some parameter combinations become more and more uncertain.
For $N_S= N_Z \geq 5$, numerical precision no longer allows to differentiate whether these modes are very uncertain or infinitely uncertain.
As before, these parameter combinations correspond to varying galaxy bias and $dn/dz$, without modifying the $\delta_i$.
}
\label{fig:fisher_toy_model_eigenvalues_gs_ns4_nz4}
\end{figure}

As the number of galaxy sample $N_S$ and the number of true redshift bins $N_Z$ increase while remaining equal, the system remains solvable if and only if galaxy clustering and shear are included.
This is illustrated in Fig.~\ref{fig:fisher_toy_model_eigenvalues_gs_ns4_nz4}.

\subsubsection{Shear multiplicative bias \& CMB lensing}

\begin{figure}[H]
\centering
\includegraphics[width=0.45\columnwidth]{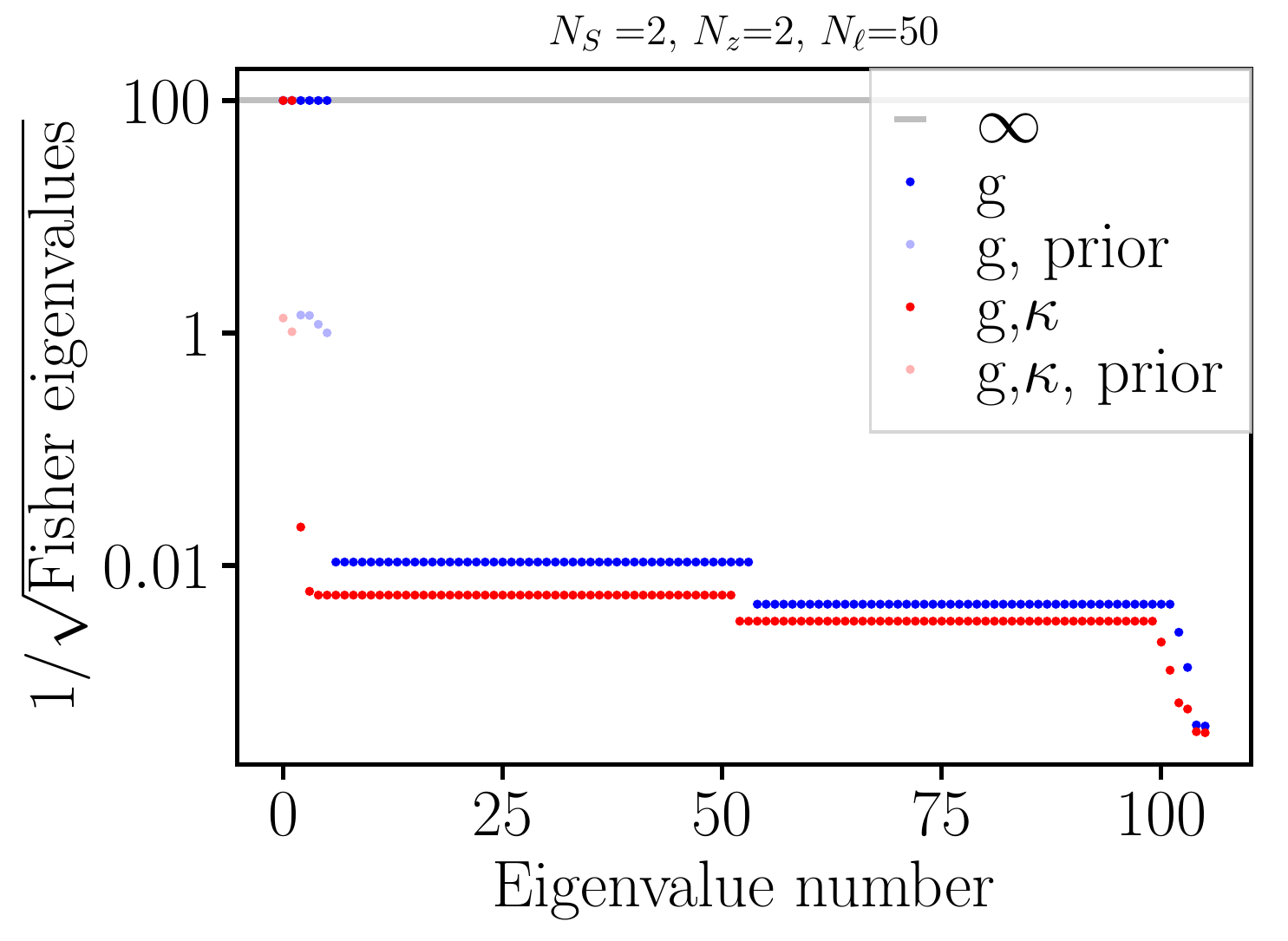}
\includegraphics[width=0.45\columnwidth]{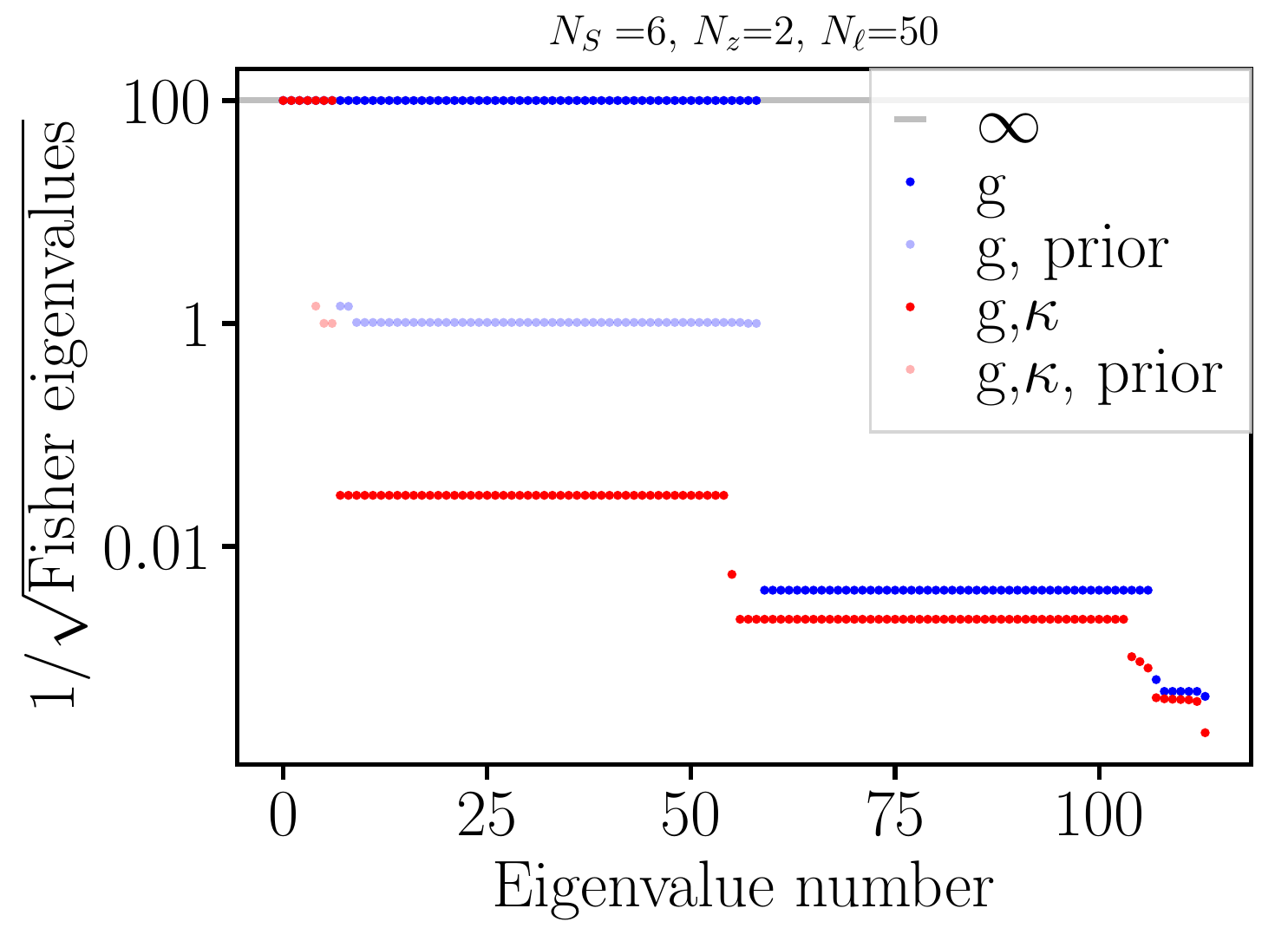}
\caption{
With clustering and galaxy lensing are included, the shear biases cannot be inferred from the data alone.
\textbf{Left:} For two tomographic samples and 2 true redshifts, one finds 2 unconstrained parameters, which are the shear biases.
\textbf{Right:} Even when $N_S \geq N_z+1$,
introducing unknown shear biases leads to unconstrained parameter combinations.
} 
\label{fig:fisher_toy_model_eigenvalues_shearbias}
\end{figure}

Unknown shear multiplicative biases add $N_S$ unknowns.
Na\"ively, one may think that this is a negligible increase in the number of unknowns as long as $N_\ell$ is large enough.
However, in the correct parameter counting described in the main text, this addition of unknowns means that the system should never be solvable. 
In particular, increasing the number $N_S$ of observed samples does not help.
This is shown in Fig.~\ref{fig:fisher_toy_model_eigenvalues_shearbias}: even when both clustering and shear are measured, the system is underconstrained.

\begin{figure}[H]
\centering
\includegraphics[width=0.45\columnwidth]{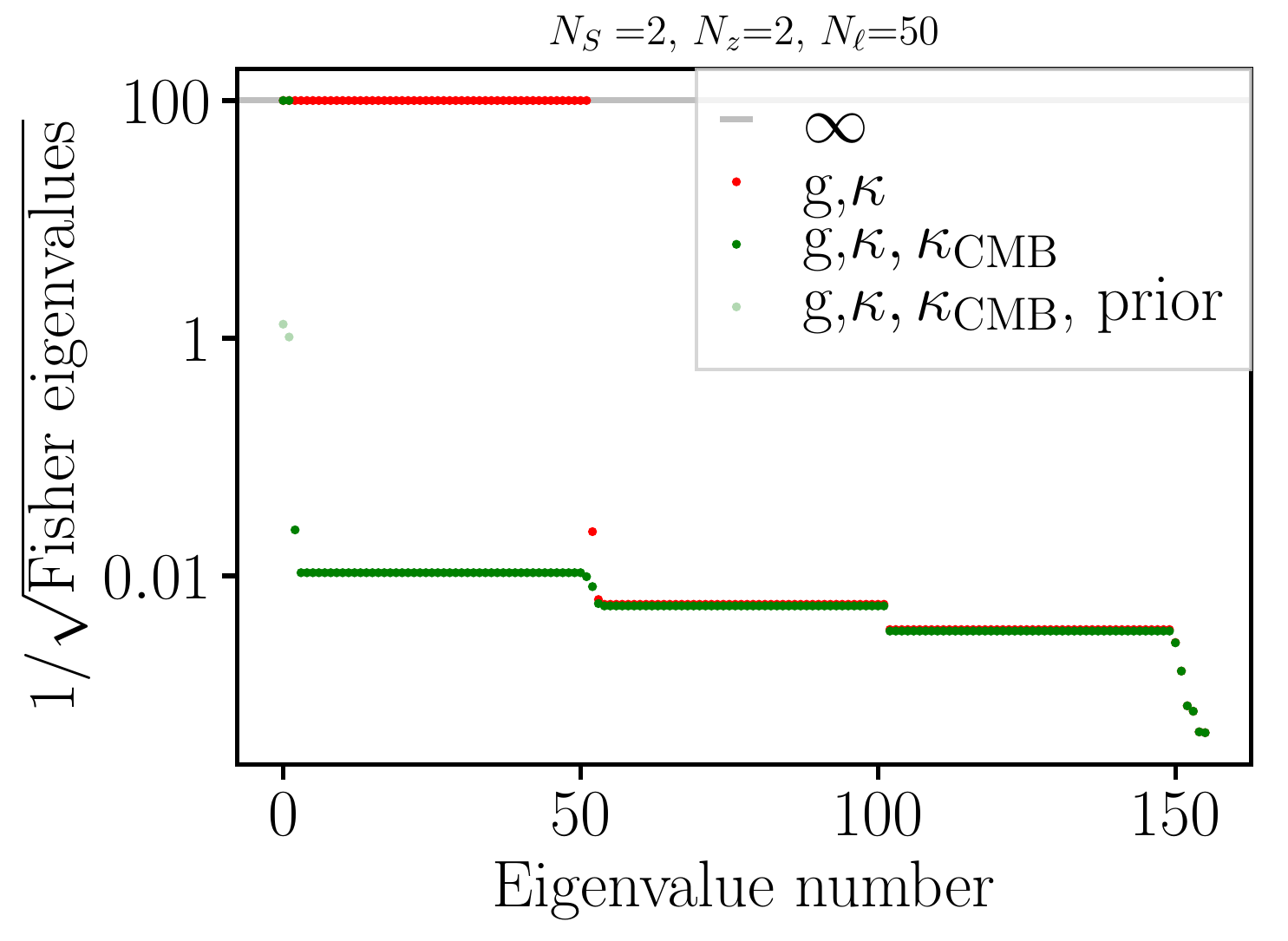}
\caption{
When only galaxy number densities and galaxy convergence are observed (red), $N_\ell + N_S$ parameter combinations are unconstrained. These correspond to the shear biases and $\delta_\text{high z}(\ell)$.
When CMB lensing is added, the $N_\ell$ parameter combinations related to $\delta_\text{high z}(\ell)$ become constrained, but the unknown $N_S$ shear biases remain unconstrained.
If a prior is added to describe the decorrelation between $\delta_\text{high z}(\ell)$ and the low redshift matter density fields (pale green dots), the shear biases become constrained.
}
\label{fig:fisher_toy_model_eigenvalues_gks_shearbias}
\end{figure}

We then add the CMB lensing field, as described in our toy model. As mentioned in the main text, this adds $N_\ell$ observables from the $\kappa_\text{CMB}(\ell)$, but it adds the same number of unknowns, in the form of the $\delta_\text{high z}(\ell)$.
If the shear biases are known, the CMB lensing field then constrains the $\delta_\text{high z}(\ell)$ field.
But when the shear biases are left free, we find that CMB lensing does not constrain them, as Fig.~\ref{fig:fisher_toy_model_eigenvalues_gks_shearbias} shows.

\end{document}